\documentclass[3p,times,twocolumn,sort&compress]{elsarticle}
\usepackage{ecrc}
\volume{00}
\firstpage{1}
\journalname{Nuclear Physics B Proceedings Supplement}
\runauth{Boettcher, Pawlowski, Diehl}
\jid{nuphbp}
\jnltitlelogo{Nuclear Physics B Proceedings Supplement}

\usepackage[figuresright]{rotating}

\usepackage{amssymb}
\usepackage{amsmath}
\usepackage{graphics}
\usepackage{graphicx}
\usepackage{psfrag}
\usepackage{hyperref}
\usepackage{bbold}
\usepackage{color}

\usepackage{verbatim}

\newcommand{\vphi}{\varphi}
\newcommand{\vare}{\varepsilon}
\newcommand{\rmi}{{\rm i}}

\def\eq#1{(\ref{#1})}
\def\Eq#1{Eq.~(\ref{#1})}

\begin{document}

\begin{frontmatter}

\dochead{}

\title{Ultracold atoms and the Functional Renormalization Group}

\author[add1]{Igor Boettcher}
\author[add1,add2]{Jan M. Pawlowski}
\author[add3,add4]{Sebastian Diehl}

\address[add1]{Institute for Theoretical Physics, University of Heidelberg, D-69120 Heidelberg, Germany}
\address[add2]{ExtreMe Matter Institute EMMI, GSI Helmholtzzentrum f\"ur Schwerionenforschung mbH, Planckstr. 1, D-64291 Darmstadt, Germany}
\address[add3]{Institute for Theoretical Physics, University of Innsbruck, A-6020 Innsbruck, Austria}
\address[add4]{Institute for Quantum Optics and Quantum Information of the Austrian Academy of Sciences, A-6020 Innsbruck, Austria}

\begin{abstract}
\noindent We give a self-contained introduction to the physics of ultracold atoms using functional integral techniques. Based on a consideration of the relevant length scales, we derive the universal effective low energy Hamiltonian describing ultracold alkali atoms. We then introduce the concept of the effective action, which generalizes the classical action principle to full quantum status and provides an intuitive and versatile tool for practical calculations. This framework is applied to weakly interacting degenerate bosons and fermions in the spatial continuum. In particular, we discuss the related BEC and BCS quantum condensation mechanisms. We then turn to the BCS-BEC crossover, which interpolates between both phenomena, and which is realized experimentally in the vicinity of a Feshbach resonance. For its description, we  introduce the Functional Renormalization Group approach. After a general discussion of the method in the cold atoms context, we present a detailed and pedagogical application to the crossover problem. This not only provides the physical mechanism underlying this phenomenon. More generally, it also reveals how the renormalization group can be used as a tool to capture physics at all scales, from few-body scattering on microscopic scales, through the finite temperature phase diagram governed by many-body length scales, up to critical phenomena dictating long distance physics at the phase transition.\\
The presentation aims to equip students at the beginning PhD level with knowledge on key physical phenomena and flexible tools for their description, and should enable to embark upon practical calculations in this field. 
\end{abstract}

\begin{keyword}
Ultracold atoms, Functional Renormalization Group, BCS-BEC crossover


\end{keyword}

\end{frontmatter}



\tableofcontents

\section{Introduction}
\label{sec:introduction}

Cold atomic many-body systems make up a young and rapidly evolving
area of modern physics. The field was born in 1995, where an
almost pure and weakly interacting Bose--Einstein condensate (BEC) was
created in alkaline atomic vapors \cite{pethick-book,pitaevski-book}.
This state of matter exhibits macroscopic phase coherence as 
a manifest feature of many-body quantum mechanics. Soon after, further
milestones were achieved -- among them, the observation of the quantum phase
transition from a Mott insulator to a superfluid in optical
lattices, to which atoms are confined due to their interaction with
light \cite{lewenstein-rmp-56-135,bloch-review}, and the
implementation of the BCS-BEC crossover realizing strongly
interacting fermion ensembles
\cite{RevModPhys.80.1215,gurarie-review}.

These achievements reflect both the high degree of control in the
manipulation of the atomic constituents, as well as the remarkable
tunability of the scales and interactions governing the
system. From a practical point of view, this prepares the ground for
efficient \emph{quantum simulations} using cold atomic gases. 
A prominent example is provided by the determination of the ground
state of the two-dimensional Fermi--Hubbard model, which remains a challenge
to theory \cite{RevModPhys.84.299}, but can be
implemented in cold atomic gases with high accuracy as a direct
simulator to gain insight into its low temperature quantum physics, once sufficient cooling is achieved
\cite{EsslingerRev10}. With increasing control over the microscopic
constituents in optical lattices, a long term goal is to go
beyond such special task quantum simulators, and to build truly
programmable, universal quantum simulation devices
\cite{T_feynman-IJTPhys-21-467,T_buluta-science-326-108,BlochDalibard12,Cirac12}.
From a theoretical point of view, cold atomic samples offer a unique
\emph{testbed for modern nonperturbative quantum field theoretical
  approaches}: While a precise microscopic understanding is often not
available in complex many-body quantum systems realized in condensed
matter or high energy physics, the ability to experimentally probe a
cold atomic sample at all scales, and to manipulate its microphysics
in a controlled way, allows for a direct comparison of experiment and
theory in a strongly interacting context. The transition from microscopic 
simplicity to macroscopic complexity, performed in a
given theoretical approach, can be benchmarked in a direct way.

Beyond these general considerations, cold atomic gases also
offer a number of \emph{physical situations which do not have an
  immediate counterpart in other branches of physics}. Without aiming
at completeness, but for the sake of giving a flavor, we briefly point
out two directions which recently attracted interest. The first one
harnesses quantum optical manipulation tools for constructing
\emph{microscopic Hamiltonians with exotic interactions}. For instance, 
long range interactions such as $1/r^3$, in
part with strong spatial anisotropy, are obtained in the context of dipolar
atoms and polar molecules
\cite{
  njp-focus-polarmolecules,baranov-review,baranov12} or  Rydberg dressed
atoms \cite{RevModPhys.82.2313}. Moreover, it is
possible to realize multicomponent interactions with high degree of
symmetry such as ${\rm SU}(N)$. This is achieved by controlling and addressing internal
states of atoms, which possess a rich level structures, like earth alkaline-like atoms
\cite{Ye27062008,Gorshkov10,daley-review}. Reliably extracting the
many-body physics of each of these systems poses its own specific
challenges to theory. A second direction is given by \emph{non-equilibrium
  physics with cold atoms}. This comprises, on the one hand, the
dynamics of \emph{closed systems}, where key questions are related to the
relaxation dynamics towards thermodynamic equilibrium
\cite{PhysRevA.72.063604,PhysRevLett.100.030602,Rigol2008,Wenger06,Hoffer07},
the behavior of a many-body system following a quench of
microscopic parameters
\cite{PhysRevLett.96.136801,PhysRevLett.98.180601,GreinerM02,Sadler06},
or dynamical phenomena such as the propagation of local perturbations
through the system \cite{Cheneau12}. On the other hand, this concerns
\emph{open systems}, where many-body ensembles are driven far from
thermodynamic equilibrium by coupling to effective external reservoirs
-- which may occur naturally, or via specific reservoir
engineering. Such systems may exhibit \emph{stable non-equilibrium
  stationary states} with intriguing quantum mechanical properties and
rich phase diagrams
\cite{diehl08,verstraete08,PhysRevLett.105.015702,TDGA2010}. In the
non-equilibrium context, it is a first challenge for both experiment
and theory to identify situations which exhibit a sufficient degree of
universality, i.e. with a phenomenology which occurs in classes of
systems and settings beyond a particular realization. Furthermore,
this calls for the development of flexible theoretical tools -- both
numerical and analytical -- to describe them efficiently.

The aim of these lecture notes is to act as a door-opener to this
exciting field. It should equip students at the beginning of their
theoretical PhD studies with the knowledge of key physical many-body
phenomena in ultracold quantum gases in the spatial continuum, as well
as with flexible functional techniques for their description. In
particular, we introduce the concept of the effective action, which
generalizes the classical action principle to full status and is an
intuitive and versatile tool for practical calculations. We describe
the physics of weakly interacting degenerate bosons and fermions in
this framework, and discuss the related BEC and BCS quantum
condensation mechanisms. We also describe a Functional Renormalization
Group (FRG) approach to ultracold atoms, for reviews see
\cite{Blaizot:2008xx,Diehl:2009ma,Scherer:2010sv,Floerchinger:2011yv,RevModPhys.84.299},
and extract the full finite temperature phase diagram for the BCS-BEC
crossover within a simple approximation for the running effective
action. This discussion not only provides the physical mechanism
underlying this phenomenon, which interpolates between the above two
cornerstones of quantum condensation phenomena with a particularly
challenging strongly interacting regime in between. It also
illustrates how the concept of the renormalization group can be used
in practice beyond the realm of critical phenomena at very long
distances. In fact, it provides a powerful tool to smoothly perform
the transition from micro- to macrophysics. This comprises the
quantitative description of few-body scattering in the physical vacuum
at short distances, the phase diagram and thermodynamics governed by
many-body scales such as interparticle spacing and de Broglie
wavelength, and finally also includes the description of critical
behavior at large wavelength, in one unified framework.

The lecture notes are organized as follows. In Sec. \ref{Sec2}, we
discuss the basic microscopic and many-body scales present in
ultracold gases, and construct an effective Hamiltonian which
universally describes the physics of bosonic and fermionic alkali
gases at low energies, as appropriate for ultracold experiments. For
the sake of a self-contained presentation, we also provide brief
reminders on thermodynamics and the quantum statistical mechanics of
noninteracting bosons and fermions. In Sec. \ref{Sec3}, we introduce
the functional integral representation of the quantum partition
function (with technical details on the derivation provided in the
appendix), and switch to a more intuitive object encoding the same
information, the effective action. We describe the key phenomenon of
spontaneous symmetry breaking, and how to extract thermodynamic
information from this object. We apply these concepts to weakly
interacting bosons and fermions, with an emphasis on the condensation
phenomenon and the nature of the low energy excitations in both
cases. In Sec. \ref{Sec4}, we introduce Wetterich's FRG framework for
the effective action, which encodes the full information on the
many-body problem in terms of an exact functional differential
equation, and is ideally suited for the implementation of practical
approximation schemes (truncations) beyond mean field theory. After a
brief general discussion on the application of this framework to
ultracold atoms systems, we apply it to the BCS-BEC crossover
problem. The emphasis is on a detailed presentation of a simple
truncation, which is able to produce the full finite temperature phase
diagram and already at this level demonstrates a number of
improvements compared to extended mean field theories, due to a
consistent inclusion of bosonic degrees of freedom. Since, on a
technical level, this covers both the treatment of interacting
fermionic and bosonic theories, it will enable the reader to embark
upon practical calculations in this field. We provide a (subjective)
list of challenges for the future.

This work is based on lectures delivered by S. D. at the
49th Schladming Graduate School for Theoretical Physics.

\section{Basics of ultracold atomic physics}
\label{Sec2}

\subsection{Scales and interactions}
\label{scales} 
\noindent The physics of ultracold quantum gases is governed by the
interplay of several scales. Tuning their relative size, it is
possible to access different regions of the phase diagram of a given
system, in this way exploring its physics. In this section, we show
which scales are relevant in the context of alkali atoms. In
particular, we will discuss the conditions under which we have an
ultracold quantum gas. These model-independent considerations will
also reveal why it is possible to formulate a simple effective
Hamiltonian, described by a few experi\-mentally measurable parameters
only, which governs all alkali (single valence electron) atoms.

Given a homogeneous gas of atoms with density $n$ in $d$ spatial
dimensions, we may write
 \begin{equation}
 \label{sc1} n = \ell^{-d},
 \end{equation}
 with $\ell$ being the \emph{interparticle spacing}. Indeed, consider
 a homogeneous system in a box of volume $V$. We divide this volume
 into cells of size $\ell^d$ each. Putting exactly one atom into each
 cell, it is possible to distribute $N = V/\ell^d$ particles. Thus, we
 arrive at the density $n = N/V = \ell^{-d}$.

 Experiments on cold atoms are performed in either magnetic or optical
 traps (see \cite{pethick-book,ketterle-review,grimm-review} and
 references therein). Therefore, the ground state of the many-body
 system will not be homogeneous. In particular, the density depends on
 space. However, there are many cases where the picture of a locally
 homogeneous system is still valid and useful
 \cite{RevModPhys.71.463}. In order to understand this, we consider a
 time-independent external trapping potential of harmonic shape. We
 have
\begin{equation}
 \label{sc2} V_{\rm ext}(\vec{x}) = \frac{m}{2} \omega_0^2 r^2
\end{equation}
with $r=|\vec{x}|$ and $m$ being the mass of the atoms. The potential
is characterized by the trapping frequency $\omega_0$. Equivalently,
we may write $V_{\rm ext}(\vec{x}) = \frac{\hbar \omega_0}{2}
(r/\ell_{\rm osc})^2$ with the \emph{oscillator length}
\begin{equation}
 \label{sc4} \ell_{\rm osc} = \left(\frac{\hbar}{m\omega_0}\right)^{1/2}.
\end{equation}
Thus, $\hbar \omega_0$ and $\ell_{\rm osc}$ are the characteristic
energy and length scales of the trap, respectively. We will later see
that in a typical situation, $\ell_{\rm osc}$ constitutes the by far
largest length scale in the system. (It will, however, act as an
infrared (IR) cutoff for very long wavelength fluctuations present
e.g. at a critical point.) Accordingly, $\hbar \omega_0$ usually
provides the smallest energy scale of the problem.

If the physics under consideration takes place on much shorter distances
than $\ell_{\rm osc}$, we can use this separation of scales to work in a so-called
\emph{local density approximation}: Consider the density
at points $\vec{x}_1$ and $\vec{x}_2$, respectively. We can then
expand $n(\vec{x}_1) = n(\vec{x}_2)(1 +
O(|\vec{x}_1-\vec{x}_2|/\ell_{\rm osc}))$. Obviously, for both points
being close to each other we can neglect the correction and assume the
density to be locally constant. In particular, for large values of
$\ell_{\rm osc}$ this may hold for subvolumes of the trapped cloud
which contain many particles. The rules of thermodynamic equilibrium
can then be applied to these small, homogeneous subvolumes. We will come back to this point in
the section on thermodynamics of cold quantum gases.

The statistical behavior of our trapped cloud is determined by the
ratio between the interparticle spacing and the so-called
\emph{thermal} or \emph{de Broglie wavelength}. To get an intuition
for the latter quantity, consider a gas of atoms coupled to a heat
bath of temperature $T$. The nonvanishing temperature induces a
nonzero average kinetic energy $\langle p^2 \rangle_T/2m$ per spatial
direction of the particles. The de Broglie wavelength is the length
scale associated to this energy, according to $\lambda_T = h/\langle
p^2 \rangle_T^{1/2}$. More precisely, using $p = \hbar k = h/\lambda$
($k$ and $\lambda$ the wavenumber and -length, respectively), we
define $\lambda_T$ as the de Broglie wavelength of a particle with
kinetic energy $p^2/2m = \pi k_{\rm B}T$. (The factor of $\pi$ is
purely conventional but standard.) This leads to
\begin{equation}
 \label{sc6} \lambda_T = \left(\frac{2\pi\hbar^2}{m k_{\rm B} T}\right)^{1/2}.
\end{equation}
Note that $\lambda_T \sim T^{-1/2}$ becomes large for decreasing
temperature. The quantities $\ell=\ell(n)$ and
$\lambda_T=\lambda_T(T)$ constitute the many-body length scales of the
system due to nonzero density and temperature, respectively.

Now we compare the length scales set by the interparticle spacing and
the de Broglie wavelength. Thinking of particles as being represented
by wavepackets rather than pointlike objects, $\lambda_T$ determines
the spread of these lumps. The ratio $\ell/\lambda_T$ is large if the
wavepackets of the individual particles are widely separated and do
not overlap. In this case the quantum nature of the particles does not
play a role. Indeed, we may follow the trajectory of an individual
particle by subsequent images, because position and momentum are
determined simultaneously, i.e. the gas can be described
classically. However, for $\ell/\lambda_T \lesssim 1$, we are dealing
with wavepackets which strongly overlap. The gas is then called
\emph{quantum degenerate}, or \emph{ultracold}. Clearly, it is then no
longer possible to distinguish the single atoms and their
trajectories. In this case we rather have to deal with the whole
many-body quantum system. The behavior is then determined by quantum
mechanics, with statistics resulting from the spin of the
constituents; ultracold atoms allow for exploring both degenerate Bose
and Fermi gases. 

\begin{figure}[t!]
 \centering
 \includegraphics[scale=0.9,keepaspectratio=true]{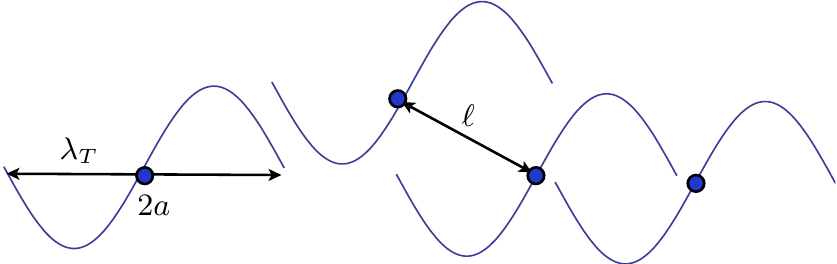}
 \caption{Quantum degeneracy is reached when the thermal wavelength
   $\lambda_T$ is of the same order as the interparticle spacing
   $\ell=n^{-1/d}$. In this regime, it is important for the statistics
   whether the particles are identical or not, leading to quantum
   many-body phenomena such as Bose condensation or a Fermi
   surface. We also indicate a typical order of magnitude for the
   scattering length $a$, which corresponds roughly to the radius of
   equivalent hard-core particles with contact interactions and cross
   section $\sigma \propto a^2$.}
 \label{FigEllLambda}
\end{figure}

The transition from the classical to the quantum
degenerate regime occurs for $ n \lambda_T^d \simeq 1$, i.e.
\begin{equation}
\ell / \lambda_T \simeq  1.
\end{equation} 
We visualize this situation in Fig. \ref{FigEllLambda}.
The combination
\begin{equation}
 \bar{\omega}=n \lambda_T^d=(\lambda_T/\ell)^d
\end{equation}
is called the \emph{phase space density}. It indicates the number of particles contained 
in a cube with linear extension set by the de Broglie wavelength.

\vspace{5mm}
\begin{center}
 \emph{Interactions and effective Hamiltonian}
\end{center}

\begin{figure}[tb!]
 \centering
 \includegraphics[scale=0.85,keepaspectratio=true]{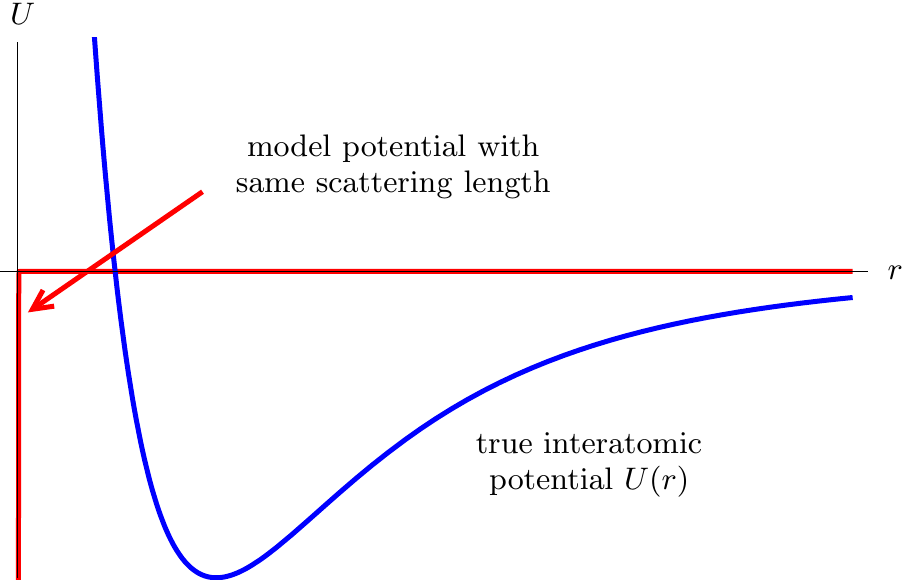}
\caption{The interatomic potential $U(r)$ between two neutral atoms
   is of the Lennard--Jones type, with an attractive van der Waals tail
   $\sim 1/r^6$ at large separations. From $U(r)$ we can calculate the
   scattering length $a$, which is the only parameter relevant for low
   energy scattering. The $\delta$-like potential from
   Eq. (\ref{sc12}), which is shown here in red, is an equally good
   description (and more handy for practical calculations), as long as
   it has the same scattering length. The reason is that under
   ultracold conditions, the short distance details of $U(r)$ are
   never resolved.}
\label{FigVDW}
\end{figure}

\noindent So far, our considerations did not depend on the
interactions of the particles. The alkali atoms used in ultracold gas
experiments are neutral and interact electromagnetically through van
der Waals forces. A typical interaction potential $U(r)$ of two atoms
separated by a distance $r$ has a strongly repulsive part for small
$r$. The physical origin of the latter is Pauli's principle which
forbids the electron clouds of the two atoms to overlap. This
repulsive part can typically be modeled by a term $U(r) \sim
1/r^{12}$, but a hard core repulsion with infinite strength works as
well. For larger distances, two atoms experience an
attraction due to mutual polarization of the electron clouds. Each
atom then acts as a small induced dipole, and they attract each other
according to a van der Waals interaction $U(r) \sim -1/r^6$. (We show
the generic shape of the total interatomic potential, the
Lennard--Jones potential, in Fig. \ref{FigVDW}.) We thus approximate
the microscopic interaction potential to be
\begin{equation}
 \label{sc7} U_{\rm vdW}(r) = \left\{ \begin{array}{cc} \infty& \hspace{3mm} (r \leq r_0),\\
                                       -C_6/r^6& \hspace{3mm} (r > r_0).
                                      \end{array}\right.
\end{equation}
We can use this expression to provide a typical length scale, the
\emph{van der Waals length}, which characterizes the interactions. A
typical length scale for zero (total) energy scattering is obtained
from equating kinetic energy of a particle with momentum
$p=\hbar/\ell_{\rm vdW}$, and potential energy $U_{\rm vdW}(\ell_{\rm
  vdW})$, resulting in
\begin{equation}
 \label{sc8} \ell_{\rm vdW} = \left(\frac{m C_6}{\hbar^2}\right)^{1/4}.
\end{equation}

For typical values of $C_6$, we find that $\ell_{\rm vdW} = (50
... 200) a_0$ ($a_0 = 5.3\times 10^2\text{nm}$ the Bohr
radius), which crucially is much smaller than the interparticle
spacing and the thermal wavelength (cf. Tab. \ref{TabScales}),
\begin{equation}\label{eq:mbvsmic}
\ell, \lambda_T \gg \ell_\text{vdW}.
\end{equation}
The many-body effects in an ultracold gas we are interested in thus never
resolve physics beyond the van der Waals length. As a consequence, we
will be able to specify an \emph{effective low energy Hamiltonian},
valid on length scales $\gtrsim \ell_{\rm vdW}$, as the microscopic
starting point of our
calculations.

After indicating the rough scale associated to interactions, we now
identify the relevant physical parameter which can be extracted from
scattering experiments, the \emph{scattering length} $a$
\cite{grimm-review}. This length scale characterizes two-body
collisions and emerges universally as the \emph{sole} parameter
characterizing low energy collisions in potentials of sufficiently
short range, such as $1/r^6$ as we deal with here. To see this, let us
consider low energy elastic scattering of two particles in a quantum
mechanical framework. (As we explain below, we can assume only elastic
two-body processes to be relevant; further note that our meaning of
``low energies'' is quantified by Eq. \eqref{eq:mbvsmic}.) Restricting
ourselves to three dimensions for concreteness, the relative wave
function of two quantum particles colliding along the $z$-axis in a
short range potential can be written as
\begin{equation}
 \label{sc9} \psi_p(\vec{x}) = e^{\rmi pz/\hbar} + f(p,\theta) \frac{e^{\rmi pr/\hbar}}{r}.
\end{equation}
The scattering amplitude $f(p,\theta)$ depends on the center of mass
energy $p^2/m_{\rm r}$ ($m_{\rm r}$ is the reduced mass) and the
scattering angle $\theta$. Solving the scattering problem for a
particular potential $U(r)$ consists in determining $f(p,\theta)$ or,
equivalently, all partial wave scattering amplitudes $f_l(p)$ in the
expansion $f(p,\theta) = \sum_{l=0}^\infty (2 l+1) f_l(p) P_l(\cos
\theta)$ with Legendre polynomials $P_l$. A nonvanishing relative
angular momentum $l$ of the scattering particles introduces a
centrifugal barrier term $\hbar^2l(l+1)/2m_{\rm r}r^2$ in the
Schr\"{o}dinger equation of relative motion. As a good estimate for
the corresponding energy, we can replace $r^2 \rightarrow \ell_{\rm
  vdW}^2$ and find that this barrier is far too high for particles
with energies $p^2/2m_{\rm r} \ll \hbar^2/\ell_{\rm
  vdW}^2$. Therefore, only isotropic s-wave-scattering ($l=0$) occurs
in ultracold alkali quantum gases.

With $p=\hbar k$, the low momentum expression of the s-wave scattering amplitude is given by
\begin{equation}
 \label{sc10} f_{0}(p) = \frac{1}{-\frac{1}{a} + r_{\rm e} k^2/2-\rmi k + \dots}.
\end{equation}
In this expansion, $a$ is the scattering length anticipated above, and
constitutes the most important parameter quantifying scattering in
ultracold quantum gases in three dimensions. The coefficient $r_{\rm
  e}$ is referred to as effective range. It represents a correction
which for the available $k$ in ultracold gases is subleading, and thus
we work with $f \simeq -a$. From Eq. (\ref{sc9}) we then have
$\psi(\vec{x}) \sim - a/r$ for large $r$ and low momenta.

The limitation to s-wave scattering has drastic consequences for
ultracold gases of identical fermions. They are necessarily
noninteracting. Collisions would only be possible in the p-wave or
higher channels, but these cannot be reached due to the low
energies. In order to have interactions between ultracold alkali
fermions, we therefore always need at least two different species.

From a low energy expansion of the s-wave scattering amplitude in one
and two dimensions, respectively, it is possible to derive parameters
similar to $a$ which quantify scattering in reduced
dimensionality. Such low dimensional geometries can be designed in
experiments by choosing a highly anisotropic harmonic potential with
strong confinement in either one or two directions \cite{hazibabic06}.

Let us briefly comment on the role of inelastic collisions. For
collisions which do not change the spin of the particles, the most
important inelastic mechanism is the formation of a molecule: If two
atoms come close to each other, there may be energetically lower
lying bound states and it is desirable for both atoms to build a
molecule. However, without a third partner which allows for
conservation of energy and momentum in this process, the excess energy
from binding cannot go anywhere. Therefore, in two-body processes
molecule formation is ruled out. If a third atom is involved, we end
up with a high kinetic energy of both the third atom and the
molecule. These fast particles are then expelled from the trap. This
three-body loss results in a finite lifetime of the gas. Due to
diluteness and the contact interaction nature of ultracold atoms, such
processes are suppressed and we find stable gases even at extremely
low temperatures, where solidification would be expected. Increasing
the density, we have to ensure the typical time scale of three-body
recombination to be much larger than the experimental time of
observation.

Equipped with the length scale characterizing interactions, we give a concrete meaning to the notion of ``weak'' interactions by requiring the scattering length to be much smaller than the interparticle spacing in this case. This is equivalent to the \emph{gas parameter} $a n^{1/3}$ being small. The criterion for weak interactions
\begin{equation}
 \label{sc11} a n^{1/3} \ll 1
\end{equation}
is often referred to as \emph{diluteness condition}. This interpretation is motivated by the fact that the scattering length provides the typical extent of a particle as far as its collisional properties are concerned. We indicate this in Fig. \ref{FigEllLambda}. In the dilute regime, it is possible to perform a perturbation theory in the gas parameter.

\begin{table*}[tb!]
\begin{center}
\begin{tabular}{|c|c|c|c|}
  \hline
  \hspace{3mm} Scattering length \hspace{3mm} & \hspace{3mm} Interparticle spacing\hspace{3mm}  &\hspace{3mm}  de Broglie wavelength \hspace{3mm} & \hspace{6mm} Trap size\hspace{6mm} \\
  \hline
  $a/a_{\rm 0}$ & $\ell/a_{\rm 0}$ & $\lambda_T/a_{\rm 0}$ & $\ell_{\rm osc}/a_0$\\
  \hline
  $(0.05\dots0.2)\times10^3$ & $(0.8\dots3)\times10^3$ & 
  $(10\dots40)\times10^3$ & $(3\dots300)\times10^3$\\
  \hline
\end{tabular}
\end{center}
\caption{Standard scale hierarchy in ultracold quantum gases with typical values in units of the Bohr radius $a_0=0.53 \times 10^{-10}$m. The ratios of scales have the following physical meaning:  $a/\ell \ll 1$ -- weakly interacting or dilute; $\ell/\lambda_T \ll 1$ -- ultracold. As long as $\ell_{\rm osc}$ is the largest length scale, the local density approximation is valid (except for long distance physics in the vicinity of a critical point).}
\label{TabScales}
\end{table*}

\begin{figure*}[tb!]
 \centering
 \includegraphics[scale=0.45,keepaspectratio=true]{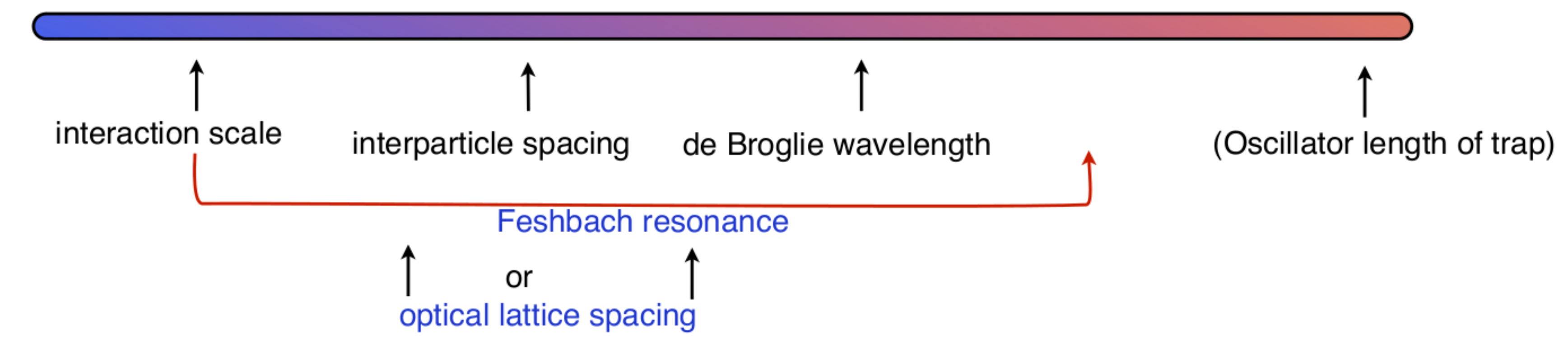}
\caption{Violations of the scale hierarchy, which do not invalidate
   the effective Hamiltonian, since all length scales are much larger
   than $\ell_{\rm vdW}$}
\end{figure*}

For short range interaction potentials and low energy scattering, the s-wave scattering length can be calculated within the Born approximation. It is then given by the Fourier transform of the interaction potential at zero wave vector \cite{pethick-book},
\begin{equation}
\label{born}a = \frac{m}{4 \pi \hbar^2} \int \mbox{d}^3x U(r).
\end{equation} 
In particular, this formula can be applied to the Lennard--Jones potential for cold atoms introduced above. Importantly, from Eq. (\ref{born}) we learn (i) that value and sign of
the scattering length may depend sensitively on the short distance
physics of the interatomic potential and (ii) that we do not need to
know these details, since very different shapes of the interaction
potential will have the \emph{same} scattering length, i.e. the same
low energy scattering behavior. Quite remarkably, it is therefore
possible to replace the microscopic Lennard--Jones potential by any
other model potential producing the same scattering length
(cf. Fig. \ref{FigVDW}). For practical reasons, it is often convenient
to work with completely local \emph{contact potentials}
\begin{equation}
 \label{sc12} U(r) = g_\Lambda \delta(\vec{x}).
\end{equation}
This simple model potential needs regularization at short distances $L
= \hbar /\Lambda$, and a subsequent renormalization
procedure. We remind to this fact with the index referring to an
ultraviolet cutoff $\Lambda$, and refer to Sec. \ref{ManyBodyFRG} for a detailed
presentation of the procedure. The cutoff-independent renormalized
coupling constant $g$ is related to the physically measured scattering
length by the simple formula
\begin{equation}
 \label{sc13} a = \frac{m}{4 \pi \hbar^2}g.
\end{equation}

The above considerations on ultracold atoms can be summarized in the
\emph{effective Hamiltonian}
\begin{equation}
 \label{sc13b}\hat{H} = \int_{\vec{x}} \left( \hat{a}^\dagger(\vec{x})\left(-
     \frac{\nabla^2}{2m}+V_{\rm ext}(\vec{x})\right)\hat{a}(\vec{x}) + g_\Lambda \hat{n}(\vec{x})^2\right),
\end{equation}
where the operators $\hat{a}^\dagger(\vec{x})$ and $\hat{a}(\vec{x})$
create and annihilate an atom at point $\vec{x}$, respectively, and
$\hat{n}(\vec{x})=\hat{a}^\dagger(\vec{x})\hat{a}(\vec{x})$ is the
local particle density operator. Note that the power of two in the
interaction term $\sim g \hat{n}^2$ stems from the fact that two
particles have to meet at one point in order to interact. The trapping
potential $V_{\rm ext}(\vec{x})$ lifts the energy of the particles at
the point $\vec{x}$ and thus this term is proportional to
$\hat{n}(\vec{x})$. This rather universal Hamiltonian provides an
accurate description for all ultracold alkali atoms.

It is a key feature of ultracold quantum gases that they are
accurately described by effective microscopic Hamiltonians which
depend only on a few system parameters. The latter can be measured in
experiments to a high precision, e.g. by spectroscopic methods or by
measurement of the collisional cross sections \cite{grimm-review},
without the need to resolve the full interatomic potentials. This
situation is very distinct from condensed matter systems, where the
underlying microscopic model is not known to such precision, and often
has to be approximated by an educated guess. Moreover, realizations of
ultracold quantum gases allow to change the system parameters
continuously and thus to understand their influence on the many-body
state.

We summarize our discussion by indicating the standard scale hierarchy
in Tab. \ref{TabScales}, which is built up from the scattering length
$a$, the interparticle distance (density) $\ell$, the thermal
wavelength (temperature) $\lambda_T$ and the oscillator length
$\ell_{\rm osc}$. Moreover, the system has a natural UV cutoff
$\Lambda^{-1} \ll \ell_{\rm vdW}$. Microscopic details on shorter length
scales are irrelevant for our purposes because none of the many-body
length scales can resolve the underlying physics.

It is both experimentally and theoretically appealing that ultracold
atoms can be tuned such that they \emph{violate the scale hierarchy},
allowing to reach strongly interacting regimes -- crucially, without
loosing the validity of the above discussion. One way is provided by
\emph{Feshbach resonances} of the scattering length
\cite{grimm-review}. Here, we can loosen the condition $ a \ll \ell$
and explore new regimes of the many-body system which are not captured
by mean field theory or perturbative expansions. Such resonances are
realized in cold atoms if a bound state is located close to the zero
energy scattering threshold, and is tuned to resonance due to the
variation of an external magnetic field. From this we infer that a
Feshbach resonance is a result of a specific fine-tuning of the
microphysics. The scattering length can then be parametrized according
to
\begin{equation}
 \label{sc14} a = a_{\rm bg}\left(1  - \frac{\Delta B}{B-B_0}\right),
\end{equation}
where $a_{\rm bg}$, $\Delta B$ and $B_0$ are background scattering
length, width and position of the resonance, respectively. Obviously,
approaching $B_0$, we can obtain an anomalously large scattering
length, meaning that, by virtue of fine-tuned microphysics, it greatly
exceeds the generic scale set by the van der Waals length,
\begin{equation}
 \label{sc14b} |a| \gg \ell_{\rm vdW}.
\end{equation}
We will discuss the issue of Feshbach resonances in more detail in
Sec. \ref{SecFesh} when introducing our effective model for the BCS-BEC
crossover.

Another way to reach an interaction dominated regime is by
superimposing an \emph{optical lattice}
\cite{lewenstein-rmp-56-135,bloch-review}. This is a standing wave of
counterpropagating laser beams in each spatial direction, which
provides a conservative periodic potential landscape for the atoms via
the AC Stark effect. Tuning the depth of the lattice wells via the
laser intensity, we can withdraw the kinetic energy more strongly then
the interaction energy and thus arrive at a strongly correlated
system. The lattice spacing is related to the wavelength of the light
used for the optical lattice. By engineering neighboring sites close
to each other, we can reach high densities (``fillings''). Each of
these effects enhances the correlations in the system.

Recall that the validity of the effective Hamiltonian in
Eq. (\ref{sc13b}) is restricted to length scales sufficiently larger
than $\ell_{\rm vdW}$. Since the mentioned scale violations happen at
larger scales, the faithful microscopic modeling is not
touched. Therefore, the pointlike description of the interactions is
also applicable in the dense and strong coupling regimes.

For a more detailed presentation of low energy universality in atomic
few-body systems and from the viewpoint of quantum field theory, we
refer to \cite{braaten-review}.

\subsection{Thermodynamics}
\noindent In this section we review a few thermodynamic concepts which are of relevance for experiments with ultracold atoms. We derive general thermodynamic statements, which hold independently of the particular system under consideration. We will see that the phase diagram and the equation of state encode important, experimentally accessible information about a system and thus are desirable quantities to be computed from first principles. This also serves as one of the motivations to investigate cold atoms with the Functional Renormalization Group.

For thermodynamics to be applicable, we require the internal processes
of a many-body system to be such that the system is in equilibrium on
the time scale of observation. Strictly speaking, thermodynamic
statements and, in particular, the theory of phase transitions are
only valid in infinitely large systems. But this requirement is less
severe as it might seem at first sight, because any thermodynamic
relation can be expressed in terms of intensive quantities only, like
particle density, entropy density, or magnetization per
particle. Taking these densities to be local quantities, we can apply
the laws of thermodynamics locally for small \emph{subsystems} of
finite volume and particle number. This procedure works perfectly at
room temperature with large particle numbers $N \sim 10^{23}$, and is
still justified for trapped gases with typically $N \sim 10^4 -
10^7$. Let us add that in addition, it turns out that systems with low
atom loss rate and long lifetime can indeed be assumed to be
thermodynamically equilibrated over the period of observation. Such a
system is provided by two-component fermions in the BCS-BEC crossover.

We recall that the full thermodynamic information of a system is
stored in the \emph{equation of state} $P(\mu,T)$, which can be
expressed in terms of the pressure as a function of chemical potential
and temperature. Using the Gibbs--Duhem relations $\mbox{d}P = n
\mbox{d} \mu + s \mbox{d}T$ and $\vare = Ts-P+\mu n$ we can calculate
all other intensive thermodynamic quantities from the pressure. Here,
$n=N/V$, $s=S/V$ and $\vare=E/V$ are the densities of particle number,
entropy and energy, respectively. The chemical potential $\mu$ is a
parameter which determines the particle number $N(\mu)$ for a given
temperature. Eliminating $\mu$ for the density $n(\mu,T)$, the
equation of state can also be formulated in terms of the free energy
density $f(n,T)$, which is the Legendre transform of the pressure
according to $f = \mu n - P$.

In order to understand the influence of a trap, we consider a cloud in
a time-independent external potential $V_{\rm ext}(\vec{x})$ which
varies on much larger length scales than the typical atomic ones
(e.g. interparticle spacing and scattering length). Picking two
neighboring small but yet macroscopic subvolumes $V_1$ and $V_2$ of
the cloud, thermal and chemical processes between them will result in
the equality of their temperature and chemical potential. Since the
subvolumes were arbitrary, we conclude that temperature and
\emph{full} chemical potential are constant inside the trap. However,
from the Gibbs-Duhem relation we infer that the full chemical
potential corresponds to the Gibbs free energy $G=F+PV$ per particle,
$\mu = G/ N$. The latter is spatially inhomogeneous due to the trap
and we find $\mu = \mu_{\rm hom}(n(\vec{x}),T) + V_{\rm ext}(\vec{x})
= \mbox{const}.$ In this formula, $\mu_{\rm hom}(n,T)$ is the chemical
potential obtained from a calculation in a homogeneous setting, e.g. a
box of volume $V$ containing $N$ particles.

We conclude that a system where the thermodynamic quantities are replaced according to
\begin{equation}
 \label{td4} P(\mu,T) \rightarrow P(\mu-V_{\rm ext}(\vec{x}),T)
\end{equation}
behaves like a system trapped in a potential of large spatial
extent. This prescription is called \emph{local density approximation}
(LDA). The above derivation provides an intuitive understanding why
this procedure should give reasonable results. Of course, if we cannot
pick small, yet macroscopic subvolumes, the argument breaks down. The
applicability of LDA is therefore limited to systems, where the trap
$\ell_{\rm osc}$ provides the largest length scale. This agrees with
our earlier considerations. From a field theory perspective it is very
promising that properties of homogeneous systems can be obtained from
trapped gases and, indeed, there have already been beautiful
measurements of the equation of state of the BCS-BEC crossover to a
high precision using LDA \cite{Nascimbene10,Navon10,Ku12}.

\begin{figure}[t!]
 \centering
 \includegraphics[scale=0.8,keepaspectratio=true]{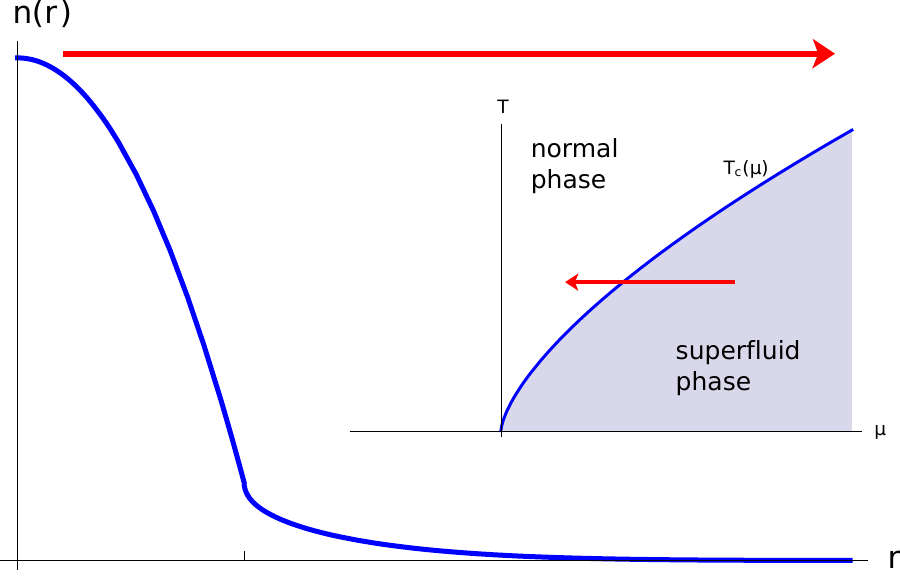}
 \caption{Within LDA, the inner regions of the trapped cloud correspond to higher chemical potentials: $\mu_{\rm loc}(r)=\mu - V_{\rm ext}(r)$. We show here the density profile of a weakly interacting Bose gas in an external harmonic confinement. The inset shows a typical phase diagram in the $(\mu,T)$-plane, where the blue region represents the superfluid phase. We cross the critical line of the superfluid phase transition for fixed temperature $T$ at a certain chemical potential $\mu_{\rm c}(T)$. This corresponds to a kink in the density profile at a critical value $n_{\rm c}(T)$. Note, however, that LDA breaks down in the outer regions of the cloud, where the gas is extremely dilute.}
 \label{LDA}
\end{figure}

The equation of state also contains information about possible phase transitions appearing in the many-body system. Phases consist of extended parameter regimes which can be distinguished from each other by macroscopic observables. As an example, we consider the element iron. Despite the difference of solid, liquid and gaseous phase we can independently also distinguish the ferromagnetically  from the antiferromagnetically ordered phase, or furthermore the crystal structures of $\gamma$-Fe and $\alpha$-Fe. Phase transitions manifest themselves through kinks and jumps in the thermodynamic functions, typically in the higher derivatives of $P(\mu,T)$. These root in non-analyticities contained in the partition function. It is easy to see that true phase transitions need a continuum of degrees of freedom, i.e. occur only in the thermodynamic limit. Indeed, the partition function is $Z = {\rm Tr} e^{-\beta H} = \sum_n e^{-\beta E_n}$, $E_n$ the eigenenergies of the system. Each of the contributions is analytic. Non-analyticities can only be generated in the case of infinitely many states entering the sum.

More formally, we distinguish two phases by an \emph{order parameter} $\rho_0(\mu,T)$, which depends on the thermodynamic variables. In different phases, it is either zero or nonzero, which gives rises to the \emph{phase diagram} in the $(\mu,T)$-plane. For a fixed value of the chemical potential, we define the critical temperature $T_{\rm c}(\mu)$ via the relation $\rho_0(\mu,T_{\rm c}(\mu))=0$. Of course, we can also fix the density $n$ to obtain the critical temperature $T_{\rm c}(n)$ as a function of $n$.

In the regime where LDA is applicable, the local chemical potential $\mu_{\rm loc}(\vec{x}) = \mu - V_{\rm ext}(\vec{x})$ has its largest value at the minima of the trapping potential. Accordingly, an increase of the potential reduces $\mu_{\rm loc}$. For this reason, we can \emph{scan the phase diagram} over a certain region from a density image in a harmonic potential $V_{\rm ext}(\vec{x}) = \frac{m}{2}\omega_0^2r^2$, see Fig. \ref{LDA}. From our above considerations we conclude that the corresponding path in the $(\mu,T)$-plane is an isothermal line. In particular, we may cross the phase boundary when the local chemical potential reaches the critical value $\mu_{\rm c}(T)$. For this reason, we can have a superfluid gas in the inner regions of the cloud, whereas the outer shell is in its normal phase. The lobes in the phase diagram of the Bose-Hubbard model lead to a wedding cake structure of the density profile \cite{PhysRevLett.97.060403}.

\subsection{Noninteracting Bose and Fermi gases}
\label{ideal}

\noindent After these general remarks on thermodynamics we turn our attention to degenerate, noninteracting Bose and Fermi gases. The notions of Bose--Einstein condensation and Fermi surfaces are introduced. They constitute the two cornerstones of quantum statistical phenomena and are crucial for understanding interacting gases.

The state of a single particle can be addressed by its momentum $\vec{p}$ and spin-projection $\sigma$. The corresponding occupation numbers $n_{\vec{p},\sigma}$ are restricted to $0,1$ for fermions due to Pauli's principle, whereas they can have arbitrary integer values $0,1,2,\dots$ for bosons. As is known from statistical mechanics, we then find for the equation of state
\begin{equation}
 \label{td4b} P(\mu,T) = \mp g k_{\rm B} T \int \frac{\mbox{d}^dp}{(2\pi\hbar)^d} \log \left( 1 \mp e^{-\beta(\varepsilon_p-\mu)}\right),
\end{equation}
where $\varepsilon_p=\vec{p}^2/2m$ and $g$ is the spin degeneracy of the momentum states. We have $g=1$ for spinless bosons considered here, and $g=2$ for spin-$1/2$ fermions. The upper (lower) sign in Eq. (\ref{td4b}) holds for bosons (fermions). As we will see below, for bosons, this expression is valid in the absence of a condensate only.

\vspace{5mm}
\begin{center}
 \emph{Free bosons and Bose--Einstein condensation}
\end{center}

\noindent To understand the appearance of condensation as a purely quantum statistical effect, we consider an ideal gas of identical bosons. At zero temperature, we expect all bosons to be in the single particle state with energy $\vare=0$. In particular, this means that the occupation number $N_0$ of that state is extensive, $N_0\sim V$. We say that the zero mode $\vare=0$ is occupied \emph{macroscopically}. At low nonzero temperature, some particles will be thermally excited into the higher states. At very high temperatures, we approach the Boltzmann limit, where all occupation numbers are small (in particular, none of them is occupied macroscopically) and  the distribution function $n(\vare)$ is very broad. Therefore, there must be a critical temperature $T_{\rm c}$ below which macroscopic occupation of the single particle ground state sets in. Since this particular behavior is due to quantum statistics and absent in a classical gas, we can estimate the critical temperature very roughly to occur for $\lambda_{T_{\rm c}} \simeq \ell$.

Starting from Eq. (\ref{td4b}) for the pressure, we obtain the particle number in a three-dimensional box of volume $V$ by virtue of a $\mu$-derivative as to be given by
\begin{equation}
 \label{td9} N(T,V,\mu) = \sum_{\vec{q}} \langle \hat{a}^\dagger_{\vec{q}}\hat{a}_{\vec{q}}\rangle =\frac{V}{\lambda_T^3} \frac{1}{\Gamma(3/2)} \int_0^\infty \frac{\mbox{d}\varepsilon \sqrt{\varepsilon}}{e^{\varepsilon-\beta\mu}-1}.
\end{equation}
 For fixed temperature and volume, this formula has a maximum $N_{\rm max}$ at $\mu=0$. However, if we decide to put more than $N_{\rm max}$ particles into the box, the expression necessarily becomes invalid. The critical temperature $T_{\rm c}(n)$ where this happens is determined by a critical phase space density
\begin{equation}
 \label{td10} \bar \omega_{\rm c} = n\lambda_{T_{\rm c}}^3 = (\lambda_{T_{\rm c}}/\ell)^3 = \zeta(3/2) \simeq 2.612,
\end{equation}
i.e. $\lambda_{T_{\rm c}}/\ell = O(1)$ as anticipated above.

Since our starting point was physically sound,  but we ended up with an unphysical result, we must have made an error. This led Einstein and Bose to treating the zero momentum mode separately \cite{einstein24,einstein25}. Indeed, in Eq. (\ref{td9}) we did not appropriately incorporate the states with $\varepsilon=0$:   Replacing the quantized momenta  of the finite system $\vec{p}_{\vec{n}}=2\pi\hbar\vec{n}/L$ in the naive continuum limit
\begin{equation}
 \label{td8b} \frac{1}{V} \sum_{\vec{n}\in\mathbb{Z}^3} \rightarrow \int \frac{\mbox{d}^3p}{(2 \pi \hbar)^3} \sim \int_0^\infty \mbox{d} \varepsilon \sqrt{\varepsilon},
\end{equation}
we multiply their contribution with $\varepsilon=0$ (or equivalently
$p^2 =0$). This corresponds to a vanishing occupation of the single
particle ground state, which constitutes a bad approximation, as is
apparent from our above considerations.

Therefore, the situation at temperatures $T< T_{\rm c}(n)$ is as
follows. Formula (\ref{td9}) with $\mu=0$ describes the excited
particles in the states with $\varepsilon >0$. The remaining $N_0(T) =
N- N_{\rm ex}(T)$ particles are condensed to the zero energy state,
leading to its macroscopic occupation. This resolves the puzzle from
above. If we put more than $N_{\rm max}$ particles into the box, they
will add to the condensate. The particle number below $T_{\rm c}$ is
given by
\begin{equation}
  \label{td11} N(T,V) = \langle \hat{a}^\dagger_{\vec{0}}\hat{a}_{\vec{0}}
  \rangle  + \frac{gV}{\lambda_T^3} \zeta(3/2).
\end{equation}
Obviously, $N_0(T) = \langle
\hat{a}^\dagger_{\vec{0}}\hat{a}_{\vec{0}}\rangle \sim V$ is
extensive. The condensate fraction $N_0(T)/N$ is an order parameter
for the Bose--Einstein condensation phase transition. From
Eq. (\ref{td10}) we conclude
\begin{equation}
 \label{tf12} \frac{N_0(T)}{N} = 1 - \left(\frac{T}{T_{\rm c}}\right)^{3/2} \text{ for } T \leq T_{\rm c}.
\end{equation}
It vanishes continuously for $T \rightarrow T_{\rm c}$, which signals
a second order phase transition.

\begin{figure}[tb!]
 \centering
 \includegraphics[scale=0.8,keepaspectratio=true]{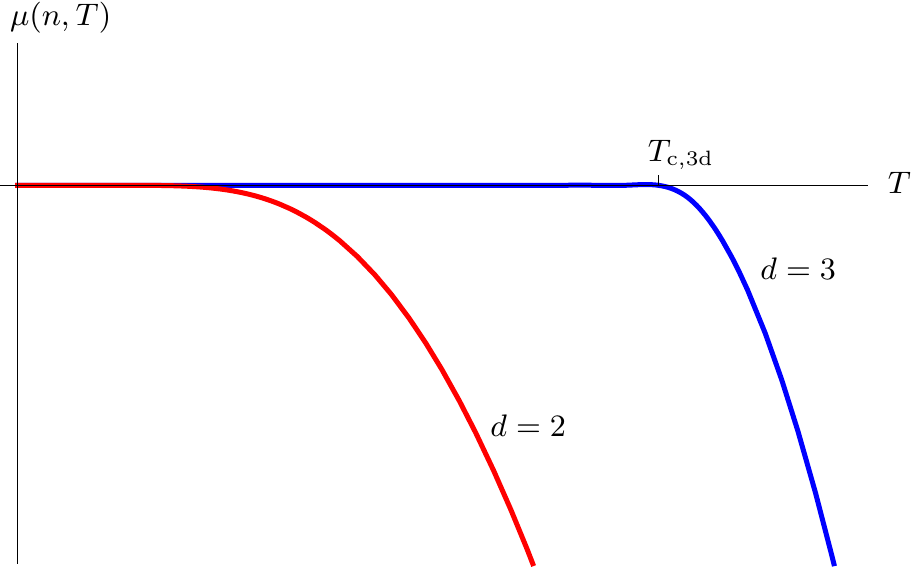}
 \caption{We plot the chemical potential $\mu(n,T)$ from
   Eq. (\ref{td4b}) for an ideal Bose gas at fixed density $n$. In
   three spatial dimensions, the function hits zero at $T_{\rm c}(n)
   >0$. Since Eq. (\ref{td4b}) cannot be applied for positive $\mu$,
   the chemical potential remains zero and condensation sets in. In
   contrast, the chemical potential in two dimensions is negative for
   $T>0$ and thus there are always enough thermally excited states and
   condensation is absent.}
\label{FigBECmu}
\end{figure}

In Eq. (\ref{td8b}) we used the three-dimensional density of states
$\rho(\vare) \sim \sqrt{\vare}$ to show why condensation appears. In
$d$ spatial dimension, we have $\rho(\vare) \sim \vare^{d/2-1}$ and
the ground state contribution is not multiplied by zero for $d \leq
2$. Indeed, a similar calculation shows that for one- and
two-dimensional systems the particle number $N(T,V,\mu)$ does not have
a maximum at nonzero temperatures and thus Bose--Einstein condensation
is absent. For $d=1$ this also holds at zero temperature. In
Fig. \ref{FigBECmu}, we plot the chemical potential as a function of
temperature. Whereas $\mu(T_{\rm c})=0$ for a nonzero $T_{\rm c}$ in
three dimensions, we find $T_{\rm c}=0$ for $d=2$. Our finding for
noninteracting particles is a special case of the generally valid
\emph{Mermin--Wagner theorem} \cite{amit-book}, which states that
there is no spontaneous breaking of a continuous symmetry in $d \leq
2$ (noncompact) dimensions. The ingredients to this theorem are the
locality of the underlying Hamiltonian, and the universal relativistic
long-wavelength form of the dispersion relation. The long-range order
is then destroyed by fluctuations with very long wavelengths. However,
in atomic gas experiments the trap provides the largest length scale
$\ell_{\rm osc}$, such that these fluctuations are not present and
condensation can be observed in lower-dimensional geometries.

\vspace{5mm}
\begin{center}
 \emph{Free fermions and Fermi surface}
\end{center}

\noindent Whereas the appearance of a Bose--Einstein condensate is
closely related to the fact that identical bosons can have arbitrarily
large occupation numbers, the notion of a \emph{Fermi surface} is a
consequence of Pauli's principle for many-fermion systems.

To get an intuition, we consider an ideal gas of $N$ identical
fermions. What will be the ground state of the quantum many-body
system? (This state is realized at zero temperature.) Obviously, each
of the particles seeks to minimize its energy. But since every single
particle state can only be occupied by at most one fermion, the ground
state will be such that precisely the $N$ energetically lowest lying
states are occupied. Equivalently, due to rotation symmetry, all
states with momenta inside a sphere of radius $p_{\rm F}$ in momentum
space will be occupied. Restricting to three dimensions, we can count
states by dividing the classical phase space into cells of volume
$h^3$. This yields
\begin{equation}
 \label{td5} N \stackrel{!}{=} \frac{gV}{(2\pi \hbar)^3} \frac{4\pi}{3}  p_{\rm F}^3.
\end{equation}
We call $p_{\rm F} = \hbar k_{\rm F}$ the Fermi momentum and deduce
\begin{equation}
 \label{td6}k_{\rm F}(n) = (6 \pi^2 n/g) ^{1/3}.
\end{equation}
Up to a prefactor of order unity, $k_{\rm F}(n)$ equals the inverse
interparticle spacing $\ell^{-1} = n^{1/3}$. Eq. (\ref{td6}) can be
used to express thermodynamic quantities as a function of $k_{\rm
  F}(n)$ instead of the density $n$. In this case, $k_{\rm F}(n)$ is
not bound to the presence of a Fermi surface. From $p_{\rm F}$ we
construct the \emph{Fermi energy} and \emph{temperature},
$\varepsilon_{\rm F}=\vare_{p_{\rm F}}=p_{\rm F}^2/2m$ and $T_{\rm F}
= \varepsilon_{\rm F}/k_{\rm B}$, respectively.

\begin{figure}[tb!]
 \centering
 \includegraphics[scale=0.8,keepaspectratio=true]{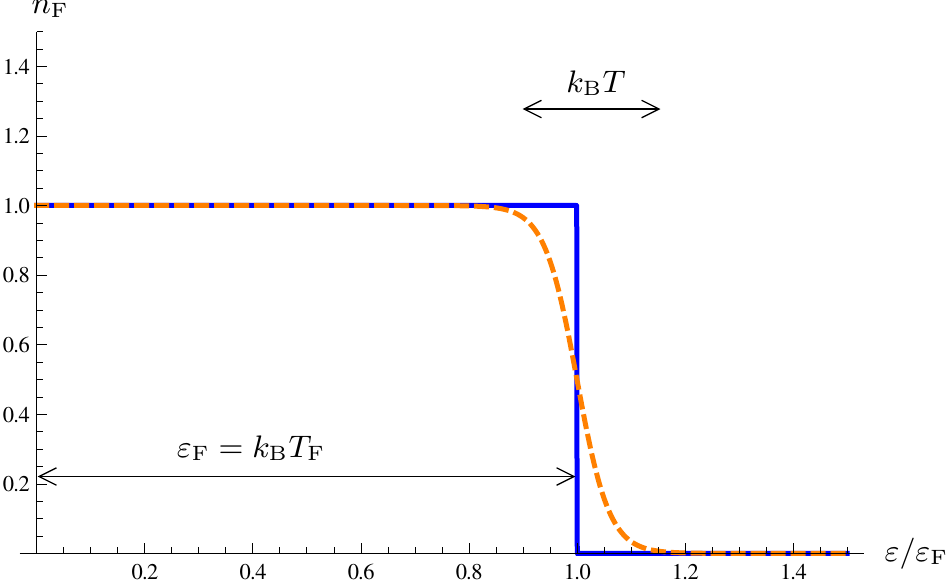}
 \caption{The Fermi--Dirac distribution at zero temperature (solid
   line) constitutes a step function. For $T>0$ (dashed line),
   broadening appears in a region of width $k_{\rm B}T$ around the
   Fermi edge located at $\vare=\vare_{\rm F} = k_{\rm B}T_{\rm
     F}$. If the distance of the edge from the origin is much larger
   than the area of broadening, the distribution function still
   displays the characteristic step-like behavior. Clearly, this
   picture is valid for the dimensionless parameter $T/T_{\rm F}$
   being small.}
\label{FermiDirac}
\end{figure}

Our simple picture of the many-body ground state is correct, because
from Eq. (\ref{td4b}) we have
\begin{align}
  \nonumber n &= \frac{\partial P}{\partial \mu} = g \int \frac{\mbox{d}^3p}{(2\pi\hbar)^3} \frac{1}{e^{\beta(\varepsilon_p-\mu)}+1}\\
 \label{td7}&\xrightarrow{T\rightarrow 0} g \int\frac{\mbox{d}^3p}{(2\pi\hbar)^3} \theta(\varepsilon_p-\mu) = \frac{g}{6 \pi^2 \hbar^3} (2 m \mu)^{3/2}.
\end{align}
On the other hand, from the zero temperature limit of the Fermi--Dirac
distribution we infer that the highest energy present in the system is
$\varepsilon_{\rm F} = \mu$ and thus we find $p_{\rm F} = (2
m\mu)^{1/2}$. As before, we finally arrive at $k_{\rm F} = (6 \pi^2
n/g)^{1/3}$. The Fermi--Dirac distribution at zero temperature is
shown in Fig. \ref{FermiDirac}.

What happens to this picture at nonzero temperature? The Fermi--Dirac
distribution $n_{\rm F}(\varepsilon) = (e^{(\varepsilon-\mu)/k_{\rm
    B}T}+1)^{-1}$ is no longer a sharp step function but rather smears
out around $ \epsilon = \mu$. Nevertheless, the smeared out region is
of order $k_{\rm B}T$, whereas the distance of the edge from
$\varepsilon=0$ is of order $\mu \simeq \varepsilon_{\rm F} = k_{\rm
  B}T_{\rm F}$. Therefore, as long as $T/T_{\rm F} \ll 1$, the
distribution function looks approximately like a step function. We
visualize this situation in Fig. \ref{FermiDirac}. For $T>0$, there
are thermally excited particles with energies close to the chemical
potential. We conclude that the low lying excitations of a Fermi gas
are not at zero momentum but rather at momenta close to the Fermi
surface, which consists of the momenta $|\vec{p}| = p_{\rm F}$.

\section{Functional methods for interacting bosons and fermions}
\label{Sec3}

\subsection{Functional integral and effective action}
\label{SecFun}
\noindent In this section, the quantum field theoretical formulation
of interacting cold atoms is put forward. Starting from the functional
integral representation of the quantum partition function $Z$, we
introduce the effective action $\Gamma$ \cite{amit-book}. The latter
stores the same information as the partition function, however in a
way that is more intuitive. In particular, it naturally provides the
classical limit. The effects of both quantum and thermal fluctuations
on physical observables can be derived from it in the few- and
many-body context. Moreover, it allows for a transparent discussion of
spontaneous symmetry breaking, and allows to leverage the power of
symmetry considerations from the classical action over to the full
quantum theory. We set $\hbar = k_{\rm B}=1$. For the moment, we keep
the nonrelativistic mass $M$ in our formulation, but later we will set
$2M=1$ in the same spirit as for the fundamental constants.

\vspace{5mm}
\begin{center}
 \emph{Functional integral}
\end{center}

\noindent As we have shown in section \ref{scales}, a system of ultracold atoms is accurately described by the effective Hamiltonian
\begin{align}
 \label{Func1}\hat{H} = \int_{\vec{x}} \left( \hat{a}^\dagger(\vec{x})\left(-\frac{\nabla^2}{2M}+V_{\rm ext}(\vec{x})\right)\hat{a}(\vec{x}) + \frac{g}{2} \hat{n}(\vec{x})^2\right),
\end{align}
where $\hat{a}^\dagger(\vec{x})$ and $\hat{a}(\vec{x})$ are operators which create and annihilate an atom at position $\vec{x}$, respectively. Depending on whether we consider bosons or fermions, these operators satisfy commutation or anti-commutation relations. The density operator is given by $\hat{n}(\vec{x})=\hat{a}^\dagger(\vec{x})\hat{a}(\vec{x})$.

The Hamiltonian in Eq. (\ref{Func1}) defines a quantum field theory with operators $\hat{a}$ and $\hat{a}^\dagger$ on each point of space. Physical observables are derived from expectation values of functions of these operators. However, the corresponding quantum field theory can also be formulated in terms of a functional integral. The latter does no longer depend on the notion of operators. In the context of quantum many-body systems, a possible derivation starts from the grand canonical partition function
\begin{equation}
 \label{Func2} Z(\mu,T) = \mbox{Tr} \Bigl( e^{-\beta(\hat{H}-\mu \hat{N})}\Bigr),
\end{equation}
where the trace is taken over Fock space. This trace can be represented in the basis of so-called coherent states, which are eigenstates of the annihilation operator $\hat{a}(\vec{x})$. We then obtain 
\begin{equation}
 \label{fun1} Z(\mu,T) = \int \mbox{D}\varphi^*\, \mbox{D}\, \varphi e^{-S[\varphi^*,\varphi]}.
\end{equation}

The expression in Eq. (\ref{fun1}) is called a \emph{functional} or \emph{path integral}. It contains the microscopic action $S[\vphi^*,\vphi]$ of a field theory, which in our case is nonrelativistic. It is related to the normal ordered\footnote{Normal ordering is a prerequisite to the construction of the coherent state functional integral, cf. \ref{AppFun}. Starting from Eq. \eqref{Func1}, normal ordering introduces a shift $\mu \to \mu - g/2$, which we simply absorb into a redefinition of the chemical potential.} Hamiltonian $\hat{H}=H[\hat{a}^\dagger({\vec{x}}),\hat{a}({\vec{x}})]$ according to
\begin{align}
 \nonumber S[\vphi^*,\vphi] =\int_0^\beta &\mbox{d}\tau \int_{\vec{x}} \vphi^*(\tau,\vec{x})(\partial_\tau-\mu) \vphi(\tau,\vec{x}) \\
 \label{Func3}&+\int_0^\beta \mbox{d}\tau H[\vphi^*(\tau,\vec{x}),\vphi(\tau,\vec{x})].
\end{align}
For the particular choice of the effective Hamiltonian in Eq. (\ref{Func1}), we have
\begin{align}
 \nonumber S[\varphi^*,\varphi] = \int_0^\beta \mbox{d}\tau\int_{\vec{x}} &\Bigl(\varphi^*(\tau,\vec{x})\Bigl(\partial_\tau - \frac{\nabla^2}{2M}  - \mu\Bigr) \varphi(\tau,\vec{x})\\
 \label{Func4} &+\frac{g}{2} (\varphi^*(\tau,\vec{x})\varphi(\tau,\vec{x}))^2 \Bigr).
\end{align}

The explicit construction of the functional integral representation of the partition function for a generic many-body Hamiltonian is carried out in \ref{AppFun}. We summarize here the two main findings.
\begin{itemize}
 \item[1)] Bosonic atoms are represented by complex fields $\vphi(\tau,\vec{x})$, whereas fermions are described in terms of Grassmann valued fields $\psi(\tau,\vec{x})$.
 \item[2)] The non-commutativity of operators introduces the \emph{imaginary time} $\tau$, which is restricted to the interval $[0,\beta]$. Bosonic fields are $\beta$-periodic in time, i.e. $\vphi(\beta,\vec{x})=\vphi(0,\vec{x})$. In contrast, fermionic fields satis\-fy $\psi(\beta,\vec{x}) = - \psi(0,\vec{x})$.
\end{itemize}
The second property implies that the Fourier transform of the fields $\vphi$ and $\psi$ in imaginary time direction reduces to a Fourier series with discrete \emph{Matsubara frequencies}
\begin{equation}
 \label{Func5} \omega_n = \begin{cases} 2 \pi n T &({\rm bosons})\\ 2 \pi(n+1/2)T & ({\rm fermions})\end{cases}, \hspace{5mm} n \in \mathbb{Z}.
\end{equation}
We say that the imaginary time direction is compactified to a torus of circumference $\beta$. In fact, introducing a chemical potential $\mu$ and compactifying the time direction to $0 \leq \tau \leq \beta$, we can describe any euclidean quantum field theory at nonzero density and temperature. For example, this procedure can be applied to quantum chromodynamics.

\vspace{5mm}
\begin{center}
 \emph{Generating functional and effective action}
\end{center}

\noindent Starting from the functional integral representation for the partition function $Z$, we now construct the corresponding effective action. The procedure outlined here focuses on the application to systems of ultracold atoms. However, additional insights into these concepts can be obtained from a comparison to classical Ising magnets on a discrete lattice. For this reason, we included a detailed discussion of the latter system in \ref{AppIsing}. There, we also perform the continuum limit and review the notions of functional differentiation and integration. A dictionary for the translation between ultracold bosonic atoms and the classical Ising model is given in Tab. \ref{TabDictionary}. For simplicity, we mostly restrict to the bosonic case in this section. There are only minor modifications for fermions, which are discussed at the end of the section.

\begin{table}[t]
\begin{center}
\begin{tabular}{|c|c|}
\hline
 \hspace{3mm} \small{\textbf{Ising magnets}} \hspace{3mm}  &\hspace{3mm}  \small{\textbf{Bosonic atoms}} \hspace{3mm} \\
\hline
 \small{lattice sites} $\vec{x}_i$ &  \small{space-time points} $X=(\tau,\vec{x})$ \\
\hline
  \small{magnetic moment} $m_i = \pm 1$ & \small{complex field} $\vphi(X)$ \\
\hline
 \small{magnetic field} $h_i$ & \small{external source} $j(X)$ \\
\hline
 \small{mean field} $\bar{m}_i = \langle m_i \rangle_h$ & $\phi(X)=\langle \vphi(X)\rangle_j$ \\
\hline
  \small{partial derivative} $\frac{\partial}{\partial h_i}$ & \small{functional derivative} $\frac{\delta}{\delta j(X)}$ \\
\hline
  \small{summation over sites} $\sum_i$ & \small{functional integral} $\int \mbox{D}\vphi^*\mbox{D}\vphi$ \\
\hline
  \small{functional measure} $\prod_i {\rm d}m_i$ &  $\mbox{D}\vphi^*\mbox{D}\vphi = \prod_X \frac{{\rm d}\vphi^*(X){\rm d}\vphi(X)}{\mathcal{N}}$ \\
\hline
 \small{partition function} $Z(\{h_i\})$ & \small{generating functional} $Z[j^*,j]$ \\
\hline
  $\bar{m}_i = \frac{\partial \log Z}{\partial h_i}$ & $\phi(X) = \frac{\delta \log Z}{\delta j^*(X)}$ \\
\hline
 $\langle m_i m_j \rangle_h - \bar{m}_i\bar{m}_j = \frac{\partial^2\log Z}{\partial h_i \partial h_j}$ & $\langle \vphi_X^* \vphi_Y\rangle_j - \phi^*_X \phi_Y =\frac{\delta^2 \log Z}{\delta j_X\delta j^*_Y}$ \\
\hline
 \small{effective action} $\Gamma(\{\bar{m}_i\})$ & $\Gamma[\phi^*,\phi]$ \\
\hline
  $h_i = \frac{\partial \Gamma}{\partial \bar{m}_i}$ & $j(X) = \frac{\delta \Gamma}{\delta \phi^*(X)}$ \\
\hline
\end{tabular}
\end{center}
\caption{Correspondence of quantities for classical Ising magnets and cold atoms. The magnet case on a lattice is discussed in \ref{AppIsing}.}
\label{TabDictionary}
\end{table}

The bosonic functional integral in Eq. (\ref{fun1}) allows for the
definition of a probability measure on the set of fields
$\vphi$. Given an observable $\mathcal{O}(\{\vphi\})$ which depends on
the field, we define
\begin{equation}
 \label{Func6}\langle \mathcal{O} \rangle = 
 \frac{1}{Z} \int \mbox{D}\,\vphi^*\mbox{D}\vphi\, \mathcal{O}(\{\vphi\}) e^{-S[\vphi^*,\vphi]}.
\end{equation}
Herein, the action $S$ acts as a weight. For instance, from
$\mathcal{O} = \vphi(X)$ or $\mathcal{O} = \vphi^*(X) \vphi(Y)$ we
obtain the one- and (disconnected) two-point correlation functions of
the theory. More generally, we obtain averages of observables by
introducing a complex source field $j(X)$ according to
\begin{equation}
  \label{eff16} Z[j^*,j] = \int \mbox{D} \varphi^* \,\mbox{D} \varphi 
  \,e^{-S[\varphi^*,\varphi] + \int_X j^*_X\varphi_X + \int_X \varphi^*_X j_X}.
\end{equation}
We call $Z[j^*,j]$ the \emph{generating functional} and have
\begin{align}
 \label{Func7a} &\phi(X) = \langle \vphi(X)\rangle_j = \frac{\delta \log Z}{\delta j^*(X)},\\
 \label{Func7b} &\langle \vphi^*(X)\vphi(Y) \rangle_j= \frac{1}{Z} \frac{\delta^2 Z}{\delta j(X) \delta j^*(Y)},
\end{align}
etc. The subscript $j$ indicates that the external source is not yet
set to zero. Generalizing Eqs. (\ref{Func7a}) and (\ref{Func7b}), we
find the representation of a general expectation value
\begin{equation}
  \label{Func8} \langle \mathcal{O} \rangle_j = \frac{1}{Z} \mathcal{O}\Bigl(
\Bigl\{\frac{\delta}{\delta j}\Bigr\}\Bigr) Z[j^*,j].
\end{equation}
We conclude that all correlation functions of interest can be obtained
from the generating functional $Z$ or
\begin{equation}
 \label{Func9} W[j^*,j] = \log Z[j^*,j].
\end{equation}
The latter quantity is called the Schwinger functional. It constitutes
the generating functional of connected $n$-point functions. For
instance, we find for the connected two-point function
\begin{align}
 \nonumber W^{(2)}[j](X,Y) &= \frac{\delta^2 W}{\delta j^*(X)\delta j(Y)}\\
 \nonumber &= \langle \vphi(X)\vphi^*(Y)\rangle_{j} - \phi(X)\phi^*(Y)  \\
 \label{eq:w2}&= \langle \vphi(X)\vphi^*(Y)\rangle_{j,\rm c},
\end{align}
where the index refers to ``connected''. This object is also called
(time ordered) \emph{Green's function} or \emph{propagator} of the
theory. Imposing the time-ordering for the propagators leads to
time-ordered general correlation functions \eq{Func8}.

The field expectation value carries a direct physical
significance. For example, in a homogeneous situation, $\phi(X) =
\phi_0$ describes the Bose--Einstein condensate.  It therefore seems
desirable to implement it into the theory in a more direct way. In
fact, by the aid of a Legendre transformation, this is possible and
gives rise to the \emph{effective action}, see e.g.\ \cite{amit-book}.  Similar
to our considerations for the Ising magnets in \ref{AppIsing}, we
introduce the latter as the generating functional which depends on the
\emph{mean} or \emph{classical field} $\phi$ defined by
\begin{equation}
 \label{eff19} \phi(X) = \langle \varphi(X) \rangle_j = \frac{\delta W}{\delta j^*(X)}.
\end{equation}
Assume we have solved this equation. We can then construct the
effective action according to the Legendre transformation
\begin{equation}
 \label{eff20} \Gamma[\phi^*,\phi] = \int_X (\phi^*_X j_X + j^*_X \phi_X) - W[j^*,j],
\end{equation}
where $j$ and $j^*$ are defined implicitly by Eq. (\ref{eff19}). Note
that while the active variable for the partition function is the
source, $Z= Z[j]$, the active variable for the effective action is the
field expectation value, $\Gamma = \Gamma[\phi]$. The effective action
is thus parametrized directly in terms of a physical
observable. Applying the chain rule for functional differentiation we
find
\begin{equation}
 \label{eff21} \frac{\delta \Gamma}{\delta \phi(X)}[\phi^*,\phi] = j^*(X).
\end{equation}
For the derivation of this relation in the case of discrete variables,
see Eq. (\ref{eff6}).

Technically speaking, the effective action is the generating
functional of \emph{one-particle irreducible (1PI) correlation
  functions}. They can be obtained from $\Gamma$ by taking successive
functional derivatives with respect to $\phi(X)$ and
$\phi^*(X)$. Diagrammatically the 1PI correlation functions are
  given by all diagrams that cannot be split by cutting one (internal)
  line, hence the name.  Physically, such inhomogeneous mean fields
can be obtained by applying external sources, namely by choosing $j$
such that $\delta \Gamma/\delta \phi=j$. Often, we are mainly
interested in the situation of vanishing source. Then, given the
effective action, we have to solve the \emph{equations of motion}
\begin{equation}
 \label{eff22} \frac{\delta \Gamma}{\delta \phi(X)}[\phi^*_0,\phi_0] =0
\end{equation}
to obtain the thermodynamic equilibrium state $\phi_0 = \langle
\vphi\rangle_{j=0}$ of the theory. The reference to the external field
is no longer present and also not needed, because it is already
included in $\Gamma[\phi^*,\phi]$. Typically, the solution $\phi_0$ to
Eq. (\ref{eff22}) is constant in space-time. This will be explained in
more detail below Eq. (\ref{SSB2}). However, in general there also
might be inhomogeneous solutions $\phi_0$ to the nonlinear partial
differential equation Eq. (\ref{eff22}), such as instantons, solitons
or vortices (see Sec.  \ref{SecSSB}). 

Higher derivatives of the effective action with respect to the fields
$\phi,\phi^*$, denoted with $\Gamma^{(n)}$ for the $n$th
derivative, provide the one-particle-irreducible vertices.  The second derivative of the 
effective action, 
\begin{equation}
 \label{eff22b} \Gamma^{(2)}(X,Y) = \frac{\delta^2 \Gamma}{\delta \phi^*(X) \delta \phi(Y)},
\end{equation}
plays a special r$\hat{\rm o}$le, as it is the
\emph{inverse propagator}. This is easily proven by 
\begin{eqnarray}\nonumber 
&& \hspace{-1cm}  \int_Z \Gamma^{(2)}(X,Z)  W^{(2)}(Z,Y)\\[1ex]\nonumber 
&&=\int_Z\left( \frac{\delta j^*(Z)}{\delta \phi^*(X)}  \frac{\delta \phi^*(Y)}{\delta j^*(Z)}+ \frac{\delta j(Z)}{\delta \phi^*(X)}
  \frac{\delta \phi^*(Y)}{\delta j(Z)}\right)\\[1ex]
&&=  \delta (X-Y)\,,
\label{eff22c}\end{eqnarray}
where we have used \eq{eff19},\eq{eff21} and the completeness relation
  of derivatives with respect to $j\,, j^*$. 

In principle,
  Eq. (\ref{eff21}) can be taken as a starting point to calculate the
  effective action $\Gamma[\phi^*,\phi]$ in certain
  approximations. However, the definition of $\Gamma$ implies an exact
  identity, which is equivalent to Eq. (\ref{eff21}), but more
  useful. Applying Eqs. (\ref{eff20}), (\ref{eff21}) and $W = \log Z$,
  we arrive at
\begin{align}
  \nonumber &e^{-\Gamma[\phi^*,\phi]} = e^{-\int_X(j^*\phi + \phi^* j) + W} \\
  \nonumber &= e^{-\int_X(j^* \phi+\phi^* j)}  \int \mbox{D}\varphi^*\mbox{D} \varphi\, e^{-S[\varphi^*,\varphi] + \int_X(j^* \varphi+\varphi^* j)}\\
  \nonumber &= \int \mbox{D}\varphi^*\,\mbox{D} \varphi \,e^{-S[\varphi^*,\varphi] + \int_X(j^* (\varphi-\phi)+(\varphi^*-\phi^*) j)}\\
 \label{eff23} &=\int \mbox{D}\varphi^*\,\mbox{D} \varphi \,e^{-S[\varphi^*,\varphi] + \int_X(\frac{\delta \Gamma}{\delta \phi}[\phi]\cdot (\varphi-\phi)+(\varphi^*-\phi^*)\cdot \frac{\delta \Gamma}{\delta \phi^*}[\phi])}.
\end{align}
This equation is called the \emph{background field identity} for the
effective action. For $\phi=\phi_0$ with $\phi_0$ satisfying
Eq. (\ref{eff22}), we recover
\begin{equation}
 \label{Func13b} Z(\mu,T) = e^{-\Gamma[\phi_0^*,\phi_0]},
\end{equation}
i.e. the effective action then corresponds to the grand canonical
potential. Furthermore, by performing a shift of the integration
variable, we rewrite Eq. (\ref{eff23}) as
\begin{align}
 \nonumber &e^{-\Gamma[\phi^*,\phi]} \\
 \label{Func18} &= \int \mbox{D}\delta\varphi^*\mbox{D} \delta\varphi
 e^{-S[\phi^*+\delta \vphi^*,\phi+\delta\vphi] + \int_X(\frac{\delta
     \Gamma}{\delta \phi}[\phi]\cdot \delta \vphi+\delta \vphi^*\cdot
   \frac{\delta \Gamma}{\delta \phi^*}[\phi])}.
\end{align}
This \emph{functional integral representation of the effective action}
gives rise to the intuitive picture that the effective action encodes
the complete information on the euclidean field theory by means of
summing over all possible field configurations $\delta\vphi$ deviating
from the classical one, $\phi$.

We now show that in the \emph{classical limit}, the effective action
and the classical action coincide. Reintroducing Planck's constant
$\hbar$, we have $\Gamma/\hbar$ and $S/\hbar$ appearing in
Eq. (\ref{Func18}). The classical limit is obtained for $\hbar \rightarrow
0$ at fixed $\Gamma$.  The integrand is then sharply peaked around the
solution to the classical equations of motion $\delta S/\delta \vphi
=0$. This results in $\Gamma = S$, which physically is the
\emph{classical
  approximation}.

It is clear that the effective action lends itself ideally for
semiclassical approximations, and also systematic improvements
thereon. From Eqs. (\ref{eff23}) or \eq{Func18}, we can
easily go one step beyond the classical approximation, by expanding
the exponent in the functional integral around its minimum value
$\vphi_0$ determined by
\begin{equation}
  \label{Func14} - \frac{\delta S}{\delta \vphi^*}[\vphi_0] + 
  \frac{\delta \Gamma}{\delta \phi^*}[\phi] = 0 = - \frac{\delta S}{\delta \vphi}[\vphi_0] 
  + \frac{\delta \Gamma}{\delta \phi}[\phi].
\end{equation}
For the particularly simple case where $\phi = \vphi_0$, the linear
derivatives cancel and we obtain
\begin{equation}
  \label{Func17} e^{-\Gamma[\vphi_0]} \simeq e^{-S[\vphi_0]}\int \mbox{D}\vphi^*
  \mbox{D}\vphi e^{-\frac{1}{2}\int (\vphi,\vphi^*)\cdot S^{(2)}[\vphi_0]\cdot \binom{\vphi}{\vphi^*}}
\end{equation}
with $S^{(2)}$ the second functional derivative of $S$ with respect to
$\vphi,\vphi^*$. 

More generally, with the help of formula (\ref{ising5}), the Gaussian approximation to Eq. (\ref{Func18}) can be evaluated at any field $\phi$ which ensures the path integral to be dominated by small fluctuations $\delta \vphi$. This leads to the so-called \emph{one-loop formula}
\begin{equation}
 \label{Func19} \Gamma[\phi^*,\phi] \simeq S[\phi^*,\phi] + \frac{1}{2} \mbox{Tr} \log S^{(2)}[\phi^*,\phi].
\end{equation}
In this order of approximation, the linear derivative terms cancel due
to the tree level relation $\Gamma \simeq S$. Note that the effective
action equals the classical action also in the case of a free,
noninteracting theory. Expanding the Tr log expression in powers of
the field, we generate one-loop perturbation theory. We may therefore
expect Eq. (\ref{Func19}) to give good results in the perturbative
regime of small coupling.

Our considerations can easily be extended to fermions as well. We
introduce independent Grassmannian source terms $\eta(X)$ and
$\eta^*(X)$ into the generating functional $Z[j^*,j,\eta^*,\eta]$,
which couple linearly to the fields $\psi^*(X)$ and $\psi(X)$,
respectively. The effective action is defined in the same manner as
before via the Legendre transformation of $\log Z$ with respect to the
mean fields. The ground state of the theory necessarily satisfies
$\langle \psi(X) \rangle_{\eta=0} = \langle \psi^*(X)
\rangle_{\eta=0}=0$, since Pauli's principle forbids macroscopic
occupation of fermionic states. However, the generating functional
$\Gamma$ depends on nonvanishing fermionic ``mean fields''. Such
fields $\bar{\psi}(X)$ can be constructed by applying a source
$\eta(X)=\delta \Gamma/\delta\bar{\psi}(X)$. They must not be regarded
as physical objects, but rather as bookkeeping parameters used to
generate the 1PI correlation functions via Grassmannian functional
differentiation.

Eq. (\ref{Func19}) is also valid for fermionic fields, but with an
additional minus sign in front of the trace. For a mixed theory of
both bosons and fermions we introduce the so-called \emph{supertrace},
$\mbox{STr}$, which takes into account this sign for fermionic
terms. Thus, we arrive at the one-loop formula
\begin{equation}
 \label{Func20} \Gamma[\phi,\bar{\psi}] \simeq S[\phi,\bar{\psi}] 
+ \frac{1}{2} \mbox{STr} \log S^{(2)}[\phi,\bar{\psi}].
\end{equation}

More detailed presentations on functional integrals can be found in
\cite{negele-book,altland-book}.

\subsection{Effective potential and spontaneous symmetry breaking}
\label{SecSSB}
\noindent In this section, we discuss how phase transitions and
spontaneous symmetry breaking (SSB) find their natural description in terms
of the effective potential $U(\rho)$. The latter is the part of the
effective action which does not contain derivatives of the field. It
includes both quantum and thermal fluctuations, and typically changes
its shape by tuning the system parameters like temperature, chemical
potential or interaction strength. For parameter regions where the
minimum of the effective potential is nonzero, small perturbations can
drive the system into an equilibrium ground state which does not
respect the symmetry of the underlying physical theory. The symmetry
is \emph{spontaneously} broken. We exemplify this important concept of
many-body physics on systems with $\mathbb{Z}_2$- and ${\rm
  U}(1)$-symmetry, respectively.

An intuitive picture of SSB is provided by a
simple daily life observation. Suppose a pencil is balanced on its tip
to stand upright. Due to the cylindrical symmetry, the pencil should
stay in this position. Indeed, the underlying physics, here given by
the gravitational force pointing downwards, does not prefer any
direction. However, if there is a small perturbation of this symmetry
due to the environment, the pencil will immediately fall to the side
and thereby minimize its energy. Even if the perturbation is removed
now, the pencil will remain in the horizontal position.

\vspace{5mm}
\begin{center}
 \emph{Thermodynamics from the effective action}
\end{center}

\noindent In order to study the properties of the thermodynamic
equilibrium state, we consider a system of bosons. We assume the
trapping potential $V_{\rm ext}(\vec{x})$ to vanish and the external
source to be constant, $j(X)=j$. Hence, the setting is homogeneous in
space-time. We learned in Eq. (\ref{Func13b}) that the grand canonical
partition function $Z(\mu,T)$ is related to the effective action
according to
\begin{equation}
 \label{SSB1} \Gamma[\phi_0] = - \log Z(\mu,T).
\end{equation}
Herein, the field expectation value $\phi_0(X)$ minimizes the
effective action, as can be seen from Eqs. (\ref{eff20}) and
(\ref{eff22}). The effective action $\Gamma[\phi]$ has the structure
\begin{equation}
  \label{SSB2} \Gamma[\phi] = \int_{X} \Bigl(\text{terms containing derivatives} 
  \Bigr) + \int_{X} U(\phi(X))\,. 
\end{equation} 
If the part containing derivatives can be expanded in orders of the
derivatives, it is necessarily non-negative for the sake of
stability. Otherwise, $\Gamma[\phi]$ would not possess a minimum,
because we could arbitrarily decrease its value by creating heavily
oscillating fields. Then, $\phi_0$ is a constant field
which additionally minimizes the \emph{effective potential} $U(\phi)$
according to
\begin{equation}
 \label{SSB3} U(\phi_0) = \min_{\phi}\Bigl[U(\phi)\Bigr].
\end{equation}
Since the effective potential depends on both the external parameters
$\mu$ and $T$, the same will be true for the field expectation value
$\phi_0=\phi_0(\mu,T)$. In the presence of a nonvanishing background
source field, we also have an explicit dependence on $j$. Note that
the above argument does not exclude the existence of inhomogeneous
ground states as they cannot be expanded in terms of derivatives.  The
existence of such inhomogeneous ground states is common in low
dimensions, in particular in 1+1 dimensions, see e.g.\
\cite{Thies:2006ti} for the class of models under discussion here.

Using Eq. \eqref{Func13b}, the effective potential at its minimum
value is related to the pressure according to
\begin{equation}
 \label{SSB4} P(\mu,T) = - U(\phi_0,\mu,T).
\end{equation}
This constitutes the \emph{equation of state} of the system. Often, we
are mainly interested in the density $n(\mu,T)$, which is found from
$\mbox{d}P=n\mbox{d}\mu+s\mbox{d}T$. The relevant thermodynamic
information contained in the effective potential can thus be
summarized in the two equations
\begin{align}
 \label{SSB5b} \frac{\partial U}{\partial \phi}(\phi_0,\mu,T) &= 0 \hspace{5mm} (\text{gap equation}),\\
 \label{SSB5}  \frac{\partial U}{\partial \mu}(\phi_0,\mu,T) &= - n \hspace{5mm} (\text{equation of state}).
\end{align}
These equations are generally valid and constitute the main building
blocks for the evaluation of the phase diagram of the many-body
problem. In particular, the above discussion is not limited to bosons,
but can be applied to an arbitrary many-body system or quantum field
theory, since the effective action approach is applicable to all of
these system. For instance, as outlined earlier, the field $\vphi(X)$
may as well describe the degrees of freedom in a Heisenberg
ferromagnet with magnetic moments $\vec{m}_i$ and $\vec{m}(\vec{x})$
on a lattice or in the continuum, respectively. However, we keep
denoting the fields by $\vphi$ and $\phi$, which may have to be
replaced appropriately.

\vspace{5mm}
\begin{center}
 \emph{Spontaneous symmetry breaking}
\end{center}

\noindent As a preparation for the more formal discussion of SSB, we first relate symmetries of the microscopic action to those of the effective action. To this end, we recall the definition of the effective action to be
\begin{equation}
 \label{SSB6} e^{-\Gamma[\phi]} = \int \mbox{D}\vphi \,e^{-S[\vphi] + \int_X j[\phi]\cdot(\vphi-\phi)}.
\end{equation}
Setting the external source $j$ to zero, we see that any symmetry of the microscopic action which is respected by the functional measure, will also be a symmetry of the effective action. A nonvanishing source $j(X)$, instead, typically leads to terms in the effective action which \emph{explicitly} break the microscopic symmetry. This is accompanied by a nonzero expectation value $\phi(X)$, because $j(X)$ either introduces a nonhomogeneity in space-time or at least singles out a direction in field space $\vphi$.

\emph{Spontaneous} breaking of a symmetry refers to a different scenario. In this case, the external source vanishes such that the effective action manifestly shares the symmetry of the microscopic action. Nevertheless, the ground state of the theory (or, more generally, the thermodynamic equilibrium state), may spontaneously break this symmetry due to a nonzero expectation value according to
\begin{equation}
\phi_0=\langle \vphi \rangle_{j\to0} \neq 0.
\end{equation} 
The symmetry is then broken because the field expectation value transforms nontrivially under the symmetry transformation. For a more detailed discussion on the interplay between the thermodynamic limit and the limit $j \rightarrow 0$, we refer to \ref{AppIsing}.

We illustrate this discussion with examples. First, we consider classical Ising magnets on a lattice. The symmetry transformation exerted on the Ising variables $m_i$ is a global reflection, $m_i \to -m_i$ for all $i$. The Hamiltonian $ H[m] = - J \sum_{i} m_{i}m_{i+1}$ is reflection symmetric, meaning that 
\begin{equation}
 \label{SSB7} H[m]=H[-m].
\end{equation}
Since the functional measure $\int \prod_i \mbox{d} m_i \delta(m_i^2-1)$ does not break this symmetry, we have for the effective action
\begin{equation}
 \label{SSB8} \Gamma[\bar{m}] = \Gamma[-\bar{m}].
\end{equation}
We call this a $\mathbb{Z}_2$-symmetry.

Analogously, the microscopic action of cold atomic bosons given in Eq. (\ref{Func4}) has a global ${\rm U}(1)$-symmetry, meaning that it is invariant under the following global  transformation of the fields,
\begin{equation}
 \label{SSB9} \vphi \rightarrow \vphi' = e^{\rmi \alpha} \vphi, \hspace{5mm} \vphi^* \rightarrow (\vphi^*)'=e^{-\rmi \alpha} \vphi^*,
\end{equation}
with real parameter $\alpha$. In the basis of real fields, $\vphi = \vphi_1 + \rmi \vphi_2$, this corresponds to a rotation
\begin{equation}
 \label{SSB10} \left(\begin{array}{c} \vphi_1'(\vec{x})\\ \vphi_2'(\vec{x})\end{array}\right) = \left(\begin{array}{cc} \cos \alpha & - \sin \alpha \\ \sin \alpha & \cos \alpha \end{array}\right)\left(\begin{array}{c} \vphi_1(\vec{x})\\ \vphi_2(\vec{x})\end{array}\right)
\end{equation}
in field space. Since the functional measure $\int\mbox{D}\vphi^*\mbox{D}\vphi$ shares this symmetry, the effective action $\Gamma[\phi^*,\phi]$  possesses the global ${\rm U}(1)$-symmetry as well. 

By virtue of Noether's theorem, the global ${\rm U}(1)$-symmetry in conjunction with a linearly appearing time derivative in the kinetic term of the microscopic action, leads to the conservation of total particle number $N=\int_{\vec{x}} \langle \vphi^*(\vec{x})\vphi(\vec{x})\rangle$. This is a characteristic feature of nonrelativistic field theories. A brief review of Noether's theorem in the classical and quantum case is given in \ref{SymmetriesEffAct}.

The above mentioned properties of the effective action for vanishing external source have a profound consequence for the effective potential $U$. Indeed, from Eq. (\ref{SSB2}) we deduce that the latter is not a function of $\phi, \phi^*$ alone, but we rather have
\begin{equation}
 \label{SSB11} U = U(\rho),
\end{equation}
where $\rho$ is the most general combination of fields allowed by symmetry. For instance, we have
\begin{equation}
 \label{SSB12} \rho = \begin{cases} \bar{m}^2 & (\mathbb{Z}_2-\text{symmetry}),\\ \phi^*\phi & ({\rm U}(1)-\text{symmetry}).\end{cases}
\end{equation}

\begin{figure}[tb!]
 \centering
 \includegraphics[scale=0.85,keepaspectratio=true]{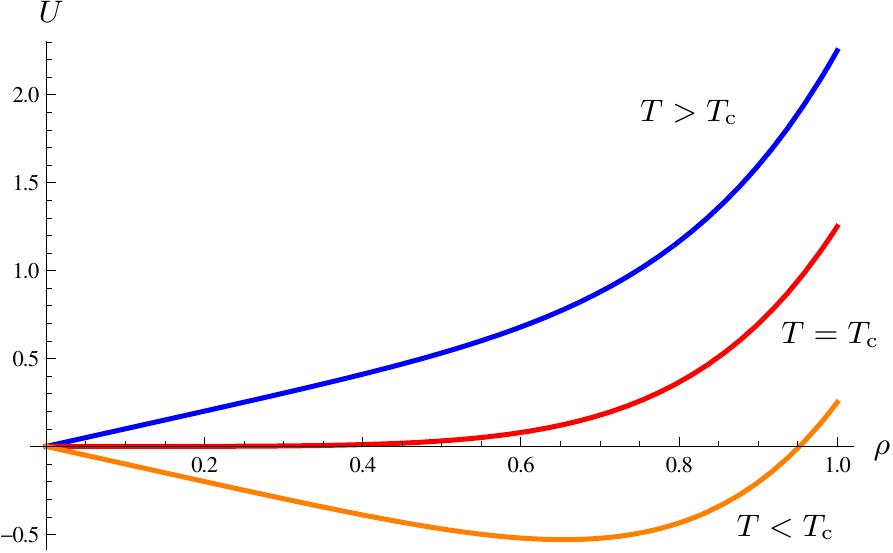}
\caption{The effective potential $U(\rho)$ for vanishing external sources is a function of the symmetry invariant $\rho$. The latter is given by $\rho = \bar{m}^2$ or $\rho=|\phi|^2$ for magnets or ultracold bosons, respectively. Throughout a second order phase transition, the location of the minimum of the effective potential changes from $\rho_0=0$ to $\rho_0>0$ in a continuous manner. We have chosen here the temperature to be the control parameter. However, since the effective potential depends on $\mu$, $T$, and the microscopic parameters of the theory (e.g. coupling constants), we may also drive the phase transition differently.}
\label{EffPot2nd}
\end{figure}

\begin{figure}[tb!]
 \centering
 \includegraphics[scale=0.85,keepaspectratio=true]{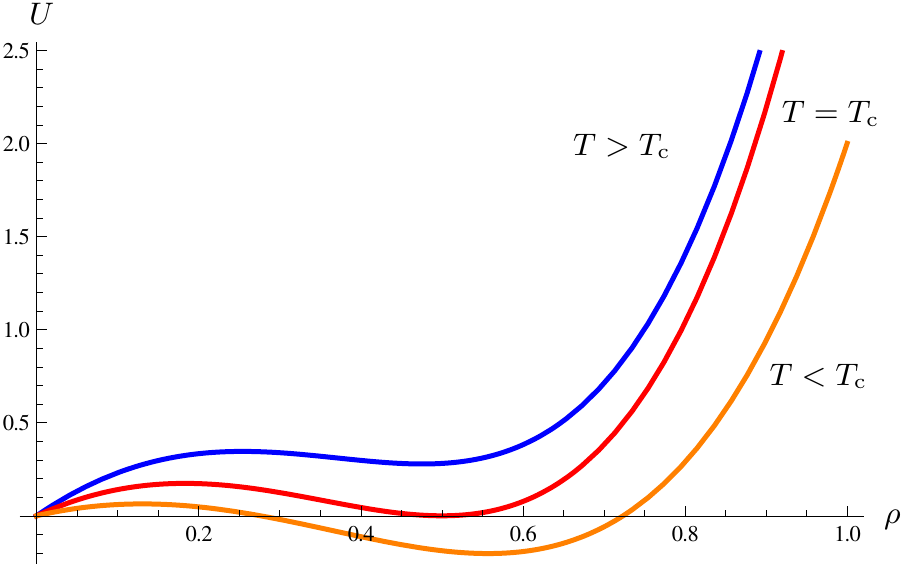}
\caption{In a first order phase transition, we have a jump in the order parameter $\rho_0$ as we cross $T=T_{\rm c}$. From the plot of the effective potential we see how this discontinuous behavior can arise, although we smoothly vary the system parameters. Note that the effective potential is actual a convex function, as the effective action originates from the Legendre transform of the convex Schwinger functional. The non-convex parts should therefore be replaced by straight lines according to the Maxwell construction, but this does not invalidate the overall picture of first order phase transitions.}
\label{EffPot1st}
\end{figure}

We plot the effective potential $U(\rho)$ for a second and first order phase transition in Figs. \ref{EffPot2nd} and \ref{EffPot1st}, respectively. The critical temperature $T_{\rm c}(\mu)$ is defined such that the location of the minimum $\rho_0(\mu,T)$ becomes zero -- either continuously or discontinuously. In particular, for a second order phase transition we distinguish the following three cases:
\begin{itemize}
 \item[(i)] $\rho_0 \neq 0$, $U'(\rho_0)=0$: phase with broken symmetry,
 \item[(ii)] $\rho_0=0$, $U'(\rho_0) =0$: critical point,
 \item[(iii)] $\rho_0=0$, $U'(\rho_0) \neq 0$: symmetric phase.
\end{itemize}

In the broken phase, the location of the minimum $\rho_0$ of the effective potential does not necessarily completely determine the thermodynamic equilibrium state of the theory. In the case of magnets, we have $\bar{m}_0^2 \neq 0$ and thus there is still the freedom to choose the sign of $\bar{m}_0$, which is a $\mathbb{Z}_2$-transformation. For the case of bosons, the condition $|\phi_0|^2 \neq 0$ only fixes the amplitude of the complex field $\phi_0 = |\phi_0|e^{\rmi\theta}$, whereas the phase $\theta$ can still be chosen arbitrarily. The possible nonequivalent choices are given by $\theta \in [0,2\pi) \simeq {\rm U}(1)$. We say that the condition on $\rho_0$ singles out a manifold of possible ground states $\phi_0$, which in our examples is given by $\mathbb{Z}_2$ and ${\rm U}(1)$, respectively. In the absence of explicit symmetry breaking terms, the precise choice of the ground state in the degenerate manifold indeed happens \emph{spontaneously} -- it is induced by fluctuations or perturbations due to the environment, which we can neither resolve nor control \cite{Nambu60}. Nevertheless, this phenomenon is ubiquitously observed experimentally; for example, spontaneous phase symmetry breaking can be detected in interference experiments of initially disjunct condensates \cite{andrews97}. 

\begin{figure}[tb!]
 \centering
 \includegraphics[scale=0.85,keepaspectratio=true]{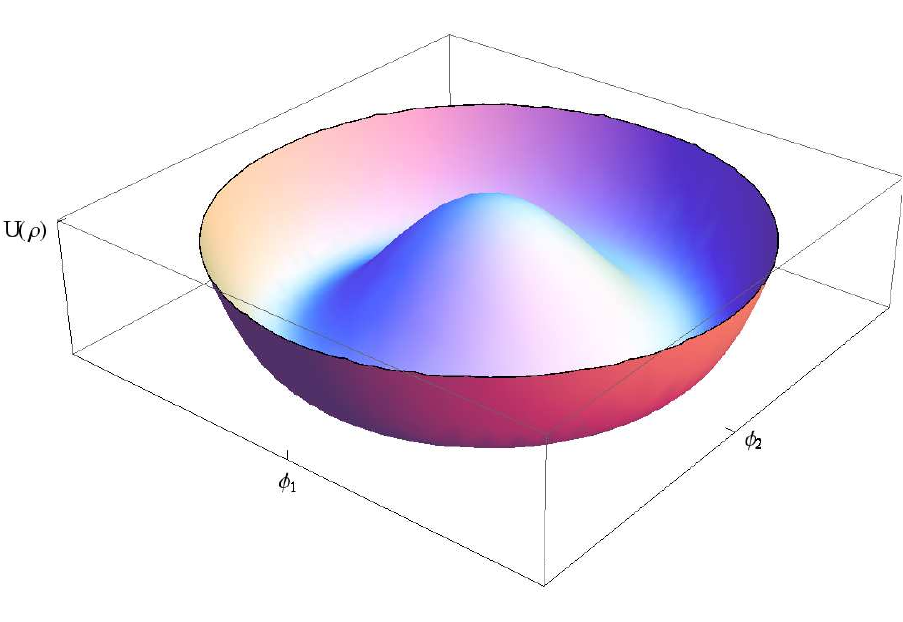}
\caption{The Mexican hat potential from Eq. (\ref{SSB13}) only depends on the amplitude of the complex field $\phi=\phi_1 + \rmi \phi_2$. Thus, it reflects the $\rm{U(1)}$-symmetry of the bosonic theory, which is invariant under phase rotations $\phi \rightarrow e^{\rmi \alpha}\phi$. The ground state of the system will, however, spontaneously break this symmetry, e.g. by choosing $\phi_0 \in \mathbb{R}$.}
\label{MexHat}
\end{figure}

In Fig. \ref{MexHat}, we plot the boson effective action in the tree level approximation $\Gamma[\phi]\simeq S[\phi]$ for a constant field in the complex $\phi$-plane. The microscopic action $S$ is given in Eq. (\ref{Func4}). We write 
\begin{equation}
 \label{SSB13} \frac{1}{\beta V}\Gamma[\phi={\rm const.}]=U(\phi) = - \mu |\phi|^2 + \frac{g}{2} |\phi|^4.
\end{equation}
For obvious reasons, $U(\phi)$ is also called the \emph{Mexican hat potential}. Without loss of generality we assume the ground state $\phi_0$ to be real, such that real and imaginary components of $\phi=\phi_1 + \rmi \phi_2$ are distinct directions in the complex plane. The ground state singles out the point $(\phi_0,0)$. Now consider the field $\vphi$ to be fluctuating around this point. As usual, we write
\begin{equation}
 \label{SSB14} \vphi(\tau,\vec{x}) = \phi_0 + \delta \vphi(\tau,\vec{x})
\end{equation}
with $\langle \delta\vphi \rangle =0$. The fluctuations $\delta \vphi$ are complex and can vary both amplitude and phase of $\vphi$. However, the fluctuations which increase the amplitude away from $\phi_0$ have to climb up the hill and thus are energetically unfavorable, i.e. they are suppressed in the functional integral by a term 
\begin{equation}
 \label{SSB15}\int \mbox{D} \vphi_1 e^{-m^2_1\delta \vphi_1^2 }.
\end{equation}
We call them radial or gapped excitations. In contrast, fluctuations of the phase are not hindered energetically, because they are along the well of the Mexican hat.

The existence of a massless or gapless mode in a symmetry broken phase observed in the example above is a general phenomenon. In fact, it is an exact property of the full theory, as has been established by Goldstone \cite{goldstone61}. More precisely, Goldstone's theorem states that any spontaneous breaking of a continuous symmetry results in the appearance of gapless modes in the excitation spectrum of the system. The proof of Goldstone's theorem is very simple in the effective action framework. Since we are interested in a statement about the masses of the theory, i.e. properties of the system in the homogeneous limit of vanishing frequencies and momenta, we can restrict ourselves to the effective potential $U$. As we have seen above, $U(\rho)$ only depends on the symmetry invariant $\rho = |\phi|^2$. Consider the field equation $\frac{\delta \Gamma}{\delta \phi_1}=0$ for the radial field. Since we assumed a homogeneous setting, this reduces to
\begin{equation}
 \label{SSB16} 0 = \frac{\partial U}{\partial \phi_1}(\phi_0) = \phi_0 U'(\rho_0) \stackrel{{\rm SSB}}{\Longrightarrow} U'(\rho_0)=0.
\end{equation}
Here, a prime denotes differentiation with respect to $\rho$. The mass of the Goldstone mode $\delta \vphi_2$ is then found to be
\begin{align}
 \nonumber m^2_{\phi_2} &= \frac{\partial^2 U}{\partial \phi_2^2}(\phi_0) = \left.\biggl( \frac{\partial^2\rho}{\partial \phi_2^2} U'(\rho) + \Bigl(\frac{\partial^2\rho}{\partial\phi_2}\Bigr)^2U''(\rho)\biggr)\right|_{\phi=\phi_0}\\
 \label{SSB17} &= \left.\Bigl(U'(\rho) + \phi_2^2 U''(\rho)\Bigr)\right|_{\phi=\phi_0} = U'(\rho_0) =0.
\end{align}
We used the continuity of the symmetry by requiring $\rho$ to depend smoothly on $\phi_{1,2}$. We have carried out the proof for the symmetry ${\rm U}(1) \simeq {\rm O}(2)$. The above steps can be performed analogously for larger symmetry groups such as ${\rm O}(N)$, leading to $N-1$ massless Goldstone modes. In the above case, the vanishing of the mass term allows for strong fluctuations of the phase field in the phase of broken symmetry. In particular, they question the assumption of small fluctuations $\delta \vphi$ in the functional integral, which lead to the one-loop formula given in Eq. (\ref{Func19}).

\subsection{Condensation of weakly interacting bosons}
\label{SecWeakBos}
\noindent 
\noindent We compute the properties of a weakly interacting Bose gas within the formalism introduced in the previous section. Mean field theory and the Gaussian approximation, which is equivalent to the Bogoliubov theory in a second quantized formulation, already capture many effects which are relevant for experiment. For reasons of stability, a Bose gas usually has to be dilute or weakly interacting such that these simple approximations work well. However, we show below that there are situations, where the Bose gas has to be treated by more sophisticated methods.

We start from the microscopic action of the weakly interacting Bose gas given in Eq. (\ref{Func4}),
\begin{align}
 \nonumber S&[\varphi^*,\varphi] = \int_0^\beta \mbox{d}\tau\int_{\vec{x}} \left\{ \varphi^*(\tau,\vec{x})\left(\partial_\tau - \frac{\nabla^2}{2M}  - \mu\right) \varphi(\tau,\vec{x})\right.\\
 \label{bog2} &+V_{\rm ext}(\vec{x}) \vphi^*(\tau,\vec{x})\vphi(\tau,\vec{x}) + \frac{g}{2} (\varphi^*(\tau,\vec{x})\varphi(\tau,\vec{x}))^2 \biggr\}
\end{align}
with $g > 0$. We recall that $\varphi$ is a complex field. As discussed in Eq. (\ref{SSB9}), the microscopic action has a global ${\rm U}(1)$-symmetry.

The goal of this section is to calculate the corresponding effective action for this problem. Due to the interaction term of fourth order in the field, the functional integral cannot be calculated in a straightforward manner and we rather have to rely on some approximative method. At sufficiently low temperatures ($\ell/\lambda_T \ll 1$), we can utilize the fact that the condensate is occupied by a macroscopic number of particles, $\phi_0 \sim \sqrt{N}$. This scaling of the field expectation value had been found for noninteracting bosons in Sec. \ref{ideal} for a sufficiently high spatial dimension. It is reasonable to assume that this scaling is also valid for the case of weak interactions. This identifies an \emph{ordering principle}, which will justify our subsequent approximations.

To understand the mechanism underlying an ordering principle of this type, we study a simple toy model: Consider a smooth, real-valued function $f(x)$ which has its minimum at $x_0$. Our goal is to evaluate the integral
\begin{equation}
 \label{bog2b} I = \int_{-\infty}^\infty \mbox{d} x e^{-N f(x)},
\end{equation}
where $N$ is large. Expanding the exponent around the minimum of $f$, we have $f'(x_0)=0$ and $f''(x_0) >0$, leading to
\begin{equation}
 \label{bog2c} I = e^{-N f(x_0)} \int_{-\infty}^\infty \mbox{d} x e^{-N \frac{1}{2} f''(x_0)(x-x_0)^2 + N O(|x-x_0|^3)}.
\end{equation}
However, performing a variable transformation $x'=\sqrt{N}x$ we find
\begin{align}
 \nonumber I= & e^{-Nf(x_0)} \int_{-\infty}^\infty\frac{\mbox{d} x'}{\sqrt{N}}  e^{-\frac{1}{2}f''(x_0)(x'-x_0')^2 + \frac{C}{\sqrt{N}}(x'-x'_0)^3+\dots}\\
 \label{bog2d} \simeq &\mbox{ } e^{-N f(x_0)} \sqrt{2\pi/Nf''(x_0)},
\end{align}
where $C$ is some constant. Here, ``$\simeq$`` means that the
correction term vanishes in the limit $N \rightarrow \infty$. If we
are only interested in $\frac{1}{N} \log I$, we can neglect the square
root term in Eq. (\ref{bog2d}). Note that we did not require the
minimum of $f(x)$ at $x_0$ to be strongly expressed, because this is
ensured by a sufficiently large $N$. This basic mechanism underlies
all large $N$ expansion strategies: after identification of a suitable
parameter $N$, the remaining functional integral becomes Gaussian,
just as in our one-dimensional toy model.

In the context of weakly interacting bosons, we want to compute the
effective action $\Gamma[\phi^*,\phi]$. We start from
Eq. (\ref{Func18}), which we write in condensed notation as
\begin{equation}
  \label{bog2e} e^{-\Gamma[\phi]} = \int \mbox{D} \delta \vphi 
  e^{-S[\phi+\delta \vphi]+\int_X \frac{\delta \Gamma}{\delta \phi}[\phi] \cdot \delta \vphi},
\end{equation}
where $\phi(\tau,\vec{x})$ is a complex field vector. Expanding the
exponent around $\phi$, the linear terms cancel and we arrive at
\begin{equation}
  \label{bog2f} e^{-\Gamma[\phi]} = \int \mbox{D} \delta \vphi 
  e^{-S[\phi]-\frac{1}{2}\int_{X,Y} \delta \vphi \cdot S^{(2)}[\phi] \cdot \delta \vphi +\dots}.
\end{equation}
Now, if the expansion point obeys $\phi=\sqrt{N} \phi'$, where
$\phi'=O(N^0)$, we find $S^{(2)}[\phi]=NS'^{(2)}[\phi']$, because the
microscopic action only contains terms up to fourth order in the
field. Accordingly, we have
\begin{equation}
  \label{bog2g} e^{-\Gamma[\phi]} = e^{-S[\phi]} \int \frac{\mbox{D}
    \delta\vphi'}{\sqrt{N}} e^{-\frac{1}{2} \delta \vphi'\cdot S'^{(2)}\cdot \delta \vphi' + O(N^{-1/2})}.
\end{equation}
We will see below that we indeed have $\phi \sim \sqrt{N}$ for the
expansion point. In comparison, the fluctuations scale with $\delta
\vphi \sim N^0$ and their contribution is negligible for large
$N$. The validity of the classical approximation $\Gamma \simeq S$,
and subsequent improvements, thus relies on the existence of a
macroscopically occupied condensate. We emphasize that the term
``classical'' here refers to the absence of fluctuations, and not to
the limit $\hbar \rightarrow 0$. In fact, in the next paragraph we
will discuss specific features which crucially build on a truly
quantum mechanical feature: macroscopic phase coherence. In addition,
we note that the approximation is not based on perturbation theory in
the coupling constant $g$. The notion of weak interactions is,
however, needed to justify the scaling $\phi\sim\sqrt{N}$ derived from
the noninteracting case.

\vspace{5mm}
\begin{center}
 \emph{Classical limit and Gross--Pitaevskii equation}
\end{center}

\noindent From Eq. (\ref{Func17}) we obtain an approximate expression for the effective action $\Gamma$, when evaluated for the solution of the classical equations of motion. It corresponds to a saddle-point approximation for the functional integral. With Eq. (\ref{Func14}) we find the condition for vanishing external sources 
\begin{align}
 \nonumber 0 =\frac{\delta S}{\delta \vphi^*(\tau,\vec{x})}[\vphi_0] = &\left( \partial_\tau - \frac{\nabla^2}{2M} -\mu+ V_{\rm ext}(\vec{x})\right)\vphi_0(\tau,\vec{x})\\
 \label{bog3} & +g \Bigl(\vphi_0^*(\tau,\vec{x})\vphi_0(\tau,\vec{x}))\Bigr) \vphi_0(\tau,\vec{x}). 
\end{align}
This equation can be analytically continued to real time $t=- \rmi \tau$ and is then known as the \emph{Gross--Pitaevskii equation} \cite{pitaevski61,gross61}. Restoring $\hbar$, we find
\begin{align}
 \nonumber \rmi \hbar \frac{\partial \vphi_0}{\partial t}(t,\vec{x}) &= \left(- \frac{\hbar^2\nabla^2}{2M} -\mu+ V_{\rm ext}(\vec{x})\right) \vphi_0(t,\vec{x}) \\
 \label{bog9} &+ g \Bigl(\vphi_0^*(t,\vec{x})\vphi_0(t,\vec{x})\Bigr) \vphi_0(t,\vec{x}).
\end{align}
For vanishing coupling $g$, this formally is a Schr\"{o}dinger equation for a single particle in an external potential. For this reason, the order parameter $\vphi_0(t,\vec{x})$ is sometimes called the \emph{macroscopic wave function} and we can expect characteristic features from quantum mechanics to be found in weakly interacting Bose--Einstein condensates. For instance, effects of phase coherence can be observed in a condensate, which as anticipated above is possible since the classical limit considered here derives from $N \rightarrow \infty$ and not $\hbar \rightarrow 0$. For $g \neq 0$, the Gross--Pitaevskii equation is nonlinear and thus shows a richer spectrum of solutions than the Schr\"{o}dinger equation.

\begin{figure}[tb!]
 \centering
 \includegraphics[scale=0.8,keepaspectratio=true]{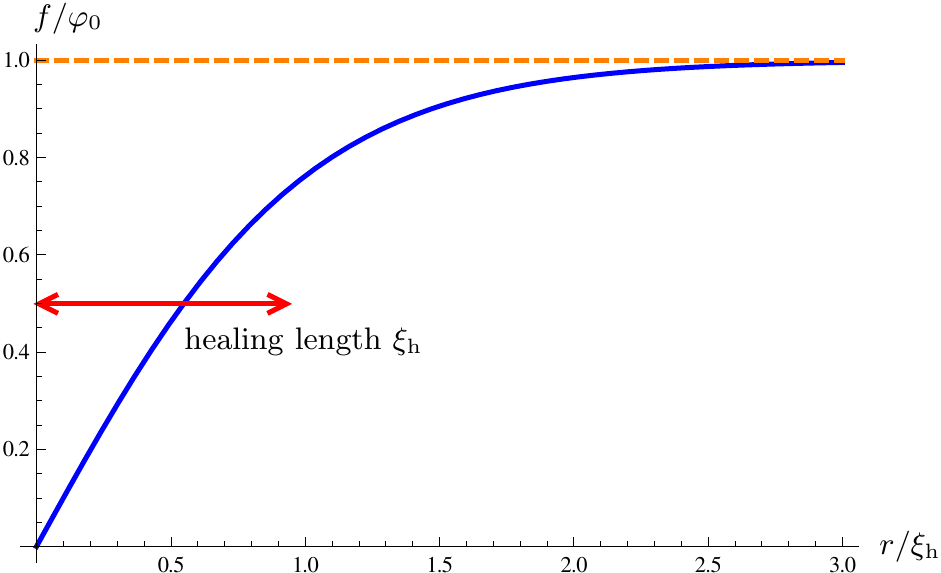}
\caption{We locate the center of the vortex solution to Eq. (\ref{bog8}) at position $r=0$ in the $(x,y)$-plane and plot the behavior of the amplitude $f(r)$. For distances which are small in comparison to the healing length $\xi_{\rm h}=\hbar/(2Mg\rho_0)^{1/2}$, we have a strong deviation from the constant solution $\vphi_0=\sqrt{\mu/g}$. For larger distance, however, the amplitude approaches this mean field value. We thus deduce the healing length to be the typical size of a vortex or, more generally, to be the characteristic length scale of possible inhomogeneities in the otherwise constant condensate amplitude.}
\label{VortexSolution}
\end{figure}

One particular example for this interplay between nonlinearity and quantum mechanics is the existence of vortex solutions for the equations of motion with quantized phase. For this purpose, consider the situation of vanishing external potential, $V_{\rm ext}(\vec{x}) =0$. We may then look for a static solution with cylindrical symmetry according to $\vphi_0(t,\vec{x}) = f(r) e^{\rmi l \theta}$, where $r=(x^2+y^2)^{1/2}$ and $\theta$ is the polar angle. The \emph{winding number} $l$ must be an integer in order to guarantee the uniqueness of the macroscopic wave function as $\theta$ wraps around the origin. Plugging this ansatz into the equations of motion, we arrive at the ordinary differential equation
\begin{equation}
 \label{bog8} 0 = - \frac{\hbar^2}{2M} \left( f'' + \frac{f'}{r} - \frac{l^2 f}{r^2}\right) - \mu f + g f^3.
\end{equation}
The corresponding field configuration $\vphi_0(\vec{x})$ is called a \emph{vortex solution}, with radial function $f$ plotted in Fig. \ref{VortexSolution}. The solution has two qualitatively distinct regimes separated by the length scale related to the nonlinearity, the \emph{healing length} 
\begin{equation}
\label{bog38} \xi_{\rm h}=\hbar/(2Mg\rho_0)^{1/2}.
\end{equation}
For length scales $r\gg \xi_{\rm h}$, the amplitude $|\vphi_0(\vec{x})|$ approaches the value $\sqrt{\mu/g}$, which is a constant solution of Eq. (\ref{bog9}). Instead, for $r\ll \xi_{\rm h}$, the condensate amplitude must vanish due to the centrifugal barrier, which in turn roots in the quantization of the phase (the term $\sim l^2 /r^2$).  Quantized vortices are a hallmark of quantum condensation phenomena, and have been observed experimentally in bosonic \cite{PhysRevLett.83.2498} and fermionic \cite{zwierlein05} condensed systems. The vortex core size $\xi_{\rm h}$ gives direct information on the interactions in the many-body problem.
The action of a vortex field configuration is larger than that of a homogenous condensate. Therefore, it must be triggered externally, such as via rotating the trap, which imprints angular momentum onto the system. We refer to \cite{cooper-review} for an extensive discussion of these topics. 

\vspace{5mm}
\begin{center}
 \emph{Effective potential and spontaneous symmetry breaking}
\end{center}

\noindent The effective potential in the classical approximation $\Gamma \simeq S$ is given by the Mexican hat potential discussed below Eq. (\ref{SSB13}). Indeed, for the homogeneous case $V_{\rm ext}=0$, the solution to Eq. (\ref{bog3}) is given by a constant $\phi_0$, which minimizes
\begin{equation}
 \label{bog6} U(\phi) = - \mu |\phi|^2 + \frac{g}{2} |\phi|^4.
\end{equation}
For $\mu <0$, we have $\phi_0=0$. For the cases of interest here, however, the chemical potential is positive and we have besides $\phi_0=0$ also the solution
\begin{equation}
 \label{bog4} |\phi_0(\tau,\vec{x})|^2 = \rho_0 = \mu/g \in \mathbb{R},
\end{equation}
which has a smaller value of the action $S[\phi_0]=\Gamma[\phi_0]$.  We draw attention to the point that the amplitude $\sqrt{\mu/g}$ was also found for the vortex solution far from the center of the vortex. The latter inhomogeneous solution necessarily has to approach this limiting value for the action to remain finite.

Note that the phase of $\phi_0$ is not specified by the action principle $\delta S /\delta \vphi^* =0$. Indeed, the global ${\rm U}(1)$-symmetry of the microscopic action $S$ is carried over to the equations of motion derived from it. Clearly, the actual ground state of the system must have a particular value of the phase and without loss of generality we assume $\phi_0$ to be real, i.e.
\begin{equation}
 \label{bog5} \phi_0 = \sqrt{\rho_0}.
\end{equation}
As discussed above, this is a manifestation of spontaneous symmetry breaking.

From the effective potential $U(\phi)$, we can deduce the phase diagram and the equation of state using Eqs. (\ref{SSB5b}) and (\ref{SSB5}). We  find in the classical approximation
\begin{align}
 \label{bog7b} &0 = \frac{\partial U}{\partial \phi^*} (\phi_0) = \bigl(-\mu + g |\phi_0|^2\bigr)\phi_0, \hspace{5mm} \text{(gap equation)}\\
 \label{bog7c} &n(\mu) = - \frac{\partial U}{\partial \mu}(\phi_0) = |\phi_0|^2 = \rho_0, \hspace{5mm} \text{(equation of state)}.
\end{align}
The first equation yields the constant solutions from Eq. (\ref{bog4}). From the equation of state $n(\mu)=\rho_0$ we infer that all particles are condensed in the classical approximation at zero temperature. In particular, the phase diagram consist of the two regions, $\mu>0$ and $\mu<0$, with and without particles, respectively. Furthermore, since the classical approximation we applied so far does not include any thermal fluctuations, we are restricted to zero temperature. To get a more physical picture of the phase structure and the thermodynamics, we have to improve formula (\ref{bog6}) for the effective potential by including fluctuations.

\vspace{5mm}
\begin{center}
 \emph{Quadratic fluctuations and excitation spectrum}
\end{center}

\noindent We now go one step beyond the classical limit and include quadratic fluctuations in $\vphi$ in the effective action. The treatment of quadratic (or Gaussian) fluctuations is often referred to as mean field theory, although sometimes this term is also used for the classical approximation $\Gamma \simeq S$. The Gaussian corrections to the classical formulae for a weakly interacting Bose--Einstein condensate have first been derived by Bogoliubov in the second quantized formalism \cite{bogoliubov47}. We are aiming at a discussion of this contribution with the help of the functional integral techniques developed so far.

The correction to the effective action due to quadratic fluctuations is summarized in the one-loop formula (\ref{Func19}),
\begin{align}
 \label{bog9c}\Gamma[\phi^*,\phi] = S[\phi^*,\phi] + \frac{1}{2} \mbox{Tr}\log S^{(2)}[\phi^*,\phi],
\end{align}
which is valid for small fluctuating fields. Here, $\phi(\tau,\vec{x})$ is an arbitrary complex field. After the trace has been evaluated, the ground state configuration $\phi_0$ is found from the minimum of the full (one-loop) effective potential. However, to get a first idea of the underlying physics, we approximate the full ground state to be given by the classical solution $\vphi_0$ from Eqs. (\ref{bog3}) and (\ref{bog4}). In this case, Eq. (\ref{Func17}) yields
\begin{align}
 \label{bog8b} e^{-\Gamma[\vphi_0]} \simeq e^{-S[\vphi_0]}\int \mbox{D}\vphi^*\,\mbox{D}\vphi\, e^{-\frac{1}{2}\int (\vphi,\vphi^*)\cdot S^{(2)}[\vphi_0]\cdot \binom{\vphi}{\vphi^*}}.
\end{align}
It is favorable to work in the real basis for the fluctuating field with transformation specified by $ \vphi = \frac{1}{\sqrt{2}}(\vphi_1 + \rmi \vphi_2)$, and to switch to momentum space (using translation invariance of the investigated situation). The quadratic fluctuations then take the form
\begin{equation}
 \label{bog8c} \frac{1}{2} \int_Q (\vphi_{1,-Q}, \vphi_{2,-Q}) \left(\begin{array}{cc} \vare_q + 2 g \rho_0 & - \omega_n\\ \omega_n & \vare_q\end{array}\right)\left(\begin{array}{c}  \vphi_{1,Q} \\ \vphi_{2,Q}\end{array}\right),
\end{equation}
where $\vare_q = \vec{q}\,\,^2/2M$, $Q=(\omega_n,\vec{q})$ and $\omega_n$ is a bosonic Matsubara frequency, see Eq. (\ref{Func5}). Moreover, we used $\rho_0 = |\vphi_0|^2 = \mu/g$.

Eq. (\ref{bog8c}) fully confirms our picture of fluctuations in the Mexican hat potential. Indeed, the real field $\delta \vphi_1$ constitutes the radial mode, which is gapped, i.e. it has a mass term $2 g \rho_0$. This mass term suppresses the corresponding \emph{amplitude fluctuations} of the field. The second kind of fluctuations, $\delta \vphi_2$, however, is not gapped and there can be arbitrary many excitations of this mode. It corresponds to \emph{phase fluctuations}, which take place in the valley of the Mexican hat. It constitutes the Goldstone mode associated to the spontaneous breaking of the ${\rm U}(1)$-symmetry.

Next we wish to calculate the excitation spectrum. To this end, we first note that the matrix appearing in Eq. (\ref{bog8c}) constitutes the full inverse propagator, or inverse Green's function, $G^{-1}(Q)$, within our simple approximation (cf. the discussion of the effective action in Sec. \ref{SecFun}). More precisely, it is defined according to $\Gamma^{(2)}(Q',Q)[\phi_0]=G^{-1}(Q)\delta(Q+Q')$ and thus coincides with the classical inverse propagator $S^{(2)}[\vphi_0]$ at this level of approximation. The excitation spectrum of the system, i.e. the dispersion relation, is obtained from the poles of the full propagator $G(\omega,\vec{q})$ after analytic continuation to real-time frequencies $\omega$, which correspond to real times $t$. 

In order to find the right prescription for the analytic continuation, we consider a field $\phi(\tau,\vec{x})$ in euclidean time $\tau$. Its Fourier representation is given by $\phi(\tau,\vec{x}) = \int_Q e^{\rmi(\vec{q}\cdot\vec{x}+\omega_n\tau)}\phi(Q)$. The sign of the temporal term in the exponent reflects the fact that time and space are treated equally in euclidean space. For real times $t$, we expect a reversed sign instead, describing wave propagation. Thus, using $\tau = \rmi t$, we find from $e^{\rmi(\vec{q}\cdot\vec{x}+\omega_n\tau)} = e^{\rmi(\vec{q}\cdot\vec{x}-\omega t)}$ the rule
\begin{equation}
 \label{bog8d} \omega_n = \rmi \omega.
\end{equation}
The poles of the full propagator at frequencies $\omega=E_q$ are then obtained via
\begin{equation}
 \label{bog8e} 0 = \mbox{det}\Bigl(G^{-1}(\omega_n=\rmi E_q,\vec{q})\Bigr) = (\vare_q+2g\rho_0)\vare_q - E_{q}^2.
\end{equation}
This leads to the Bogoliubov excitation spectrum
\begin{equation}
 \label{bog31} E_q = \sqrt{\vare_q(\vare_q+2g\rho_0)}.
\end{equation}

\begin{figure}[tb!]
 \centering
 \includegraphics[scale=0.85,keepaspectratio=true]{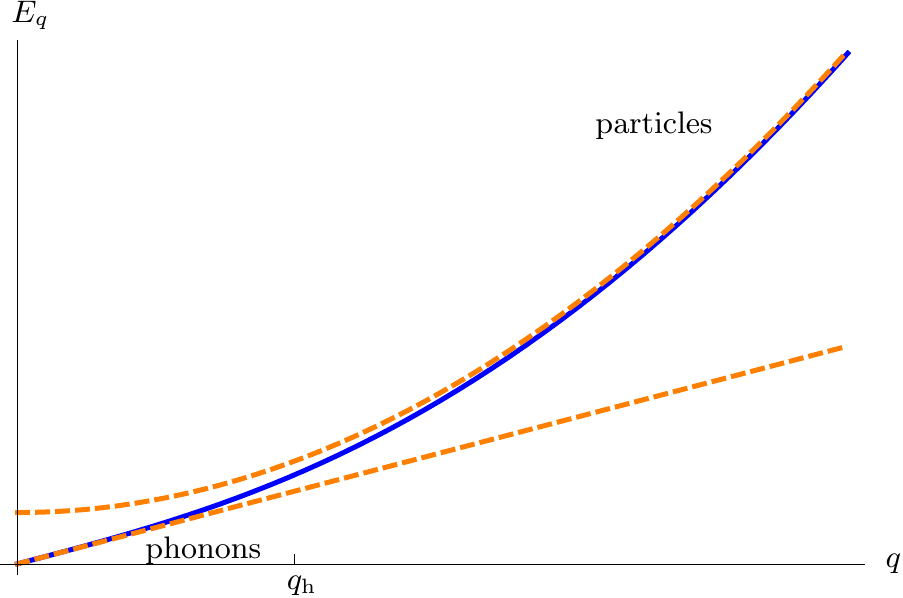}
\caption{The energy spectrum of Bogoliubov excitations consists of a phonon-like part for momenta $q \lesssim q_{\rm h}$, where $q_{\rm h} = \hbar/\xi_{\rm h}$ is associated to the healing length $\xi_{\rm h}$. For larger momenta, the excitations show a quadratic momentum dependence, which is typical for nonrelativistic particles.}
\label{BogDispersion}
\end{figure}

The form of the spectrum of elementary excitations (\ref{bog31}) has important physical consequences. For small momenta $\vec q\,^2\ll Mg\rho_0$, we have a linear and gapless dispersion 
\begin{equation}
 \label{bog36} E_q \approx c | \vec q| , \hspace{5mm} c=\sqrt{\frac{g\rho_0}{M}},
\end{equation}
which is characteristic for phonons with velocity of sound $c$, whereas the spectrum gets quadratic and thus particle-like for high momenta $q^2 \gg Mg \rho_0$,
\begin{equation}
 \label{bog37} E_q \approx \vare_q = \frac{\vec q\,^2}{2M}.
\end{equation}
Thus, for long wavelength excitations (low momenta) the linear part of the spectrum dominates, whereas on short scales (high momenta) the particle nature of the system is still visible. The fact that the low-momentum degrees of freedom have a phonon-like dispersion hints at the typical collective behavior of many-body systems, where the effective quasiparticles do no longer coincide with the microscopic particle-like degrees of freedom. 

Below, we will see that the phenomenon of superfluidity is intimately connected to this modification of the spectrum at low momenta. We can get more insight into the nature of these fluctuations by integrating out the massive modes $\vphi_1$ in the functional integral Eq. (\ref{bog8b}). (This can be done by completing the square in the exponent). This procedure is useful if we focus on momenta with energies below the gap, $ \epsilon_q \ll 2g\rho_0$. It produces a ``renormalized'' low energy theory for the massless excitations $\vphi_2$, which reads
\begin{equation}
 S[\vphi_2] = \frac{1}{4g\rho_0} \int_Q  \vphi_{2,-Q} (\omega^2 + c^2 \vec q\,^2 )\vphi_{2,Q}.
\end{equation}
This low energy action reveals that the linear dispersion is due to the fluctuations of the phase. 

The characteristic momentum scale $q_{\rm h}$ which separates both regimes is given by the inverse \emph{healing length}, $q_{\rm h} = \hbar/\xi_{\rm h}$, which we already encountered as the characteristic size of a vortex inside an otherwise homogeneous condensate. The Bogoliubov dispersion and its limits are shown in Fig. \ref{BogDispersion}.

To complete the predictions available from Bogoliubov theory, we indicate the condensate depletion. Physically, this depletion is an observable effect of quantum fluctuations related to the presence of a finite interaction strength $g $, and results from scattering processes out of and back into the condensate. In particular, it occurs in the absence of thermal fluctuations at $T=0$. One finds
\begin{align}\label{bog28ba}
 \nonumber n  &= \rho_0 + \frac{1}{2} \int \frac{\mbox{d}^3q}{(2\pi)^3} \biggl( \frac{\vare_q+g \rho_0}{E_q} - 1 \biggr).
\end{align}
The calculation is rather lengthy, but instructive as we can learn from it how to deal with ultraviolet divergences in nonrelativistic quantum field theory. It is presented in \ref{AppOneLoopBos}.

\vspace{5mm}
\begin{center}
 \emph{Superfluidity and Landau criterion}
\end{center}

\noindent The excitation spectrum $E_q=\sqrt{\vare_q(\vare_q+2g\rho_0)}$ allows for superfluidity. This phenomenon was first observed in liquid $^4$He and manifests itself, for example, in the frictionless flow through small slits or capillaries. The onset of superfluid behavior constitutes a phase transition, which is of second order for the weakly interacting Bose gas in three dimensions. The hydrodynamic description of a superfluid system is altered by the separation of the macroscopic motion into normal fluid flow and superfluid flow, the latter being frictionless, irrotational and entropy conserving. The corresponding two-fluid hydrodynamic equations have been developed by Landau and coworkers (cf. \cite{landau-book}). Above the critical temperature of superfluidity, which coincides with Bose condensation here, the two-fluid equations turn into the hydrodynamics of a one-component fluid.

Condensation and superfluidity are related phenomena, but they do not necessarily coincide. Condensation refers to the macroscopic \emph{occupation} of a single mode and thus is a statistical effect. In contrast, superfluidity refers to the \emph{response} of a given system as we will see below. The fact that these are independent concepts becomes particularly important in spatial dimension less than three, where the Mermin--Wagner theorem forbids the existence of a condensate. In contrast, superfluidity is present below a critical temperature. The related phase transition is known as the Berezinski--Kosterlitz--Thouless transition. For more information on this subject, we refer to the literature (e.g. \cite{altland-book}).

\begin{figure}[tb!]
 \centering
 \includegraphics[scale=0.55,keepaspectratio=true]{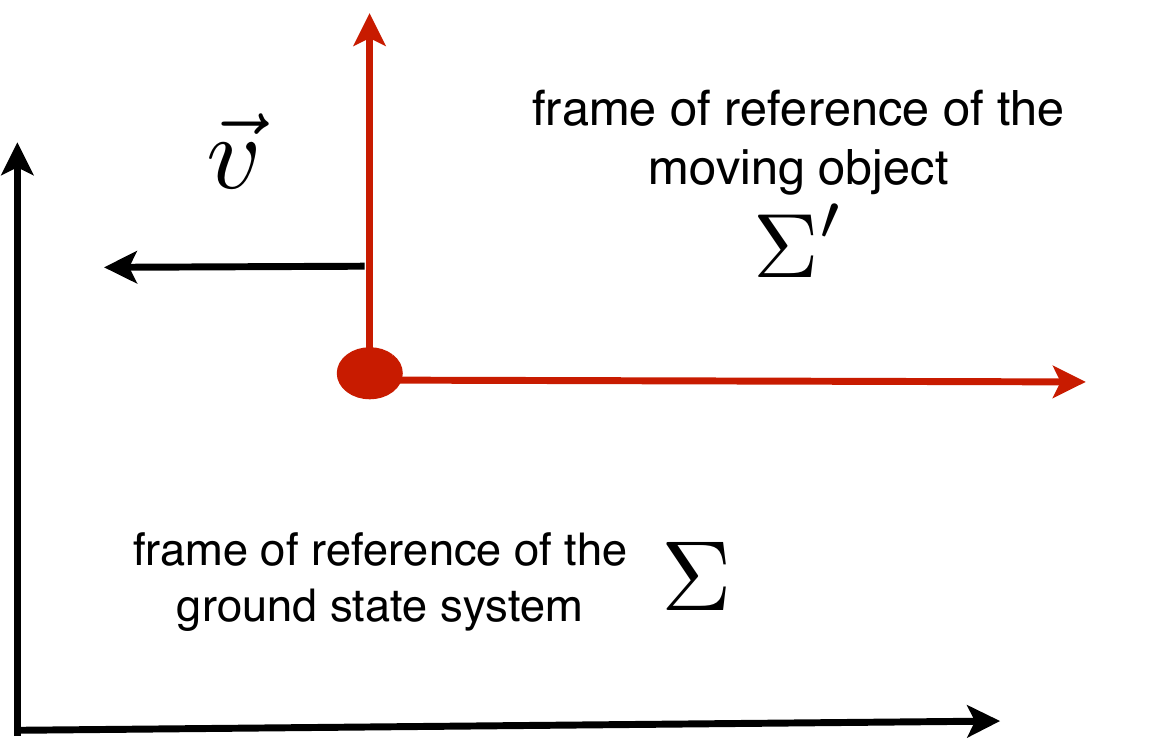}
\caption{Frames of reference for Landau's gedankenexperiment to identify superfluid flow.}
\label{fig:landau}
\end{figure}

Landau established a beautiful criterion for the existence of superfluidity, identified via the property of frictionless flow, from a simple yet general kinematic argument. We consider an object uniformly moving in the liquid equilibrium state of a system at velocity $\vec v$, and ask when it is favorable to create an excitation in that state, which leads to friction. To describe the situation, we introduce two frames of reference: $\Sigma'$ is comoving with the object, $\Sigma$ is the frame of reference for the liquid. (This is illustrated in Fig. \ref{fig:landau}.) The general transformation of energy and momentum under a Galilean boost with velocity $\vec v$ in these two frames is given by
\begin{align}
\nonumber \Sigma &:\mbox{ } E,\quad  \vec p,\\
\Sigma' &: \mbox{ }E' = E - \vec p \cdot \vec v + \tfrac 1 2 M \vec v ^2 ,\quad \vec p'= \vec p - M\vec v,
\end{align}
where $M$ is the total mass. We apply this transformation to the calculation of the total energy and momentum in the following two situations:
\begin{itemize}
\item[(i)] the ground state of the liquid
\begin{align}
\nonumber\Sigma &: \mbox{ }E_0,\quad  \vec p_0 =\vec{0},\\
\Sigma' &: \mbox{ }  E'_0 = E_0  + \tfrac 1 2 M \vec v ^2 ,\quad \vec p'_0=  - M\vec v
\end{align}
\item[(ii)] the ground state of the liquid plus an excitation with momentum $\vec p$ and energy $\epsilon_p$
\begin{align}
\nonumber\Sigma :\mbox{ }  E_{\text{ex}} &= E_0 + \epsilon_p ,\quad  \vec p_\text{ex} =\vec p ,\\
\nonumber \Sigma' : \mbox{ }  E'_\text{ex} &= E_0  +\epsilon_p - \vec p \cdot \vec v + \tfrac 1 2 M \vec v ^2 ,\\
 \vec p'_\text{ex} &= \vec p  - M\vec v
\end{align}
\end{itemize}
The creation of an excitation is \emph{unfavorable} if $E_\text{ex}' -E'_0  = \epsilon_p -\vec{p} \cdot \vec v  \geq \epsilon_p -|\vec p|| \vec v| >0$. Thus, no excitations occur, and we have frictionless transport of the object,  as long as the velocity is smaller than a \emph{critical velocity},
\begin{equation}\label{eq:landaucrit}
|\vec  v| < v_\text{c} = \min_{\vec{p}} \frac{\epsilon_ p}{|\vec p|}.
\end{equation}
Moving the object with a larger velocity leads to friction. Let us apply this criterion to two particularly important systems. First, for a  weakly interacting Bose gas, the low momentum linear dispersion is $E_p = c |\vec p|$. Thus, the system is superfluid with a critical velocity $c$. Second, in a free Bose gas, the scaling of energy with momentum is quadratic, $E_p = \vec p^2/2M$. In consequence, a free Bose--Einstein condensate is not a superfluid; its critical velocity $v_\text{c} =0$. This is another example where the concepts of superfluidity and condensation do not coincide.

The Landau criterion can also be established from a field theoretical perspective. In order to study this problem, we first need to analytically continue back to real times. The euclidean microscopic action $S_{\rm E}$ and the real time action $S$ are connected by the requirement that $e^{-S_{\rm E}} = e^{\rmi S}$ appears in the path integral. Thus, starting from Eq. (\ref{bog2}), we employ $\tau = \rmi t$ to arrive at the real time microscopic action
\begin{equation}
 \label{RealTimeAction} S[\vphi] = \int \mbox{d} t \int_{\vec{x}} \biggl\{ \vphi^*\Bigl(\rmi \partial_t + \frac{\nabla^2}{2M} + \mu\Bigr) \vphi - \frac{g}{2}(\vphi^*\vphi)^2\biggr\}.
\end{equation}
Working in the Bogoliubov approximation, we expand this action to quadratic order in the fluctuations about the solution $\vphi_0=\sqrt{\rho_0}=\sqrt{\mu/g}$ of the classical equations of motion. Writing $\vphi = \vphi_0 + \frac{1}{\sqrt{2}}(\delta \vphi_1 + \rmi \delta \vphi_2)$, the corresponding real time action for the Bogoliubov excitations is found to be given by
\begin{align}
 \nonumber &S_{\rm bog}[\delta \vphi_1,\delta \vphi_2] = \frac{1}{2}  \int\mbox{d} t \int_{\vec{x}}g \rho_0^2\\
 \label{BogAction} &+\frac{1}{2} \int \mbox{d} t \int_{\vec{x}} (\delta \vphi_1,\delta \vphi_2) \left( \begin{array}{cc} \frac{\nabla^2}{2M}-2 g \rho_0 & - \partial_t \\ \partial_t & \frac{\nabla^2}{2M} \end{array}\right) \left( \begin{array}{c} \delta \vphi_1 \\ \delta \vphi_2 \end{array}\right).
\end{align}

In view of identifying a critical velocity, we ask whether this action describing the excitations on top of the condensed ground state is stable under the transformation of the field
\begin{equation}
 \label{Lan1} \vphi(t,\vec{x}) \rightarrow \vphi'(t,\vec{x}) = e^{\rmi (\vec{p}\cdot \vec{x} - E t)} \vphi(t,\vec{x}).
\end{equation}
This describes the imprint of a plane wave with momentum and velocity $\vec p = M\vec v $ on the fields, and we comment on the temporal phase rotations below. In order to understand this transformation, we first decompose $\vphi(t,\vec{x}) = \phi_0 + \delta\vphi(t,\vec{x})$, where $\phi_0$ describes the condensed ground state, and $ \delta\vphi(t,\vec{x})$ small fluctuations around it. In particular, we see that under the transformation, the ground state picks up a position dependent phase $\phi_0 \to  \phi'_0(t,\vec{x}) = \phi_0e^{\rmi (\vec{p}\cdot \vec{x} - E t)}$ and thus carries a supercurrent $\vec j = \tfrac{\mathrm i}{2M} [\nabla \phi^{' *}_0(t, \vec{x})  \phi'_0(t, \vec{x}) - \phi^{' *}_0(t, \vec{x}) \nabla \phi'_0(t, \vec{x}) ]  = \phi_0^*\phi_0 \tfrac{\vec p}{M} =\phi_0^*\phi_0 \vec v$. We will now ask whether this supercurrent (superfluid flow) is persistent or stable, and thus the system a true superfluid.\footnote{We note that this is \emph{not} a Galilei transformation describing the change of frame of reference, given by $\vphi(t,\vec{x}) \rightarrow \vphi'(t,\vec{x}) = e^{\rmi (\vec{p}\cdot \vec{x} - E t)} \vphi(t,\vec{x} - \vec v t)$ (the full microscopic action must be invariant under such transformations). It rather is a local gauge transformation.} To this end, we apply the transformation (\ref{Lan1}) to the microscopic action in Eq. (\ref{RealTimeAction}), and find
\begin{align}
 &S[\vphi] \to S[\vphi] + \int \mbox{d} t \int_{\vec{x}} \vphi^* \Bigl( E -  \rmi \vec{v} \cdot \nabla - \frac{\vec{p}^2}{2M}\Bigr) \vphi.
\end{align}
The temporal component of the transformation describes an adjustment of the zero of energy, and, choosing $E=\vec{p}^2/2M$, we shift the latter back to its original value.

In the momentum representation of the fields, $\vphi(\omega,\vec{q}) = \int e^{-\rmi(\vec{q}\cdot \vec{x} - \omega t)}\vphi(t,\vec{x})$, we find the quadratic part of the transformed Bogoliubov action in Eq. (\ref{BogAction}) to be given by
\begin{align}
 \label{TrfProp} - \left( \begin{array}{cc} \vare_q + 2 g \rho_0 & - \rmi ( \omega + \vec{v} \cdot \vec{q})\\ \rmi ( \omega +  \vec{v} \cdot \vec{q}) & \vare_q \end{array}\right).
\end{align}
The system is stable under the transformation in Eq. (\ref{Lan1}), and therefore supports superfluid flow, if the excitation energies resulting from the matrix in Eq. (\ref{TrfProp}) are positive -- in this case, the system is located at a (local) minimum in energy. Diagonalization shows that the effect of the above transformation is to shift the Bogoliubov excitation energies according to 
\begin{equation}
E_q \to E_q' = E_q - \vec v\cdot \vec q \geq E_q - |\vec v| |\vec q| \stackrel{!}{>} 0.
\end{equation}
We thus recover the Landau criterion Eq. (\ref{eq:landaucrit}) for the critical velocity: Below the critical velocity for the perturbation, the excitation energies $E_q'$ for all modes are positive, and thus the equilibrium state $\phi'_0(t,\vec{x})$ carrying superfluid flow is stable. Above $v_\text{c}$, however, there exist unstable fluctuations which ultimately tend to destroy the superfluid flow, driving the system to a state which no longer is described by $\phi'_0(t,\vec{x})$. We note that the considerations  also hold for the noninteracting case, where $E_q = \vec{q}^2/2M$, resulting in a vanishing superfluid velocity.

\vspace{5mm}
\begin{center}
 \emph{Validity of Bogoliubov theory}
\end{center}

\noindent To close this section, we discuss the validity of Bogoliubov theory. In particular, our analysis will reveal why many experimental observations on cold trapped bosons are captured within this framework.

We have seen in Eq. (\ref{bog2g}) that the validity of Bogoliubov theory is related to the ordering principle of a macroscopically occupied condensate, which allows for an approximate evaluation of the path integral. Obviously, such a procedure breaks down if no condensate exists. This situation is found in two-dimensional systems at nonzero temperature and always in one-dimensional systems. (See our discussion of the Mermin--Wagner theorem and its relevance for cold atoms in Sec. \ref{ideal}.) In these lower-dimensional settings, one necessarily has to rely on nonperturbative approaches. 

Even in three dimensions it is questionable whether an expansion in powers of the fluctuating field $\delta \vphi$ is valid for low momenta. Indeed, from Eq. (\ref{bog8c}) for the  classical inverse propagator we find for the occupation of the $\vec{q}$-mode
\begin{equation}
 \label{bog44} n_{\vec{q}} = \int_{\omega} \langle \delta \vphi^*_Q \delta \vphi_Q \rangle \stackrel{q\rightarrow 0}{\sim} \frac{1}{E_q} \sim \frac{1}{|\vec{q}|}.
\end{equation}
The high, diverging occupation of low momentum modes, allowing to roughly count $\delta \vphi_{\vec{q}} \sim |\vec{q}|^{-1/2}$, questions the validity of a simple ordering principle set by the macroscopic condensate occupation $\phi_0 \sim \sqrt{N}$. 

\begin{figure}[tb!]
 \centering
 \includegraphics[scale=0.65,keepaspectratio=true]{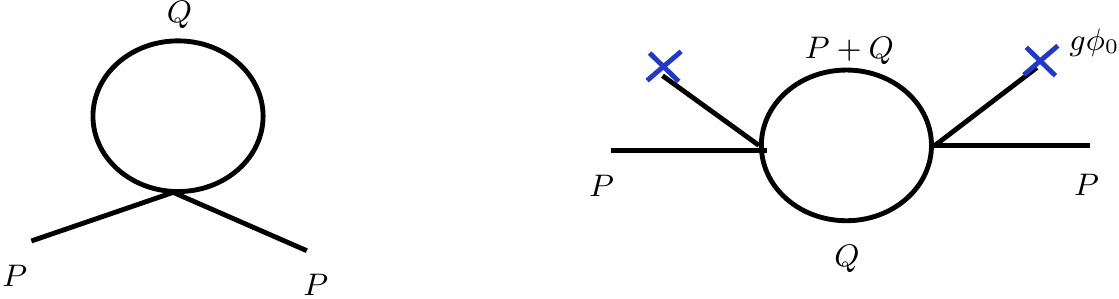}
\caption{First order perturbative corrections to the self-energy. Whereas the diagram to the left is infrared regular, the one to the right is divergent for low momenta $P$, see Eq. (\ref{bog48b}).}
\label{FigBreakdownReg}
\end{figure}

In view of estimating the momentum scale where Bogoliubov theory breaks down, we study the perturbative effects on the self-energy $\Sigma$ for weakly interacting bosons at zero temperature \cite{PhysRevA.80.043627}. The full inverse propagator in the $(\phi,\phi^*)$-basis is given by
\begin{align}
 \nonumber &G^{-1}(P) \\
 \label{bog45} &= \left(\begin{array}{cc} \Sigma_{\rm an}(P) & - \rmi \omega_n +\vare_p -\mu +\Sigma_{\rm n}(P) \\ \rmi \omega_n +\vare_p -\mu +\Sigma_{\rm n}(P) & \Sigma_{\rm an}(P) \end{array}\right).
\end{align}
We may regard Bogoliubov theory as the tree-level self-energies
\begin{equation}
 \label{bog46} \Sigma^{(0)}_{\rm n}(P) = 2 g \rho_0, \hspace{5mm} \Sigma^{(0)}_{\rm an}(P) = g \rho_0.
\end{equation}
The leading perturbative corrections are shown diagrammatically in Fig. \ref{FigBreakdownReg}. The second diagram has an infrared divergence, which is logarithmic in $d=3$ spatial dimensions and polynomial for $d <3$. Indeed, the low momentum contribution to the corresponding loop integrals is given by
\begin{equation}
 \label{bog47} \Sigma^{(1)}_{\rm n}(P) \sim \Sigma^{(1)}_{\rm an}(P) \sim - g^2 \rho_0 \int_Q G_{22}(Q)G_{22}(P+Q)
\end{equation}
with
\begin{equation}
 \label{bog48}G_{22}(P) = \frac{2 g \rho_0}{\omega^2+c^2\vec{p}^2}
\end{equation}
and $c=(g\rho_0/M)^{1/2}$. For very low momenta we find
\begin{align}
 \nonumber \Sigma^{(1)}(P \rightarrow 0) &\sim g^2 \rho_0 M^2  c \int_{p}^\Lambda \mbox{d}^{d+1}\bar{q} \frac{1}{q^4} \\
 \label{bog48b}&\sim g^2 \rho_0 M p_{\rm h} \begin{cases} \log(\Lambda/p) & (d=3)\\ p^{d-3} & (d<3)\end{cases},
\end{align}
where $\bar{q}=(\omega/c,\vec{q})$ is a $(d+1)$-dimensional vector and $q=|\bar{q}|$. In this regime we can use $\Lambda =p_{\rm h}$ with $p_{\rm h} =(M g \rho_0)^{1/2}=Mc$ being the momentum scale associated with the healing length, discussed earlier in this section.

We expect perturbation theory to break down when both zeroth and first order corrections to the self-energy become of comparable size, i.e.
\begin{equation}
 \label{bog49} |\Sigma^{(0)}(P_{\rm np})| \simeq |\Sigma^{(1)}(P_{\rm np})|.
\end{equation}
From this relation we can deduce a characteristic momentum scale $p_{\rm np}$ where the superfluid becomes \emph{strongly correlated} and has to be described nonperturbatively. We arrive at
\begin{equation}
 \label{bog49b} \frac{p_{\rm np}}{p_{\rm h}} = \begin{cases} \exp(-1/g M p_{\rm h}) & (d=3)\\ g M & (d=2) \\ (g M/p_{\rm h})^{1/2} & (d=1)\end{cases}.
\end{equation}
In three dimension we have $g \propto a$. Together with $\rho_0 \approx n$ we find $gp_{\rm h} \propto (a^3n)^{1/2}$. Thus, in the dilute regime of small gas parameter, the nonperturbative physics happen at exponentially small momenta. In two dimensions instead, where the coupling constant is dimensionless, the condition $p_{\rm np} \approx p_{\rm h}$ is reached for $g$ being of order unity.

The finding of Eq. (\ref{bog49b}) can also be expressed as
\begin{equation}
 \label{bog50} \frac{p_{\rm np}}{p_{\rm h}} = \begin{cases} \exp(-1/\tilde{g}^{3/2}) &(d=3) \\ \tilde{g}^{\frac{d}{2(3-d)}} & (d <3)\end{cases}.
\end{equation}
The dimensionless quantity $\tilde{g}$ constitutes the ratio of interaction versus kinetic energy in the nonrelativstic superfluid,
\begin{equation}
 \label{bog51} \tilde{g} = \frac{E_{\rm int}}{E_{\rm kin}} = \frac{ g \rho_0}{1/(M\ell^2)} = g M \rho_0^{1-2/d} \sim (p_{\rm h}\ell)^2.
\end{equation}
We used here again $\rho_0 \sim n$ for the weakly interacting condensate and $\ell = n^{-1/d}$ is the interparticle spacing.

\begin{figure}[tb!]
 \centering
 \includegraphics[scale=0.85,keepaspectratio=true]{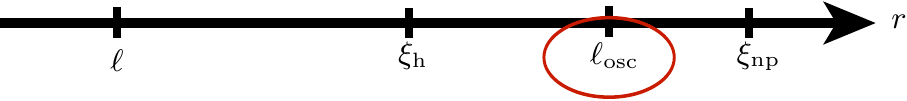}
\caption{The hierarchy of length scales is cut off by the trapping potential with characteristic size $\ell_{\rm osc}$. Fluctuations with larger wavelength do not appear in the system due to the finite extent. Since perturbation theory breaks down beyond that scale only, nonperturbative effects are difficult to observe in experiments and Bogoliubov theory is a sufficient approximation. The healing length $\xi_{\rm h}$, which is the characteristic size of a vortex, is smaller than $\ell_{\rm osc}$. The observations of vortices and vortex lattices in trapped systems confirm this picture \cite{PhysRevLett.83.2498,zwierlein05}.}
\label{InfraredTrap}
\end{figure}

Accordingly, superfluids can be classified as \cite{PhysRevA.80.043627}
\begin{itemize}
 \item[(1)] \emph{weakly correlated:} We have $\tilde{g} \ll 1$ and therefore $p_{\rm np} \ll p_{\rm h} \ll \ell^{-1}$ from Eqs. (\ref{bog50}) and (\ref{bog51}). Bogoliubov theory is valid for a large part of the spectrum, namely all modes with momenta $|\vec{q}| \gtrsim p_{\rm np}$.
 \item[(2)] \emph{strongly correlated:} For $\tilde{g} \simeq 1$ we find $p_{\rm np} \simeq p_{\rm h} \simeq \ell^{-1}$. Bogoliubov theory breaks down.
\end{itemize}
In typical traps, the oscillator length $\ell_{\rm osc}$ is smaller
than the scale $\xi_{\rm np} = 1/p_{\rm np}$. Therefore, the external
potential provides an \emph{infrared cutoff} towards the strongly
correlated regime. In the spatial continuum discussed here, the
effects of fluctuations with very large wavelengths are not
encountered in a finite trapping geometry. The observation of
vortices, which are of size $\xi_{\rm h}$ separating the particle-
from the phonon-like regime, shows that $\ell^{-1} \ll \xi_{\rm h} \ll
\ell_{\rm osc}$. We note, however, that the strongly correlated regime
can be reached by reducing the kinetic energy with respect to the
potential energy, thus decreasing $\tilde{g}$. This happens for bosons
on a lattice close to the phase transition between a Mott insulator
and a superfluid \cite{PhysRevB.83.172501,PhysRevA.85.011602}. The
scale hierarchy for bosons in a harmonic trap is summarized in
Fig. \ref{InfraredTrap}.

We conclude that Bogoliubov theory provides a good description for
many experimental settings. It may break down for special geometries
like optical lattices and lower-dimensional traps, which then allow to
study nonperturbative effects beyond the mean field approximation in
experiment.

What happens beyond the scale $\xi_\text{np}$, where perturbation
theory is plagued with infrared divergences? It has been recognized a
long time ago that a phase-amplitude (hydrodynamic) description,
useful for the regime of distances larger than $\xi_\text{np}$, does
not have these problems \cite{pata73}. Based on this, an exact
argument properly taking into account the coupling of phase and
amplitude mode has been given that there must be \emph{two} massless
modes \cite{Nepo75,Nepo78,popov-book}. This is in stark contrast to
the naive expectation from Bogoliubov theory, which predicts one
massless (transverse) and one massive (longitudinal) mode as we have
seen above. In a renormalization group language, this behavior is
reflected in a fixed point of the scale dependent interaction coupling
$g_k$ at zero, such that the longitudinal mass $\sim g_{k\to 0} \rho$
vanishes
\cite{PhysRevB.69.024513,PhysRevLett.78.1612,PhysRevB.77.064504}. In
addition, a symmetry enhancement from Galilean to Lorentz symmetry has
been observed. This is reflected by the coefficient of the linear
frequency term $\omega$ taking fixed point at zero, while the
coefficient of the term $\omega^2$ arrives at a finite value
\cite{PhysRevB.77.064504}. This is possible since Galilean invariance
is broken in a many-body system even at zero temperature due to the
presence of a condensate. The full crossover from the particle-like to
the phonon-like region at momentum scale $1/\xi_\text{h}$, and then
from the phonon-like to the hydrodynamic regime at momentum scale
$1/\xi_\text{np}$, has been followed continuously in unified
approaches based on FRG techniques
\cite{PhysRevB.77.064504,PhysRevA.77.053603,PhysRevA.79.013601,%
  PhysRevLett.102.120601,PhysRevLett.102.190401,PhysRevA.82.063632,PhysRevA.80.043627}.

The vanishing of the longitudinal mass (or the divergence of the
longitudinal static susceptibility $\chi (\omega =\vec q =0) \sim
1/g\rho$) also has interesting physical consequences. For example, it
has been shown that at zero temperature and in spatial dimension $d=2$
the (real frequency domain) spectral function possesses a critical
continuum which starts directly above the spin wave pole at $\omega =
c |\vec q|$ \cite{PhysRevB.59.14054,PhysRevLett.92.027203,%
  PhysRevLett.102.120601,PhysRevA.80.043627}. Physically, the large
spectral weight stems from the possibility of the (naively gapped)
amplitude mode breaking into a pair of spin waves for any $\omega
>c|\vec q|$.

\subsection{Superfluidity of weakly attractive fermions}
\label{SecBCS}
\noindent In this section we discuss the Bardeen--Cooper--Schrieffer
(BCS) theory of weakly interacting two-component fermions applied in a
cold atoms context. After a brief survey on the main statements of BCS
theory we solve the theory in a Gaussian approximation. On the
technical side we perform a Hubbard--Stratonovich transformation in
order to introduce effective bosonic degrees of freedom into the
fermionic theory. While the boson field here plays the role of an
auxiliary degree of freedom describing a Cooper pair condensate, it
will acquire a more direct meaning when considering the BCS-BEC
crossover in an FRG framework, where it will play the role of a
molecular bound state in the BEC regime of this problem. Many of the
formulae we derive in this section will be useful for the analysis of
the strongly interacting case. The bosonic degrees of freedom also
play an important role in view of structuring the phase diagram, and
we will be able to apply our knowledge on phase transitions for Bose
systems here. We will, in part, also adopt a ``purely fermionic point
of view'', and show that the BCS instability is related to a
divergence of the fermionic four-point vertex.

BCS theory has been developed for superconductors and thus is
originally a theory of electrons, which are two-component fermions due
to the spin $1/2$. In ultracold quantum gas experiments, two-component
fermions are realized by two distinct hyperfine states, for example of
either $^6$Li or $^{40}$K. A balanced mixture of the spin components
is not fundamental in this context, but can easily be achieved via
proper spin polarization.

We describe the system by an effective Hamiltonian with local
interactions. As in the case of bosons, for fermionic isotopes of
alkali atoms, it is given by
\begin{align}
 \label{bcs0} \hat{H} = \int \mbox{d}^3x \left\{ \hat{a}^\dagger(\vec{x}) \biggl(-\frac{\nabla^2}{2M}\biggr)\hat{a}(\vec{x}) + \frac{\lambda}{2} \hat{n}(\vec{x})^2\right\}.
\end{align}
The creation and annihilation operators satisfy anti-commutation relations. The corresponding action reads
\begin{align}
  \nonumber S[\psi^*,\psi] = \int_0^\beta \mbox{d}\tau\int \mbox{d}^3 x &\left\{ \psi^\dagger(\tau,\vec{x})(\partial_\tau - \frac{\nabla^2}{2M}  - \mu) \psi(\tau,\vec{x})\right.\\
 \label{bcs0b} &+\frac{\lambda}{2} (\psi^\dagger(\tau,\vec{x})\psi(\tau,\vec{x}))^2 \Biggr\}
\end{align}
with independent Grassmann fields $\psi = (\psi_1,\psi_2)^t$ and $\psi^*=(\psi_1^*,\psi_2^*)^t$. (A shift in $\mu$ from normal ordering, a prerequisite for the construction of the functional integral, has been absorbed.) We formally defined the operation $^\dagger = (^*)^t$. The microscopic coupling constant $\lambda$ will receive contributions from fluctuations. Including them leads to the renormalized coupling $\lambda_{\rm R}$, which is then related to the scattering length measured in experiments by relation (\ref{sc13}), i.e. $\lambda_{\rm R} = 4 \pi \hbar^2 a/M$.

We begin with some qualitative remarks. The expectation value of the fermion field vanishes due to Pauli's principle, $\langle \psi_i \rangle =0$ for $i=1,2$. However, there is no fundamental principle preventing a nonzero pairing correlation
\begin{equation}
 \label{bcs3}  \langle \psi_{1} \psi_{2}\rangle \neq 0.
\end{equation} 
As we will see below, this behavior is indeed found for low temperatures and \emph{attractive} interactions. A description of this phenomenon within the BCS framework is possible for 
\begin{equation}
 \label{bcs1} a <0, \hspace{5mm} | k_{\rm F} a| \simeq |a/\ell| \ll 1.
\end{equation}
It is a key feature of BCS theory that the small interaction scale cannot substantially modify the Fermi sphere which we encountered in the discussion of noninteracting fermions in Sec. \ref{ideal}. In particular, $\mu =\vare_{\rm F}(n)$ remains valid.

The nonvanishing correlation (\ref{bcs3}) is equivalent to a condensation of bosonic quasiparticles in their lowest energy state. Indeed, suppose two fermions build a composite bosonic state. The relative energy of the partners will be minimized for a spin singlet and vanishing center of mass energy. Thus, the total momentum of the boson is zero and the momenta of the partners are opposite. Due to the fermionic origin of the excitations, the relative momenta lie on antipodal points of the Fermi surface. We arrive at a pairing
\begin{equation}
 \label{bcs2} \langle \psi_1(\epsilon_{\rm F},\vec{p}) \psi_2(\epsilon_{\rm F},-\vec{p})\rangle \neq 0,
\end{equation}
which is local in momentum space. This is a very small effect as we will quantify below, since it only occurs in the vicinity of the Fermi surface. The composite bosons just described are called \emph{Cooper pairs}. Note that Eq. (\ref{bcs2}) results in a ground state of the many-body system which breaks the global ${\rm U}(1)$-symmetry of the action (\ref{bcs0b}).

\vspace{5mm}
\begin{center}
 \emph{Pairing field and Hubbard--Stratonovich transformation}
\end{center}

\noindent When looking at the action in Eq. (\ref{bcs0b}), we may wonder whether the quartic Grassmann field term is identically zero. Indeed, this would be the case for a one-component Fermi gas (which thus does not have local interactions). However, for our two-component spinors we have
\begin{align}
 \nonumber (\psi^\dagger \psi)^2 &= \left[ (\psi_1^*,\psi_2^*) \left(\begin{array}{c}\psi_1\\ \psi_2\end{array}\right) \right]^2 = (\psi_1^*\psi_1 + \psi_2^*\psi_2)^2 \\
 \label{bcs4} &= - 2 \psi_1^*\psi_2^*\psi_1\psi_2.
\end{align}
With the fully anti-symmetric tensor in two dimensions $\vare_{ij} = - \vare_{ji}$, $\vare_{12}=1$, i.e.
\begin{equation}
 \label{bcs5} \vare = \left(\begin{array}{cc} 0 & 1 \\ -1 & 0 \end{array}\right),
\end{equation}
we easily find
\begin{equation}
 \label{bcs6} (\psi^\dagger \psi)^2 = - \frac{1}{2} (\psi^\dagger \vare \psi^*) ( \psi^t \vare \psi).
\end{equation}
This rewriting is called a Fierz transformation. Note that $(\psi^t\vare \psi)^* = - \psi^\dagger \vare \psi^*$.

From Eq. (\ref{fun27}) we deduce the simple identity
\begin{equation}
 \label{bcs7} \int \mbox{D} \varphi^* \,\mbox{D} \vphi \,\exp\left\{-\int_X m^2 (\varphi^*-\vphi_{\rm a})(\vphi-\vphi_{\rm b})\right\} =\mathcal{N}.
\end{equation}
Herein, $\vphi_a(X)$ and $\vphi_b(X)$ can be chosen arbitrarily, because they can be eliminated by a shift of the corresponding integration measure. Note however that $m^2$ has to be positive for the integral to converge. For this reason, we wrote it in a suggestive manner as a square. $\mathcal{N}$ is related to the determinant of the identity operator, but irrelevant for our purposes since we later take the logarithm of the corresponding expression.  The idea behind the \emph{Hubbard--Stratonovich transformation} is to insert this unity into the path integral $\int \mbox{D}\psi e^{-S}$ and then choose the free parameters in such a way that the action gets more suitable for subsequent approximations. In our case, we want to eliminate the quartic fermionic interaction term $\sim \lambda (\psi^\dagger\psi)^2$; it is traded for the complex field $\vphi$ as additional bosonic degree of freedom. As it turns out, this is actually a very physical effect. 

We choose
\begin{equation}
 \label{bcs8} \vphi_{\rm a} = \frac{h}{2m^2} (\psi^t\vare\psi)^*, \hspace{5mm} \vphi_{\rm b}= \frac{h}{2m^2} (\psi^t \vare \psi),
\end{equation}
where the physical meaning of the constants $m^2$ and $h$ will become clear below. This leads to
\begin{align}
 \nonumber &\mathcal{N} = \int \mbox{D} \vphi^* \mbox{D} \vphi \\
 \nonumber &\times\exp\left\{-\int_X m^2 (\varphi^*-\frac{h}{2m^2} (\psi^t\vare\psi)^*)(\vphi-\frac{h}{2m^2} (\psi^t \vare \psi))\right\}\\
 \nonumber &=\int \mbox{D} \vphi^* \mbox{D} \vphi \\
 \nonumber &\times\exp\left\{-\int_X m^2 (\varphi^*+\frac{h}{2m^2} (\psi^\dagger\vare\psi^*))(\vphi-\frac{h}{2m^2} (\psi^t \vare \psi))\right\}\\
 \nonumber &=\int \mbox{D} \vphi^* \mbox{D} \vphi \exp \Biggl\{-\int_X \biggl(m^2 \vphi^*\vphi\frac{}{}\\
 \label{bcs9} &\left.+ \frac{h}{2}\left(\vphi \psi^\dagger\vare\psi^*-\vphi^*\psi^t\vare\psi\right)-\frac{h^2}{4m^2}(\psi^\dagger\vare\psi^*)(\psi^t\vare\psi)\right)\Biggr\}.
\end{align}
Obviously, inserting this into the coherent state path integral for the fermions
\begin{equation}
 \label{bcs10} \int \mbox{D} \psi^* \mbox{D} \psi \,e^{-S[\psi^*,\psi]}
\end{equation}
is equivalent to having a theory with both fermions and bosons and action
\begin{align}
 \nonumber &S[\psi^*,\psi,\vphi^*,\vphi] = \int_X \left\{ \psi^\dagger\Bigl(\partial_\tau - \frac{\nabla^2}{2M}  - \mu\Bigr)\psi\right.\\
 \nonumber &+ m^2 \vphi^*\vphi + \frac{h}{2}\left(\vphi \psi^\dagger\vare\psi^*-\vphi^*\psi^t\vare\psi\right)\\
 \label{bcs11} &\left.-\frac{1}{4} \left(\lambda +\frac{h^2}{m^2}\right) (\psi^\dagger\vare\psi^*)(\psi^t\vare\psi)\right\}.
\end{align}

Before proceeding, we remark on the strategy of the outlined procedure. The rewriting of the purely fermionic theory in terms of a theory of both fermions and bosons is exact and did not involve any approximation. On the other hand, we obviously did not gain anything so far, because the functional integral still has to be evaluated. However, we need to recognize that we will essentially never be able to perform the full functional integral. Therefore, it is reasonable to reformulate the theory in such a way that already the leading order captures the physically most relevant phenomena. Based on physical insight, this is achieved by introducing the proper collective degrees of freedom. We found above that for attractive fermions, condensation of pairs should be the relevant mechanism. Our choice $\vphi_{\rm a}^* = \vphi_{\rm b} \sim \psi^t \vare\psi$ in Eq. (\ref{bcs8}) effectively substitutes $\vphi$ for $\psi^t\vare\psi = 2 \psi_1 \psi_2$ in the action. Thus, $\vphi$ is directly related to the \emph{pairing order parameter} $\langle \psi_1\psi_2\rangle$. A condensation of $\vphi$ will then yield a nonvanishing pairing correlation. Therefore, we can hope that already the introduction of $\vphi$ and a Gaussian approximation to the path integral can be sufficient to describe the theory.

We emphasize, however, that the choice of the bosonic degree of freedom, is not unique from a mathematical point of view. In fact, the ``wrong'' choice of $\vphi_{\rm a}$ and $\vphi_{\rm b}$ does not lead to a satisfying result. For example, if we had chosen $\vphi_{\rm a}$ and $\vphi_{\rm b}$ as hermitean bilinears, such that $\vphi \sim \psi_1^*\psi_2$, then this would describe well the features of a theory with particle-hole-pairing, but fail here poorly, because the instability occurs in the particle-particle-channel.

For the case of \emph{attractive} interactions, $\lambda <0$, the action in Eq. (\ref{bcs11}) can be simplified by choosing
\begin{equation}
 \label{bcs12} \lambda = - \frac{h^2}{m^2}.
\end{equation}
Thus, only the ratio $h^2/m^2$ is a physical quantity. In particular, the sign of $h$ is irrelevant. This can also be inferred from Eq. (\ref{bcs8}), because any phase of $h$ can be absorbed into a redefinition of $\vphi_{\rm a}$ and $\vphi_{\rm b}$. By a rescaling of the bosonic field according to $\vphi \rightarrow h \vphi$, we arrive at the microscopic action
\begin{align}
 \nonumber &S[\psi^*,\psi,\vphi^*,\vphi] = \int_X \left\{ \psi^\dagger\Bigl(\partial_\tau - \frac{\nabla^2}{2M}  - \mu\Bigr)\psi\right.\\
 \label{bcs13} &\left.+ \frac{1}{|\lambda|} \vphi^*\vphi + \frac{1}{2}\left(\vphi \psi^\dagger\vare\psi^*-\vphi^*\psi^t\vare\psi\right)\right\},
\end{align}
which only depends on the single parameter $\lambda$.

\vspace{5mm}
\begin{center}
 \emph{One-loop effective potential}
\end{center}

\noindent We perform the calculation of the one-loop effective potential analogously to the case of weakly interacting bosons in \ref{AppOneLoopBos}. Starting from the action in Eq. (\ref{bcs13}), we introduce \emph{Nambu spinors}
\begin{equation}
 \label{bcs14} \Psi = \left(\begin{array}{c} \psi_1\\ \psi_2^* \end{array}\right), \hspace{5mm} \Psi^\dagger = ( \psi_1^*, \psi_2).
\end{equation}
The action can then be expressed as
\begin{equation}
 \label{bcs15} S[\Psi^\dagger,\Psi,\vphi^*,\vphi] = -\frac{1}{\lambda} \int_X \vphi^* \vphi + \int_X \Psi^\dagger \left(\begin{array}{cc} P & \vphi \\  \vphi^* & P' \end{array}\right) \Psi,
\end{equation}
which is manifestly \emph{quadratic in the fermions}, with
\begin{equation}
 \label{bcs16} P = \partial_\tau - \frac{\nabla^2}{2M} -\mu, \hspace{5mm} P' = \partial_\tau + \frac{\nabla^2}{2M}+\mu.
\end{equation}
Using the generalization of Eq. (\ref{eff23}) for the effective action $\Gamma[\bar{\Psi}^\dagger,\bar{\Psi},\phi^*,\phi]$, we can now evaluate the Gaussian integral over the fermionic fields $\Psi^\dagger$ and $\Psi$. This results in a theory $\Gamma[\phi^*,\phi]$ of interacting bosons.

The BCS approximation consists of integrating out the fermions and \emph{neglecting the boson field fluctuations}. Therefore, it is a mean field approximation for the bosonic degrees of freedom. We choose a constant background field $\phi$, which is not yet evaluated at its equilibrium value. The second functional derivative of the action is found from Eq. (\ref{bcs15}) to be given by
\begin{align}
 \nonumber &S^{(2)}_{\Psi^\dagger\Psi}[\phi^*,\phi](Q',Q) = \frac{\stackrel{\rightarrow}{\delta}}{\delta \Psi^{\dagger}(Q')} S \frac{\stackrel{\leftarrow}{\delta}}{\delta \Psi(Q)}\\
 \label{bcs17} &= \delta(Q+Q') \left(\begin{array}{cc} \rmi \omega_n +\vare_q -\mu & \phi \\ \phi^* & \rmi \omega_n -\vare_q +\mu \end{array}\right).
\end{align}
The matrix appearing in this expression is the inverse classical propagator $G^{-1}_{\Psi^\dagger \Psi}(Q)$. For the effective potential $U = \Gamma/\beta V$ we obtain
\begin{align}
 \nonumber &U(\rho=\phi^*\phi) = -\frac{1}{\lambda} \phi^*\phi - \frac{1}{2} \mbox{Tr} \log S^{(2)}[\phi^*,\phi]\\
 \nonumber& = -\frac{1}{\lambda}\rho  - \int_Q \log \det G^{-1}_{\Psi^\dagger \Psi}(Q)\\ 
 \nonumber &= -\frac{1}{\lambda}\rho - \int_Q \log\Bigl(\omega_n^2 +(\vare_q-\mu)^2+\rho\Bigr)\\
 \label{bcs18} &= -\frac{1}{\lambda}\rho - 2T \int \frac{\mbox{d}^3q}{(2\pi)^3} \log \cosh \Bigl(\frac{E_q(\rho)}{2T}\Bigr).
\end{align}
We used $\sum_n \log(1+\frac{x^2}{\pi^2(n+1/2)^2})=2 \log \cosh x$ and dropped an overall constant in the last line, which is irrelevant for the thermodynamics. We also introduced
\begin{equation}
 \label{bcs19} E_q(\rho) = \sqrt{(\vare_q-\mu)^2+\Delta^2}
\end{equation}
with $\Delta^2=\rho$.

\vspace{5mm}
\begin{center}
 \emph{Excitation spectrum}
\end{center}

\noindent The order parameter $\rho_0(\mu,T)=\phi^*_0\phi_0$ is found from the gap equation
\begin{equation}
 \label{bcs20}0 = \frac{\partial U}{\partial \phi^*}(\phi_0) = \phi_0 \cdot U'(\rho_0).
\end{equation}
The three types of solutions to this equation have been discussed in Sec. \ref{SecSSB}. Given a nonvanishing order parameter $\rho_0 >0$ in the phase with spontaneously broken symmetry, the fermion excitation spectrum found from $\det G^{-1}_{\Psi^\dagger\Psi}(\omega_n=\rmi E_q,\vec{q})=0$ acquires a gap,
\begin{equation}
 \label{bcs20b} E_q = \sqrt{(\vare_q - \mu)^2 + \Delta_0^2}.
\end{equation}
Due to the appearance of the gap, even the excitation of fermions with momenta at the Fermi surface is suppressed. (Note that the fermion dispersion vanishes for $\vare = \vare_{\rm F}$ in the symmetric regime.) The absence of single particle fermion excitations at sufficiently low energies has an important implication: The lowest-lying modes are bosonic phonons with linear dispersion. This results in \emph{superfluidity} of the system according to Landau's criterion discussed in Sec. \ref{SecWeakBos}.

\begin{figure}[tb!]
 \centering
 \includegraphics[scale=0.78,keepaspectratio=true]{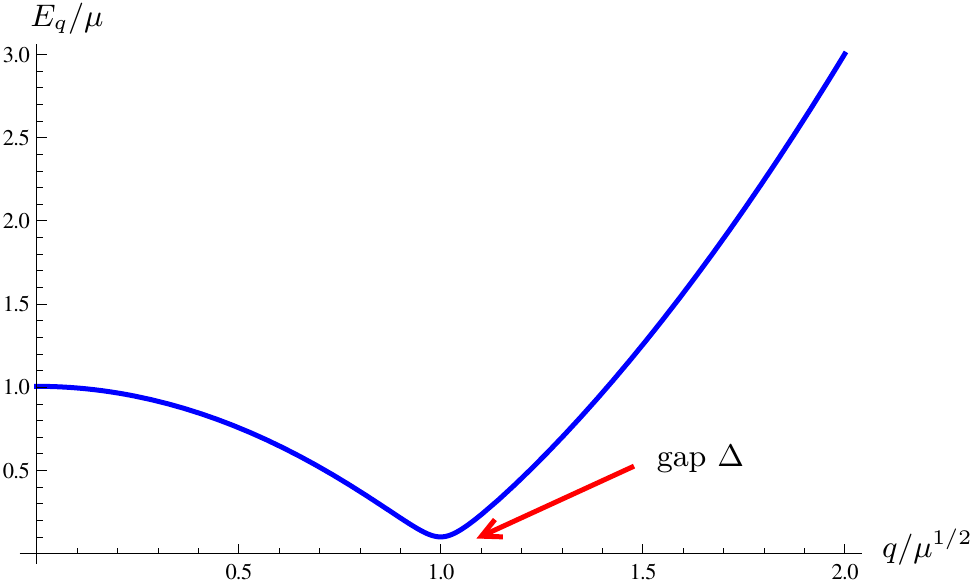}
\caption{We plot the function $E_q =\sqrt{(\vare_q-\mu)^2+\Delta^2}$ from Eq. (\ref{bcs19}) for a nonzero gap $\Delta$. For $\Delta =0$, the branch of the dispersion relation hits zero at $q=\sqrt{\mu}$ and, therefore, it is even at very low energies possible to excite particles with momenta at the Fermi surface. For $\Delta >0$, these excitations are costly and at very low energies or temperatures, we cannot excite them. In the BCS case, the gap is exponentially small.}
\label{BCSDispersion}
\end{figure}

\vspace{5mm}
\begin{center}
 \emph{Ultraviolet renormalization}
\end{center}

\begin{figure}[tb!]
 \centering
 \includegraphics[scale=0.75,keepaspectratio=true]{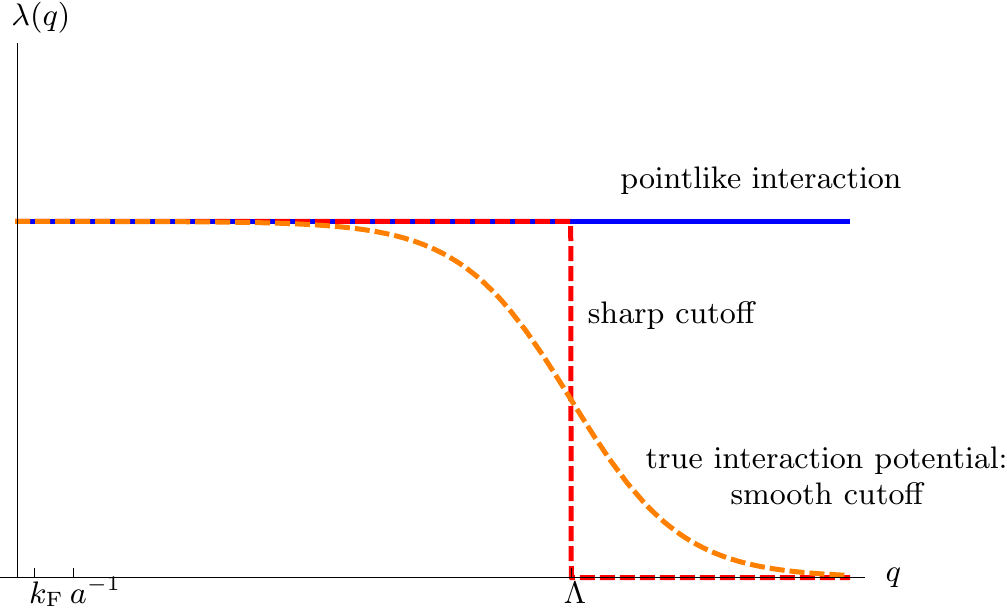}
\caption{The effective Hamiltonian in Eq. (\ref{bcs0}) assumes a pointlike interaction, which is valid for the scales $k_{\rm F},a^{-1},\lambda_T^{-1},\ell_{\rm osc}^{-1}$ we encounter in ultracold atom experiments. However, when calculating the effective action in perturbation theory, we are confronted with divergences when integrating over all momenta. These singularities arise because of momenta $\vec{q}^2 \gtrsim \Lambda^2$. At these scales, the microscopic details of the interatomic potential can be resolved and we cannot rely on a pointlike approximation, see Fig. \ref{FigVDW}. We cure the problem by observing that the true coupling $\lambda(q)$ is derived from a more realistic potential and falls off smoothly in the UV. This cannot be described in the pointlike approximation, but is taken into account by introducing a sharp cutoff at $\Lambda$.}
\label{FigUVcoupling}
\end{figure}

\noindent The onset of superfluidity is determined by the conditions $\rho_0 =0$ and $U'(\rho_0)=0$, which correspond to the critical point. Using Eq. (\ref{bcs18}) for the effective potential, we conclude from the gap equation that this particular point is given by 
\begin{align}
 \nonumber 0 &= - \frac{1}{\lambda} - \int \frac{\mbox{d}^3q}{(2\pi)^3} \frac{\partial E_q}{\partial \rho}\Bigr|_{\rho=0} \tanh\Bigl(\frac{E_q(\rho=0)}{2T_{\rm c}}\Bigr)\\
 \label{bcs20c} &= - \frac{1}{\lambda} - \frac{1}{4\pi^2} \int_0^\infty \mbox{d} q \frac{q^2}{\vare_q-\mu} \tanh \Bigl( \frac{\vare_q-\mu}{2T_{\rm c}}\Bigr).
\end{align}
The integral is linearly divergent as the integrand tends to unity for $q \rightarrow \infty$. This UV divergence is due to our simplifying model assumption that the interactions are pointlike and thus constant in momentum space for arbitrarily large momenta. In reality, the coupling $\lambda(q)$ is cut off smoothly at large momenta. However, as we have seen in Sec. \ref{scales}, the details of the interatomic potential are not essential for low energy collisions, which are solely determined by the scattering length. We visualize this situation in Fig. \ref{FigUVcoupling}.

We cure the divergence in Eq. (\ref{bcs20c}) by a proper \emph{UV renormalization}, where the goal is to trade the ``bare'' coupling constant $\lambda$ against a physically observable one, in this way eliminating the UV divergence. The procedure consists of two steps: (i)  We \emph{regularize} the momentum integral by the introduction of a sharp UV (high momentum) cutoff $\Lambda$. $\lambda$ is then interpreted as a bare coupling, which depends explicitly on the cutoff $\Lambda$. (ii) Next we perform the \emph{renormalization}. To this end, we observe that in experiments, we actually measure a renormalized coupling $\lambda_{\rm R}$ at low energies, which necessarily includes the effects of quantum fluctuations. Therefore, the bare coupling is not accessible to us and might have a very large or very small value. Expressing everything in terms of the renormalized coupling $\lambda_{\rm R}$, the cutoff $\Lambda$ will eventually drop out of the results. To compute $\lambda_{\rm R}$, we consider the vacuum situation where $T=\mu=\rho_0=0$ and the excitation of bosons is suppressed. The renormalized boson mass $U'(\rho_0)=m^2_\vphi$ should then be positive and equal to $-1/\lambda_{\rm R}$, see Eq. (\ref{bcs12}) (with $h$ absorbed into the fields). In the vacuum limit, we deduce from the explicit form of the effective potential in Eq. (\ref{bcs18}) the relation
\begin{equation}
 \label{bcs20d} -\frac{1}{\lambda_{\rm R}} \stackrel{!}{=} -\frac{1}{\lambda} - \frac{1}{4 \pi^2} \int_0^\Lambda\mbox{d} q \frac{q^2}{\vare_q} = - \frac{1}{\lambda} - \frac{M \Lambda}{2\pi^2}.
\end{equation}
Inserting this expression into Eq. (\ref{bcs20c}), we can send $\Lambda \rightarrow \infty$. 

This yields the \emph{renormalized gap equation} which determines the critical temperature $T_{\rm c}(\mu,\lambda_{\rm R})$,
\begin{align}
 \label{bcs20e}  0&= - \frac{1}{\lambda_{\rm R}} - \frac{2M}{4\pi^2} \int_0^\infty \mbox{d}q \left(\frac{\vare_q}{\vare_q-\mu}\tanh \biggl( \frac{\vare_q-\mu}{2T_{\rm c}}\biggr)-1\right).
\end{align}
We note a difference to the usual condensed matter argumentation for BCS theory in solids. There, the validity of the approximation of pointlike attraction is restricted to energies in the vicinity of the Fermi surface, where the attraction is mediated by phonons. Therefore, the cutoff (Debye frequency) is much closer to the Fermi surface, cf. e.g. \cite{altland-book}. Nevertheless, observables such as the critical temperature display the same functional dependencies, as they are dominated by effects very close to the Fermi surface as we will see next.

\vspace{5mm}
\begin{center}
 \emph{Critical temperature and paired state}
\end{center}

\noindent We already mentioned that the weak interactions do not
significantly modify the Fermi surface and we can assume $\mu =
\vare_{\rm F}$. This will also be justified below by a direct
computation of the equation of state. Rescaling the momenta in
Eq. (\ref{bcs20e}) with the Fermi momentum, $\hat{q} = q/k_{\rm F}$,
and using $\vare_{\rm F} = T_{\rm F} = k_{\rm F}^2/2M$, we arrive at
the condition determining the critical temperature $T_{\rm c}$
\begin{equation}
  \label{bcs20f} 0 = -\frac{1}{k_{\rm F}a} - \frac{2}{\pi}\int_0^\infty 
  \mbox{d} \hat{q} \left(\frac{\hat{q}^2}{\hat{q}^2-1}\tanh \biggl(
    \frac{\hat{q}^2-1}{2 T_{\rm c}/T_{\rm F}}\biggr)-1\right).
\end{equation}
We introduced the scattering length for distinguishable fermions
according to $\lambda_{\rm R} = 4 \pi a/M<0$. For $T_{\rm c}/T_{\rm
  F}\rightarrow 0$, the integral develops a logarithmic singularity at
the Fermi surface, where $\hat{q}=1$. Thus, we obtain a solution of
Eq. (\ref{bcs20f}) for arbitrarily weak interactions $a$. The
corresponding BCS critical temperature is
\begin{equation}
 \label{bcs20g} \frac{T_{\rm c}}{T_{\rm F}} = \frac{ 8 e^{\gamma-2}}{\pi} e^{-\frac{\pi}{2|k_{\rm F}a|}}.  
\end{equation}
Here, $\gamma=0.577$ is Euler's constant such that the overall
prefactor becomes $0.61$. Note that $T_{\rm c}/T_{\rm F}$ is
exponentially small.

For $T < T_{\rm c}$, a gap $\rho_0 >0$ develops, which cures the
logarithmic divergence. In particular, we find from the gap equation
at zero temperature
\begin{equation}
 \label{bcs20h} \frac{\rho_0(T=0)}{T_{\rm c}} = \frac{\pi}{e^\gamma} = 0.57.
\end{equation}

In the spirit of Landau's theory of second order phase transitions, we
expand the effective potential (i.e. the grand canonical potential)
around its minimum $\rho_0$. Since the effective potential depends on
the renormalized quantities including all fluctuations, we have
\begin{align}
 \label{bcs23} U(\rho,\mu,T) &=-P(\mu,T) + m_{\phi,\rm R}^2 (\rho - \rho_0) + \frac{u_{\phi,\rm R}}{2}  (\rho-\rho_0)^2.
\end{align}
In the symmetric phase, we have $\rho_0 =0$ and, accordingly,
$m_{\phi,\rm R}^2 =\frac{1}{|\lambda_{\rm R}|}>0$. The order parameter
$\rho_0$ acquires a nonzero value when either $m_{\phi,\rm R}^2$
becomes zero or $\lambda_{\rm R}$ diverges. The fact that the ``mass'' $m_{\phi,\rm R}^2$
of the bosons  becomes zero when going from the
symmetric to the broken phase will later be important in our
renormalization group study. In particular, we will employ a
truncation of the effective potential which closely resembles the
shape of Eq. (\ref{bcs23}).

Let us finally comment on the relation to a ``purely fermionic''
approach which does not introduce bosonic degrees of freedom. To this
end, we compare Eqs. (\ref{bcs20d}) and (\ref{bcs20e}). We have
interpreted the left hand side of Eq. (\ref{bcs20d}) as the inverse vacuum interaction
parameter, which includes the effects of vacuum quantum fluctuations
-- in an effective vacuum action, it would appear as the renormalized
four-fermion (two-body) interaction term
$\tfrac{\lambda_R}{2}(\psi^\dag\psi)^2$. In full analogy, we can
interpret the left hand side of Eq. (\ref{bcs20e}) as the inverse ``many-body''
interaction parameter $1/\lambda_{\text{mb}}(T,\mu)$, which \emph{in
  addition} to the vacuum quantum fluctuations also includes many-body
fluctuation effects related to the many-body scales $\ell^{-1} =
k_\text{F} \sim \sqrt{\mu}, \lambda_T \sim \sqrt{T}$. (Note that
$\lambda_{\text{mb}}(T=0,\mu=0) = \lambda_\text{R}$.) The zero of the inverse many-body interaction strength, i.e. the divergence of the many-body interaction vertex, 
signals the onset of new physics, more precisely an instability
towards a state which needs a qualitative modification of the
theoretical approach for its proper description (such as a nontrivial
minimum in its effective potential). If this phase is understood
qualitatively, and supposed to be described quantitatively, the
approach via Hubbard--Stratonovich decoupling is preferable. On the
other hand, the purely fermionic approach is less biased and may serve
as a guide to identify the proper decoupling channel. We emphasize
that the possibility of reducing the question of stability to a single
parameter is remarkable in a many-body context: The full four-fermion
vertex of our problem is, in principle, a tensor $\lambda_{Q_1, ... ,
  Q_4}$ depending on four 4-momenta. This highly nontrivial reduction
of complexity in the vicinity of the Fermi surface has been shown by
Shankar \cite{shankar91}, and is reviewed in \cite{RevModPhys.66.129}.

\vspace{5mm}
\begin{center}
 \emph{Equation of state}
\end{center}

\noindent Besides the phase structure, we also obtain the
thermodynamics from the one-loop effective
potential. Eq. (\ref{bcs18}) implies for the density
\begin{align}
 \nonumber n(\mu,T) &= - \frac{\partial U}{\partial \mu}(\rho_0)= 
-2 \int \frac{\mbox{d}^3q}{(2\pi)^3} \tanh \Bigl( \frac{E_q}{2T}\Bigr)\\
\label{bcs20i} &= 2 \left( \int \frac{\mbox{d}^3q}{(2\pi)^3}
  \frac{1}{e^{E_q/T}+1} - \int \frac{\mbox{d}^3q}{(2\pi)^3}
  \frac{1}{2}\right).
\end{align}
In the second line, we have split up the integral into a physical
contribution and an artefact from the functional integral. It accounts
for the relation between field expectation values and operator
expectation values $\langle \delta \psi^*\delta \psi \rangle = \langle
\hat{a}^\dagger \hat{a}\rangle - \frac{1}{2}$ for each momentum
mode. (We refer to \ref{AppOneLoopBos} and Eq. (\ref{bog28b}) for a
more detailed discussion.) We thus arrive at
\begin{equation}
 \label{bcs20j} n(\mu,T) = 2 \int \frac{\mbox{d}^3q}{(2\pi)^3} \frac{1}{e^{E_q/T}+1}.
\end{equation}

From Eq. (\ref{bcs20j}) we find our above considerations to be
consistent: The Fermi surface is clearly expressed for small
temperatures and an at best exponentially small gap or order parameter
identified above, cf. Eqs. (\ref{bcs20g}) and (\ref{bcs20h}), and we
have $\mu \simeq \vare_{\rm F}$. Despite its smallness, the nonzero
gap of the dispersion relation around the Fermi energy results in an
important qualitative effect, namely the superfluidity of the system.

\vspace{5mm}
\begin{center}
 \emph{Validity and experimental (ir)relevance of BCS superfluidity} 
\end{center}

\noindent We close this section by an estimate of the temperature
regime which has to be reached in order to observe BCS superfluidity
in cold atom experiments. We found in Eq. (\ref{bcs20g}) that for a
given density (i.e. Fermi momentum $k_{\rm F}(n)$) we have
\begin{equation}
 \label{bcs24} \frac{T_{\rm c}}{T_{\rm F}(n)} = 0.61 e^{-\pi/2|k_{\rm F}a|}.
\end{equation}
Due to $|k_{\rm F}a| \ll 1$, we have an exponential suppression of the
critical temperature. To experimentally reach this exponentially low temperature
regime is difficult. 

We compare to a bosonic system. For particle number density
$n$ and boson density $n_{\rm B}$, we can formally define a Fermi
temperature via $T_{\rm F} = (6 \pi^2n/g_{\rm F})^{2/3}/2M$. Together with
the critical temperature for Bose condensation found from $g_{\rm
  B}\zeta(3/2)=n_{\rm B}\lambda_{T_{\rm c}}^3 = n_{\rm B}(2\pi/M_{\rm
  B}T_{\rm c})^{3/2}$, we get
\begin{align}
 \nonumber \frac{T_{\rm c}}{T_{\rm F}(n)} &= \frac{4\pi M}{M_{\rm B}}
 \biggl(\frac{n_{\rm B}/n}{3 \pi^2 \zeta(3/2)}\biggr)^{2/3} \\
 \label{bcs25}&= \begin{cases} 0.692 & (n_{\rm B}=n, M_{\rm B}=M) \\
   0.218 & (n_{\rm B} = n/2, M_{\rm B}=2 M)\end{cases},
\end{align}
where we assumed boson and fermion degeneracy to be $g_{\rm B}=1$ and
$g_{\rm F}=2$, respectively. The second case corresponds to a gas of
composite bosons made of two fermions each. Since the ratio in
Eq. (\ref{bcs25}) is of order unity, it is easier (yet nontrivial) to access bosonic
superfluidity in ultracold quantum gases. In addition, due to Pauli
blocking, the cooling of degenerate fermions is more challenging than
for bosons.

From Eq. (\ref{bcs24}) we see that there is an exponential increase in
$T_{\rm c}$ for rising $k_{\rm F}a$, i.e. towards strong
interactions. We may ask how this trend continues, and if it can help
to achieve fermion pair condensation. More generally, one can imagine
to tune the interaction parameter $k_\text{F}a$ entering
Eq. (\ref{bcs20f}) from negative values through a resonance to
positive ones. Such a knob indeed exists in cold atom experiment
thanks to Feshbach resonances. Not unexpectedly, the critical
temperature doe not rise indefinitely when cranking up the
interactions; in addition, for large positive values of $k_\text{F}a$
we will smoothly approach the limit of weakly interacting pointlike
bosons. This is the BCS-BEC crossover. However, the one-loop formula
for the effective action used so far does not capture the crossover
problem, because it relies on weak coupling and small amplitudes of
the field. A more complete description is needed.

\section{Strong correlations and the Functional Renormalization Group}
\label{Sec4}
\noindent A full, quantitative grip on the strongly-correlated physics
discussed in the previous chapter requires the use of nonperturbative
techniques. Various methods have been applied to the physics of
ultracold gases, ranging from quantum Monte-Carlo and diagrammatic
Monte-Carlo Simulations to functional methods and resummation schemes
such as Dyson--Schwinger and Kadanoff--Baym equations as well as the
Functional Renormalization Group (FRG). The latter allows to access
the whole phase diagram of ultracold gases in a unified approach, and,
in particular, is also applicable in strongly correlated or strongly
coupled regimes.  Its setup and applications are described in the
present chapter. This work does not aim at a fully detailed
introduction to the FRG. This has been done in various topical and
general reviews and lecture notes, and we refer the interested reader
to introductory and advanced general reviews
\cite{Aoki:2000wm,Bagnuls:2000ae,Berges:2000ew,Salmhofer:2001tr,%
  Polonyi:2001se,Pawlowski:2005xe,Delamotte:2007pf,Rosten:2010vm,%
  Braun:2011pp}, and to more topical reviews on nuclear and atomic
physics
\cite{Blaizot:2008xx,Diehl:2009ma,Scherer:2010sv,2010arXiv1012.4914B,%
Floerchinger:2011yv,RevModPhys.84.299,2012arXiv1201.2510F,2012arXiv1203.1779F},
non-equilibrium RG \cite{2009EPJST.168..179S,2011arXiv1112.1375T,Berges:2012ty}, on gauge
theories, QCD and QCD effective models,
\cite{Litim:1998nf,Gies:2006wv,Schaefer:2006sr}, and quantum gravity
\cite{Litim:2008tt,Percacci:2007sz,Reuter:2012id}.

\subsection{Flow equation}\label{sec:flow}
\noindent We begin with an introduction of the basic concepts of the
FRG. It is based on the continuum version of Kadanoff's block-spinning
transformations on the lattice \cite{Kadanoff:1966}, and has been
formulated for the continuum by Wilson
\cite{Wilson:1971bg,Wilson:1971dh}.  Its modern functional form for
the effective action used in the present work has been put forward by
Wetterich in \cite{Wetterich:1992yh}.

For the description of ultracold atom experiments, the action $S$
derived from the Hamiltonian in Eq. (\ref{sc13b}) is a microscopic
starting point. It is related to an ultraviolet momentum scale
$\Lambda \gg \ell_{\rm vdW}^{-1}$.  The relevant physics, however,
takes place at momentum scales $k$ far smaller than $\Lambda$, and the
respective quantum and thermal fluctuations have to be included. In
the FRG framework, these fluctuations are included successively at a
given momentum scale $k$ starting at $\Lambda$ with the action
$\Gamma_\Lambda=S$, leading to an effective action $\Gamma_k$. The
latter already includes all quantum and thermal fluctuations above the
momentum scale $k$. It can be interpreted as a microscopic action for
the physics below the scale $k$ in the very same way $S$ has been
introduced as the microscopic action of ultracold gases. After the
inclusion of all fluctuations we arrive at the full effective action
$\Gamma$,
\begin{align}
 \label{FRG2} \Gamma_{k=\Lambda} &= S,\\
 \label{FRG3} \Gamma_{k=0} &= \Gamma.
\end{align}
The effective action $\Gamma_k$ interpolates smoothly between the
microscopic (or initial effective) action $\Gamma_\Lambda$ and the
full effective action $\Gamma=\Gamma_{k=0}$. An infinitesimal change
of the effective action with the scale $k$ is described by a \emph{flow
equation} for $\partial_k \Gamma_k$, which depends on the correlation
functions of the theory at the scale $k$ as well as the specific way
the infrared modes with momenta smaller than $k$ are suppressed. Such
an RG-step has similarities to a coarse graining where details on short
distances are continuously washed out, the difference being that the
effective action $\Gamma_k$ still keeps the information about the
fluctuations between $\Lambda$ and $k$. At the end of the process, for
$k\rightarrow 0$, we include fluctuations with large wavelength. These
are the problematic modes which cause infrared divergences in other
approaches. Due to the stepwise inclusion of fluctuations, the
renormalization group procedure is not plagued by such divergences. In
conclusion, a given initial effective action $\Gamma_\Lambda$ and the
flow equation Eq.~(\ref{FRG1}) define the full quantum theory
analogously to the setting with classical action and the path integral.

In this section, we derive the flow equation for $\Gamma_k$ and discuss
its practical solution. To that end, we specify a suppression
of low momentum fluctuations $\omega,\vec{q}^2 \leq k^2$. This is most
easily achieved via a mass-like infrared modification of the
dispersion relation, while the ultraviolet modes should remain
unchanged: we add a \emph{regulator} or \emph{cutoff} term $\Delta
S_k[\vphi]$ to the microscopic action $S[\vphi]$ which is quadratic in
the fields, 
\begin{equation}
 \label{FRG4} S[\vphi] \rightarrow S[\vphi] + \Delta S_k[\vphi].
\end{equation}
The field $\vphi$ is general and may be a collection of fields. For
concreteness, we will use a notation analogous to ultracold bosons and
write $\phi(X) = \langle \vphi(X)\rangle$. We have 
\begin{align}
 \label{FRG5} \Delta S_k[\vphi]&= \int_Q \vphi^*(Q) \,R_k(Q)\, \vphi(Q).
\end{align}
The requirement of the suppression of low momentum modes entails that $R_k(Q\rightarrow 0) \neq 0$. 
In turn, for large momenta (in comparison to $k$), the regulator has to
vanish,  $R_k(Q\to\infty)\to 0$. These properties can be summarized in the conditions
\begin{align}
  \label{FRG6J} \lim_{\vec q^2/k^2\to 0} R_k(Q)=k^2\,, \quad
  \lim_{\vec q^2/k^2 \to\infty} R_k(Q)=0\,.
\end{align}
For the sake of simplicity, we have restricted ourselves in
Eq.~(\ref{FRG6J}) to regulators that only depend on $\vec q$ and have a
standard normalization $R_k(0)=k^2$ in the infrared. The extension to
general regulators is straightforward.

If we interpret the action in Eq. (\ref{FRG4}) as the microscopic action
of a theory, it has a trivial infrared sector: The fields are gapped
with gap $k^2$. The generating functional of this theory is given by
\begin{equation}\label{FRGZk}
Z_k[j] = \int \mbox{D} \vphi \,e^{-S-\Delta S_k + \int
  j\cdot \vphi}\,. 
\end{equation}
From Eq.~(\ref{FRG6J}), we infer that $Z[j]=Z_{k=0}[j]$ is the full
generating functional of the theory introduced in Eq.~(\ref{fun1}). For
$k\to\Lambda$, the regulator term dominates the path
integral as all physical scales are far smaller and we are left with a
trivial Gaussian integral. Moreover, for a given $k$, the correlation
functions $\langle \varphi(Q_1)\cdots \varphi(Q_n)\rangle$ tend
towards the full correlation functions for $Q_i^2\gg k^2$ for all
$i=1,...,n$. In turn, for $Q_i^2\ll k^2$, the correlation functions are
trivial, as the fields are gapped.

For explicit computations, it is more convenient to deal with the
effective action $\Gamma_k$, which is obtained via a modified Legendre
transform according to
\begin{equation}\label{Gk}
\Gamma_k[\phi] = \int j \cdot \phi -
\log Z_k[j] - \Delta S_k[\phi]\,,
\end{equation}
where $j=j_k[\phi]$ satisfies $(\delta \log Z_k/\delta j)[j] =\phi$. We have
already shown that the effective action has the simple physical
interpretation of a free energy in a given background
$\phi$. Diagrammatically, it generates all one-particle irreducible
diagrams. As in the case without regulator term, $\Gamma_k$ satisfies
a functional integro-differential equation similar to Eqs. \eq{eff23},
\eq{Func18}. Applying the definitions of $Z_k$ and $\Gamma_k$ we find
\begin{align} 
  \nonumber &\exp\{-\Gamma_k[\phi]\} \\
  \label{FRG10} &=\int \mbox{D} \vphi\, \exp\left\{-S[\phi+\vphi]
    -\Delta S_k[\vphi]+\int_X\,\frac{\delta\Gamma_k}{\delta\phi}\cdot
    \vphi\right\},
\end{align}
where we have used the condensed notation introduced in Eq. \eq{bog2e}, as well as 
\begin{equation} \label{J}
j[\phi]=\frac{\delta (\Gamma_k+\Delta S_k)}{\delta\phi}\,,
\end{equation} 
following from the definition of the Legendre transform \eq{Gk}. 
\Eq{FRG10} makes the suppression of the fluctuations even more
apparent. Note first that the action $S$ in the exponent depends on
the sum $\phi+\varphi$, whereas the cutoff term only depends on the
fluctuation $\varphi$. Hence, for large cutoff scales
$k\to\Lambda$, the functional integral in Eq. \eq{FRG10} gets
Gaussian and the effective action tends towards the microscopic
action, $\Gamma_{k\to\Lambda}\to S$. For $k\to 0$, the
regulator vanishes, $R_k\to 0$, and we are left with Eq. \eq{Func18}.

For a successive integration of momentum modes we need to know the 
flow $\partial_k \Gamma_k$. Applying the $k$-derivative to Eq. \eq{Gk} leads to 
\begin{equation}\label{dkGk}
\partial_k \Gamma_k[\phi] = -
 \partial_k\bigr|_j \log Z_k[j] - \partial_k \Delta S_k[\phi].
\end{equation}
The notation signals that $j$ is
$k$-dependent, but the terms proportional to $\partial_k j$ cancel. We have
$\partial_k \Delta S_k[\phi] = \int_Q \partial_k R_k(Q) \phi(Q)\phi^*(Q)$. The
generating functional $Z_k$ only depends on $k$ via the cutoff term
$\Delta S_k$. Taking the $k$-derivative of Eq. \eq{FRGZk}, we can compute $\partial_k|_j \log Z_k$ to arrive at
\begin{equation}\label{dkGk1}
  \partial_k \Gamma_k[\phi] =\int_Q \partial_k R_k(Q)\left[\Bigl\langle  
\vphi(Q)  \vphi^*(Q) \Bigr\rangle_k-\phi(Q)  \phi^*(Q)\right].
\end{equation}
Herein, we have restricted ourselves to bosonic fields $\varphi$.
In the case of fermions, a global minus sign occurs due to the
Grassmann nature of the fermions. The expression in the square bracket
in Eq. \eq{dkGk1} is the full, field-dependent propagator, which reads in
terms of the effective action 
\begin{equation}
 \label{FRG6}\Bigl\langle  
 \vphi(Q)  \vphi^*(Q) \Bigr\rangle_k-\phi(Q)  \phi^*(Q)=  
\frac{1}{\Gamma^{(2)}_k+R_k}(Q,Q)\,.
\end{equation}
In Eq. \eq{FRG6}, we have used the property of Legendre transforms that
the second derivatives of a functional and its Legendre transform are
inversely related. In the present case, we note that the Legendre transform of $\log Z_k$
is $\Gamma_k+\Delta S_k$, as defined in Eq. \eq{Gk}. Hence, we are led schematically to 
$\delta^2 \log Z_k/\delta
j^2\cdot (\Gamma^{(2)}+R_k)=1$, which we have used in Eq. \eq{FRG6}. 

The momentum integral in Eq. \eq{dkGk1} can be conveniently written in
terms of a trace. Including also the possibility of internal indices
and different species of fields, we are led to the final expression for
the flow equation for $\Gamma_k$,
\begin{equation}
 \label{FRG1} \partial_t \Gamma_k = \frac{1}{2} \mbox{STr}\, 
\left[\frac{1}{\Gamma_k^{(2)}+R_k}\, \partial_t R_k\right]\,, 
\end{equation}
the Wetterich equation. The supertrace includes the momentum
integration, and a summation over internal indices and field species,
see also Eq. (\ref{Func20}). In Eq. \eq{FRG1} we have introduced the
RG-time $t=\log{k/k_0}$ with some reference scale $k_0$, typically
being either the ultraviolet scale, $k_0=\Lambda$, or some physical
scale. For a given quantity $\mathcal{O}_k$, the logarithmic scale derivative 
$\partial_t \mathcal{O}_k = k \partial_k \mathcal{O}_k$ has the same properties under RG-scaling
as the quantity itself. It also is
convenient as one usually integrates the flow over several orders of
magnitude in the momentum scale $k$. Henceforth we shall use the
standard choice $t=\log k/\Lambda$.

Above, we have argued that regulators with the properties \eq{FRG6J}
lead to a suppression of the infrared physics of the theory. Moreover,
since the finite initial effective action $\Gamma_\Lambda$ at the
initial scale $\Lambda$ already includes all fluctuations of momentum
modes with momenta larger than $\Lambda$, no ultraviolet divergences
should be present. These properties have to be reflected in the flow
equation \eq{FRG1}: It has to be both infrared {\it and} ultraviolet
finite. Here, we show this explicitly for the case of bosonic fields.
For low momenta, the regulator adds a positive mass to
$\Gamma_k^{(2)}$ in the denominator. The typical size of this mass is
$k^2$, in Eq. \eq{FRG6J} we have normalized it to $k^2$. For the sake
of simplicity, consider a classical dispersion $\Gamma_k^{(2)}\simeq
\rmi\omega+\vec q^2$ (with $2M=1$) for small momenta which tends to
zero for vanishing momentum. Schematically, we have for small momenta
\begin{equation}
 \label{FRG7} \frac{1}{\Gamma^{(2)}_k(Q)+R_k(Q)} \stackrel{\vec q^2\to 0}{
\longrightarrow} 
\frac{1}{\rmi \omega+\vec q^2+k^2},
\end{equation}
which is finite for $Q \rightarrow 0$. For fermions, the
infrared singularities arise close to the Fermi surface. Accordingly,
the propagators have to be regularized there. In summary, this implies
infrared safe flows. 

\begin{figure}[tb!]
 \centering
 \includegraphics[width=.9\columnwidth,keepaspectratio=true]{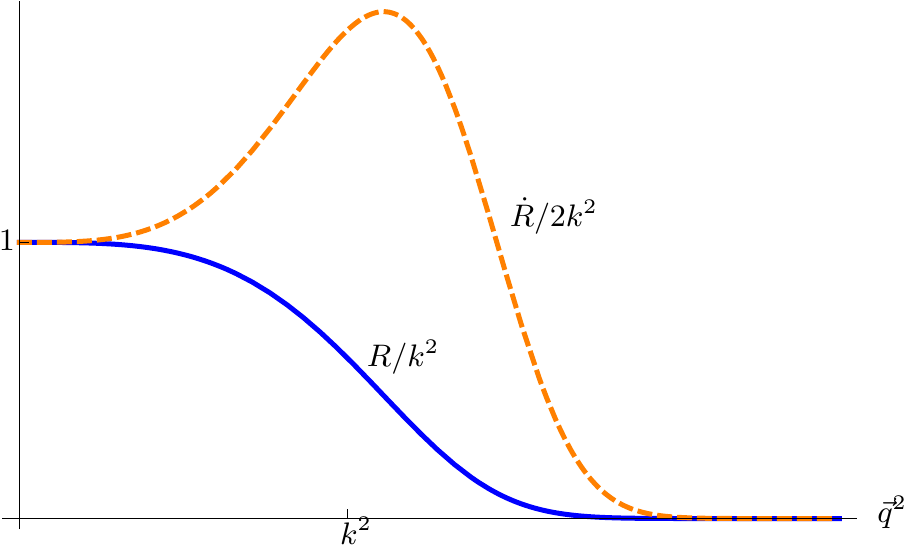}
 \caption{We plot a typical cutoff function $R_k(Q) = R_k(\vec{q}^2)$,
   which only depends on the spatial momentum. The function is nonzero
   for $ \vec{q}^2 \lesssim k^2$ and thus provides an infrared cutoff
   for the propagators. For large momenta, it falls off rapidly, thus
   becoming inactive in the UV. The scale derivative $\dot{R}_k(Q) =
   k \partial_k R_k(Q)$ is sharply peaked at $\vec{q}^2 \approx
   k^2$. For this reason, the loop integral on the right hand side of
   the flow equation is dominated by these modes. This provides the
   mechanism how momentum shells are successively integrated out in
   the FRG framework.}
\label{CutoffPlot}
\end{figure}

In turn, for large momenta, the scale-derivative $\partial_t R_k(Q)$
vanishes due to Eq. \eq{FRG6J}. If this happens sufficiently fast, 
\begin{equation}\label{UVsafe} 
\lim_{\vec q^2/k^2\to\infty} \vec q^2\partial_t R_k(Q)\to 0\,, 
\end{equation}
the momentum integral in Eq. \eq{FRG1} is finite. In the following, we
shall show results for regulators that satisfy Eq. \eq{UVsafe}. We also
remark that mass-like regulators, i.e.\ $R_k=k^2$, do not satisfy
Eq. \eq{UVsafe} and hence require UV renormalization. The related flows
are functional Callan--Symanzik equations as first derived in
\cite{Symanzik:1970rt}. They are sometimes used due to computational
simplicity, see e.g. \cite{Diehl:2009ma}. The generic shape of a cutoff
is shown in Fig. \ref{CutoffPlot}.
    
\begin{figure}[tb!]
 \centering
 \includegraphics[width=.8\columnwidth,keepaspectratio=true]{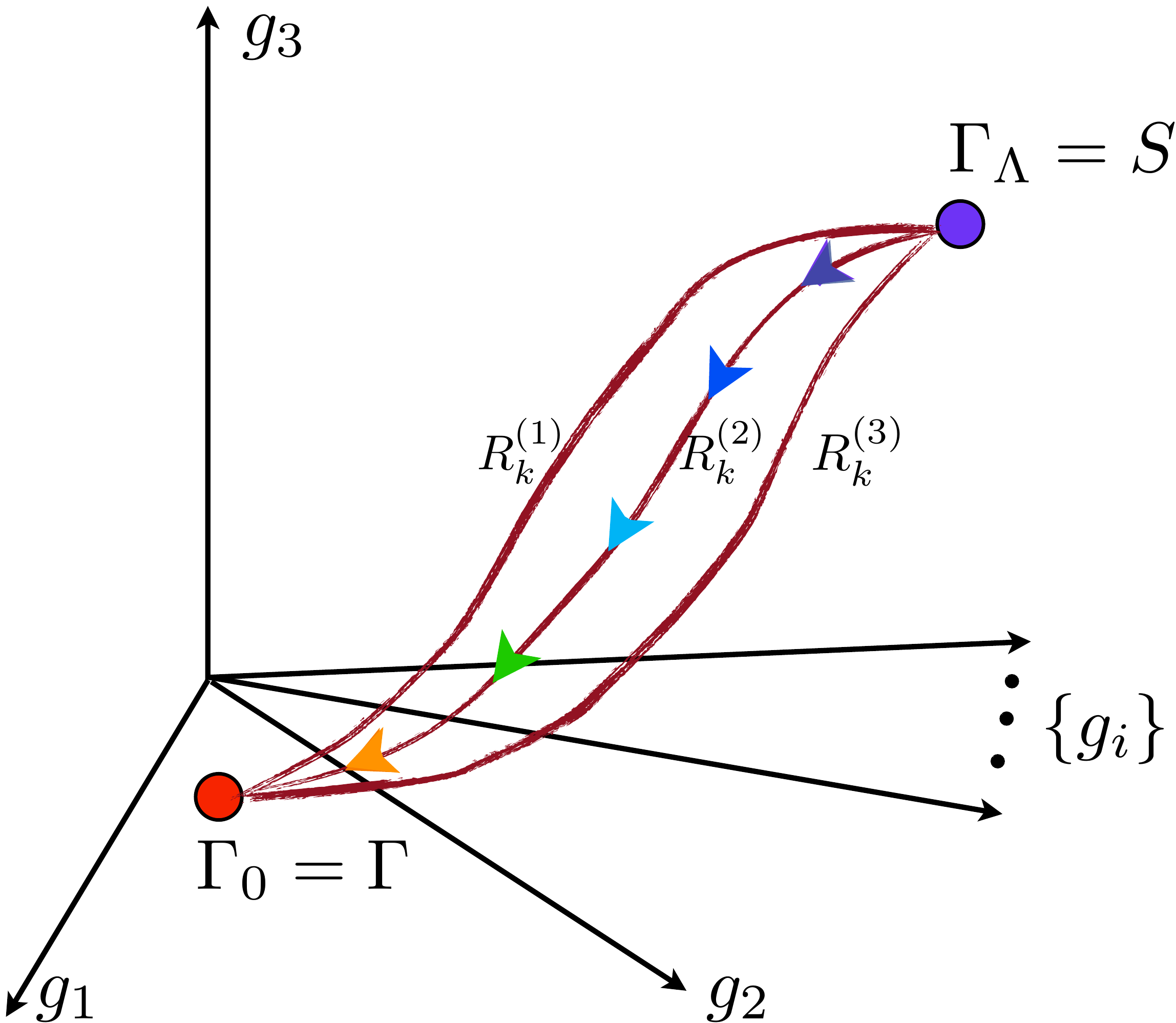}
 \caption{The flow of $\Gamma_k$ connects the microscopic action to
   the effective action in the \emph{theory space} of all possible
   action functionals. The latter is of infinite dimension since the
   effective action is characterized by an infinite set of couplings
   (or correlation functions). This is indicated here schematically by
   the couplings $g_1, g_2, g_3$ and $\{g_i\}$. For different choices
   of the regulator $R_k^{(i)}$, the trajectories in theory space
   differ, as is indicated in the figure. At $k=0$, however, the
   particular paths merge again and eventually arrive at the full
   effective action.}
\label{ThSpaceCutoffs}
\end{figure}

It is apparent from the derivation that $\Gamma_k[\phi]$ depends on
the shape of the regulator. This regulator-dependence disappears for
$k\to 0$, hence physical observables are independent of the choice of
$R_k$, but the trajectory $\Gamma_k$ from $k=\Lambda$ to $k=0$ depends
on $R_k$, see. Fig. \ref{ThSpaceCutoffs}.  This leaves us with some freedom for the choice of
the regulator. Indeed, its choice can be optimized to the
approximation under investigation
\cite{Litim:2000ci,Litim:2001up,Litim:2001fd,Pawlowski:2005xe}. In general, such
a choice is further guided by computational simplicity, as in
complicated systems the computational costs can be high. 
Typical choices are functions $R_k(Q)$ which decay
exponentially or even vanish identically for high $Q$. A slight
complication for nonrelativistic system is provided by the fact that
frequencies and spatial momenta appear differently. The Galilei
symmetric combination is given by $\rmi \omega + \vec{q}^2$, in contrast to 
the ${\rm O}(4)$-symmetric combination $(q_\mu)^2 = \omega^2+\vec{q}^2$ for
relativistic systems. At nonvanishing temperature, Galilei
symmetry is broken. In the Matsubara formulation used in the present
work, the coupling to the heat bath leads to periodicity in the imaginary
time $\tau$ with period $\beta$. Therefore, we may also choose
a regulator which only depends on frequency or momentum
space. Moreover, we may sacrifice Galilei symmetry in order to obtain
simpler expressions for the flow equation.

Here, we briefly discuss some common choices and their pros and cons.
This should give the reader the chance to embark on first computations
on their own while being aware of the advantages and limitations of
the choices. A detailed discussion, however, is beyond the scope of the present work. 

For instance, exponential cutoffs for ultracold bosons are given by
\begin{equation}
 \label{FRG11} R_k(Q) = \frac{k^2}{e^{(\vec{q}^2/k^2)^n}+1}, 
\qquad \frac{k^2}{e^{(\omega/k^2)^n}+1}.
\end{equation}
The power of $n$ can be chosen such that the cutoff falls off
sufficiently fast for high $Q$. We remark that despite its exponential
nature, one has to choose $n>1$ in order to have flows which are local
in momentum space, see \cite{Fister:2011uw}. This is mirrored in the
property that for $n=1$ the cutoff insertion $\partial_t R_k(\vec
q^2)$ is not peaked at about $\vec q^2\approx k^2$ but is a 
monotonously decaying function. Only for larger $n$ do we get peaked
cutoff insertions. We also remark that rapid decay of $R_k$ as a
function of $\omega$ poses problems for computing thermodynamic
quantities, as they lead to oscillations in the flow. These
oscillations mirror the property that the periodicity in $\beta$, even
though present for smooth regulators, is lost for a sharp cutoff in
$\omega$. 

A particularly useful cutoff is the
three-dimensional Litim cutoff \cite{Litim:2001up,Litim:2006ag}, 
\begin{equation}
 \label{FRG12} R_k(Q) = (k^2 - \vec{q}^2) \theta(k^2-\vec{q}^2),
\end{equation}
which effectively reduces the momentum integral to $\vec{q}^2\leq k^2$
and replaces $\vec{q}^2$ by $k^2$. It facilitates the analytic
derivation of flow equations for correlation functions in the
derivative expansion, and hence leads to important computational
simplifications. Its analytic property also allows an easy access to
the structure and interrelation of the flows (and hence the
correlation functions). These properties make it the standard
choice within (lower orders of) the derivative expansion. Moreover, in
three-dimensional theories, the cutoff in Eq. (\ref{FRG12}) provides an
optimal choice \cite{Litim:2000ci,Pawlowski:2005xe} within the lowest
order of the derivative expansion scheme as will be employed below. In
$(3+1)$-dimensional theories, it still shares some of the stability
features it has in three dimensions \cite{Litim:2006ag}, but loses
locality in momentum space necessary for full quantitative precision
\cite{Fister:2011uw}. 

 A manifestly Galilei symmetric regulator is provided by \cite{Floerchinger:2011sc},
\begin{equation}
 \label{FRG13} R_k(Q) = \frac{k^2}{1 + c\bigl(\frac{\rmi \omega +\vec{q}^2}{k^2}\bigr)^n},
\end{equation}
where $n$ determines the algebraic decay for large momenta and $c$ is
a prefactor of order unity.  \Eq{FRG13} can be extended to more
general rational functions in the Galilei invariant $ \rmi \omega
+\vec{q}^2$. Its key advantages are its Galilei invariance as well
as its analytic structure. The latter allows to continue the results to
real time, and hence may give access to transport properties or more
generally dynamics of ultracold gases. Similar choices in relativistic
theories can be used for computing decay properties
\cite{Floerchinger:2011sc}.

Note that the flow equation (\ref{FRG1}) has a one-loop structure,
which can be traced back to the quadratic form of the regulator in
Eq. (\ref{FRG5}). Indeed, we may rewrite Eq. \eq{FRG1} as 
\begin{equation}
 \label{FRG14} \partial_t \Gamma_k[\phi] = \frac{1}{2} \mbox{STr} \,\, 
 \tilde{\partial}_t \log\, \Bigl(\Gamma^{(2)}_k[\phi]+R_k\Bigr)\,,
\end{equation}
where the derivative $\tilde{\partial}_t$ only acts on the $k$-dependence of the
regulator, i.e. we have
\begin{equation}
 \label{FRG15} \tilde{\partial}_t =\partial_t\Bigr|_{\Gamma_k^{(2)}}.
\end{equation}
In Eq. \eq{FRG14}, we identify the one-loop formula for the effective action
(\ref{Func20}) on the right hand side. Therein, we have to substitute
$S^{(2)}$ with the full two-point function $\Gamma_k^{(2)}+R_k$. 

Eq.~(\ref{FRG14}) is a very convenient form of Eq. \eq{FRG1} for deriving
the flows for correlation functions, e.g. $\partial_t \Gamma_k^{(n)}$.
It also allows for an easy access to the fluctuation-dependence of
specific correlation functions without performing any calculation. First we note 
that the flow equation for any correlation function (e.g. a
coupling constant) can be obtained by writing down all one-loop
diagrams which contribute to the expression, replace the propagators
and vertices by full ones, 
\begin{equation}\label{replace}
  \frac{1}{S^{(2)}}\to \frac{1}{\Gamma_k^{(2)}+R_k}\,,
  \qquad S^{(n)}\to \Gamma_k^{(n)}\,,
\end{equation}
and then take the $\tilde{\partial}_t$-derivative. Remarkably, this
renders the one-loop expression an exact (flow) equation. We emphasize
that this only holds true for additive IR regularizations of the
one-loop formula for the effective action (\ref{Func20}), see
\cite{Litim:2002xm}.  Note also that one also has to take into account
perturbative one-loop diagrams where the vertices involved vanish
classically, $S^{(n)}=0$. Still this one-loop structure is very
useful: If the loop expansion of a given correlation function does not
exhibit a one-loop diagram, this correlation function is not sensitive
to quantum fluctuations. This either happens due to internal
symmetries or the pole structure of the diagrams. The latter is
specific to nonrelativistic theories and, as we will see below, in the
case of ultracold atoms it leads to strong simplifications for vacuum
scattering properties (cf. \ref{AppVac}).

The Wetterich equation \eq{FRG1} is an equation for a functional and
thus may be evaluated for any (possibly inhomogeneous) mean field
$\phi(X)$. It is a functional integro-differential equation and its
full solution is, in most theories, beyond reach. Instead, one has to
use approximation schemes to the full effective action $\Gamma_k$,
which include the physics at hand already at a low order of the
approximation. The systematics of a given approximation scheme and the
control of the related systematic error is of chief importance when it
comes to the discussion of the reliability of the results. A
discussion of this interesting point is beyond the scope of the
present work, and is tightly linked to the discussion of optimal
choices of regulators mentioned above.

Here, we briefly discuss the most important approximation schemes which
cover (in variations) all approximation schemes used in the
literature. The most important scheme, which is partially behind all
approximations used, is the {\it vertex expansion} about a specific
background $\bar\phi$, schematically written as
\begin{equation}\label{vertexex}
  \Gamma_k[\phi]=\sum_n \frac{1}{n!} \int
  \Gamma_k^{(n)}[\bar \phi](X_1,...,X_N) \prod_{i=1}^n\left[\phi(X_i)-\bar\phi(X_i)\right].
\end{equation}
The information about the effective action is encoded in the vertices
$\Gamma_k^{(n)}$. The related flow can be derived from that of the
effective action as 
\begin{equation}
 \label{FRG16} \partial_t \Gamma^{(n)}_k[\bar\phi](X_1,\dots,X_n) = 
\frac{\delta^n}{\delta \phi(X_1)\dots\delta \phi(X_n)} \partial_t 
\Gamma_k[\bar\phi]\,.
\end{equation}
On the right hand side, we have to take the $n$th derivative of the one-loop diagram in Eq. \eq{FRG1},
\begin{equation}
 \label{FRG16r} 
\frac{\delta^n}{\delta \phi(X_1)\dots\delta \phi(X_n)} \frac{1}{2} \mbox{STr}\, 
\left[\frac{1}{\Gamma_k^{(2)}+R_k}\, \partial_t R_k\right]\,,
\end{equation}
which produces all possible one-loop diagrams with cutoff
insertions. Evidently, the diagrams for the flow of $\Gamma_k^{(n)}$
depend on $\Gamma_k^{(m)}$ with $m\leq n+2$.  Hence, within the vertex
expansion described above, we arrive at an infinite hierarchy of
equations, because the flow equation for $\Gamma^{(n)}$ requires input
from $\Gamma^{(n+1)}$ and $\Gamma^{(n+2)}$. The flow of the latter two
quantities depends again on higher correlation functions and
eventually the system never closes. We should not be surprised about
this, as the effective action necessarily contains infinitely many
independent terms, and we have just rewritten the functional
integro-differential equation in terms of infinitely many partial
integro-differential equations. In most interesting cases, it will not
be possible to derive a closed expression for the functional
$\Gamma[\phi]$. Practically, one \emph{truncates} the hierarchy of
flow equations at a given order $n$, i.e. approximates
$\Gamma_k^{(m>n)}\approx 0$, and solves the restricted, finite set of
partial integro-differential equations for $\Gamma_k^{(m\leq
  n)}(Q_1,...,Q_m)$. Examples for this scheme can be found in e.g.\
\cite{RevModPhys.84.299,Benitez:2011xx,Husemann:2011wr,Fister:2011um,%
  PhysRevLett.102.190401,PhysRevA.80.043627,PhysRevLett.102.120601,PhysRevA.82.063632,PhysRevA.83.063620}. The
self-consistency of this approximation can be checked by computing the
flow $\partial_t\Gamma_k^{(m>n)}$ as a function of $\Gamma_k^{(m\leq
  n)}$. This provides some error control.

A further important approximation scheme is the {\it derivative
expansion}. Formulated in momentum space, it is an expansion of the
vertices in powers of the momenta (derivatives). Its $n$th order
relates to the $n$th-order in $\rmi \omega +\vec q^2$. In contrast to the
vertex expansion, all vertices are present
already at the lowest order of the derivative expansion. Here, we exemplify this expansion for the
case of the effective action $\Gamma_k$ of a Bose gas. An often used ansatz for this theory is provided by
\begin{equation}
 \label{FRG17} \Gamma_k[\phi^*,\phi] = \int_X \Bigl(\phi^*
(Z_k \partial_\tau - A_k \nabla^2)\phi + U_k(\phi^*\phi)\Bigr)\,.  
\end{equation}
Herein, $U(\phi^*\phi)$ is the full effective potential. It is a
general function of $\rho=\phi^*\phi$. Accordingly, we have
$U^{(n)}\neq 0$ and thus vertices
$\Gamma_k^{(n)}$ to all orders in $n$. For $Z_k=A_k \equiv 1$, the ansatz in Eq. \eq{FRG17} has
the same momentum dependence as the classical action and is the lowest
(or zeroth) order in the derivative expansion. For flowing $Z_k, A_k$,
one goes beyond the lowest order. Note however that it is not the
full first order in the derivative expansion, as this requires
field-dependent $Z_k(\rho), A_k(\rho)$. The derivative
expansion and in particular the above ansatz \eq{FRG17} assumes lower
orders of the differential operators to be more relevant than the
higher ones. In the presence of a mass gap $m_{\rm gap}$, this is expected 
to be valid in the infrared,
because 
\begin{equation}\label{IRgap}
  (\partial_\tau/m^2_{\rm gap} )^n \sim (\nabla^2/m^2_{\rm gap})^n 
  \sim |\vec{q}/m_{\rm gap}|^{2n}
  \rightarrow 0\,.
\end{equation} 
Hence, within the derivative expansion, we make an expansion of the
effective action about the low energy effective action. From the more
technical point of view, we project the flow of $\Gamma_k$ onto a
subspace of functionals, which mimic the shape of the microscopic
action. The latter is given by $Z_\Lambda=1$, $A_\Lambda=1$, and the
effective potential $U_\Lambda(\rho)=-\mu \rho + \frac{g}{2}
\rho^2$. These values constitute the initial conditions for the flow
equation. During the flow, dressed quantities $Z_k$, $A_k$ and
$U_k(\rho)$ emerge. Besides the ansatz in Eq. (\ref{FRG17}), we also
have to specify a projection description which determines the flow
equations $\dot{Z}_k$, $\dot{A}_k$ and $\dot{U}_k(\rho)$ from
Eq. (\ref{FRG1}). Examples for the full lowest order derivative
expansion in bosonic as well as mixed fermionic-bosonic theories can
be found in e.g.\
\cite{Berges:2000ew,Schaefer:2006sr,Schaefer:2004en}. 

Most applications to ultracold atoms discussed in the present work are
done within low orders of the derivative expansion within an
additional field expansion of the effective potential $U_k$ up to the
$n$th order of the fields. Of course, such an approximation can also be 
interpreted as the $n$th order of the vertex expansion with an additional
expansion in powers of momenta and frequencies. Indeed, as has been
mentioned before, any approximation scheme used in the literature can
be seen as combination, deformation or further approximation of the
vertex expansion and the derivative expansion.  In any case, when
using such an approximation, we restrict the space of functionals. For
this reason, although we started from an exact flow equation, we may
accumulate errors. In particular, given the exact flow equation, every
regulator satisfying the mentioned properties should give the same
result. But since we never integrate the full flow, we may end up at
different ``effective actions'' $\Gamma_{k=0}$ if we used two
different regulators. This regulator dependence can be applied to
partially check this source of uncertainty. The approximate
independence of the results at vanishing cutoff, $k=0$, guarantees the
self-consistency of the approximation.  In turn, the independence of
$\Gamma_{k=0}$ of the chosen regulator $R_k$ or the chosen trajectory
in theory space can be utilized for devising regulators that are
best-suited (optimal) for the given order of a given approximation
scheme at hand, see
\cite{Liao:1999sh,Latorre:2000qc,Litim:2000ci,Litim:2001up,Litim:2001fd,Canet:2002gs,Pawlowski:2005xe,Salmhofer:2006pn}.

In summary, the Functional Renormalization Group approach for the
effective action constitutes a fully \emph{nonperturbative} approach
to quantum field theory. In fact, the functional differential equation (\ref{FRG1}) may be seen as an alternative but equivalent formulation to the functional integrals (\ref{eff16}) or (\ref{Func18}) in Sec. \ref{SecFun}. In particular, it is neither restricted to
small couplings nor to small amplitudes. For this reason, it can be
applied to many strongly coupled and/or strongly correlated systems
such as superconductors, superfluids, quantum chromodynamics,
quantum gravity, or -- in our case -- the unitary Fermi gas.

\subsection{The many-body problem in ultracold atoms from the FRG
  point of view}
\label{ManyBodyFRG}

\noindent We are now ready to proceed to the application of the FRG to
the physics of ultracold atoms. Particular emphasis is put on the
difference between the vacuum and the many-body limit of the system,
which can both be accessed in experiments and thereby allow for a high
precision comparison between experiment and theory. For this reason,
cold quantum gases also provide an ideal testing ground for different
approximation and truncation schemes within the FRG approach. Our
considerations build the basis for the detailed analysis of the
BCS-BEC crossover of two-component fermions in the subsequent section,
but are generally applicable to all ultracold atom settings. In
addition, it will provide an RG point of view on UV renormalization
discussed in Sec. \ref{Sec3}.

The microscopic action $\Gamma_\Lambda=S$ of an ultracold Bose gas at
the initial ultraviolet scale $\Lambda$ is given
by
\begin{align}
  \label{MB1}S[\vphi^*,\vphi] = \int_X \Bigl(
  &\vphi^*\bigl(\partial_\tau - \frac{\nabla^2}{2M} -\mu\bigr) \vphi +
  \frac{g_\Lambda}{2} (\vphi^*\vphi)^2 \Bigr)\,.
\end{align}
As an effective action, this expression is only valid in the UV, which is
given here by the van der Waals length. Going to the low energy scales
realized in experiments with ultracold atoms, fluctuations are
included and the parameter $g_\Lambda$ gets replaced by a dressed quantity
$g$. The corrections to the UV value arise from quantum and
thermal fluctuations.

Assume we want to measure the scattering length. Say, we perform a
scattering experiment between two atoms such that there is no
influence from other particles. Hence, there are no effects which are
associated to nonzero temperatures or densities in this
setting. Therefore, we call such processes to take place in the 
\emph{vacuum}, because they could also be observed if we had nothing
but the two colliding partners. In practice, this situation is achieved in ultracold experiments for $T\to0$, 
but with vanishing phase space density $\bar \omega =(\lambda_T/\ell)^d\ll1$ (cf. the discussion around Eq. (\ref{td10})). 

The scattering length $a$ of two identical bosons is related to the
coupling constant according to
\begin{equation}
 \label{MB2} g = 8 \pi \hbar^2 a/M\,.
\end{equation}
It is important to realize that the dressed or \emph{renormalized}
coupling constant $g$ appears in this equation. Although the
scattering takes place in the absence of statistical many-body
effects, we still have quantum fluctuations, which are always
present. For this reason, the {\it measured} coupling constant $g$ does not
coincide with the \emph{bare} coupling constant $g_\Lambda$ of the
microscopic action.

In fact, the experimental relevance of the bare parameters in Eq. \eq{MB1}
is at best indirect. Our microscopic action is not a realistic
description for atoms at high energies, but rather a simple model with
the same low energy physics as a more elaborate description. Put
differently, we utilize here that only the renormalized parameters are
important for observations in cold atom experiments.

It is a key property of ultracold quantum gases that the interaction
parameters of the atoms can be measured without reference to the
many-body system. In our case, we know the value of the scattering
length $a$ and can use this as an input for many-body predictions. For
instance, we may express the equation of state at zero temperature as
a function of $a$ and the chemical potential $\mu$, $P=P(\mu,a)$, and
verify experimentally the predicted dependence on both $\mu$ and
$a$. This optimal situation is not met generically in condensed matter
systems like e.g.\ solids, where the parameters of the model
Hamiltonian are not known and have to be adjusted according to the
observed many-body physics. In addition, a systematic investigation of
interaction effects is often not possible in solid state physics,
because the parameters are fixed by the sample and cannot be tuned
arbitrarily.

The ``dictionary'' between bare and dressed parameters of the
  microscopic action in vacuum is sometimes referred to as the UV
  renormalization. We already encountered such a procedure in the
 treatment of weakly interacting bosons and
fermions in Sec. \ref{Sec3}. Within the FRG approach, the notion of UV renormalization of
microscopic parameters arises naturally.

The microscopic action in Eq. (\ref{MB1}) enters the flow equation
(\ref{FRG1}) for the effective action. Depending on how we
choose the initial parameters $\mu$ and $T$ of the microscopic action,
we will arrive for $k=0$ either in vacuum or at a many-body
system. Herein, the vacuum effective action is defined through a
diluting procedure $n, T \rightarrow 0$, such that the system is kept
above criticality, $T/T_{\rm c}(n) >1$. In this way, condensation is excluded and we end up in the physical vacuum of a few scattering particles. An equivalent way to express the same idea is to take the limit $T\to0$, and at the same time sending the phase space density $\bar \omega =(\lambda_T/\ell)^d \to 0$ ($n=\ell^{-d},\lambda_T\sim 1/\sqrt{T}$, cf. Eqs. (\ref{sc1}) and (\ref{sc6})). In other words, the system gets dilute faster than it gets cold. Indeed, the vacuum scattering experiments work precisely in this ultracold limit at low phase space density.  These considerations are summarized with the definition
\begin{equation}
 \label{MB3} \Gamma_{\rm vac} = \Gamma\bigr|_{T\to 0, \mbox{ } \bar\omega=n \lambda_T^d\to 0}.
\end{equation}
Hence, in this case, $\Gamma_k$ is in the
symmetric phase for all $k$ and tends towards that at $n=T=0$ for
$k=0$.

\begin{figure}[tb!]
 \centering
 \includegraphics[width=.9\columnwidth,keepaspectratio=true]{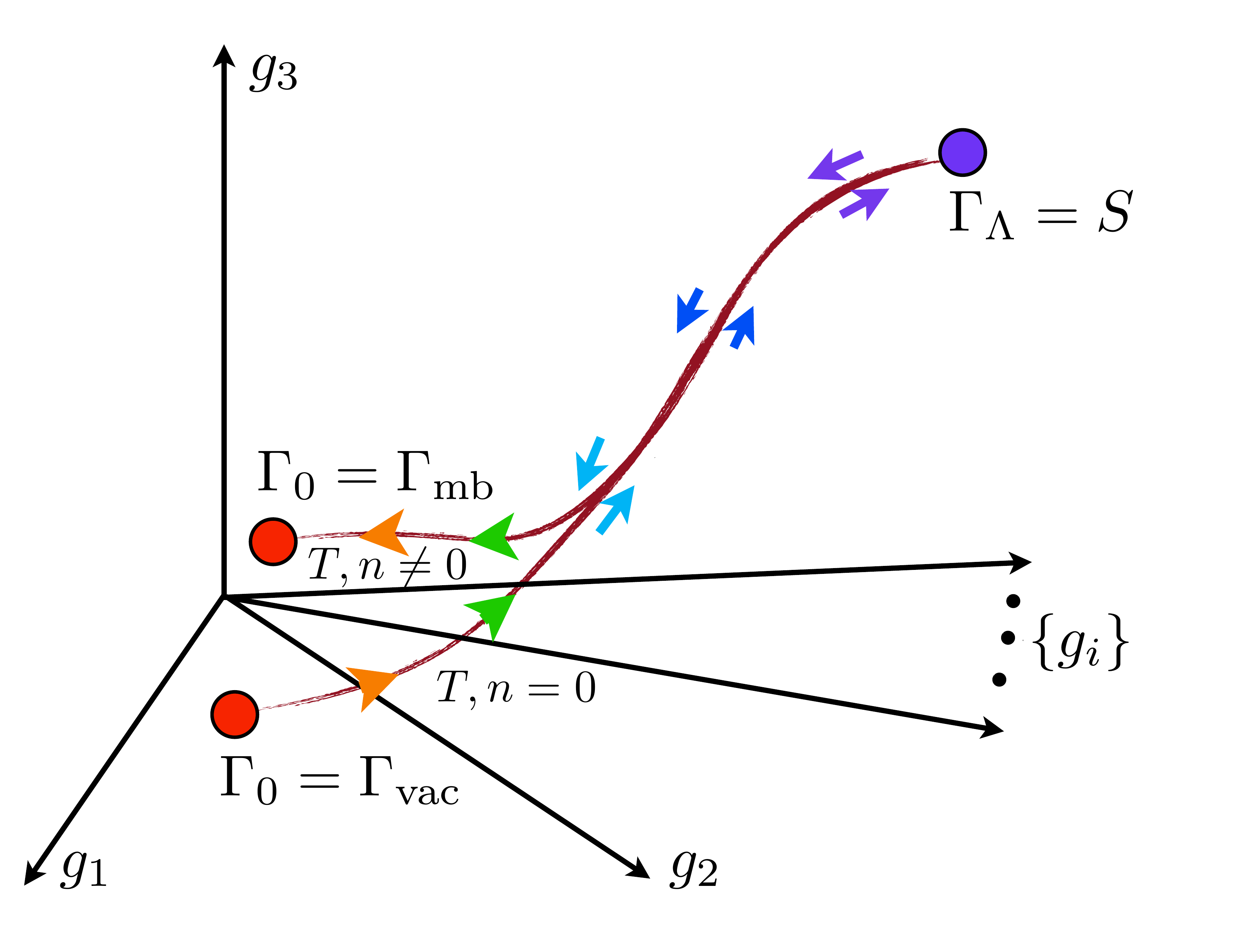}
 \caption{For given $\mu$ and $T$, the effective action
   $\Gamma_{k=0}=\Gamma$ will either describe a few-body or a
   many-body system. We denote these two cases by $\Gamma_{\rm vac}$
   and $\Gamma_{\rm mb}$ in the plot, respectively. To compute
   observables for ultracold quantum gases we need both: First, we
   solve the vacuum case, which provides us with the renormalized
   parameters of the microscopic action. Then, in a second step, we
   switch on the many-body scales $k_{\rm mb} =
   \ell^{-1},\lambda_T^{-1}$ (inverse interparticle spacing and de
   Broglie wavelength), which influences the trajectory of $\Gamma_k$
   once $k \approx k_{\rm mb}$.}
\label{TheoryFlow}
\end{figure}

This allows us to set-up the solution of the many-body problem with
the FRG in a two-step procedure:
\begin{itemize}
\item[(1)] \emph{Solving the flow in vacuum}\\ Choosing $\mu$
  and $T$ such that $\Gamma_{k=0}=\Gamma_{\rm vac}$, we start from a
  given set of bare parameters and compute the resulting renormalized
  ones via a successive inclusion of quantum fluctuations. The
  renormalized coupling can be measured and tuned in practice, and thus allows for a direct matching of theory and experiment.

\item[(2)] \emph{Solving the flow for the many-body system}\\
  We set the chemical potential and the temperature such that we arrive at a desired
  $n>0$ and $T \geq 0$ for $k=0$. 
\end{itemize}
Let us now follow the integration of the flow equation at finite
density and temperature. For large $k$ with $T/k^2,\mu/k^2\ll 1$, the
flow agrees with the vacuum flow in the symmetric regime up to
subleading contributions. Once the flow parameter $k$ reaches the
many-body momentum scales $\mu^{1/2}, T^{1/2}$, the flow
deviates from the vacuum trajectory and eventually arrives at the
effective action $\Gamma$ of the many-body system. Having started with
a given set of bare couplings, we know from step (1) what the
corresponding renormalized couplings are. Thus, we can express the
many-body observables in terms of $n$, $T$ and physical microscopic
couplings. The situation is illustrated in Fig. \ref{TheoryFlow}.

A simple example for this two-step procedure is provided by the
microscopic action for bosons given in Eq. (\ref{MB1}). Herein,
$g_\Lambda$ is the only bare parameter which receives substantial renormalization. In order to compute the
renormalized coupling, we follow the flow for $T=n=0$ and
$g_\Lambda>0$ down to $k=0$. The renormalized coupling
$g=g(g_\Lambda)$ can then be read-off from the vacuum effective action
$\Gamma_{k=0}$. This completes step (1). Now, we again solve the flow
equation but with $n>0$ and $T\geq0$. For instance, we may compute
the equation of state $n(\mu,T,g_\Lambda)$. From step (1) and
Eq. (\ref{MB2}) we can replace $g_\Lambda$ by the scattering length
and obtain the equation of state in the experimentally accessible form
$n(\mu,T,a)$.

\subsection{BCS-BEC crossover and unitary Fermi gas}

\noindent In this section, we investigate the BCS-BEC crossover of
two-component ultracold fermions. Applying the FRG in a simple
truncation, we capture both the weakly and the strongly interacting
regime. In particular, we show how the BCS and BEC ground states are
linked by the unitary Fermi gas.  Furthermore, we are able to resolve
physics at all scales, from the few-body physics over the many-body
sector down to very long wavelengths, relevant for critical
behavior. One of the goals of this section is to show how the FRG
provides a unified approach which captures all effects of the system
within the same description.

\vspace{5mm}
\begin{center}
 \emph{Feshbach resonances and microscopic model}
\end{center}
\label{SecFesh}

\noindent We have discussed the two cornerstones of quantum
condensation phenomena in the weak coupling regimes: On the one hand,
attractive interactions lead to superfluidity of two-component
fermions via the formation of Cooper pairs. The momenta of two
fermions constituting such a pair are located on opposite points of
the Fermi surface. This locality in momentum space implies that the
spacing between them may be large in position space. On the other
hand, we discussed Bose condensation of weakly repulsive bosons, which
are microscopically pointlike objects, localized in position
space. Tightly bound pairs of two fermions could effectively realize
such bosons.

There exists an experimental knob to connect these two scenarios. It
is provided by the Feshbach resonance \cite{grimm-review}, which
allows to change the scattering length through the variation of an
external magnetic field $B$. We write
\begin{equation}
 \label{Fesh1} a(B) = a_{\rm bg} \left( 1 - \frac{\Delta B}{B-B_0}\right)\,,
\end{equation}
where $a_{\rm bg}$, $\Delta B$ and $B_0$ are parameters which can be determined
experimentally. This formula is a decent parametrization in a range
of order $\Delta B$ around $B \approx B_0$. In particular, at $B =
B_0$, the scattering length changes sign and becomes anomalously
large, $|a| \gg \ell_{\rm vdW}$. Recall that this does not invalidate
our effective Hamiltonian and the fact that scattering can be assumed
to be pointlike, see the discussion at the end of Sec \ref{scales}.

Sufficiently stable ultracold quantum gases of two-component fermions
are built from either $^6$Li or $^{40}$K, which are alkali
atoms. Their internal structure is relevant in order to have fermionic
s-wave interactions at all, but also manifests itself in the
appearance of Fesh\-bach resonances. To understand this, we consider a
single alkali atom. We can approximate the system to consist of an
atomic core and a valence electron. The ground state of the system is
given by the electron being in the s-orbital. Correspondingly, the
orbital angular momentum of the valence electron vanishes and thus a
fine structure does not appear. However, the electron spin $S$ couples
to the spin of the nucleus $I$. The resulting quantum number $F$
introduces a (tiny) hyperfine splitting of the ground state. Since
$S=1/2$, the value of $F$ is given by $F=I\pm1/2$. In addition, every
hyperfine state has a $(2F+1)$-fold degeneracy $m_F = -F, \dots,
F$. Thus, alkali atoms in their electronic ground state can be
distinguished according to their hyperfine state $|F,m_F\rangle$.

\begin{figure}[tb!]
 \centering
 \includegraphics[scale=0.85,keepaspectratio=true]{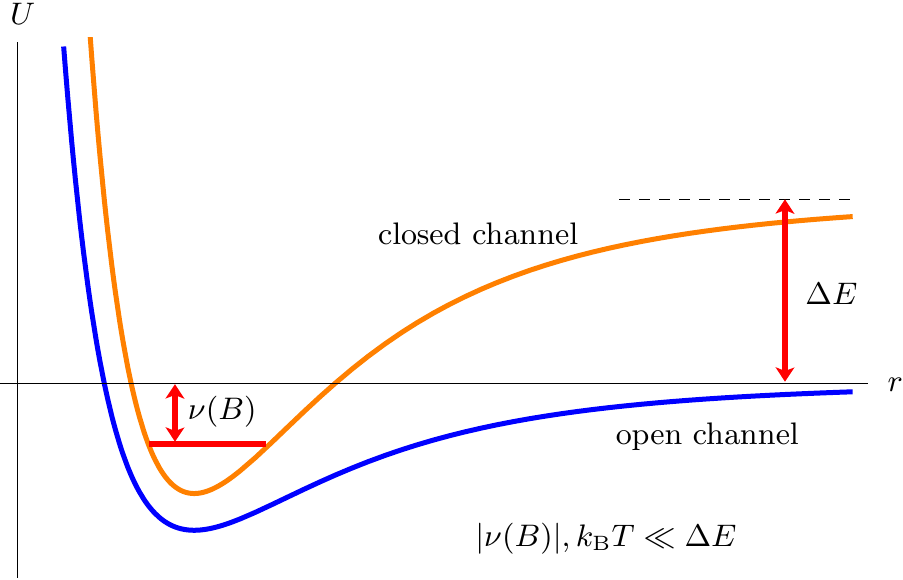}
 \caption{Interatomic potential $U$ between two fermions in distinct
   hyperfine states separated by a distance $r$. The closed channel
   consists of bound states and low energy scattering can only take
   place in the open channel. However, by changing the external
   magnetic field $B$, we can drive one of the bound states close to
   the scattering threshold $U(r \rightarrow \infty)=0$. The energy
   distance related to this particular bound state is denoted as
   $\nu(B) = \Delta \mu\cdot(B-B_0)$. The resulting scattering length
   $a=a(B)$ is parametrized according to Eq. (\ref{Fesh1}). For $B
   \approx B_0$, it becomes anomalously large, and thus it can largely exceed
   the interparticle spacing: $|a| \gg \ell_{\rm vdW} \Rightarrow |(k_{\rm
     F}a)^{-1}| \lesssim 1$.}
\label{Feshbach}
\end{figure}

Now suppose that two atoms in different hyperfine states scatter off
each other. Due to the internal structure of the colliding partners,
we call this a multichannel scattering. The two-body system of atoms
will be in a superposition of the singlet and the triplet
state. Depending on the species of atoms, the former will have a
higher or a lower energy than the latter, while the first option is
more generic. Moreover, due to the hyperfine coupling there will in
general be a mixing between both states. For our purpose it is enough
to restrict our considerations to two relevant channels, an
\emph{open} and a \emph{closed channel}, which have different magnetic
moments. We normalize the potential such that two atoms in the open
channel at infinite distance have zero energy; this sets the
\emph{scattering threshold}. The closed channel is separated from the
open one by a large energy gap $\Delta E$
(cf. Fig. \ref{Feshbach}). It thus cannot be accessed by atoms in the
lower channel. The relevant feature of the closed channel is a bound
state lying close to the open channel scattering threshold. It is
evident that this situation is not particularly generic since typical
bound state level spacings are much larger than typical collision
energies in ultracold gases, and thus requires specific, fine tuned
conditions.

Because of the difference $\Delta \mu$ in magnetic moment, open and
closed channel couple differently to an external magnetic field
$B$. For this reason, the difference in energies between both channels
can be tuned according to $\Delta E \rightarrow \Delta E + \Delta \mu
\cdot B$. Consider a particular bound state from the closed
channel. Its energetic distance from the scattering threshold $E=0$ is
called the \emph{detuning}
\begin{equation}
 \label{Fesh2} \nu(B) = \Delta \mu \cdot(B-B_0)\,.
\end{equation}
Due to second order processes, where two colliding atoms virtually
enter the closed channel and then leave it again, a bound state with
small $\nu$ affects the scattering properties of the alkali atoms. In
particular, changing the magnetic field such that $\nu \rightarrow 0$,
both channels become resonant and we obtain a strongly interacting
system. The corresponding field dependent scattering length $a(B)$ in
Eq. (\ref{Fesh1}) constitutes the most important application of such a
\emph{Feshbach resonance}
\cite{pethick-book,pitaevski-book,chin-review} in ultracold atomic
physics.

We now incorporate the physics of a Feshbach resonance for
two-component fermions on the level of the microscopic action. Recall
from Eq. (\ref{bcs0b}) that the action for a fermionic theory with
pointlike interactions is given by
\begin{align}
  \label{Fesh3} S_\psi[\psi^*,\psi] = \int_X \biggl( \psi^\dagger
  \Bigl(\partial_\tau-\frac{\nabla^2}{2M}\Bigr)\psi+
  \frac{\lambda_{\rm bg}}{2}(\psi^\dagger\psi)^2\biggr)\,,
\end{align}
with Grassmann fields $\psi=(\psi_1,\psi_2)$. Eq. (\ref{Fesh3})
constitutes a single-channel model. In order to emphasize the
structure of this and the following expressions, we have dropped the
chemical potential for the moment. It will be reintroduced below.

The closed channel is included explicitly in terms of a microscopic
bosonic field $\vphi$, which constitutes a composite degree of freedom
resulting from the interconversion of two fermions into a closed
channel molecule. The action for this boson field is modeled as
\begin{equation}
  \label{eq:bosact} S_\vphi[\vphi^*,\vphi] = \int_X \vphi^*
  \Bigl(\partial_\tau - \frac{\nabla^2}{4M} +\nu\Bigr)\vphi\,.
\end{equation}
The most important term is the detuning $\nu$, which acts as a mass
term for the bosons. In addition, we allow for a Galilean invariant
kinetic term, where the prefactor of $1/4M$ is related to the mass
$2M$ of the composite object. As we will see in a moment, the
microscopic kinetic term is, however, unimportant for the case of broad
Feshbach resonances which are studied here, and could be equally well
omitted. The full microscopic action from which we will extract the
physics of the BCS-BEC crossover is then given by
\cite{PhysRevLett.87.120406}
\begin{align}
  \nonumber &S[\psi^*,\psi,\vphi^*,\vphi] = \int_X \biggl(
  \psi^\dagger \Bigl(\partial_\tau -\frac{\nabla^2}{2M}\Bigr)\psi +
  \frac{\lambda_{\rm bg}}{2}(\psi^\dagger\psi)^2\\
\label{Fesh4} &\mbox{ }+\vphi^*\Bigl(\partial_\tau -
\frac{\nabla^2}{4M} +\nu\Bigr)\vphi- h\bigl(\vphi^* \psi_1\psi_2-\vphi
\psi_1^*\psi_2^*\bigr) \biggr).
\end{align}
As anticipated above, the Yukawa-type cubic coupling $\sim h \vphi^*
\psi_1 \psi_2$ (called \emph{Feshbach coupling} in the cold atom
context) allows for the interconversion of two fermions of opposite
spin into one molecule.

The parameters $\lambda_{\rm bg}$, $\nu$ and $h$ of the microscopic
action can be measured in experiment. We show here that they
correspond to the three parameters in Eq. (\ref{Fesh1}) for the
scattering length $a(B)$ across a Feshbach resonance in the broad
resonance limit. For this purpose, we consider the functional integral
$Z =\int \mbox{D} \vphi \mbox{D} \psi\, e^{-S[\psi,\vphi]}$. For fixed
$\psi$ and $\psi^*$, we can perform the Gaussian integral in $\vphi^*$
and $\vphi$. This is equivalent to the saddle-point approximation about
the solution of the mean field equations of motion
\begin{equation}
  \label{Fesh5} \frac{\delta S}{\delta \vphi^*} = 0 \hspace{3mm}\Rightarrow\hspace{3mm}
  \vphi= \frac{h}{\partial_\tau - \nabla^2/4M+\nu} \psi_1\psi_2\,.
\end{equation}
As the action is quadratic in $\varphi$, the saddle-point
approximation is exact.  If we formally insert this into the partition
function and integrate out the bosonic fields, we arrive at the action
\begin{equation}
  \label{Fesh6}S[\psi^*,\psi] = S_\psi + \int_X \psi_1 \psi_2
  \frac{h^2}{\partial_\tau-\nabla^2/4M+\nu}\psi_1^*\psi_2^*\,,
\end{equation}
with $S_\psi$ from Eq. (\ref{Fesh3}). We emphasize that the procedure
described here corresponds to reversing a Hubbard--Stratonovich
transformation explained in Sec. \ref{SecBCS}, for a slightly more
complicated inverse boson propagator; this is possible due to the fact
that in Eq. \eqref{eq:bosact} we work with a quadratic bosonic
action. We now take the \emph{broad resonance limit}, where $h,\nu
\rightarrow \infty$ with $h^2/\nu$ kept fixed. Then, we can neglect the
derivatives corresponding to frequency and momentum dependence of the
effective four-fermion vertex. More precisely, we scale $\nu \sim h^2$
for $h\to \infty$, while leaving the derivative coefficients of order
unity. We then obtain the action
\begin{equation}
  \label{Fesh7} S[\psi^*,\psi] = S_\psi[\psi^*,\psi] - \frac{1}{2}
 \frac{h^2}{\nu} \int_X (\psi^\dagger\psi)^2\,.
\end{equation}
Apparently, this coincides with a purely fermionic action with an
effective coupling
\begin{equation}
  \label{Fesh8} \lambda_{{\rm eff}} = \lambda_{\rm bg} - \frac{h^2}{\nu}\,.
\end{equation}
We conclude that in the broad resonance limit the two-channel and the
single-channel model become equivalent
\cite{Diehl:2005an,Diehl:2005ae,gurarie-review}. The single channel
model, however, acquires an additional effective contribution to the
coupling constant. The Feshbach resonances in $^6$Li and $^{40}$K are
broad. The narrow resonance limit is conceptually interesting, as it
can be solved exactly \cite{Diehl:2005an}. Moreover, recently, examples
of narrow resonances have been studied experimentally
\cite{Kohstall11}. In an RG language, it corresponds to a Gaussian
fixed point, while the broad resonances are governed by an interacting
(Wilson--Fisher) fixed point \cite{Diehl:2007ri}. While the macroscopic
physics depends on microscopic details of the closed channel, the
broad resonance fixed point is characterized by a large degree of
universality, i.e. a pronounced independence on the microscopic
details of the closed channel, as is plausible from the above scaling
\cite{PhysRevA.75.033608,Diehl:2007ri}.

Assuming the couplings in Eq. (\ref{Fesh8}) to be the renormalized
ones, we can relate them to the scattering length according to
$\lambda = 4 \pi a/M$.\footnote{Note the difference in convention to
  identical bosons, where $\lambda = 8 \pi a/M$.} We find
\begin{equation}
 \label{Fesh9} a = \frac{M}{4\pi} \left(\lambda_{\rm bg}-\frac{h^2}{\nu}\right)\,.
\end{equation}
Comparing this to Eqs. (\ref{Fesh1}) and (\ref{Fesh2}), we find that
indeed $\nu = \Delta \mu\cdot (B-B_0)$ corresponds to the detuning
from resonance. The four-fermion coupling $\lambda_{\rm bg}$ is related to the
background scattering length in the usual manner via $a_{\rm bg} = M
\lambda_{\rm bg}/4\pi$. With $\Delta B = h^2 \Delta\mu/\lambda_{\rm
  bg}$, the Yukawa/Feshbach coupling $h$ is seen to determine the
\emph{width of the resonance}.

For magnetic fields close to $B_0$, the scattering length becomes
anomalously large and the background scattering length can be
neglected. In what follows, we assume $a_{\rm bg}=0$ throughout the
whole crossover.

As anticipated above, only the value of the scattering length plays
the role of a relevant parameter for the crossover in the broad
resonance limit. Given the density $n$ of atoms, we build the
dimensionless parameter $(k_{\rm F}a)^{-1}$. Since the interparticle
spacing is given by $\ell \approx k^{-1}_{\rm F}$, we find the following
scheme for the crossover:
\begin{itemize}
 \item[1)] $k_{\rm F}a \rightarrow -\infty$ : weakly interacting fermions,\\[-1em]
 \item[2)] $|(k_{\rm F}a)^{-1}| \leq 1$: strong interactions, dense regime,\\[-1em]
 \item[3)] $(k_{\rm F}a)^{-1} \rightarrow \infty$: weakly interacting molecules.
\end{itemize}
The regions $a<0$ and $a>0$ are called BCS and BEC side of the
crossover, respectively.

Region 2) is often referred to as the \emph{unitary Fermi gas}. The
origin of this term is the following.  The cross section of two-body
scattering in the $s$-wave channel is given by $\sigma_{l=0} = 4 \pi
|f_{l=0}|^2 $. For the perturbative regions with $p |a| \ll 1$, where
$p$ is the relative momentum of scattering particles, we then find
$\sigma = 4\pi a^2$. Naively extrapolating this to the resonance
$|a|^{-1} \rightarrow 0$, this would imply a divergent cross section,
which is excluded from the fact that the scattering matrix is unitary.
Recalling however Eq. (\ref{sc10}), we find in the latter limit that
$\sigma \simeq 4 \pi | (-{1}/{a} - \rmi p)|^{-2} \rightarrow {4
  \pi}/{p^2}$. Since scattering is meaningful only for nonzero relative
momenta, the expression on the right hand side constitutes the upper
limit on possible s-wave scattering; in the unitary Fermi gas, the
typical scale for the scattering momentum are
$k_\text{F},\sqrt{T}$. This effect has been observed in
\cite{gupta03}. We note that exactly at the unitary point $a^{-1} =0$,
the scale associated to interactions drops out and the only remaining
scales are interparticle spacing and temperature. This hints at highly universal
properties in this regime \cite{PhysRevLett.92.090402} (distinct from
the broad resonance universality described above). However, at this
point and in its vicinity where $|k_\text{F}a|^{-1} \ll 1$, the gas
parameter Eq. \eqref{sc11} is large and cannot be used to as a control
parameter for systematic expansions. In this regime, where the
interaction length scale greatly exceeds the interparticle spacing, we
deal with a strongly coupled and dense quantum system.

In the perturbative regime of small $a$, we have a second order phase
transition towards the superfluid phase. We will see that a second
order transition is found for all values of $k_{\rm F} a$. The value
of the critical temperature $T_{\rm c}/T_{\rm F}$ as a function of
$k_{\rm F}a$ is particularly interesting. Deep in the BEC regime, we
expect the noninteracting formula (\ref{td10}) to hold, with a shift
due to the small, but nonvanishing diluteness parameter $k_{\rm
  F}a$. On the BCS side, Eq. (\ref{bcs25}) will turn out to be
insufficient, because it is lowered by a factor of approximately two
due to particle-hole fluctuations. A great challenge in many-body
theory is the calculation of $T_{\rm c}/T_{\rm F}$ at unitarity from
first principles.

\vspace{5mm}
\begin{center}
 \emph{Ansatz for the effective action}
\end{center}

\noindent After these preliminaries we turn our attention to an FRG
study of the crossover. Since this approach is not limited to weak
coupling, it can be applied to the limiting BCS and BEC regimes as
well as to the unitary Fermi gas. Our particular interest lies in the
transition from micro- to macrophysics.

We restrict our analysis to the three dimensional case. However,
performing the same calculations with general dimension $d$ of the
loop integral, we can also analyze lower-dimensional systems. The
scattering properties have to be adjusted appropriately. We expect
quantum effects to be more pronounced in reduced dimensionality,
because long-range effects are more significant there. This is
reflected in more severe infrared singularities in the respective
loop integrals. We also restrict to the balanced case of equal
densities of spin up and down atoms. For the imbalanced case, a first
order phase transition is expected and thus its implementation is more
demanding. The latter corresponds to unequal chemical potentials
$\mu_1\neq\mu_2$ for the fermions. We set $2M=1$ with the fermion mass
$M$.

The chemical potential which enters Eq. (\ref{Fesh4}) can be found
from the following consideration. Since every bosonic molecule $\phi$
consists of two fermionic atoms $\psi$, we have for the total number of atoms
$N = N_\psi + 2 N_\phi$. For this reason, what appears in the grand
canonical partition function is the combination $H - \mu N = H-
\mu(N_\psi + 2N_\phi)$. On the other hand, we can also regard the
system to be composed of two species with individual chemical
potential, i.e. $H -\mu_\psi N_\psi - \mu_\phi N_\phi$ enters the
partition function. Of course, both expression have to be equal, which
yields $\mu_\psi = \mu$ and $\mu_\phi=2\mu$. The full microscopic
action $\Gamma_\Lambda=S$ thus reads
\begin{align}
  \nonumber  S[\vphi,\psi] = \int_X & \mbox{ }
  \biggl( \psi^\dagger\Bigl(\partial_\tau -\nabla^2 -\mu\Bigr)\psi + 
\frac{1}{2}\lambda_\Lambda(\psi^\dagger\psi)^2\\
  \nonumber &+\vphi^*\Bigl(Z_\Lambda \partial_\tau - 
\frac{1}{2}A_\Lambda\nabla^2 +\nu_\Lambda-2\mu\Bigr)\vphi\\
\label{Fesh10} &- \frac{h_\Lambda}{2}(\vphi^* \psi^t \vare\psi-\vphi
\psi^\dagger\vare\psi^*) \biggr)\,.
\end{align}
The bosonic mass is related to the detuning from the Feshbach resonance as explained in the last section. 

Already a simple approximation for the effective action allows for a
qualitative study of the full crossover. Since the key to the BCS-BEC
crossover consists in the formation of a bound state, the possibility of describing a bosonic
molecule must be contained in any reasonable approximation. We thus
employ the following ansatz for the effective action,
\begin{align}
  \nonumber \Gamma_k[\phi,\psi] = &\mbox{ }\int_X 
\biggl( \psi^\dagger\Bigl(\partial_\tau -\nabla^2 -\mu\Bigr)\psi \\
  \nonumber &\mbox{ }+\phi^*\Bigl(Z_k \partial_\tau-
\tfrac{1}{2} A_k\nabla^2\Bigr)\phi +U_k(\rho)\\
\label{Fesh11} &-\frac{h_k}{2}\Bigl(\phi^*\psi^t\vare\psi-
\phi\psi^\dagger\vare\psi^*\Bigr)\biggr).
\end{align}
As explained in Sec. \ref{sec:flow}, this approximation, up to the
wave function renormalizations $Z_k,A_k$, constitutes the
lowest order in the \emph{derivative expansion}. 

The ansatz in Eq. (\ref{Fesh11}) forces the effective action
into a particular form. The general fluctuation-dependence of the vertices
$\Gamma_k^{(n)}$ is replaced by the flow of $Z_k$, $A_k$, $h_k$ and
$U_k(\rho)$. This simplified picture, however, already encodes the
most relevant physics of the theory. For instance, the fermion
propagator is considered classical in this approximation. The
inclusion of the full fermion dynamics yields higher quantitative
precision, but is not required to obtain a qualitative picture of the
phase transition, which is driven by the particle-particle loop
contribution to the boson propagator as we will see below. It is in this sense that a
successive improvement of the truncation of the effective action can
lead to new physical insights.

The scale dependent effective potential $U_k(\rho)$ can only depend on
the ${\rm U}(1)$-invariant quantity $\rho=\phi^*\phi$. This follows
from our considerations in Sec. \ref{SecSSB}. The ${\rm
  U}(1)$-symmetry of $U_k$ is not violated during the renormalization
group flow, if the regulator $\Delta S_k$ respects this symmetry. As
we employ $\Delta S_k = \int_Q \vphi^*(Q) R_k(Q) \vphi(Q) + \int_Q \psi^\dagger(Q) R_\psi(Q) \psi(Q)$, 
which is manifestly ${\rm U}(1)$-symmetric, this is the
case. It is a particular strength of the FRG approach that symmetries
of the theory are conserved if the truncation and the regulator are
chosen appropriately. The initial condition for $U_k(\rho)$ can be
deduced from Eq. (\ref{Fesh10}) and is given by
\begin{equation}
 \label{Fesh12} U_\Lambda(\rho) = (\nu_\Lambda-2\mu) \rho\,.
\end{equation}
Note that the microscopic potential does not contain a term $\sim
u_{\phi,\Lambda}\rho^2$ with dimer-dimer coupling
$u_{\phi,\Lambda}$ in accord with our discussion of the Feshbach model (\ref{Fesh4}). Despite being zero at $k=\Lambda$, this coupling will be generated during the RG flow within our truncation.

In Eq. (\ref{Fesh11}), we have neglected a possible running of
$\lambda_{\psi,k}$. Certainly, for $\lambda_{\psi,\Lambda} <0$, we can eliminate the
four-fermion vertex in the ultraviolet by choosing $h_\Lambda$
appropriately, see Eq. (\ref{bcs12}). However, during the RG flow, the
coupling $\lambda_{\psi,k}$ is generated again. By neglecting the flow of
this coupling, we simplify the flow equations but miss an important
screening effect in the many-body problem.

Another interesting feature is the irrelevance of the initial
conditions for $Z_\Lambda$, $A_\Lambda$ and $h_\Lambda$ in the broad
resonance limit. First of all, one may argue that the coupling
$A_\Lambda$ is actually not present, because it can be absorbed into
the definition of $\phi$.\footnote{One of the couplings in the inverses propagator can always be absorbed, because the field equation can be premultiplied with an arbitrary nonzero number. Here we choose $A_k$.} We apply this redefinition procedure at all
scales $k$ by introducing the renormalized field
\begin{equation}
 \label{Fesh13} \tilde{\phi} = A_k^{1/2} \phi\,.
\end{equation}
As a consequence, the effective average action is given by
\begin{align}
  \nonumber \Gamma_k[\tilde{\phi},\psi] = &\mbox{ }\int_X
  \biggl( \psi^\dagger\Bigl(\partial_\tau -\nabla^2 -\mu\Bigr)\psi \\
  \nonumber &\mbox{
  }+\tilde{\phi}^*\Bigl(S_k \partial_\tau-\frac{1}{2}
  \nabla^2\Bigr)\tilde{\phi} +U_k(\tilde{\rho})\\
  \label{Fesh14}
  &-\frac{\tilde{h}_k}{2}\Bigl(\tilde{\phi}^*\psi^t\vare\psi-
  \tilde{\phi}\psi^\dagger\vare\psi^*\Bigr)\biggr)\,,
\end{align}
where we have introduced the renormalized couplings $S_k=Z_k/A_k$ and
$\tilde{h}^2_k = h^2_k/A_k$. The designation `renormalized
quantities' is common but somewhat unfortunate, since the actual
renormalization procedure takes place in the flow with $k$. Note that
Eq. (\ref{Fesh14}) has a canonical kinetic term for the bosons without
prefactor of the $\nabla^2$-term at all scales.

\begin{figure}[tb!]
 \centering
 \includegraphics[scale=0.85,keepaspectratio=true]{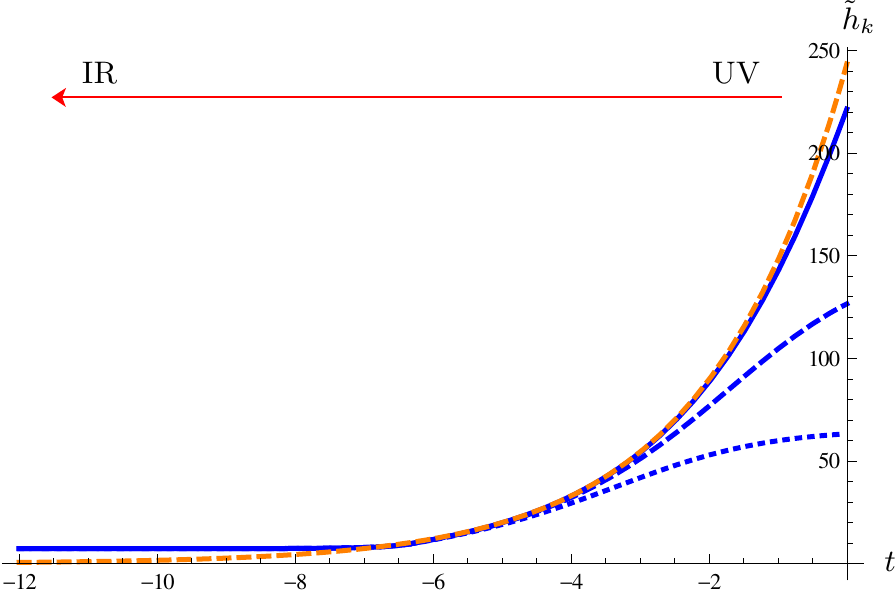}
 \caption{Flow of the Yukawa coupling $\tilde{h}_k=h/A_k^{1/2}$ for
   different initial values $\tilde{h}_\Lambda/\Lambda^{1/2}=7,4,2$ as
   a function of $t=\log(k/\Lambda)$. The curves are attracted to the
   scaling solution $\tilde{h}_k = (6 \pi^2 k)^{1/2}$ shown as an
   orange dashed line. Memory of the initial value is then lost. The
   approach to the corresponding partial fixed point of
   $\tilde{h}^2/k$ is faster for a larger initial value. In
   particular, for $\tilde{h}_\Lambda = (6 \pi^2 \Lambda)^{1/2} =7.7
   \Lambda^{1/2}$ we start the flow in the scaling regime. For this
   plot we have chosen $T=0$ and $(k_{\rm F}a)^{-1}=0.3$, although the
   situation is generic and also valid at nonzero temperature and on
   the BCS side of the crossover. For the chosen parameters, the mass
   becomes zero for $t \approx -6.5$ and the flow leaves the scaling
   regime at this point. }
\label{FPofh}
\end{figure}

As mentioned below Eq. (\ref{Fesh12}), the flow starts at $k =
\Lambda$ in the symmetric phase. In particular, at this stage of the
flow, $k$ is much larger than the many-body scales $k_{\rm F}$ and
$T^{1/2}$, which thus cannot be resolved. Additionally, bosonic
contributions to the right hand side of the flow equation vanish,
because they are proportional to the condensate $\rho_0$. Solving the
remaining flow equations, we then find the dimensionless combination
$\tilde{h}^2_k/k$ to be rapidly attracted to the (partial) fixed point
value $6 \pi^2$ \cite{Diehl:2007ri}. This means that after a few RG steps, the Yukawa
coupling shows the scaling behavior $\tilde{h}_k \sim \sqrt{6 \pi^2
  k}$ in the ultraviolet. The approach to the fixed point is faster
for a larger initial value of $h_\Lambda$. It is only a partial fixed
point, because the scaling solution becomes invalid when the scales
provided by chemical potential and temperature enter the flow. We
visualize this behavior in Fig. \ref{FPofh}.

After $\tilde{h}_k^2$ has approached its partial fixed point, the
value of $S_k$ is attracted to unity, irrespective of the initial
choice of $S_\Lambda$. Therefore, the choice of $Z_\Lambda$ and
$A_\Lambda$ is irrelevant for a sufficiently large $h_\Lambda$. This
is the case for a broad resonance. For definiteness we start with the
fixed point value $h_\Lambda = \sqrt{6 \pi^2 \Lambda}$ in the UV and
set $A_\Lambda = Z_\Lambda =1$. Even if we had set the latter two
quantities to zero, the term $S_{k \approx \Lambda}=1$ would
immediately be generated. This provides an RG argument for the
equivalence of the models with and without dynamical bosons
$\vphi$. In particular, after introducing the mass $2M=1$, we find the
boson kinetic term for $S \rightarrow 1$ to approach the form
\begin{equation}
 \label{Fesh15} \tilde{\phi}^*\Bigl(\partial_\tau -\frac{1}{4M}\nabla^2\Bigr)\tilde{\phi},
\end{equation}
which reveals the boson mass to be twice the fermion mass, $M_\phi = 2 M$.

Choosing the initial conditions such that we start at the fixed point
value of $\tilde{h}_k$, we assume the flow of the Yukawa coupling to
vanish:
\begin{equation}
 \label{Fesh15b}h_k = h_\Lambda=h\,.
\end{equation}
This is valid in the symmetric phase, but neglects a contribution
proportional to $\rho_0$ in the broken phase. The latter, however, is
expected to be subleading, and this is confirmed numerically
\cite{Diehl:2009ma}. Note that $\tilde{h}_k^2 = h^2/A_k$ has a
scale-dependence due to the running of $A_k$.

In Eq. (\ref{Fesh11}), the inverse boson propagator is assumed to be
of the simple form $\rmi Z_k \omega + \frac{1}{2}A_k \vec{q}^2$. Thus,
instead of resolving the full functional dependence of the propagator,
we only follow the flow of $Z_k$ and $A_k$. This corresponds to
replacing the partial differential equation for the propagator by a
set of ordinary differential equations. (For such a procedure to work
well, the truncation of $\Gamma_k$ has to be physically well
motivated.) The corresponding coupled equations are easily solved with
standard numerical methods -- although we are calculating the
effective action of a strongly interacting quantum field theory. We
could also expand $U_k(\rho)$ to a certain order in $\rho$ and then
consider only ordinary differential equations for the
coefficients. However, for the moment we keep the full flowing
potential, because we will recover several structures which are
already familiar from the previous sections on functional methods.

\vspace{5mm}
\begin{center}
 \emph{Flow of the effective potential}
\end{center}

\noindent Now we are aiming at a computation of the effective potential
$U(\rho) = U_{k=0}(\rho)$. For this purpose, we start from the
Wetterich equation
\begin{align}
  \nonumber \partial_k \Gamma_k[\phi,\psi] = &
\mbox{ }\underbrace{\frac{1}{2} \mbox{Tr} 
\left(\frac{\partial_k R_{\phi,k}}{\Gamma_{\phi,k}^{(2)}[
\phi,\psi]+R_{\phi,k}}\right)}_{\text{bosonic contribution}} \\
\label{Fesh16}&\underbrace{-\frac{1}{2} \mbox{Tr}
  \left(\frac{\partial_k R_{\psi,k}}{\Gamma_{\psi,k}^{(2)}[\phi,\psi]
      +R_{\psi,k}}\right)}_{\text{fermionic contribution}}\,.
\end{align}
Here, we have already performed the summation over the field indices with a
minus sign for the fermions. The field content is quite simple -
bosons $\phi$ and fermions $\psi$. So far, $\phi(X)$ and $\psi(X)$ are
arbitrary fields parametrizing the effective action. We need to keep them since we may want to take
functional derivatives of Eq. (\ref{Fesh16}). After we have derived
all flow equations $\dot{\Gamma}_k^{(n)}$ of interest, we insert physical values (possible solutions of the field equations)
$\phi={\rm const.}$ and $\psi=0$, because the ground state of the
theory will necessarily have a vanishing fermion expectation value and
the bosonic field will be constant.

Without loss of generality, we assume a non-negative real value $\phi = \sqrt{\rho}$ for the
expectation value of the complex Bose field. We emphasize that we do
not explicitly break the symmetry yet, because $\rho$ could be
zero. We rather assume that \emph{if} there is a symmetry breaking,
then the vacuum expectation value is real. The advantage of this is
that by decomposing
\begin{equation}
  \label{cr22} \phi(X) =  \phi +\frac{1}{\sqrt{2}} (\delta\phi_1(X)
 + \rmi \delta \phi_2(X))\,,
\end{equation}
into real fields $\delta \phi_i$, we can distinguish the radial mode
from the massless Goldstone mode via $\delta\phi_1$ and
$\delta\phi_2$, respectively. At the end of the calculation, we
determine the ground state value $\rho_0$ by minimizing
$U_{k=0}(\rho)$ with respect to $\rho$. If $\rho_0=0$, we arrived in
the symmetric phase above $T_{\rm c}$, whereas for $\rho_0 \neq 0$ we
have a spontaneous breaking of the ${\rm U}(1)$-symmetry and are below
$T_{\rm c}$.

The inverse propagators in the real basis have the form $\Gamma_k^{(2)}(Q',Q) =
G^{-1}(Q)\delta(Q'+Q)$, which we encountered above and are given by
\begin{align}
  \label{Fesh17} &G^{-1}_{\phi,k}= \left(\begin{array}{cc} \frac{1}{2}A_k \vec{q}^2
 +U_k'(\rho) + 2 \rho U_k''(\rho) & - Z_k \omega_n \\ Z_k \omega_n & 
\frac{1}{2} A_k \vec{q}^2 + U_k'(\rho)\end{array}\right)\,,\\
 \label{Fesh18} &G^{-1}_{\psi,k}= \left(\begin{array}{cc} - h \phi^* \vare & 
\rmi \omega_n - (\vec{q}^2-\mu)  \\ \rmi \omega_n + \vec{q}^2 -\mu &
 h \phi \vare\end{array}\right)_{4 \times 4}\,.
\end{align}
Compare this to the similar expressions in Eqs. (\ref{bog17}) and
(\ref{bcs17}). Here, $\vare$ is the antisymmetric tensor in two dimensions (cf. Eq. (\ref{bcs5})), and the unit matrix in two dimensions on the off-diagonal of Eq. (\ref{Fesh18}) is suppressed. The matrices correspond to the orderings
$(\phi_1,\phi_2)$ and $(\psi_1,\psi_2,\psi_1^*,\psi_2^*)$ of the field
variables, respectively.

The flow equation (\ref{Fesh16}) contains the regularized expression
$(\Gamma_k^{(2)}+R_k)^{-1}$. We use Litim cutoffs here. With this
choice, we effectively replace $\vec{q}^2/2\rightarrow k^2$ for the
bosons. The fermions are regularized around the Fermi surface such
that $\pm|\vec{q}^2 -\mu|$ is replaced by $\pm k^2$. Explicitly, we
have
\begin{align}
  \label{Fesh19} R_{k,\phi}(Q) = \left(\begin{array}{cc} r_\phi & 0 \\
      0 & r_\phi\end{array}\right),\hspace{5mm} R_{k,\psi}(Q) &=
  \left(\begin{array}{cc} 0 & - r_\psi \\ r_\psi &
      0\end{array}\right)_{4 \times 4}
\end{align}
with regulator functions
\begin{align}
 \label{Fesh20} r_\phi &= A_k (k^2-\vec{q}^2/2)\theta(k^2-\vec{q}^2/2)\,,\\
 \label{Fesh21} r_\psi &=
 (\mbox{sgn}(\vec{q}^2-\mu)k^2-\vec{q}^2+\mu)\theta(k^2-|\vec{q}^2-\mu|)\,.
\end{align}
Due to overall theta-functions arising from $\partial_k R_k$, the
$\vec{q}$-integration is restricted to regions where $\vec{q}^2$ is
replaced by $k^2$ such that the integration over spatial momenta
becomes trivial.

Then, the full flow equation for the effective potential is given by
\begin{align}
  \vspace{-1cm}\nonumber \partial_k U_k(\rho) &= \frac{1}{2} T \sum_{n \in \mathbb{Z}} 
\int\frac{\mbox{d}^3q}{(2\pi)^3} \mbox{tr} \Bigl( G^{-1}_{\phi,k}(Q) \partial_k R_{k,\phi}(Q)\Bigr)\\
 \label{Fesh22} & \quad -\frac{1}{2} T \sum_{n \in \mathbb{Z}} \int \frac{\mbox{d}^3q}{(2\pi)^3} 
\mbox{tr} \Bigl( G^{-1}_{\psi,k}(Q) \partial_k R_{k,\psi}(Q)\Bigr)\,.\nonumber\\
\end{align}
Besides the simple integration over spatial momenta, the Matsubara
summations can be carried out explicitly, because the $\omega_n$ are
not cut off at all. The resulting flow equation is a partial
differential equation for $U_k(\rho) = U(k,\rho)$. We find
\begin{align}
 \nonumber  \partial_k U_k(\rho) = &\mbox{ }\frac{\sqrt{2}k^6}{3\pi^2 S_k}
 \left(\sqrt{\frac{1+w_1}{1+w_2}}+\sqrt{\frac{1+w_2}{1+w_1}}\right)\\
 \nonumber &\times\left(\frac{1}{2}+\frac{1}{e^{\sqrt{(1+w_1)(1+w_2)}k^2/S_kT}-1}\right)\\
 \label{Fesh23} &-\frac{2 k^3}{3\pi^2
   \sqrt{1+w_3}}\left(\frac{1}{2}-\frac{1}{e^{\sqrt{1+w_3}k^2/T}+1}\right)\ell(\mu)\,,
\end{align}
where $w_1=U'_k(\rho)/k^2$, $w_2=(U'_k(\rho)+2\rho U''_k(\rho))/k^2$,
$w_3=h^2\rho/k^4$ and
$\ell(\mu)=(\mu+k^2)^{3/2}\theta(\mu+k^2)-(\mu-k^2)^{3/2}\theta(\mu-k^2)$. At
zero temperature, we set the Bose and Fermi distribution functions to
zero.

The flow of the effective potential in Eq. (\ref{Fesh23}) explicitly
depends on $S_k=Z_k/A_k$. Therefore, in order to close the expression,
we also need flow equations for the wave function renormalization
$Z_k$ and the gradient coefficient $A_k$. These are obtained from the
flow of the inverse boson propagator $G^{-1}_\phi$ via suitable
projection prescriptions. For example, given the flow of the boson
two-point function
$\Gamma_{k,\phi}^{(2)}(Q',Q)=\delta(Q'+Q)G^{-1}_{k,\phi}(Q)$, we
obtain the flow equation for $Z_k$ according to
\begin{align}
 \label{Fesh24} \dot{Z}_k = - \left.\frac{\partial}{\partial \omega} (\dot{G}^{-1}_{k,\phi})_{12}(\omega,0)\right|_{\omega=0}\,.
\end{align}
The flow equation for $A_k$ is derived analogously. We can then again
evaluate the Matsubara summations and $\vec{q}$-integrations. These
calculations are a little intricate, but standard and straightforward
in principle.

\vspace{5mm}
\begin{center}
 \emph{Three building blocks}
\end{center}

\noindent Approximative solutions to the flow equation (\ref{Fesh23})
for the effective potential have to fulfill three key
requirements. These are independent of the particular truncation and
summarize our earlier considerations on general properties of the
effective potential. The computation of $U_k(\rho)$ has to account for
the following three building blocks:
\begin{itemize}
\item[1)] \emph{Phase diagram $\rho_0(\mu,T)$}\\ The phase structure of
  the system under consideration is found from the minimum of the
  effective potential $U_{k=0}(\rho)$. Therefore, we have to allow for
  spontaneous symmetry breaking in the construction of the flow
  equations.
\item[2)] \emph{Equation of state $n(\mu,T)$}\\ We obtain the equation
  of state from the full effective potential via
\begin{equation}
 \label{cr26} P(\mu,T) = - U_{k=0}(\mu,T,\rho_0)\,.
\end{equation}
In most cases, we will be satisfied by having an expression for the
density $n(\mu,T) = (\partial P/\partial \mu)_T$. In particular, by
inverting the relation $n=n(\mu,T)$ for $\mu=\mu(n,T)$, we can
eliminate the chemical potential for the density.
\item[3)] \emph{Vacuum flow of the couplings}\\  As discussed earlier in this section, the microscopic
  couplings appearing in the action $S$ are not the ones we will
  measure in experiment. For this reason, we equipped them with a
  subscript $\Lambda$. In the BCS-BEC crossover, we obtain the
  equation of state and the phase diagram as a function of
  $\nu_\Lambda$ and $h$. In order to find out how these two parameters
  are connected to the scattering length $a$, we have to solve the
  vacuum flow equations. After that, we can express our observables as
  a function of $k_{\rm F}a$, which can then be compared to
  experiment.
\end{itemize}
We use the first and second points as guidelines for the construction
of a suitable truncation of $U_k(\rho)$.

In order to include the effect of spontaneous symmetry breaking, we
remark that
\begin{equation}
 \label{cr33} 0= \frac{\partial U_k}{\partial \phi^*}(\rho_{0,k}) = \phi_{0,k} 
\cdot U_k'(\rho_{0,k}) \,,
\end{equation}
has the three types of solutions discussed in Sec. \ref{SecSSB},
however, at the scale $k$. The observable phase diagram is found from
the corresponding value of $\rho_{0,k}$ in the limit $k \rightarrow
0$. Since we expect a second order phase transition here, an expansion
\begin{equation}
 \label{cr34} U_k(\rho) = m_{\phi,k}^2 (\rho -\rho_{0,k}) +
 \frac{u_{\phi,k}}{2}(\rho -\rho_{0,k})^2\,,
\end{equation}
is sufficient to get the phase structure,
cf. Eq. (\ref{bcs23}). Inserting this ansatz into Eq. (\ref{Fesh23})
we get a set of coupled ordinary differential equations for
$m_{\phi,k}^2$, $u_{\phi,k}$ and $\rho_{0,k}$. The inclusion of
higher terms yields further quantitative improvement, but can be
neglected in a first attempt. By construction, the effective potential
has a minimum at $\rho_{0,k}$. We divide the flow into two regimes:
\begin{itemize}
 \item[(i)] symmetric regime: $m^2_{\phi,k}>0$, $\rho_{0,k}=0$,
 \item[(ii)] broken regime: $m^2_{\phi,k}=0$, $\rho_{0,k} \neq0$.
\end{itemize}
There is no condensate for $k=\Lambda$, because the microscopic
potential is given by $U_\Lambda(\rho)=(\nu_\Lambda-2\mu) \rho$. We
start in the symmetric regime and follow the flow of
$m^2_{\phi,k}$. If this quantity hits zero, we switch to the flow equations
for the broken regime. The mass is then fixed to zero,
$m^2_{\phi,k}=0$, and the flow of the minimum $\rho_{0,k}$ is found
from
\begin{equation}
 \label{cr35} 0 \stackrel{!}{=} \partial_k \bigl( U_k'(\rho_{0,k}) \bigr) = 
\partial_k U_k'|_{\rho_0}(\rho_{0,k}) + U_k''(\rho_{0,k}) \partial_k \rho_{0,k}\,,
\end{equation}
to be given by
\begin{equation}
 \label{cr36} \partial_k \rho_{0,k} = - \frac{\partial_k U_k'|_{\rho_0}
(\rho_{0,k})}{u_{\phi,k}}\,.
\end{equation}
The quantity in the numerator can be obtained from a derivative of the
flow equation for $U_k$. Note that we follow the physics from small to
large scales by investigating the scale-dependent position of the
minimum $\rho_{0,k}$. If we arrive at $\rho_0 >0$ for $k=0$, we are in
the phase of spontaneous symmetry breaking. However, it may even occur
that in the broken regime of the flow the minimum becomes zero and we
have to switch flow equations again. Only the $(k=0)$-value is of
physical relevance.  We will discuss the related issue of
precondensation below.

The expansion in Eq. (\ref{cr34}) can be extended to include pressure
and density. For this purpose we write
\begin{align}
 \nonumber U_k(\rho) = &\mbox{ } -P_k  -n_k\cdot (\mu'-\mu)\\
 \label{cr37} &+ m_{\phi,k}^2 (\rho -\rho_{0,k})+\frac{u_{\phi,k}}{2}(\rho -\rho_{0,k})^2\,,
\end{align}
where $\mu'$ is an artificially introduced off-shell chemical
potential which replaces $\mu$ in all previous calculations. It is
used to generate the flow equation of $n_k$ by taking a
$\mu'$-derivative of Eq. (\ref{Fesh23}) for the effective
potential. Once we have the corresponding flow equation, we set
$\mu'=\mu$. In fact, this is the same procedure as for obtaining slope
and curvature of the effective potential ($m^2_{\phi,k}$ and
$u_{\phi,k}$) by taking derivatives with respect to $\rho$ and
setting $\rho = \rho_{0,k}$ afterwards. Again, we may include higher
orders in $\rho-\rho_{0,k}$ and $\mu'-\mu$ to obtain quantitative
improvements.

\vspace{5mm}
\begin{center}
 \emph{UV renormalization and fine-tuning of the mass}
\end{center}

\noindent The solution of the flow equations derived above results in
$n(\mu,T)$ and $\rho_0(\mu,T)$ as functions of $\nu_\Lambda$ and
$h$. For a comparison with experiment we have to relate this set of
parameters to the scattering length $a$. Another necessity for this UV
renormalization is that the values of the microscopic action are
strongly cutoff dependent. Indeed, we have a dependence $m^2_{\phi,k} \sim k$
for the running mass term for $k\rightarrow \infty$ and thus, if we slightly change $\Lambda$,
we drastically change $m^2_{\phi,\Lambda}$. Since our UV cutoff is
only roughly determined to be $\Lambda \gg \ell^{-1}_{\rm vdW}$,
the precise value of $m^2_{\phi,\Lambda}$ cannot be of any physical
relevance.

Let us briefly recapitulate the general reasoning described earlier and then apply it to the present case. The input for the flow equations is given by $\mu$, $T$ and the
microscopic couplings. As $k$ is lowered from $\Lambda$ to $0$, we
include quantum and thermal fluctuations. For high momentum scales
$k$, only quantum fluctuations are important and a few-particle system
will behave similar to a many-body system. This situations is altered
at the scales $k_{\rm F}$ and $\lambda_T^{-1}$ which are present at
nonzero densities and temperatures. When $k$ reaches these thresholds,
many-body effects set in and the flow of observables is different from
the pure vacuum flow. The FRG idea of UV renormalization is to follow
the flow \emph{twice}: First we solve the flow equations for
$\mu=\mu_{\rm v}$ and $T=0$ such that the many-body scales are never
reached. This tells us what scattering length $a$ we use. Then, in a
second attempt, we use the same initial conditions but vary
$\mu=\mu_{\rm v}+ \mu_{\rm mb}$ and $T\geq0$ to get the functions
$n(\mu,T,a)$ and $\rho_0(\mu,T,a)$.

In the ultraviolet, the effective potential is given by
\begin{equation}
 \label{cr38} U_\Lambda(\rho) = (\nu_\Lambda-2 \mu)\rho\,.
\end{equation}
For conceptual clarity we split up the chemical potential into a
vacuum and many-body part according to
\begin{equation}
 \label{cr39} \mu = \mu_{\rm v} + \mu_{\rm mb}\,.
\end{equation}
We have $\mu_{\rm v}\leq 0$ and $\mu_{\rm mb}\geq 0$. This
decomposition is, of course, artificial, because the chemical
potential is simply a parameter to change the density. However, we
will see that $\mu_{\rm v}$ is related to the binding energy of the
molecules and only appears with nonzero value on the BEC-side, whereas $\mu_{\rm mb}$
determines the density in the crossover and provides the scale $k_{\rm
  F}$.

The chemical potential appears in the inverse fermion propagator
$P_\psi =(\partial_\tau-\nabla^2-\mu)$, which is shown below not to be
renormalized in vacuum. A vanishing Fermi distribution function
$n_{\vec{q}} = \int_{\omega} P_\psi^{-1}(Q)$ thus requires $\mu\leq
0$. For the vacuum problem we set $\mu_{\rm mb}=0$. Moreover, we have
$T=0$, and since the flow will always be in the symmetric phase, we
effectively have $\rho =0$. The latter fact is related to the absence
of condensation.

The flow of the mass $m^2_{\phi,k} = U'_k(\rho_{0,k})$ is found from
Eq. (\ref{Fesh23}) to be
\begin{equation}
  \label{cr40} \partial_k m^2_{\phi,k} = 
\frac{h^2}{6\pi^2k^3} (k^2+\mu_{\rm v})^{3/2} \theta(k^2+\mu_{\rm v})\,.
\end{equation}
The right hand side is solely due to the fermionic contribution. This
equation is solved by
\begin{align}
  \label{cr41} m^2_{\phi,k} = \begin{cases} \begin{array}{l}
      m^2_{\phi,\Lambda} - \frac{h^2}{6 \pi^2}(\Lambda-k) \end{array}
    & (\mu_{\rm v} =0), \\ \begin{array}{l} m^2_{\phi,\Lambda} -
      \frac{h^2}{6\pi^2}\biggl[ \sqrt{\Lambda^2+\mu_{\rm
          v}}\Bigl(1-\frac{\mu_{\rm v}}{2\Lambda^2}\Bigr) \\ -
      \frac{3}{2}\sqrt{-\mu_{\rm v}}
      \mbox{arctan}\Bigl(\frac{\sqrt{\Lambda^2+\mu_{\rm
            v}}}{\sqrt{-\mu_{\rm v}}}\Bigr)\biggr]\end{array} &
    \left(\begin{array}{l} \mu_{\rm v} <0,\\ k \leq |\mu_{\rm
          v}|\end{array}\right) \end{cases}
\end{align}
with
\begin{equation}
 \label{cr41b} m^2_{\phi,\Lambda} = \nu_\Lambda -2 \mu_{\rm v}.
\end{equation}
Since $m^2_{\phi}(k=0)$ and $|\mu_{\rm v}|$ act as gaps for the
excitation of bosons and fermions, respectively, we can split up the
physical vacuum at $k=0$ into three sectors according to
\begin{itemize}
\item[(i)] atom sector ($a <0$): $m^2_{\phi,k=0} >0$, $\mu_{\rm v}=0$,
 \item[(ii)] resonance ($a^{-1}=0$): $m^2_{\phi,k=0} = \mu_{\rm v} =0$,
 \item[(iii)] dimer sector ($a >0$): $m^2_{\phi,k=0} =0$, $\mu_{\rm v} <0$.
\end{itemize}
The three sectors are distinguished by the corresponding type of
particles which interact in the vacuum and, in addition, they are
related to a certain scattering length. The latter statement is proven
below. Note the formal analogy to the classification of the
thermodynamic phase diagram, cf. the discussion below
Eq. (\ref{SSB12}). The positive mass term (i) corresponds to the
symmetric phase. (ii) relates to the critical point. In (iii), a scale
$\mu_\text{v}$ is generated, similarly to $\rho_0$ in the many-body
problem. Despite the fact that no symmetry is broken, the onset of the
molecular bound state shares features of a phase transition in
vacuum. Indeed, there is a (spatial) continuum of degrees of freedom.

The condition on $\mu_{\rm v}$ in (i) - (iii) is easily satisfied by
setting the chemical potential in the microscopic action to a certain
value. For the condition on $m^2_{\phi,k=0}$, we use our solution
(\ref{cr41}) of the flow equation to find a suitable choice of
$\nu_\Lambda$. Obviously, given $\mu_{\rm v}=0$, we will have
$m^2_{\phi,k=0}=0$ precisely if $ \nu_\Lambda =
\frac{h^2}{6\pi^2}\Lambda.$ In the same fashion, $\mu_{\rm v}=0$ and
$\nu_\Lambda > \frac{h^2}{6\pi^2}\Lambda$ yields $m^2_{\phi,k=0}>0$,
i.e. the atom sector of individual fermions. Note that indeed
$m^2_{\phi,\Lambda} \sim \Lambda$ strongly depends on the cutoff. If
$\mu_{\rm v}<0$, we take the second line of Eq. (\ref{cr41}) and solve
$m^2_{\phi,k=0}=0$ for $\nu_\Lambda$. This can easily be done
numerically. With this choice of $\nu_\Lambda$ we are in the molecule
phase.

If we now switch on $\mu_{\rm mb}$, we could already tell from the
choice of $\nu_\Lambda$ and $\mu_{\rm v}$ whether we are on the
BCS-side, on resonance, or on the BEC-side. However, we still need to relate Eq. (\ref{cr41b}) to the precise value of the observable fermion scattering length $a$. From Eq. (\ref{Fesh8}) we know that $\lambda_{\psi,{\rm
    eff},\Lambda} = -h_\Lambda^2/\nu_\Lambda$ is valid in the UV. A
similar equation holds for the renormalized coupling
$\lambda_{\psi,{\rm eff}}=8\pi a$. It is given by $\lambda_{\psi,\rm
  eff} = - h^2/P_\phi(\omega,\vec{q}=0,\mu_{\rm v})$, where $P_\phi$
is the boson propagator analytically continued to real time
frequencies $\omega=-\rmi \omega_n$. The expression has to be
evaluated for the on-shell condition of fermion scattering. We do not
dive into the details here. One can show that in our case we always
have
\begin{equation}
 \label{cr44} 8 \pi a = \lambda_{\psi,\rm eff} = 
- \frac{h^2}{m^2_{\phi,k=0}(\mu_{\rm v}=0)}\,.
\end{equation}

From Eq. (\ref{cr44}) we obtain the following dictionary to translate
$(h_\Lambda,\nu_\Lambda) \rightarrow a$. The fermion scattering length
$a$ is always given by
\begin{equation}
 \label{cr45} a = a(B) =  -\frac{h^2}{8 \pi \nu(B)}, \hspace{5mm} \nu(B) = \nu_\Lambda - \frac{h^2}{6 \pi^2}\Lambda.
\end{equation}
Herein, $h=h_\Lambda$ and $\nu(B)$ is the \emph{physical detuning}
from Eq. (\ref{Fesh2}). It corresponds to the energy difference
between the closed channel bound state and the open channel scattering
threshold. We write the explicit dependence of $a$ and $\nu$ on the
magnetic field $B$ to emphasize that this parameter can be tuned
experimentally. If we choose $\nu(B)>0$, the expression in the
denominator of Eq. (\ref{cr45}) is positive and we are in the atom
sector, or -- at finite density -- on the BCS side. We then have $\mu
= \mu_{\rm mb}$. If $\nu(B)<0$, the scattering length gets positive
and we are on the BEC side. Part of the chemical potential
$\mu=\mu_{\rm v}+\mu_{\rm mb}$ is then determined by the condition
that molecules in the vacuum are gapless, propagating degrees of freedom. This contribution $\mu_{\rm v}$ is found
as a solution of
\begin{align}
 \nonumber 0 \stackrel{!}{=} m^2_{\phi,k=0} = \nu_\Lambda - 
2 \mu_{\rm v}&- \frac{h^2}{6\pi^2}\biggl( \sqrt{\Lambda^2+\mu_{\rm v}}
\Bigl(1-\frac{\mu_{\rm v}}{2\Lambda^2}\Bigr)\\
 \label{cr46} &- \frac{3}{2}\sqrt{-\mu_{\rm v}} \mbox{arctan}
\Bigl(\frac{\sqrt{\Lambda^2+\mu_{\rm v}}}{\sqrt{-\mu_{\rm v}}}\Bigr)\biggr)\,,
\end{align}
for given $\nu_\Lambda$. 

We observe that only on the BEC-side the chemical potential partially
contributes to the vacuum flow. The physical interpretation of
$\mu_{\rm v}$ is very intuitive. It is \emph{half the binding energy
  of a dimer},
\begin{equation}
  \label{cr47} \mu_{\rm v} = \frac{\vare_{\rm B}}{2} \hspace{5mm} (\text{BEC side})\,.
\end{equation}
The binding energy is a negative quantity and thus the total chemical
potential $\mu = \frac{\vare_{\rm B}}{2} + \mu_{\rm mb}$ might be
\emph{negative}. In fact, in the limit $\sqrt{-\mu_{\rm
    v}}/\Lambda\rightarrow 0$ we find from Eq. (\ref{cr46})
\begin{equation}
 \label{cr48} \vare_{\rm B} = -\frac{2}{a^2} =- \frac{\hbar^2}{M a^2} 
\hspace{5mm} (\text{BEC-side})\,.
\end{equation}
This is the well known universal relation for a molecular bound state
from quantum mechanics. If $a$ gets small, $\vare_{\rm B}$ becomes
large and negative. Note that if we do not neglect higher orders in
$\sqrt{-\mu_{\rm v}}/\Lambda\rightarrow 0$, we get deviations from
Eq. (\ref{cr48}), which are related to microscopic details of the
interaction potential and thus non-universal.

We briefly comment here on the dimer-dimer scattering length. The
system of two-component fermions can be described as a gas of bosons
on the BEC side. This Bose gas has a scattering length $a_\phi$
between its constituents. The exact value is known from the quantum
mechanical calculation \cite{Petrov:2004zz} to be
\begin{equation}
 \label{cr48b} \frac{a_\phi}{a} = 0.6\,.
\end{equation}
A mean field calculation omitting bosonic fluctuations gives
$a_\phi/a = 2$. Including fluctuations of the molecules yields a
lower value. Within our truncation we find $a_\phi/a = 0.72$.

\vspace{5mm}
\begin{center}
 \emph{Crossover at zero temperature}
\end{center}

\noindent The analysis of the vacuum problem allows to interpret the
results of the flow equation for given initial conditions in terms of
the scattering length $a$. Tuning the chemical potential $\mu$, we can
now create a nonzero density. Many-body effects become important once
the flow parameter $k$ reaches $k_{\text{mb}}$. The earlier considerations on the initial conditions in the UV
regime remain valid. Making contact to the discussion at the beginning of this section, we take the limit $T\to0$, but now at fixed phase space density $\bar \omega$.

From our building blocks stated above we conclude that the solution of
the flow equation for the effective potential consists in essence in
the determination of
\begin{align}
 \nonumber &n(\mu)\hspace{5mm} (\text{equation of state})\,, \\
 \label{cr49} &\Delta(\mu) = h^2 \rho_0(\mu) \hspace{5mm} (\text{gap})\,.
\end{align}
We introduced here the gap, which has precisely the same meaning as in
the discussion of the BCS theory.

We employ the truncations of the effective action and effective
potential given in Eqs. (\ref{Fesh11}) and (\ref{cr37}) for $\mu_{\rm
  mb} = \mu - \mu_{\rm v} >0$. The resulting flow equations for $Z_k$,
$A_k$, $m^2_{\phi,k}$, $\rho_{0,k}$, $u_{\phi,k}$, $P_k$ and $n_k$ can
easily be solved numerically. The general picture is that the fermions
break the symmetry in the early stages of the flow entering the many-body regime $k\approx k_\text{mb}$, i.e. the fermionic
contributions on the right hand side of the flow equation
(\ref{Fesh23}) are responsible for a decrease of the mass
$m^2_{\phi,k}$, which ultimately hits zero at $k>0$. At this stage of
the flow, we switch flow equations and follow the flow of $\rho_{0,k}$
instead of $m^2_{\phi,k}$. The bosons were quite uninvolved in the
symmetric regime flow. Now, the bosons become very active and try to
restore the symmetry, and thus slow down the running of
$\rho_{0,k}$. We find that for $k\rightarrow 0$ the values of all
quantities saturate and we can read off the result at $k=0$. (These
observations also hold at nonzero temperature. But, depending on $\mu$
and $T$, the mass term may not be zero for $k=0$ and we end up in the
symmetric phase.)

\begin{figure}[tb!]
 \centering
 \includegraphics[scale=0.77,keepaspectratio=true]{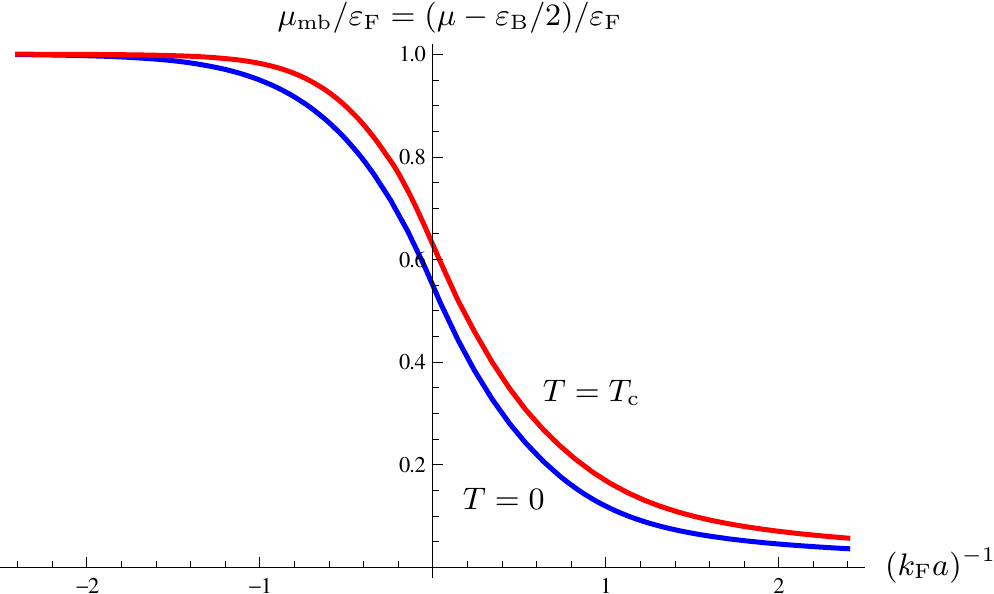}
 \caption{We plot the chemical potential (equation of state) as a
   function of $(k_{\rm F}a)^{-1}$ for $T=0$ and $T=T_{\rm c}$,
   respectively. The Fermi momentum $k_{\rm F} = (3 \pi^2 n)^{1/3}$
   corresponds to the density $n$. The binding energy of the molecular
   bound state on the BEC side for positive $a$ has been
   subtracted. The positive quantity $\mu-\vare_{\rm B}/2$ corresponds
   to $\mu_{\rm mb}$ in the notation of Eq. (\ref{cr39}). For $(k_{\rm
     F}a)^{-1} \lesssim -2$, we observe $\vare_{\rm B}=0$ and $\mu =
   \vare_{\rm F}(n)$. This is equivalent to the equation of state for
   ideal fermions. We find an explanation in the exponentially small
   gap $\Delta >0$, which opens up in the dispersion $E_q$
   (cf. Fig. \ref{BCSDispersion}.) Due to its tiny value, it does not
   influence the equation of state significantly. For zero
   temperature, we can read off the Bertsch parameter
   $\xi=\mu/\vare_{\rm F}$ at unitarity.}
\label{CrossMuMB}
\end{figure}

\begin{figure}[tb!]
 \centering
 \includegraphics[scale=0.77,keepaspectratio=true]{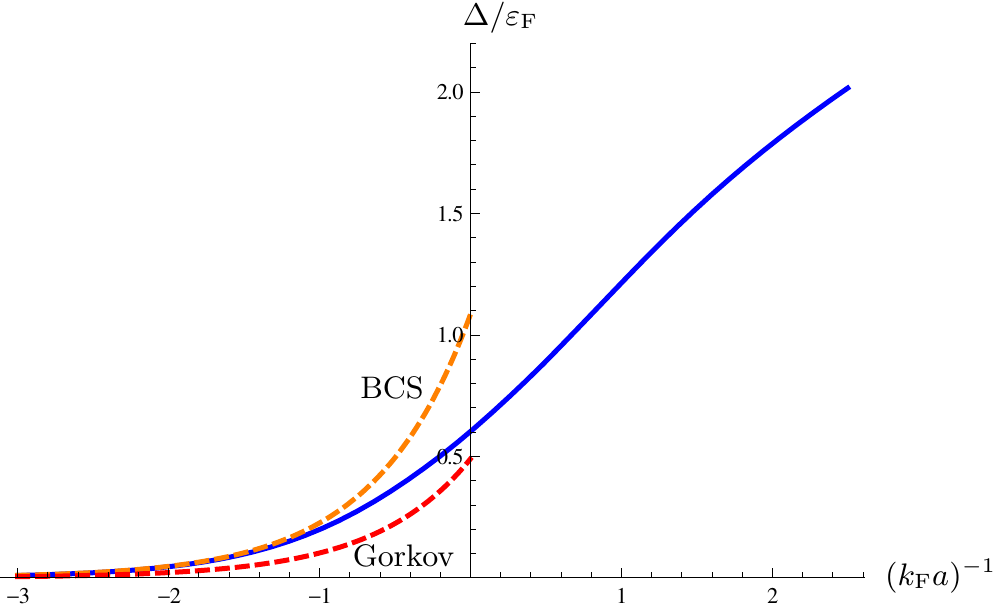}
 \caption{The gap parameter $\Delta^2 = h^2\rho_0$, shown here for
   $T=0$, constitutes the order parameter of the superfluid phase
   transition in the BCS-BEC crossover. It is nonzero for all values
   of $(k_{\rm F}a)^{-1}$ at zero temperature. We find excellent
   agreement of our truncation with the BCS result from
   Eq. (\ref{cr51}). It is well-known that there is a correction to
   the BCS result due to screening effects, which has first been
   calculated by Gorkov and Melik-Barkhudarov. We show the
   corresponding behavior of the gap $\Delta_{\rm Gorkov} =
   (2/e)^{7/3} e^{-\pi/2|k_{\rm F}a|}$, which can be captured with the
   FRG from a more elaborate truncation including particle-hole
   fluctuations.}
\label{CrossGap}
\end{figure}

The solution to the flow equation for the equation of state and the
gap parameter as a function of the crossover parameter $k_{\rm F}a$
are shown in Figs. (\ref{CrossMuMB}) and (\ref{CrossGap}). These plots
already yield a full qualitative understanding of the BCS-BEC
crossover at zero temperature. If we wish to extend our truncation, we
simply have to derive more flow equations and integrate them
numerically.

In the BCS limit ($a<0$, $|k_{\rm F}a|\ll1$), we find $\mu\rightarrow
\vare_{\rm F}(n)$. Thus we arrive at a weakly interacting Fermi gas
with a clearly expressed Fermi surface. We find good agreement with the perturbative
one-loop result from BCS theory
\begin{equation}
 \label{cr50} n = 2 \int \frac{\mbox{d}^3q}{(2\pi)^3} \frac{1}{e^{E_q/T}+1}\,,
\end{equation}
with $E_q = \sqrt{(\vare_q-\mu)^2+\Delta^2}$, see
Eq. (\ref{bcs20j}). For small values of $|k_{\rm F} a|$, the gap
parameter agrees with the BCS result
\begin{equation}
 \label{cr51} \frac{\Delta_{\rm BCS}}{\vare_{\rm F}} = \frac{8}{e^2} e^{-\pi/(2 |k_{\rm F}a|)}\,.
\end{equation}
Since only fermions around the Fermi surface contribute to the
pairing, condensation is weakly expressed. We find strong deviations
from the BCS result for $k_{\rm F}a \sim -1$.

\begin{figure}[tb!]
 \centering
 \includegraphics[scale=0.77,keepaspectratio=true]{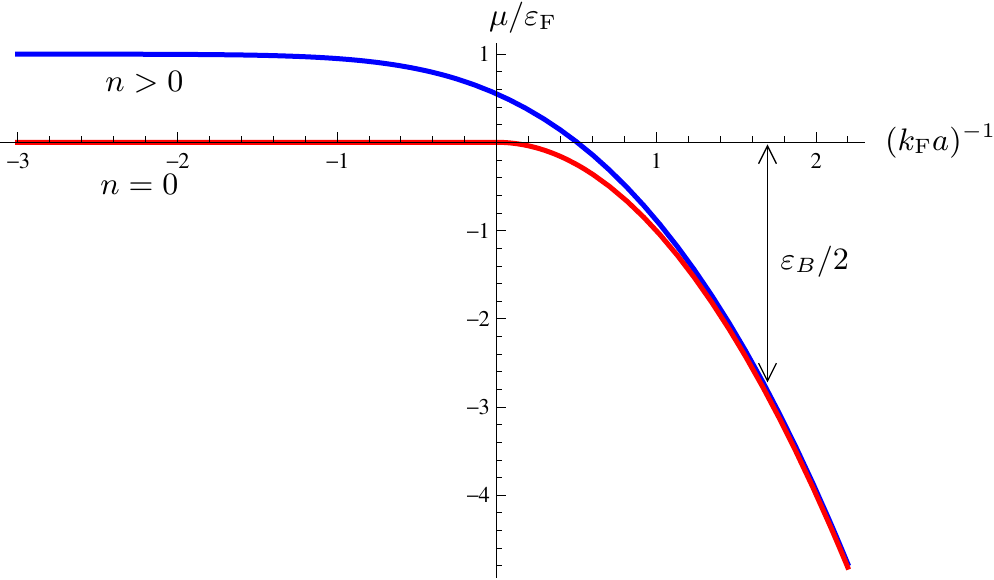}
 \caption{The fermion chemical potential $\mu$ includes the binding
   energy of the molecule on the BEC side of the crossover. It
   eventually becomes negative for $(k_{\rm F}a)^{-1} \approx 0.5$ and
   the behavior of the system is then dominated by few-body
   physics. The chemical potential in vacuum ($n=0$), which is
   equivalent to half the binding energy of the bound state, is shown
   in red. We observe that indeed both functions merge for $(k_{\rm
     F}a)^{-1} \gtrsim 1$. The smooth behavior at nonzero density
   terminates in a sharp second order phase transition in vacuum,
   which is related to the formation of a molecular bound state.}
\label{CrossMu}
\end{figure}

In the BEC limit ($a>0, k_{\rm F}a\ll1$) we find $\mu/\vare_{\rm F}
\rightarrow -\infty$. Comparing this result to the solution of the
vacuum limit, we observe that this regime is largely determined by
few-body physics. We have
\begin{equation}
 \label{cr52} \mu(n) = \vare_{\rm B}/2 + \mu_{\rm mb}(n) \rightarrow -\frac{1}{a^2}\,.
\end{equation}
The chemical potential approaches half the binding energy in this
limit and the many-body scale $\vare_{\rm F}(n) \ll \vare_{\rm B}$
drops out. In particular, we have a zero crossing of the chemical
potential at $(k_{\rm F}a)^{-1} \approx 0.5$. This is demonstrated in
Fig. \ref{CrossMu}.

The function $\mu=\mu(k_{\rm F}a)$ behaves smoothly when going from
the BCS to the BEC side, which is the manifestation of a
\emph{crossover}. In contrast, the vacuum chemical potential $\mu_{\rm
  v} = \frac{\vare_{\rm B}}{2} \theta(a\Lambda)$ has a discontinuous derivative at
$a^{-1}=0$.

In the unitary limit $a^{-1}=0$, the ratio $\mu/\vare_{\rm F}$ is a
universal number, which is also called the Bertsch parameter $\xi$. We
find here $\xi =0.55$. The experimental value is given by $\xi =
0.376(5)$ \cite{Ku12}, which shows that our simple truncation can capture the
qualitative effects but fails for quantitative precision. In
Fig. \ref{CrossMuMB}, we display in addition to the zero temperature
case the behavior of $\mu_{\rm mb}$ at $T=T_{\rm c}$ and find a larger
value of $\mu/\vare_{\rm F}$ at unitarity. This is also found in
experiments \cite{Ku12} but, again, the chemical potential obtained here exceeds the measured
value.

\vspace{5mm}
\begin{center}
 \emph{Finite temperature phase diagram}
\end{center}

\begin{figure}[tb!]
 \centering
 \includegraphics[scale=0.74,keepaspectratio=true]{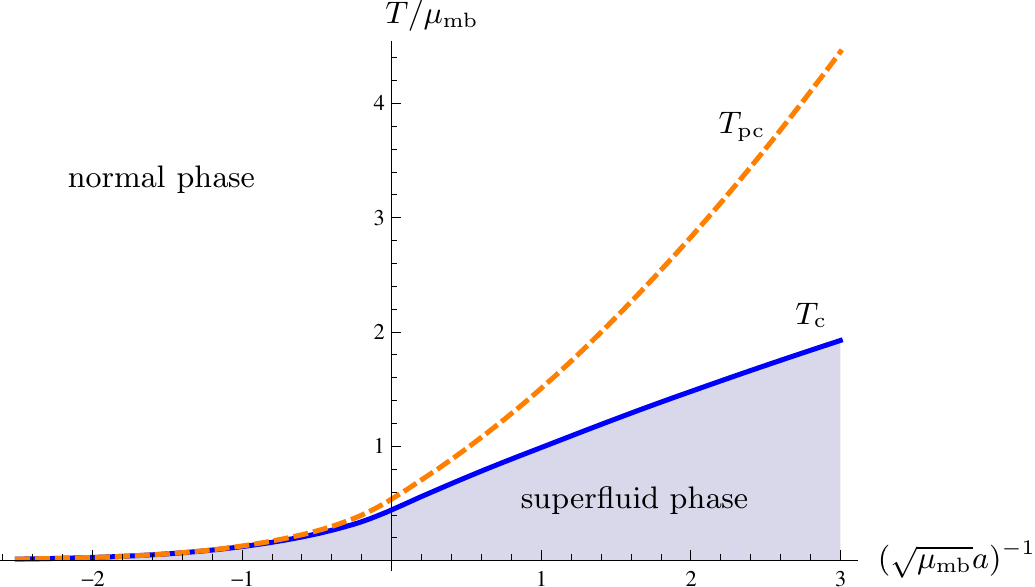}
 \caption{Phase diagram of the three-dimensional BCS-BEC crossover in
   the plane spanned by temperature, chemical potential and scattering
   length. Here, $\mu_{\rm mb}=\mu-\vare_{\rm B}/2$ is positive
   because the binding energy has already been subtracted. Thus,
   $\sqrt{\mu_{\rm mB}}a<0$ and $\sqrt{\mu_{\rm mb}}a>0$ correspond to
   the BCS and BEC sides of the crossover, respectively. The critical
   temperature separates the superfluid from the normal phase. In
   addition, there is a precondensation temperature below which a
   nonvanishing field expectation value $\rho_{0,k}$ appears at
   intermediate stages $k$ of the RG flow. In the BCS limit, $T_{\rm
     c}$ and $T_{\rm pc}$ are practically indistinguishable.}
\label{PDmu}
\end{figure}

\begin{figure}[tb!]
 \centering
 \includegraphics[scale=0.77,keepaspectratio=true]{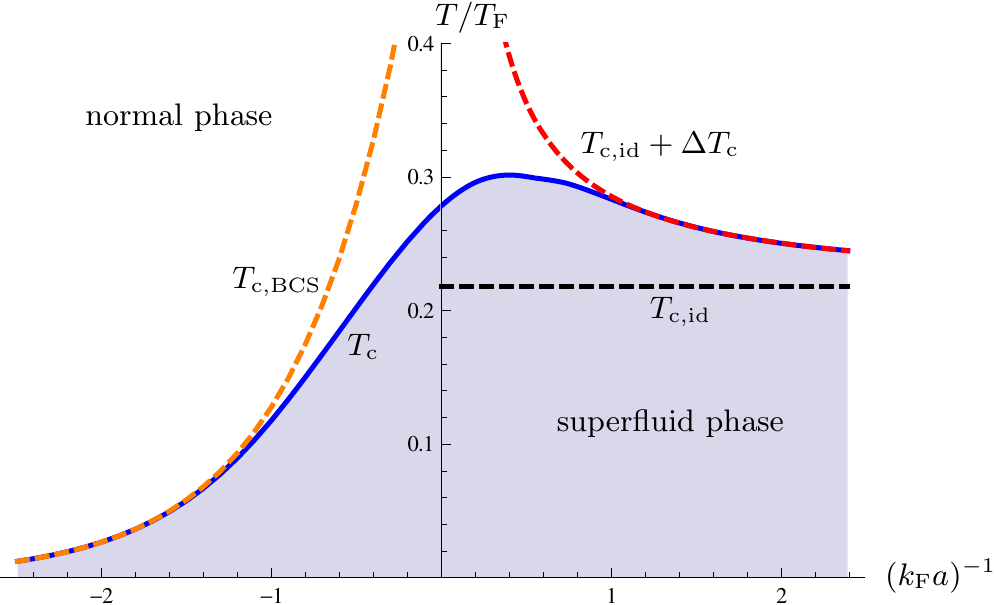}
 \caption{Combing the equation of state at $T >0$ and
   the phase diagram from Fig. \ref{PDmu}, we obtain the phase diagram
   in terms of $T_{\rm c}/T_{\rm F}$ and $k_{\rm F}a$. We indicate the
   limiting cases on the BCS and BEC sides by dashed lines. The BCS
   formula (\ref{bcs20g}) is found here to match for $(k_{\rm
     F}a)^{-1} \lesssim-1$. The condensation temperature of ideal
   bosonic dimers $T_{\rm c,id}/T_{\rm F}=0.218$ is approached on the
   BEC side with a correction proportional to the diluteness parameter
   $k_{\rm F}a$, see Eq. (\ref{cr56}).}
\label{PDkF}
\end{figure}

\begin{figure}[tb!]
 \centering
 \includegraphics[scale=0.8,keepaspectratio=true]{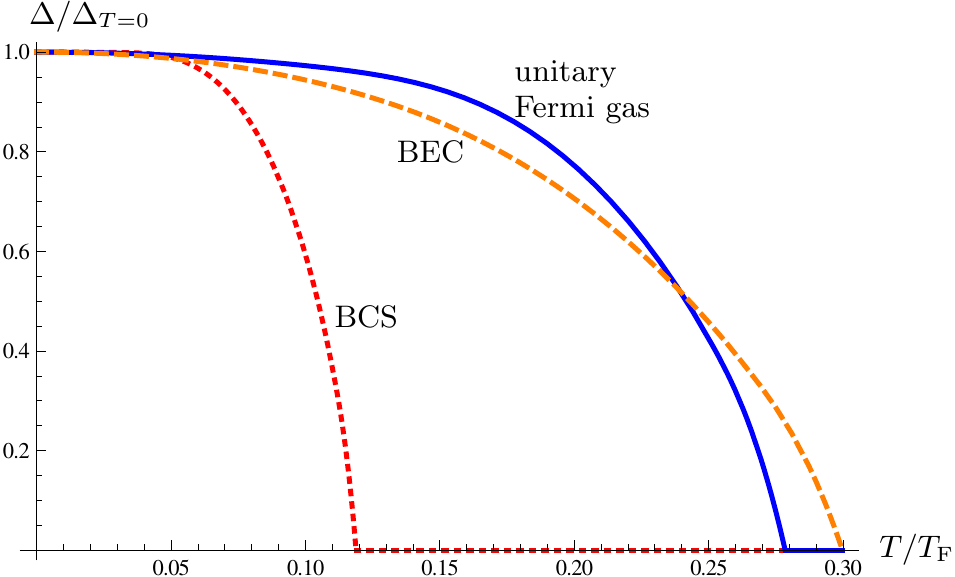}
 \caption{Temperature dependence of the gap $\Delta =h^2 \rho_0$,
   which constitutes the order parameter for the superfluid phase
   transition in the BCS-BEC crossover. The dotted, solid and dashed
   lines correspond to BCS side, unitarity and BEC side,
   respectively. We rescaled the gap by its zero temperature value to
   obtain better visibility of the exponentially small gap in the BCS
   phase. Note that $\Delta(T)$ vanishes continuously for
   $T\rightarrow T_{\rm c}$, thus revealing the phase transition to be
   of second order. It is a particular strength of the FRG that it
   properly accounts for the fluctuations at criticality, which
   diminish the order parameter and eventually drive it to
   zero. Neglecting these fluctuations, the phase transition may
   (wrongly) appear to be of first order.}
\label{TempGap}
\end{figure}

\noindent It is a particular strength of the FRG that calculations at
nonzero temperature are conceptually and technically as
straightforward as the corresponding computations at zero
temperature. We show the finite temperature phase diagram of the
crossover in Figs. (\ref{PDmu}) and (\ref{PDkF}) in terms of
$(\mu,T,a)$ and $(n,T,a)$, respectively. The curves are obtained with
the FRG from the basic truncation given in Eq. (\ref{Fesh11}). The
plots show regions, where the system is either in the normal
(symmetric) phase or in the superfluid phase with spontaneous breaking
of the global ${\rm U}(1)$-symmetry. The superfluid phase transition
is found to be of \emph{second order} throughout the whole
crossover. This justifies our truncation of the effective potential in
Eq. (\ref{cr34}) a posteriori. The temperature dependence of the gap
is shown in Fig. \ref{TempGap}.

Note that the phase diagram is equivalent to a plot of the critical
temperature $T_{\rm c}$. The value of $T_{\rm c}$ for strong coupling
is particularly interesting. Within our truncation we find $T_{\rm
  c}/\mu = 0.44$ and $T_{\rm c}/T_{\rm F} = 0.28$ at unitarity. This
has to be compared to the experimental values $T_{\rm c}/\mu =
0.32(3)$ \cite{Nascimbene10} and $T_{\rm c}/T_{\rm F} = 0.167(13)$
\cite{Ku12}. As for the Bertsch parameter, quantitative precision is
not found within this basic truncation. Part of the error in $T_{\rm
  c}/T_{\rm F}$ consists in determining $T_{\rm F}(n)=(3 \pi^2
n)^{2/3}$ at unitarity, i.e. the equation of state.

We find remarkable agreement of the value for the critical temperature
in the limiting cases of small coupling. On the BCS side, formula
(\ref{bcs20g}) is found to be valid. In particular, the exponential
vanishing of $T_{\rm c}$ with the correct exponent can be verified in
a logarithmic plot, which is not shown here. For small positive
$a=a_\psi$, we expect the system to be described by a gas of weakly
interacting bosons with scattering length $a_\phi$. Due to
interactions effects, the critical temperature deviates from the ideal
gas result $T_{\rm c, id}/T_{\rm F} = 0.218$ (cf. Eq. (\ref{bcs25}))
according to
\begin{equation}
  \label{cr55} \frac{T_{\rm c}-T_{\rm c,id}}{T_{\rm c,id}} = 
  \kappa a_\phi n_{\rm B}^{1/3} \hspace{5mm} (\text{for small }a>0)
\end{equation}
with a dimensionless constant $\kappa$. Rewriting this expression as
\begin{equation}
 \label{cr56} \frac{T_{\rm c}-T_{\rm c,id}}{T_{\rm c,id}} = \frac{\kappa}{(6\pi^2)^{1/3}} \frac{a_\phi}{a} (k_{\rm F}a),
\end{equation}
we can apply our result $a_\phi/a=0.72$ to find $\kappa=1.7$.

\begin{figure}[tb!]
 \centering
 \includegraphics[scale=0.8,keepaspectratio=true]{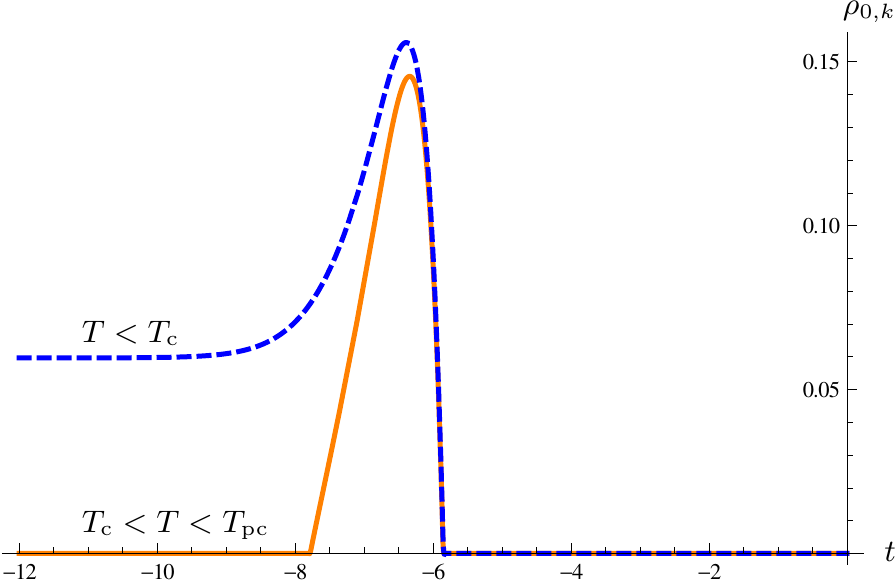}
 \caption{The precondensation temperature $T_{\rm pc}$ is defined as
   the highest temperature where a nonzero value of $\rho_{0,k}$
   appears at $k>0$. This minimum of $U_k$ is not necessarily nonzero
   for $k=0$. For this reason, there is a region in the phase diagram
   where precondensation occurs, but the system is not yet
   superfluid. The distinction between $T_{\rm c}$ and $T_{\rm pc}$ is
   illustrated in this plot. The flow parameter is parametrized
   according to $t=\log(k/\Lambda)$. We have $\rho_{0,k}\equiv0$ for
   $T > T_{\rm pc}$. In this plot, we keep $\sqrt{\mu_{\rm mb}}a$
   fixed, which corresponds to $(k_{\rm F}a)^{-1} = 0.74\mbox{ }
   (0.72)$ above (below) $T_{\rm c}$, respectively.}
\label{PreCon}
\end{figure}

In Fig. \ref{PDmu}, we show the precondensation temperature $T_{\rm
  pc}$. For $T \leq T_{\rm pc}$, there is a momentum scale $k_{\rm
  SBB}$ where a nonzero field expectation value $\rho_{0,k}$ appears
during the flow. The superfluid phase corresponds to those
temperatures, where $\rho_{0,k=0} = \rho_0$ is nonzero for $k=0$ and
thus constitutes the order parameter of the phase transition. For
intermediate temperatures $T$ such that $T_{\rm c} < T \leq T_{\rm
  pc}$, the field expectation value does not survive in the infrared
and we arrive in the symmetric phase. This precondensation region can
be viewed as a state of the system where we have correlated domains of
size $\sim k^{-d}$. We visualize the flow of $\rho_{0,k}$ in
Fig. \ref{PreCon}.

Let us pause here for a moment to discuss the relation of the truncation presented in this section to other approximations. For this purpose, we start from a more basic ansatz and then successively build in additional effects until arriving at the present truncation scheme. Quite remarkably, the qualitative physics of the crossover problem at zero temperature can be described in terms of just a BCS gap equation (\ref{bcs20f}), together with a self-consistent treatment of the equation of state (\ref{bcs20i}) with fermionic contribution alone. While in the BCS regime the equation of state is well approximated (up to exponentially small corrections) by the Fermi sphere contribution (i.e. $n = (2M\mu)^{3/2}/(3\pi^2)$ with $\mu=\epsilon_\text{F}$), ramping the inverse scattering length to large positive values leads to $n \to 2\phi^*\phi$, which describes a (renormalized) condensate without depletion. This is the original approximation of Eagels \cite{eagles69} and Leggett \cite{leggett80}. The BEC regime here is described in terms of just a condensate, but clearly there are no propagating bosonic degrees of freedom. The next stage of approximation improves on this point, and includes fermionic fluctuations which build up a bosonic propagator \cite{nozieres85,melo93}. This then allows to qualitatively describe the crossover at finite temperatures, including the phase border to the symmetric phase. This approximation yields a full, qualitatively correct finite temperature phase diagram, which often is referred to as the extended mean field or Nozi\`eres--Schmitt-Rink approximation. In the FRG language, it corresponds to keeping only the fermionic contributions to the flow equations of $U_k$, $Z_k$, and $A_k$, respectively, and no bosonic feedback. This allows to describe an effective theory of pointlike bosons in the BEC regime, which is characterized by a nonrelativistic mass $2M$ and an effective scattering length $a_\phi = 2 a$.

A consistent generalization of the extended mean field theory by
semi-analytical means is not straightforward. It has been addressed in
the frame of functional field-theoretical techniques, in particular
via $\epsilon$-expansion
\cite{PhysRevA.74.053622,PhysRevLett.97.050403,PhysRevA.75.063618,PhysRevA.75.063617,PhysRevA.75.043605,PhysRevA.75.043620},
$1/N$-expansion \cite{PhysRevA.75.033608,PhysRevD.78.125010}, t-matrix
approaches
\cite{springerlink:10.1007/BF01344058,PhysRevLett.85.2801,PhysRevB.61.15370,PhysRevLett.92.220404,PhysRevB.70.094508,PhysRevA.84.013610},
Dyson--Schwinger equations
\cite{Diehl:2005ae,Diehl:2005an,PhysRevA.77.023626}, 2-Particle
Irreducible methods \cite{PhysRevA.75.023610}, and renormalization
group flow equations
\cite{Birse2005287,Diehl:2007th,Diehl:2007ri,PhysRevLett.100.140407,Floerchinger:2008qc,PhysRevB.80.104514}. An
immediate challenge can be inferred from the fact that the inclusion
of fluctuations related to the bosonic sector leads to infrared
divergent integrals already in perturbation theory. We have met an
example of this situation in our analysis of extensions of
Bogoliubov's theory for weakly interacting bosons at zero temperature;
at finite temperature, those divergences become even more serve. In
addition, a notorious problem known for bosonic theories is the order
of the finite temperature phase transition, which typically is wrongly
found to be of first order.

Already the simple truncation scheme \cite{Diehl:2007th} advocated above, provides the means to consistently and systematically deal with these problems. The flow equation is both UV and IR finite and thus it is not plagued by the problems mentioned above. Physically, this allows to include the order parameter  fluctuations in the BCS regime, and the thermal and quantum fluctuations of the effective molecular bound state degree of freedom in a single framework. We have described a few important improvements resulting from this approach on various scales: On short distances, bosonic vacuum fluctuations renormalize the effective bosonic scattering length already close to the exact value obtained from the solution of the four-body Schr\"odinger equation. On thermodynamic scales, we observe a shift in the critical temperature, predicted for pointlike bosons from effective field theory and Monte Carlo simulations. At long distances, we correctly capture the second order phase transition throughout the whole crossover, characterized by an anomalous dimension close to the best available estimate from the $\epsilon$-expansion and numerical approaches. This discussion is summarized in Tab. \ref{TabExponents}.

\begin{table}[t]
\begin{center}
\begin{tabular}{|c|c|c|c|}
\hline
  Physical scale & MF &   FRG  &  Other \\
\hline
 \hline
{\small Microphysics,} $\frac{a_\phi}{a}$ & 2 & 0.72 & 0.6 (Ref. \cite{Petrov:2004zz})\\
 \hline
{\small  Thermodynamics,} $\kappa$ & 0 & 1.7 & 1.3 (\cite{PhysRevLett.87.120401,PhysRevLett.87.120402,Blaizot:2005wd})\\
 \hline
{\small Critical behavior,} $\eta$ & 0 & 0.05 & 0.038 (Ref. \cite{Pelissetto:2000ek})\\
 \hline
\end{tabular}
\end{center}
\caption{Quantitative precision on all scales: The BCS-BEC crossover shows a separation of scales, which can be benchmarked by several key observables. We display here a selection of representatives for each sector and its value in extended mean field theory (MF), from our FRG truncation introduced above, and other methods. The microphysics are sensitive to the scattering of composite bosons with effective molecular scattering length $a_\phi$. A nonperturbative thermodynamic effect on the BEC side consists in the shift of the critical temperature relative to the ideal gas value, $\Delta T_{\rm c,BEC}/T_{c,id} = \kappa a_\phi n_{\rm B}^{1/3}$. The critical exponent $\eta$ known from the theory of critical phenomena has to be compared to the three-dimensional ${\rm O}(2)$ universality class, because this is the symmetry of the complex order parameter $\phi_0$.}
\label{TabExponents}
\end{table}

\vspace{5mm}
\begin{center}
 \emph{Efimov physics at resonance}
\end{center}

\noindent At resonance, the gas of two-component fermions is dense and
strongly interacting. In particular, three-body processes become
relevant. Here, we would like to give a short excursion which
demonstrates that the few-body sector in the strongly interacting
regime $a^{-1}\to0$ holds interesting physics in its own right, the
qualitatively most interesting part occurring for bosons or
three-component fermions. The corresponding physics of few-particle
systems in vacuum can be investigated within an FRG framework using
vertex expansions \cite{PhysRevC.78.034001,PhysRevA.79.042705} or
refermionization techniques
\cite{PhysRevA.79.053633,PhysRevA.79.013603}. These procedures remain
tractable since massive diagrammatic simplifications occur in the
vacuum limit (\ref{MB3}), leading to a closed hierarchy of flow
equations of the $N$-body sector. (In other words, not unexpectedly,
the nonrelativistic $N$-body problem can be solved without knowledge
of the $(N+1)$-body problem.) Here we give a flavor of this physics
only, discussing the Efimov effect of three resonantly interacting
particles, which is explained by a limit cycle behavior of the RG
flow. For some further aspects of vacuum physics, we refer to
\ref{AppVac}. In particular, we point to the reviews
\cite{braaten-review,Floerchinger:2011yv}, addressing Efimov physics
from a field theoretical perspective.

Efimov found the effect named after him from a consideration of the
Schr\"{o}dinger equation for three resonantly interacting identical
bosons. He showed that this problem can be mapped to the scattering in
an inverse square potential at short distances
\cite{Efimov:1970zz,Efimov1973157}. The latter potential has a
discrete spectrum of bound states which form a geometric series,
i.e. two neighboring bound states $E_n$ and $E_{n+1}$ satisfy
\begin{equation}
 \label{vac14} \frac{E_{n+1}}{E_n} = e^{-2\pi/s_0},
\end{equation}
where $s_0 \simeq 1.00624$ is know from quantum mechanics. This
scaling of the infinite tower of bound states is known as the
\emph{Efimov effect}. With the help of Feshbach resonances, this
effect can be studied using ultracold quantum gases
\cite{Grimm1aa,Grimm2aa,zaccanti1aa,Pollack18122009,PhysRevLett.103.163202,PhysRevLett.102.165302,PhysRevLett.101.203202,PhysRevLett.103.130404,PhysRevLett.103.043201,PhysRevLett.105.103203,PhysRevLett.105.103201,PhysRevLett.105.023201,springerlink:10.1007/s00601-011-0260-7,PhysRevLett.107.120401}.

In order to describe this problem in a field theoretical FRG
framework, we need to include an atom-dimer interaction. The full
solution for the momentum dependent atom-dimer vertex can be found
within a vertex expansion of the FRG \cite{PhysRevA.79.042705}. The
dimer can consist of either two fermions or two bosons, see the
examples below. Using kinematic simplifications and projection to zero
angular momentum partial waves, the flow equation for the atom-dimer
vertex $\gamma_{3,k}$ can be reduced to a quadratic matrix
differential equation \cite{PhysRevC.78.034001,PhysRevA.79.042705}. We
will not present details on these approaches here, but rather explain
the basic phenomenon underlying the Efimov effect: a renormalization
group \emph{limit cycle}. It is a strength of the FRG that this
qualitative effect can be identified in a minimalistic truncation in
terms of a single atom-dimer interaction parameter
\cite{PhysRevA.79.042705}.

To this end we consider the following types of three-particle systems:
(i) identical bosons, (ii) two-component fermions (i.e. two hyperfine
states), and (iii) three-component fermions (i.e. three hyperfine
states). Systems (i) and (iii) are assumed to have a ${\rm SU}(2)$ and
${\rm SU}(3)$ spin symmetry, respectively. For the latter case, this
is not generic, because the resonances of the three mutual scattering
lengths $a_{12}(B)$, $a_{13}(B)$ and $a_{23}(B)$, as a function of the
magnetic field $B$, are in general not identical -- there is no
fundamental ${\rm SU}(3)$ symmetry in cold atoms. However, fine-tuning
of parameters can realize such a situation to sufficient accuracy
\cite{PhysRevLett.101.203202,PhysRevLett.105.103201}. For theoretical
work on this problem, see
Refs. \cite{PhysRevLett.103.073202,PhysRevA.79.053633,PhysRevA.79.013603,PhysRevA.81.013605,Endo:2012sa}.

We can now give a simple RG argument to identify which of the above
three-particle systems allows for the Efimov effect
\cite{PhysRevA.79.042705}. We approximate the atom-dimer-vertex matrix
by a single entry independent of momentum, i.e. a single coupling
constant $\lambda_{3,k}$. The corresponding flow equation for the
dimensionless\footnote{The canonical power counting at the unitary
  point deviates from the canonical power counting for finite
  $a^{-1}$, as can be inferred from the exact solution of the two-body
  problem, see e.g. \cite{braaten-review}.} coupling
$\tilde{\lambda}_{3,k}= \lambda_{3,k}k^2$ has the structure
\begin{equation}
 \label{vac15} \partial_t \tilde{\lambda}_3 = \alpha \tilde{\lambda}_3^2 + \beta \tilde{\lambda}_3 +\gamma,
\end{equation}
where the constants depend on the choice (i) - (iii). The solution to this equation is given by
\begin{equation}
 \label{vac16} \tilde{\lambda}_3(t) \sim \left\{\begin{array}{c} \tanh(\sqrt{D}t/2) \hspace{5mm}(D\geq 0)\\ \tan(\sqrt{|D|}t/2) \hspace{5mm} (D <0)\end{array}\right.,
\end{equation}
where $D=\beta^2-4\alpha\gamma$ is the discriminant of the beta
function in Eq. (\ref{vac15}). The first case ($D\geq 0$) is found for
system (ii). The coupling $\tilde{\lambda}_3$ then reaches an infrared
fixed point. Intuitively, this can be understood from Pauli's
principle, preventing two spin-1/2 fermions from coming too close to
each other. However, for systems (i) and (iii) it turns out that $D
<0$, and thus the solution is \emph{periodic} in RG-time
$t=\log(k/\Lambda)$. We say that the flow approaches an infrared
\emph{limit cycle}. The periodicity of the solution with time is
\begin{equation}
 \label{vac16b} T=2\pi/\sqrt{|D|}.
\end{equation}
For each divergence of the three-body scattering amplitude at zero
frequency and momentum approximated by $\lambda_{3,k}$, a new Efimov
bound state is expressed. This gives rise to an infinite number of
exponentially spaced Efimov bound states, i.e. the binding energy of
successive bound states vanishes exponentially fast. This
consideration is exact at the resonance. Away from it, due to the
scaling violations caused by $a^{-1}$, only a finite number of Efimov
states exists. Experiments with ultracold atoms have resolved the
lowest (most deeply bound) Efimov states.

In order to compute $s_0$, we perform a \emph{scale identification},
which relies on the fact that the flowing action can be seen as a
theory at momentum scale $k$. However, the subtle association of $k$
to physical scales like external momenta, bound state energies etc. is
not unique. Here, we observe that from dimensional arguments we have
$E \sim k^2$. The prefactor drops out when calculating the ratio of
two energies. We thus associate two neighboring dimer bound state
energies $E_n$ and $E_{n+1}$ with the corresponding scales $k_n$ and
$k_{n+1}$ in the limit cycle. We have $t_{n+1}=t_n -T$ and accordingly
\begin{equation}
 \label{vac17} \frac{E_{n+1}}{E_n} = \frac{k_{n+1}^2}{k_n^2} = e^{-2 T}.
\end{equation}
We compare this to Eq. (\ref{vac14}) and deduce $s_0 = \pi/T$. The
Efimov parameter is related to the periodicity of the limit cycle.

From this simple analysis including only the single parameter
$\lambda_3$, one obtains $s_0 \simeq 1.393$. This is quantitatively
not yet very accurate, but we inferred the relevant qualitative
physics. The FRG approach can be completed by solving the flow
equation for the momentum dependent vertex $\gamma_{3,k}$
numerically. The result $s_0\simeq1.0$ is compatible with the value
from other approaches \cite{braaten-review}.

\vspace{5mm}
\begin{center}
 \emph{Open challenges and further reading}
\end{center}

\noindent As discussed above, the truncation for the effective action
$\Gamma_k$ given in Eq. (\ref{Fesh11}) provides a consistent picture
of the whole crossover, interpolating smoothly between the limiting
cases of BCS superfluidity of atoms and Bose condensation of dimers,
and in particular providing a consistent treatment of the bosonic
sector, with a number of quantitative improvements. Starting from
this, open challenges remain. Some of them have been successfully
addressed and others remain for the future. For a more detailed
discussion of this subject, see the review article
\cite{Scherer:2010sv}.

\begin{figure}[tb!]
 \centering
 \includegraphics[scale=0.57,keepaspectratio=true]{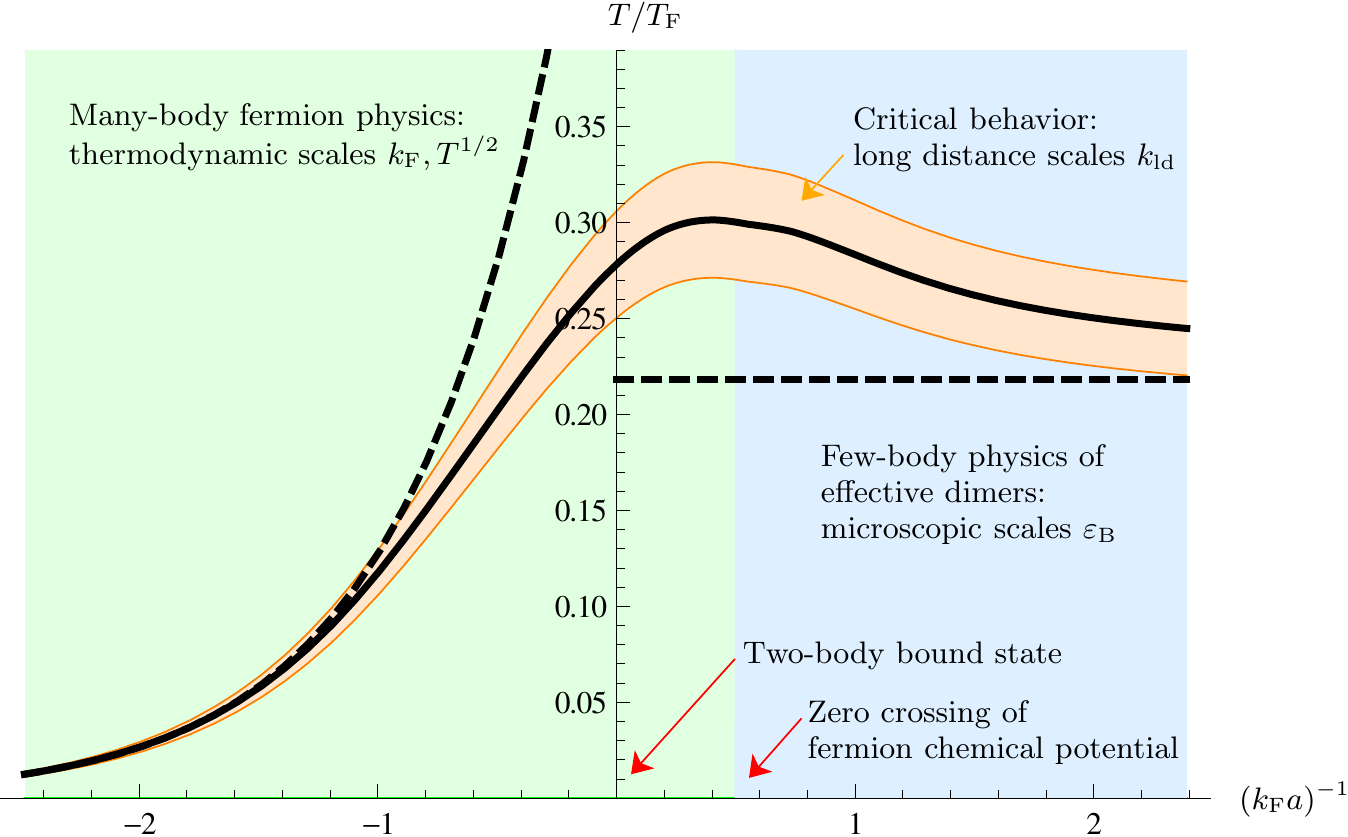}
 \caption{The phase diagram of the BCS-BEC crossover can be divided
   into regions where different momentum scales are
   dominant. Including effects beyond mean field theory provides a
   challenge in each individual sector. In addition, a unifying
   description of the whole crossover has to match the correct
   limits. The zero crossing of the fermion chemical potential at
   $(k_{\rm F}a)^{-1} \approx 0.5$ separates a region where many-body
   effects are important from a region where the main features of the
   system can be captured by an effective theory of dimers. In the
   latter, we have $|\vare_{B}| \approx \frac{2}{a^2} \gg k_{\rm
     F}^2,T$ such that microscopic scales are relevant. The situation
   is reversed to the left of the zero crossing. In the region around
   the critical line, long distance physics and infrared fluctuations
   become important.}
\label{PDScales}
\end{figure}

In order to identify these challenges, we structure the physics of
different regimes of the BCS-BEC crossover more carefully according to
the relevant length scales. This will help to find suitable
improvements for our ansatz of the effective action. We visualize the
situation in Fig. \ref{PDScales}, which serves as a guide for the
following discussion. We divide the phase diagram into three major
regions, and we will argue that the main remaining challenges are
determined by the few-body, many-body and long distance length scales,
respectively.

\begin{figure}[tb!]
 \centering
 \includegraphics[scale=0.63,keepaspectratio=true]{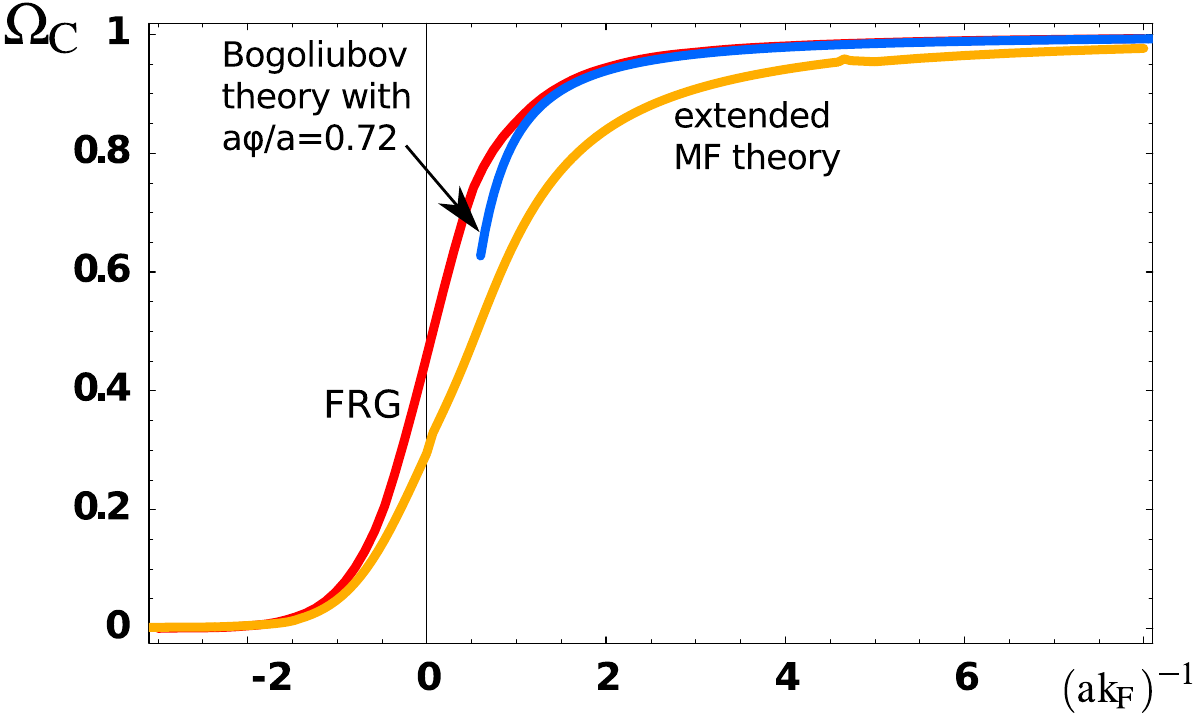}
 \caption{Condensate fraction $\Omega_{\rm C}$ at zero
   temperature. The FRG result is indicated by the red line. It agrees
   well with Bogoliubov theory of effective dimers with molecular
   scattering length $a_\phi/a=0.72$. We also show the predictions
   from extended mean field theory.}
\label{CondFrac}
\end{figure}

(i) \emph{BEC regime} -- On the BEC side of the crossover, we have a
nonvanishing binding energy $|\vare_{\rm B}| \approx 2/a^2$. In
particular, for $a \rightarrow 0$, its value exceeds all other scales
of the problem, such as the many-body scales $\vare_{\rm F}$ and
$T$. We therefore have a clear separation of scales: The physics of
the effective boson theory builds up at $k\sim 1/a$, and will not
 be influenced noticeably by the presence of the many-body scales which
become important at $k\sim k_\text{F}, \sqrt{T}$. The many-body
physics can then be described by a theory of effective bosons, cf. Fig
\ref{CrossMu}. Nevertheless, neither the physics of the microscopic bound state nor
the many-body physics of the effective bosons is trivial. The first
challenge concerns the correct description of the bosonic
self-interaction, i.e. the bosonic scattering length $a_\phi$. On
dimensional grounds, $a_\phi = \beta a$, but the determination of the
dimensionless number $\beta$ represents a genuinely nonperturbative
problem with no small expansion parameter, despite $a\to 0$. The
problem has been solved exactly from the four-body Schr\"odinger
equation \cite{Petrov:2004zz}, with the result $\beta =0.6$, see
Tab. \ref{TabExponents}. This has also been derived from
phenomenological two-loop self-consistency equations
\cite{brodsky05,PhysRevA.73.053607}. Recently, the result was also
found in a truncation of FRG equations taking into account the
feedback from an atom-dimer vertex \cite{PhysRevA.83.023621}. This
truncation could be used as a starting point for a many-body
calculation, which would provide highly accurate low-temperature
many-body results on the BEC side based on the separation of scales
and the validity of Bogoliubov theory for many-body observables. These
statements are illustrated in Fig. \ref{CondFrac}, where we plot the
condensate fraction at zero temperature as a function of $(k_{\rm
  F}a)^{-1}$ in comparison with an effective Bogoliubov theory with a
phenomenologically assumed molecular scattering length $a_\phi =0.72
a$. Deviation from the effective theory of pointlike bosons occur near
resonance, where $|\vare_{\rm B}|$ becomes of comparable size to the
thermodynamics scales. For further comparison, we plot the result from
extended mean field theory approaches, in which $a_\phi=2a$.

For higher temperatures, the infrared divergences become more severe,
and Bogoliubov theory or its finite-temperature extension known as
Popov approximation \cite{popov-book} fail to predict both the second
order phase transition as well as the shift in $T_\text{c}$
\cite{PhysRevLett.83.1703}. These genuine many-particle effects, which
can be traced back to the physics of the zero Matsubara mode only and
thus the 3D O(2) model alone, are already quite well captured in the
above truncation, see Tab. \ref{TabExponents}. Higher precision can be
achieved in an FRG treatment within higher orders of the derivative
expansion, e.g.\ \cite{Litim:2010tt,Canet:2003qd}, or by treating the full
momentum dependence, e.g.\ \cite{Benitez:2009xg,Benitez:2011xx}. 

\begin{figure*}[t!]
 \centering
 \includegraphics[scale=0.5,keepaspectratio=true]{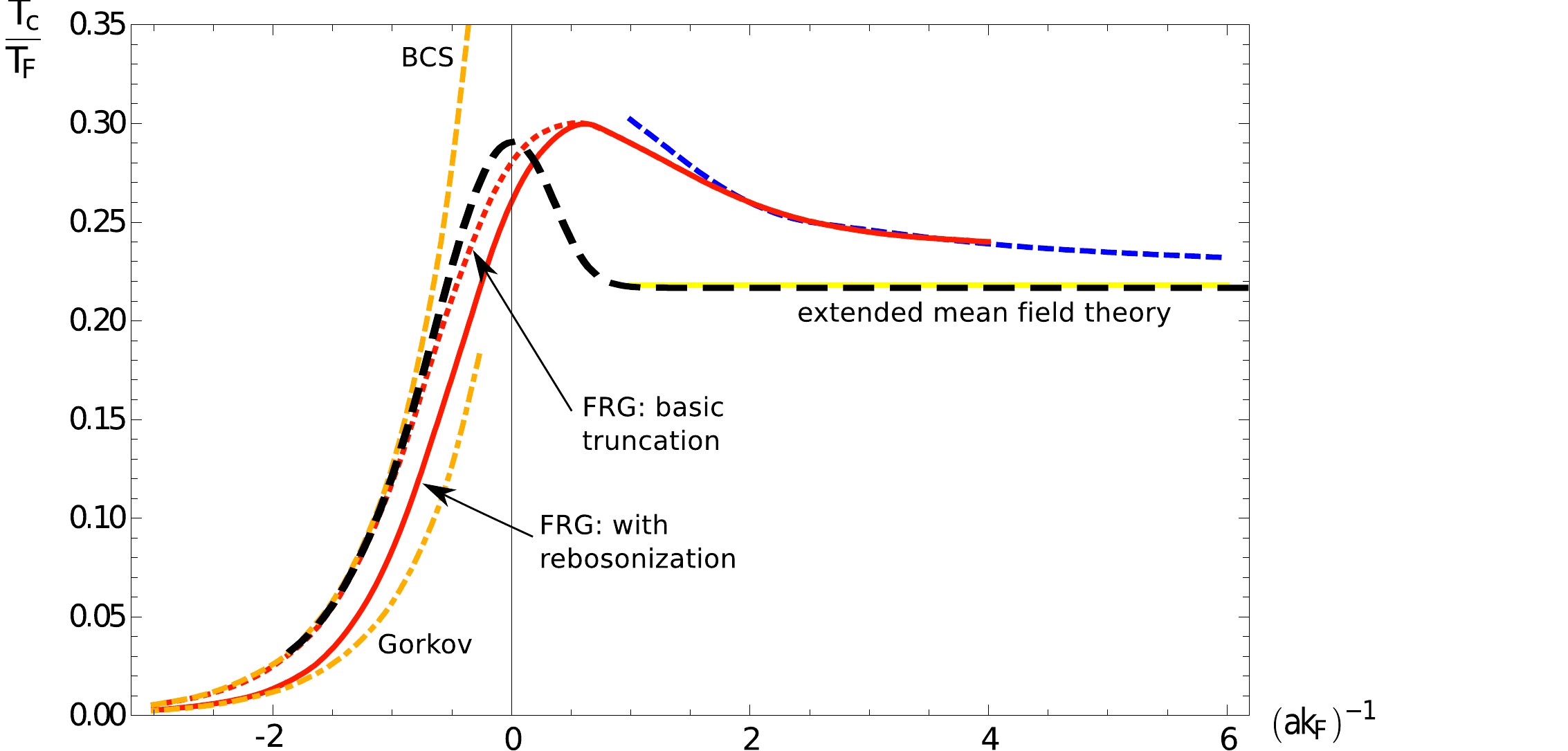}
 \caption{Critical temperature of superfluidity in the BCS-BEC
   crossover. The limiting cases of the BCS and Gorkov critical
   temperature for negative fermion scattering length are indicated as
   well as the ideal gas condensation temperature on the BEC side,
   both with and without shift $\Delta T_{\rm c, BEC}/T_{\rm c,id}
   \propto a n^{1/3}$. The basic truncation scheme for the effective
   action reproduces the correct limiting boson theory, but
   overestimates the critical temperature on the BCS side. By
   including particle-hole fluctuations within a rebosonization scheme
   \cite{Floerchinger:2008qc}, we find agreement with the Gorkov
   result on the BCS side. The corresponding many-body effect is only
   weakly expressed at resonance and vanishes at the zero crossing of
   the fermion chemical potential. We also plot the critical
   temperature obtained from extended mean field theory.}
\label{CritTemp}
\end{figure*}

(ii) \emph{BCS and unitary regime} -- No bound state is formed on the
BCS side of the crossover and thus the many-body scales $k_{\rm F}$
and $T$ dominate the physics in this region. An ordering principle is
at best given by the presence of a strongly expressed Fermi surface,
such that only modes in its vicinity can contribute to the
thermodynamics in a nontrivial way. In particular, the corrections to
the equation of state must remain exponentially small. An interesting
beyond mean-field many-body effect is present in this regime
nevertheless: While qualitatively the finite temperature transition is
governed by the $\log T$-divergence in the particle-particle channel
in the RG flow of the four-fermion vertex, there are additional
contributions from the particle-hole channel which remain regular for
$T \rightarrow 0$. They can thus be treated perturbatively, and be
taken into account via a shift of the dimensionless scattering length
according to $ak_\text{F} \to ak_\text{F} - \alpha (ak_\text{F})^2$,
with $\alpha>0$ a number of order unity. Taking this shift into
account in the exponent governing the BCS critical temperature
\eq{bcs20f}, and treating it perturbatively in $ak_\text{F}$ as
appropriate in this regime, one obtains a multiplicative correction of
the prefactor for the critical temperature
$e^{-\pi\alpha/2}=(2.2)^{-1}$, resulting in
\begin{equation}
 \label{crnew1} \frac{T_{\rm c, BCS}}{T_{\rm c, Gorkov}} = 2.2.
\end{equation}
This suppression of the critical temperature due to screening via
particle-hole fluctuations is know as Gorkov's effect
\cite{gorkov61}. The particle-hole fluctuations are tied to the
presence of a Fermi surface, around which these pairs are
created. Consequently, the Gorkov correction is absent in vacuum and
constitutes a true many-body effect.

One may now ask how to recover this effect in the FRG treatment
\cite{Floerchinger:2008qc}. Since the Hubbard--Stratonovich
transformation of the original fermionic theory has to be performed in
the particle-particle channel in order to capture the formation of the
bound state -- which is the essence of the BCS-BEC crossover -- we may
think that the particle-hole channel is lost. However, there is an
elegant trick to recover it with the help of the rebosonization
technique
\cite{Gies:2001nw,Pawlowski:2005xe,Floerchinger:2009uf,Floerchinger:2010da}. More
explicitly, we utilize the fact that, although we have chosen our
initial values such that the four-fermion coupling vanishes,
\begin{equation}
 \label{crnew2} \lambda_{\psi,\Lambda} = \lambda_{\rm bg} =0,
\end{equation}
a nonzero value of this quantity is immediately generated during the
RG flow: $\partial_k\lambda_{\psi,k} \neq 0$. The corresponding
contributions on the right hand side of this equation are precisely
the particle-hole fluctuations. Thus, an extension of our truncation
according to
\begin{equation}
 \label{crnew3} \Delta \Gamma_k = \frac{\lambda_{\psi,k}}{2} \int_X (\psi^\dagger\psi)^2
\end{equation}
captures the Gorkov correction. Spontaneous symmetry breaking during
the flow is now signaled by a divergence of the overall four-fermion
coupling $\lambda_{\psi,k} - h_k^2/m_{\phi,k}^2$. Since this is rather
difficult to resolve numerically, we conduct a Hubbard--Stratonovich
transformation on each scale $k$ to absorb the contribution from
fluctuations to $\lambda_{\psi,k}$ into the running of $h_k$. This
procedure is called rebosonization; it also allows to apply our
standard criteria for the determination of the phase diagram explained
above.

Eq. (\ref{crnew1}) suggests that our finding for $T_{\rm c}/T_{\rm F}$
at unitarity, which is above the experimental value, may be lowered
substantially by the inclusion of particle-hole fluctuations. It is a
particular strength of the FRG that it allows to answer the question
whether particle-hole screening is a relevant effect at resonance. In
particular, the feedback of the particle-hole fluctuations does not
rely on whether the effect is perturbative (as in the deep BCS regime)
or not. We simply have to improve our truncation accordingly and then
solve the flow equation. While it is clear that the Gorkov correction
is restricted to the presence of the Fermi surface $\mu>0$, and thus
has to vanish on the BEC side after the zero crossing of the fermion
chemical potential, it is still an important quantitative question to
follow its evolution into the strongly interacting unitary regime. It
has been studied in Ref. \cite{Floerchinger:2008qc}, with the result
for the critical temperature shown in Fig. \ref{CritTemp}. From this
figure we see that the relevance of the screening effect diminishes
rather rapidly as we approach resonance, and in particular is too
small to explain the large downshift in the critical temperature found
from QMC simulations
\cite{PhysRevLett.96.090404,PhysRevLett.99.120401,PhysRevA.78.023625,PhysRevLett.96.160402}.

We now turn to the present status of the crossover truncation
\cite{Floerchinger:2009pg}, which in addition to our above
improvements takes into account the renormalization of the inverse
fermion propagator generated by a diagram involving both a boson and a
fermion propagator line. As a motivation, let us summarize our finding
of the beyond mean field effects due to fluctuation effects. The boson
physics, namely the dimer-dimer interactions, drive the physics on the
BEC side of the crossover and are responsible for the shift $\Delta
T_{\rm c,BEC}$ of the critical temperature with respect to the ideal
gas value. However, bosons are massive on the BCS side (except in an
exponentially narrow vicinity of the critical point), and thus their
contribution is suppressed there. On the other hand, the Gorkov
correction is bound to the presence of a Fermi surface and vanishes on
the far BEC side: there, the fermions are gapped. In consequence, the
diagram which renormalizes the inverse fermion propagator remains
suppressed in both BCS and BEC regimes, but may be important in the
unitary regime in between.  The renormalization effect on the fermion
propagator due to the diagram containing \emph{both} bosons and
fermions is most relevant at resonance. In each of the limiting cases
of BEC and BCS, either the one or the other of them is massive. In the
unitary regime, instead, no such simple ordering principle can be
applied and a priori there is no suppression of this effect.  Again,
the influence of this effect can be investigated systematically by
improving the truncation: The fermion propagator renormalization is
taken into account by means of a derivative expansion according to
\begin{equation}
 \label{crnew4} P_{\psi,k}(Q) = Z_{\psi,k}(\rmi \omega+\vec{q}^2-\mu),
\end{equation}
see \cite{Floerchinger:2009pg}. A strong renormalization of the
propagator is indeed found in the region of anomalously large
scattering length, but nevertheless, the effect on the critical
temperature at unitarity is not strong enough to explain the
discrepancy to numerical simulations.

\begin{figure}[tb!]
 \centering
 \includegraphics[width=1\columnwidth,keepaspectratio=true]{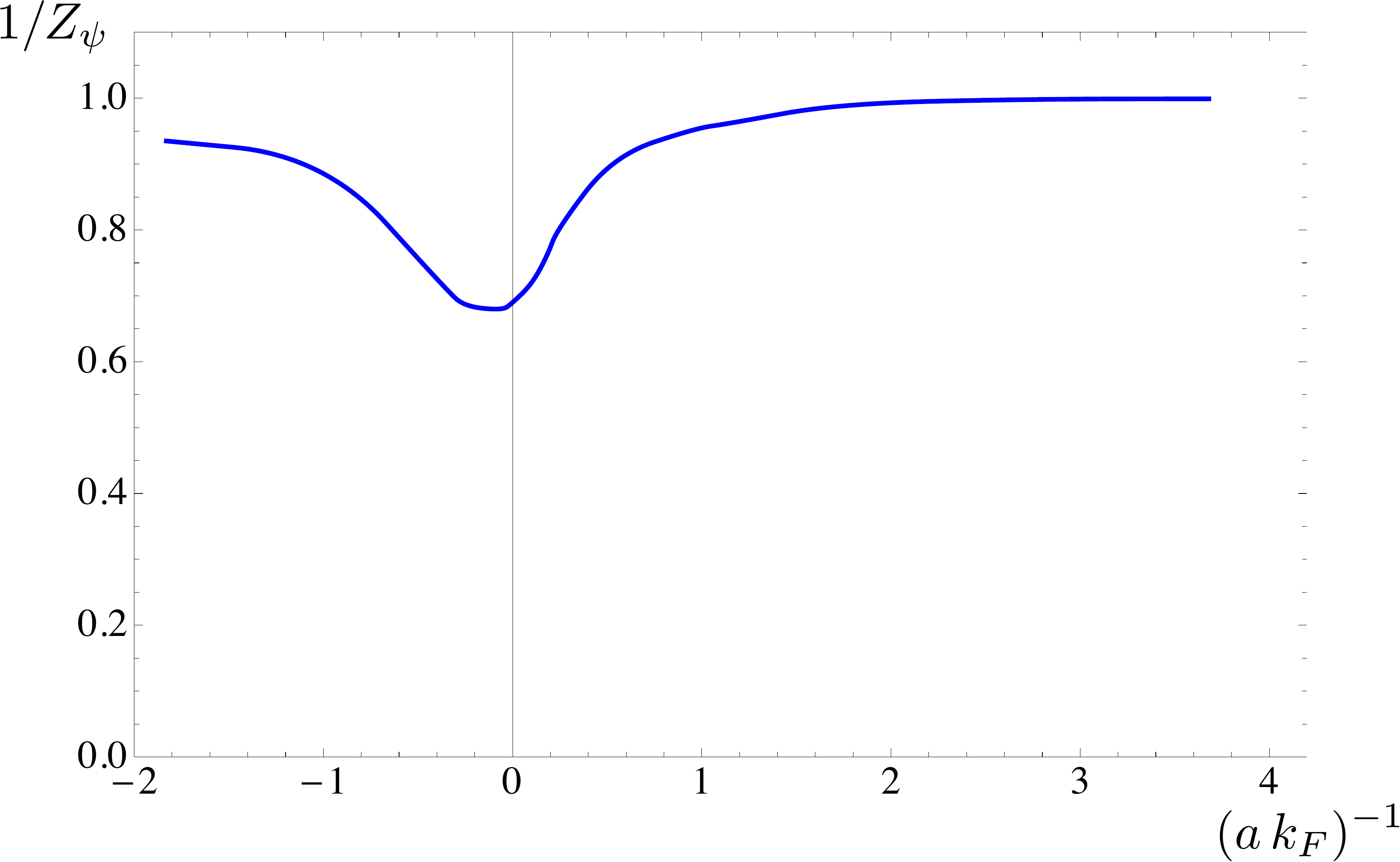}
 \caption{Infrared value of the wave function renormalization
 $Z_\psi=Z_{\psi,k=0}$ introduced in \eq{crnew4}. We observe
 strong renormalization of the fermion propagator close to
 resonance. For further details see Ref. \cite{Floerchinger:2009pg}.}
\label{RunFerm}
\end{figure}

(iii) \emph{Critical domain and universal aspects} -- In
Fig. \ref{PDScales}, we schematically indicate a third region in the
phase diagram, which is dominated by the critical fluctuations on long
distance scales $k_{\rm ld} \gg k_{\rm F}, T^{1/2}, |\vare_{\rm
  B}|^{1/2}$. While this regime is very difficult to address with
techniques other than RG approaches, it is already well under control
in the simple truncation including boson fluctuations in our
framework. We have seen that the infrared boson fluctuations drive the
order parameter to zero in a continuous manner,
cf. Fig. \ref{TempGap}. Generically, nonperturbative effects dominate
the region around the phase boundary in the phase diagram. This
regions is exponentially small in the BCS limit. From a systematic
inspection of the size of the universal region of long range
fluctuations, one finds the largest extend of the critical domain at
unitarity \cite{Diehl:2009ma}. A quantitative test for the reliability
of the critical modeling is given by the critical exponent $\eta$ of
the propagator, $G(\vec{x}) \sim r^{-(d-2+\eta)}$. In our basic
truncation, it is found to be $\eta = 0.05$ for the whole
crossover. This has to be compared to the corresponding value
$\eta=0.038$ of the three-dimensional $\rm{O}(2)$ universality class
\cite{Pelissetto:2000ek}.

Further universal aspects in the crossover phase diagram relate to the
width of the resonance. Here, we have exclusively focused on the broad
resonances \cite{PhysRevA.75.033608}. An systematic FRG investigation
has revealed the existence of two different fixed points governing the
Fermi gas with divergent scattering length -- a broad resonance,
interacting (Wilson--Fisher) fixed point and a narrow resonance,
Gaussian fixed point \cite{Diehl:2007ri}, which can be solved exactly
\cite{Diehl:2005an}. While the narrow resonance fixed point is
sensitive to microscopic details, the broad resonance fixed point is
distinguished by a pronounced insensitivity with respect to the
precise microphysics.

Finally, in order to foster comparison with QMC approaches, a finite
size study has been performed recently \cite{Braun:2011uq}, building
on finite size studies in bosonic \cite{Braun:2008sg} and fermionic
theories \cite{Braun:2010vd,Braun:2011iz}. Since QMC calculations are
performed in finite volumes and lattices, such an analysis can provide
valuable information on the domain of lattice sizes where the
extrapolation to infinite volume is justified, even if the absolute
values for the observables were not fully accurate. Complementary,
the effects of the lattice need to be negligible in order to be able
to make reliable statements on the continuum limit from QMC
simulations. For this purpose, the limit of vanishing filling has to
be taken. This has been critically examined in
\cite{PhysRevA.85.013640,chen11}.

In summary, we have seen that the BCS-BEC crossover shows important
nonperturbative effects on all scales, ranging form the microscopic
scattering physics over genuine many-body effects down to the long
distance critical physics in the vicinity of the finite temperature
phase transition. Full resolution of all of these effects requires a
unified flexible framework, which is provided by the FRG. Beyond being
free of intricate infrared divergence problems which represent a
severe obstacle to alternative many-body approaches, this setting
offers a high degree of flexibility for the inclusion of effects that
are well-understood effects in the limiting cases, together with the
possibility of following their impact when moving into the challenging
unitary regime. This provides a systematic interpolation scheme
between BCS and BEC regimes, as we have seen at the example of the
Gorkov effect. It still remains to be seen if an effect can be
identified which would be able to bridge the quantitative
discrepancies between analytical approaches and Quantum Monte Carlo
simulations
\cite{PhysRevLett.96.090404,PhysRevLett.99.120401,PhysRevA.78.023625,PhysRevLett.96.160402},
as well as recent experiments \cite{Navon10,Nascimbene10,Ku12}.

\section{Outlook}
\label{sec:outlook}

In these lecture notes, we have given an introduction to many-body
physics of ultracold atomic systems in a functional integral
framework. We have worked-out the cornerstones of quantum condensation
phenomena, Bose--Einstein condensation and the BCS mechanism in this
language. We have also seen how these phenomena are connected in the
presence of a Feshbach resonance. 

This was done by introducing and applying the concept of the
Functional Renormalization Group, which already in a simple
approximation to the full quantum theory of ultracold atoms allows to
access the complete finite temperature phase diagram. On the
technical side, we have seen on the basis of this example how the FRG
concept can be applied to fermionic and bosonic systems in the cold
atoms, i.e. nonrelativistic context. In view of potential future
applications, let us therefore come back to the discussion in the
introduction, asking to which of the challenges mentioned there the
FRG framework could usefully contribute.

\emph{Resolving physics at different scales and fostering comparison
  with experiment} -- Experimental tools such as Bragg or RF
spectroscopy provide information beyond thermodynamics, and in fact
yield detailed knowledge of e.g. the full spectral function for a
large regime of frequencies and momenta, including strongly
interacting regimes. It is therefore a pressing issue to access such
observables also theoretically with flexible tools beyond mean field
theory with quantitative precision. Important steps in the direction
of a full momentum resolution are made by taking into account higher
orders of the derivative expansion or performing vertex expansion
schemes. Both steps have not yet been applied to the full quantum
theory of ultracold atoms with fermionic and bosonic degrees of
freedom, but have been tested in various general settings, for the
higher order derivative expansion see e.g.\
\cite{Litim:2010tt,Canet:2003qd}, for vertex expansions see e.g.\
\cite{RevModPhys.84.299,Benitez:2011xx,Husemann:2011wr,Fister:2011um}. Direct
calculation of (real time domain) spectral functions have been 
performed in the context of bosonic
\cite{PhysRevLett.102.190401,PhysRevA.80.043627,PhysRevLett.102.120601,PhysRevA.82.063632}
and fermionic \cite{PhysRevA.83.063620} systems.

\emph{Lattice systems and exotic interactions} -- Lattice models have
so far been investigated using the FRG framework mainly in the
condensed matter context for fermionic systems
\cite{RevModPhys.84.299}. Optical lattices nowadays play a key role in
the physics of cold atomic systems, and in part are crucially needed
for a stable realization of many of the proposals involving long-range
and multicomponent interactions. In general, due to the possibility of
reaching high densities, lattice systems allow to access regimes of
strong correlations with relative ease, and therefore offer
particularly rich quantum phase diagrams. While the study of such
quantum phase transitions from a low energy viewpoint is interesting
in its own right \cite{PhysRevB.77.064504,PhysRevB.81.125103}, an
outstanding challenge is clearly the quantitative assessment of the
physics on various scales, such as the determination of the location
of quantum phase boundaries, where the short distance lattice physics
needs to be accounted for explicitly. First promising steps in this
direction for the conceptually simplest problem -- the Mott insulator
to superfluid phase transition in the Bose--Hubbard model -- have been
taken in
\cite{PhysRevB.83.172501,PhysRevB.84.174513,PhysRevA.85.011602}.

\emph{Non-equilibrium systems} -- The field of non-equilibrium physics
with cold atoms is only fledging but there is substantial potential
for the discovery of intriguing physics. The need for theoretical
tools for the description of out-of-equilibrium many-body systems is
thus pressing, enhanced by the fact that efficient numerical tools
comparable to Monte-Carlo simulations are scarce or only applicable to
specific circumstances -- at least in dimensions larger than one. In
the context of non-equilibrium, \emph{closed} system dynamics, in
addition to density-matrix-based approaches
\cite{Vidal2004a,Daley2004a,White2004a,Manmana2005a,1367-2630-12-5-055006}
which have proven powerful in understanding aspects of thermalization,
functional techniques \cite{Berges:2004yj,Gasenzer2009a} based on the
Keldysh real time path integral are most promising due to their high
degree of flexibility in describing physics at different scales in one
unified framework. FRG approaches to this class of problems have been
put forward in \cite{Gasenzer:2008zz,Gasenzer:2010rq} for the
investigation of the real-time evolution starting from a given initial
state and in \cite{Berges:2008sr}, applied to the analysis of strongly
nonlinear wave-turbulent states far from equilibrium.  Another
interesting direction is provided by \emph{open} systems, which in
contrast to the dynamical phenomena above can exhibit stable
\emph{non-equilibrium stationary states} -- they share exact time
translation invariance with thermodynamic equilibrium, but are
governed by different distribution functions. One can therefore hope
that FRG approaches based on the Keldysh formalism can still give
analytical insights into such problems. Theoretical approaches for
\emph{classical} non-equilibrium stationary states have been worked
out in
\cite{Canet:2003yu,Canet:2009vz,Canet:2011wf,Canet:2011ez}. Non-equilibrium
stationary states in cold atomic \emph{quantum} systems are currently
moving into the focus of research, being realized e.g. via the
competition of particle loss and repumping in physical contexts as
diverse as ensembles with optical Feshbach resonances
\cite{chin-review}, low dimensional systems of polar molecules
\cite{PhysRevLett.105.073202}, and metastable repulsive fermions
\cite{Jo09}, or in stationary states resulting from tailored
dissipation \cite{diehl08,verstraete08,muller12}.

\begin{center}
 \textbf{Acknowledgements}
\end{center}

\noindent S.~D.\ and J.~M.~P.\ thank the organizers for the
opportunity to attend and lecture/take part in the plenary discussion
at the 49th Schladming Theoretical Physics Winter School. We also thank J.~Berges, J.~Braun, S.~Fl\"orchinger,
T.~Gasenzer, H.~Gies, C.~Krahl, D.~F.~Litim, B.-J.~Schaefer, M.~Scherer, C.~Wetterich for 
discussions and collaboration on the research presented here.
 S. D. acknowledges support by the Austrian Science Fund (FWF) through
SFB FOQUS and the START grant Y 581-N16. I.~B.\ acknowledges funding
from the Graduate Academy Heidelberg. This work is supported by the
Helmholtz Alliance HA216/EMMI.

\begin{appendix}

  \section{Functional integral representation of the quantum partition
    function}
\label{AppFun}

\noindent In this appendix, we derive the functional integral
expression for the partition function of a quantum many-body
system. The construction utilizes coherent states. These are
eigenstates of the annihilation operators $\hat{a}_i$ and allow for a
parametrization of Fock space, which is different from the occupation
number representation. We show that bosonic or fermionic particles can
be formulated in terms of a nonrelativistic field theory with
euclidean time $\tau$. Whereas the former are expressed by a complex
field $\vphi_i(\tau)$, the latter correspond to Grassmann valued
fields $\psi_i(\tau)$. We give a brief introduction to Grassmann
numbers and the corresponding calculus.

To begin our analysis, we note that the Hamiltonian of a many-body
system can be expressed in terms of creation and annihilation
operators, $\hat{a}^\dagger$ and $\hat{a}$, according to
\begin{equation}
 \label{Co1} \hat{H} = H(\hat{a}^\dagger,\hat{a}) = \sum_{ij} t_{ij} \hat{a}_i^\dagger \hat{a}_j 
+ \sum_{ijkl} V_{ijkl} \hat{a}^\dagger_i\hat{a}^\dagger_j\hat{a}_k\hat{a}_l + \dots\,.
\end{equation}
As we have discussed in section \ref{scales}, this generic form of the
Hamiltonian is important for cold atoms, where only two-body
interactions (i.e. $V_{ijkl}$) are relevant. In Eq. (\ref{Co1}), all
annihilation operators are to the right of the creation operators. If
this is the case, we say that the operator is \emph{normal ordered}.
The labels $i,j,\dots$ run over a particular choice of single particle
states. Typically, we think of spin and momentum or points on a
lattice. For a finite system, this enumerates a discrete and finite
set. However, it is also common to write $\hat{a}_{\vec{p}, \sigma}$
or $\hat{a}_{\vec{x}}$ with continuous variables $\vec{p}$ and
$\vec{x}$, keeping in mind that one has to go back to a discretized
formulation if problems should occur.

The creation and annihilation operators satisfy commutation
(anti-commutation) relations for bosons (fermions). We have
\begin{align}
  \nonumber [\hat{a}_i,\hat{a}_j^\dagger] &= \hat{a}_i \hat{a}_j^\dagger - 
\hat{a}_j^\dagger \hat{a}_i = \delta_{ij}, \hspace{2mm} [\hat{a}_i,\hat{a}_j]
=0 \hspace{2mm} ({\rm bosons}),\\
\label{Co2} \{\hat{a}_i,\hat{a}_j^\dagger\} &= \hat{a}_i
\hat{a}_j^\dagger + \hat{a}_j^\dagger \hat{a}_i = \delta_{ij},
\hspace{2mm} \{\hat{a}_i,\hat{a}_j\}=0 \hspace{2mm} ({\rm fermions}).
\end{align}
The particle number operator can be expressed as
\begin{equation}
 \label{Co3} \hat{N} = N(\hat{a}^\dagger,\hat{a}) = \sum_i \hat{a}^\dagger_i \hat{a}_i.
\end{equation}
We assume that there is a vacuum state $| \rm vac\rangle$ which does not contain any excitations. Consequently, we have $\hat{a}_i | {\rm vac}\rangle=0$. Since the states in occupation number representation form a basis of the many-body Fock state, we can write the partition function $Z = \mbox{Tr} e^{-\beta(\hat{H}-\mu\hat{N})}$ as
\begin{equation}
 \label{Co4} Z(\mu,T) = \sum_{|n_1n_2\dots\rangle} \langle n_1n_2\dots| e^{-\beta(\hat{H}-\mu \hat{N})}|n_1n_2\dots\rangle.
\end{equation}

From the last formula we easily obtain Eq. (\ref{td4b}) for the pressure of a noninteracting gas. However, for an interacting system, it will in general not be possible to exactly calculate the partition function. Therefore, we aim to rewrite Eq. (\ref{Co4}) in terms of a functional integral, which is particularly well-suited for treating interaction effects in a systematic fashion.

\vspace{5mm}
\begin{center}
 \emph{Coherent states for bosons}
\end{center}

\noindent For a functional integral representation, we would like to substitute the operators $\hat{a}_i^\dagger$ and $\hat{a}_i$ for corresponding classical fields $\varphi_i$, which are then quantized. We first restrict ourselves to the \emph{bosonic case}. The natural way to replace an operator by a number is to let it act on an eigenstate. We ask whether there are states $|\varphi_1\varphi_2\dots\rangle$ with complex numbers $\varphi_i \in \mathbb{C}$ such that
\begin{equation}
 \label{Co5} \hat{a}_i|\varphi_1\varphi_2\dots\rangle = \varphi_i |\varphi_1\varphi_2\dots\rangle
\end{equation}
for every $i$. We call such a state a \emph{coherent state}. The
$\varphi_i$ are not necessarily real, because $\hat{a}_i$ is not
selfadjoint. The creation operator $\hat{a}_i^\dagger$ cannot have an
eigenstate. Indeed, assume there was an eigenstate $|\Psi\rangle$ of
$\hat{a}_i^\dagger$. Then, as every Fock space state, $|\Psi\rangle$
could be expressed in the occupation number representation as a linear
combination of several basis states $|n_1n_2\dots\rangle$. In this
linear combination, one basis state has the smallest particle number
$\sum_i n_i$. Now, letting $\hat{a}_i^\dagger$ act on $|\Psi\rangle$,
the particle number of every basis state in the superposition gets
increased by one. Due to the fact that this also increases the minimal
particle number by one, $\hat{a}_i^\dagger|\Psi\rangle$ cannot be
proportional to $|\Psi\rangle$.

We claim that, for every collection $\{\varphi_i\}$ of complex numbers, 
\begin{equation}
  \label{Co6} |\varphi\rangle = |\varphi_1\varphi_2\dots\rangle = e^{\sum_i \varphi_i \hat{a}_i^\dagger}|\rm vac\rangle
\end{equation}
constitutes an eigenstate of $\{\hat{a}_i\}$ with eigenvalues
$\{\varphi_i\}$. Applying $\hat{a}_i$ to $|\varphi\rangle$, it will
commute with all $\hat{a}_j^\dagger$ for $i \neq j$. We can then --
for fixed $i$ -- use the relation $[\hat{a},(\hat{a}^\dagger)^n]=n
(\hat{a}^\dagger)^{n-1}$. This yields
\begin{align}
 \nonumber \hat{a}|\varphi\rangle &= \hat{a} e^{\varphi \hat{a}^\dagger}|{\rm vac}\rangle = [\hat{a}, e^{\varphi \hat{a}^\dagger}]|{\rm vac}\rangle \\
 \nonumber &= \sum_{n=0}^\infty  \frac{\varphi^n}{n!}[\hat{a},(\hat{a}^\dagger)^n] |{\rm vac}\rangle \\
 \label{Co7} &= \varphi \sum_{n=1}^\infty  \frac{\varphi^{n-1}}{(n-1)!}(\hat{a}^\dagger)^{n-1} |{\rm vac}\rangle = \varphi|\varphi\rangle,
\end{align}
which proves $\hat{a}_i|\varphi\rangle = \varphi_i
|\varphi\rangle$. Via complex conjugation we find $\langle
\varphi|\hat{a}_i^\dagger = \langle\varphi|\varphi_i^*$.

Clearly, for every choice of complex numbers $\{\varphi_i\}$, we can
construct a coherent state. In contrast, the occupation number basis
$|n_1n_2\dots\rangle$ of the Fock space was limited to integers
$\{n_i\}$. This increase of degrees of freedom results in the fact
that coherent states are over-complete. They do not represent an
orthonormal basis of the Fock space, but nevertheless any state can be
expressed as a superposition of coherent states. With the explicit
formula in Eq. (\ref{Co6}) we find for the overlap of two coherent
states
\begin{equation}
 \label{Co8} \langle \varphi'|\varphi\rangle = \exp\Bigl\{\sum_i \varphi_i'^*\varphi_i\Bigr\}.
\end{equation}

\begin{center}
 \emph{Functional integral}
\end{center}

\noindent One can show that the weighted sum
\begin{equation}
 \label{Co9}\int \mbox{D} \varphi^*\mbox{D} \varphi  e^{-\sum_i \varphi_i^*\varphi_i} |\varphi\rangle \langle \varphi| = \mathbb{1}
\end{equation}
constitutes the unit operator in Fock space. The functional measure is defined as
\begin{equation}
 \label{Co10} \mbox{D} \varphi^*\mbox{D}\varphi  = \prod_i \frac{\mbox{d} \varphi_i \mbox{d} \varphi_i^*}{2\pi {\rm i}} = \prod_i\frac{\mbox{d} (\mbox{Re}\varphi_i)\mbox{d} (\mbox{Im}\varphi_i)}{\pi}.
\end{equation}
The integration over all values of $\varphi_i$ is similar to the
summation over all occupation numbers, but due to the
over-completeness we have to suppress the individual contributions by
a weighting factor.

Since we assumed the Hamiltonian to be \emph{normal ordered}, we have 
\begin{align}
 &\nonumber \langle\varphi|\left(\hat{H} - \mu \hat{N}\right)|\varphi\rangle = \langle\varphi|\Bigl(H(\hat{a}^\dagger,\hat{a}) - \mu N(\hat{a}^\dagger,\hat{a})\Bigr)|\varphi\rangle\\
 \label{Co11} &= \Bigl(H(\varphi'^*,\varphi) - \mu N(\varphi'^*,\varphi)\Bigr) e^{\sum_i \varphi_i'^*\varphi_i}.
\end{align}
Obviously, we arrived at our goal to substitute the operators for
complex numbers. There are no operators appearing in the functional
integral. We could now insert the identity operator into $Z=\mbox{Tr}
e^{-\beta(\hat{H}-\mu\hat{N})}$ and express the partition function as
a path integral. However, there is one complication: The exponential
of the normal ordered operator $\hat{H}-\mu \hat{N}$ is no longer
normal ordered. Denoting normal ordering by double dots we have
\begin{equation}
 \label{Co12} e^{: \hat{A} :} \neq \mbox{ } : e^{\hat{A}} :.
\end{equation}
The situation is less severe if we introduce a small parameter $\varepsilon$, because
\begin{equation}
 \label{Co13} e^{: \varepsilon \hat{A} :} = \mbox{ }: e^{\varepsilon \hat{A}} : + \mbox{ }O(\varepsilon^2).
\end{equation}
We therefore divide the inverse temperature $\beta$ into $M$ small intervals of length $\varepsilon$ such that $\beta = M \varepsilon$. We then insert $M-1$ resolutions of the identity. This introduces a discrete label $s=1,\dots,M$ according to $\vphi_i \rightarrow \{\vphi_{si}\}_s$. It will later be associated with imaginary time. We arrive at
\begin{align}
 \nonumber &Z(\mu,T) = \mbox{Tr} e^{-\beta(\hat{H}-\mu \hat{N})} \\
 \nonumber &= \int \mbox{D}\varphi_0^{*}\mbox{D}\varphi_0 e^{-\sum_i\varphi_{0i}^*\varphi_{0i}} \langle \varphi_0| e^{-\beta(\hat{H}-\mu \hat{N})} | \varphi_0\rangle\\
 \nonumber  &= \int \mbox{D}\varphi_0^{*}\mbox{D}\varphi_0 e^{-\sum_i\varphi_{0i}^*\varphi_{0i}}\langle \varphi_0| e^{-\varepsilon(\hat{H}-\mu \hat{N})} \mathbb{1} \dots \mathbb{1} e^{-\vare(\hat{H}-\mu\hat{N})}| \varphi_0\rangle\\
 \nonumber  &= \int \mbox{D}\varphi_{M-1}^{*}\mbox{D}\varphi_{M-1} \dots \mbox{D}\varphi_0^{*}\mbox{D}\varphi_0e^{-\Bigl(\sum_i\varphi_{M-1,i}^*\varphi_{M-1,i} + \dots + \sum_i \vphi^*_{0i}\vphi_{0i}\Bigr)} \\
 \nonumber &\hspace{5mm} \times \langle \varphi_{0}| e^{-\varepsilon(\hat{H}-\mu \hat{N})} | \varphi_{M-1}\rangle\dots \langle \varphi_{1}| e^{-\varepsilon(\hat{H}-\mu \hat{N})} | \varphi_{0}\rangle\\
 \nonumber &= \int_{\varphi_{Mi}=\varphi_{0i}} \mbox{D} \varphi^* \mbox{D} \varphi \exp \biggl\{-\vare \sum_{s=1}^M \Bigr[ \sum_i \vphi_{si}\Bigl(\frac{\vphi_{si}-\vphi_{s-1,i}}{\vare}\Bigr)\\
 \nonumber &\hspace{5mm} +\Bigl(H(\vphi^*_s,\vphi_{s-1})-\mu N(\vphi_{s}^*,\vphi_{s-1})\Bigr)\Bigl]\biggr\}\\
 \label{Co14} &= \int_{\vphi_{Mi}=\vphi_{0i}} \mbox{D}\vphi^* \mbox{D}\vphi e^{-S_\vare[\vphi^*,\vphi]}.
\end{align}
In the second to last line we introduced a new complex field $\varphi_{s i}$ which depends on both $i$ and $s=1,\dots,M=\beta/\varepsilon$. The functional measure is naturally extended to be
\begin{equation}
 \label{Co15} \mbox{D}\varphi^* \mbox{D} \varphi = \prod_{s,i} \frac{\mbox{d} \varphi_{s i} \mbox{d} \varphi_{s i}^*}{2\pi {\rm i}}.
\end{equation}
For these fields we introduced the action
\begin{align}
 \nonumber S_\vare[\varphi^*,\varphi] = \varepsilon \sum_{s=1}^{M} &\Bigl[ \sum_i\varphi_{s i}^*\Bigl(\frac{\varphi_{s i}-\varphi_{s-1,i}}{\varepsilon}\Bigr)\\
 \label{Co16} & + \Bigl(H(\varphi_{s}^*,\varphi_{s-1})-\mu N(\varphi_{s}^*,\varphi_{s-1})\Bigr)\Bigr].
\end{align}
The condition $\varphi_{Mi} = \varphi_{0i}$ for all $i$ originates from the fact that the partition function is a trace.

Sending $M \rightarrow \infty$ and $\vare\rightarrow 0$ while keeping $\beta = M\vare$ fixed, the discrete variable $\vare s =\tau$ becomes continuous. The partition function then acquires the form
\begin{equation}
 \label{Co16b} Z(\mu,T) = \int_{\vphi_i(\beta)=\vphi_i(0)} \mbox{D}\vphi^*\mbox{D}\vphi e^{-S[\vphi^*,\vphi]},
\end{equation}
with microscopic action
\begin{align}
 \nonumber S[\varphi^*,\varphi] = \int_0^{\beta} \mbox{d}\tau\Bigl( &\sum_i \varphi_i^*(\tau)(\partial_\tau-\mu) \varphi_i(\tau) \\
 \label{Co17} & + H(\varphi^*(\tau),\varphi(\tau))\Bigr).
\end{align}
Recall that $i$ can be a continuous variable, too. Typically we are
interested in $i = \vec{x}$ and fields $\varphi(\tau,\vec{x})$. If an
expression happens to be not well-defined during a calculation, we can
always go back to the discretized form of the action in
Eq. (\ref{Co16}). The condition $\varphi_i(\beta) = \varphi_i(0)$
restricts the functions $\varphi$ and $\varphi^*$ to be periodic
functions in $\tau$.

To summarize, we found a functional integral representation of a
generic many-body system of bosons. The non-commutativity of operators
introduced the time variable $\tau$.

\vspace{5mm}
\begin{center}
 \emph{Coherent states for fermions}
\end{center}

\noindent How can these considerations be extended to include
fermions? Most of the formulae from above remain valid or only receive
corrections due to some signs. However, there is one important
conceptual difference in the path integral representation for
fermions.

We already noticed that no operators appear in the functional integral
formulation. In the bosonic case, complex numbers took the positions
of the (normal ordered) annihilation and creation operators. In order
to satisfy the anti-commutation relations for fermionic operators
instead, the eigenvalues of the $\{\hat{a}_i\}$ cannot be complex
numbers. Indeed, if $|\varphi\rangle = | \varphi_1
\varphi_2\dots\rangle$ was a coherent state of arbitrary complex
numbers, we would arrive at the contradiction
\begin{equation}
 \label{Co19} 0 = \{ \hat{a}_i,\hat{a}_j \} |\varphi \rangle = (\varphi_i \varphi_j + \varphi_j \varphi_i) |\varphi\rangle \neq 0.
\end{equation}

Complex numbers thus cannot be applied for constructing coherent states of fermions. However, there are objects which can formally be multiplied and obey anti-commutation relations. They are called Grassmann variables (or Grassmannians) and a calculus can be developed for them. We will reduce our discussion of this issue to a minimum and refer the reader to the textbooks \cite{negele-book,altland-book} for more details. Since any two Grassmannians $\psi$ and $\eta$ satisfy
\begin{equation}
 \label{Co20} \psi \eta = - \eta \psi,
\end{equation}
the construction of functions of them is pretty simple. Assume we have a polynomial expression $f(X) = \sum_i f_i X^i$ with $f_i \in \mathbb{C}$. We then define
\begin{equation}
 \label{Co21} f(\psi) = f_0 + f_1 \psi,
\end{equation}
which incorporates the condition $\psi^2 = \psi \psi = - \psi \psi \Rightarrow \psi^2 =0$. Functions $f(X_1,\dots,X_n)$ of more than one variable, and analytic functions which allow for a series expansion, can be extended to Grassmann-valued arguments as well. The associated series will always terminate at a finite order and we never have to bother about convergence problems. We will often have to deal with expression of the form
\begin{equation}
 \label{Co22} e^{ c \psi} = 1 + c \psi
\end{equation}
for $c$ being either complex or Grassmann valued. In the first (second) case, we have $f(X) = e^{cX}$ ($f(X,Y) = e^{XY}$).

The product $\pi$ of an even number of Grassmannians satisfies $\pi \psi = \psi \pi$ for any Grassmannian $\psi$, because the minus signs from anti-commuting through $\psi$ cancel each other. We refer to such products as being \emph{even}. Complex numbers are trivially even. The remaining Grassmannians are then \emph{odd}.

We have just seen that functions of Grassmann variables can be linearized and it is thus possible to define a differentiation rule, although there is no actual way in which a Grassmannian is either small or large. (In fact, they do not have particular values like complex numbers, but will rather only serve for generating correlation functions.) We introduce a left- and right-derivative according to
\begin{align}
 \nonumber \frac{\stackrel{\rightarrow}{\partial}}{\partial \psi} ( f_0 + \psi f_1) &= f_1,\\
 \label{Co23} (f_0 + \bar{f}_1 \psi) \frac{\stackrel{\leftarrow}{\partial}}{\partial \psi} &= \bar{f}_1.
\end{align}
Note that $f_1 = \bar{f}_1$ ($f_1 = - \bar{f}_1$) if $f_1$ is even (odd).

Integration of Grassmann functions is defined to be linear and to satisfy
\begin{equation}
 \int \mbox{d} \psi 1 = 0, \hspace{5mm} \int \mbox{d} \psi \psi =1.
\end{equation}
Thus, it coincides with left-differentiation. Note however that $\mbox{d} \psi \mbox{d} \eta = - \mbox{d}\eta \mbox{d} \psi$. 

We can now formulate the coherent state path integral for fermions in terms of Grassmann fields $\psi_i(\tau)$. It is given by
\begin{equation}
 \label{Co27} Z(\mu,T) = \int_{\psi_i(\beta) = - \psi_i(0)} \mbox{D} \psi^* \mbox{D} \psi e^{-S[\psi^*,\psi]}.
\end{equation}
The action $S[\psi^*,\psi]$ is constructed from the Hamiltonian analogous to the bosonic case in Eq. (\ref{Co17}). The functional measure reads
\begin{equation}
 \label{Co27b} \mbox{D} \psi^* \mbox{D} \psi = \prod_{s,i} \mbox{d} \psi_{s,i}^* \mbox{d} \psi_{s,i}.
\end{equation}
In contrast to the bosonic case, the partition function is now restricted to anti-periodic functions in time-direction, satisfying
\begin{equation}
 \label{Co28} \psi_i(\beta) = - \psi_i(0)
\end{equation}
for all $i$. 

The (anti-)periodicity reduces the Fourier transformation of the fields in time direction to a Fourier series with discrete frequencies. Indeed, for a function $f(x)$ satisfying $f(x+L)=f(x)$ we have
\begin{equation}
 \label{Co29} f(x) = \frac{1}{L} \sum_{n \in \mathbb{Z}} e^{2\pi \rmi nx/L}.
\end{equation}
Thus, in our case, we arrive at
\begin{equation}
 \label{Co30} \vphi(\tau,\vec{x}) = T \sum_{n \in \mathbb{Z}} \int \frac{\mbox{d}^dq}{(2\pi)^d} e^{\rmi(\vec{q}\cdot \vec{x} + \omega_n \tau)} \vphi_n(\vec{q})
\end{equation}
with Matsubara frequencies
\begin{equation}
 \label{Co31} \omega_n = \begin{cases} 2 \pi n T &({\rm bosons}),\\ 2 \pi(n+1/2)T & ({\rm fermions}).\end{cases}
\end{equation}
In the zero temperature limit, the stepsize $\Delta \omega = 2 \pi T$ gets infinitesimally small and the sum is replaced by a Riemann integral over the continuous variable $\omega$. More explicitly,  we have
\begin{align}
 \label{Co32} T \sum_{n \in \mathbb{Z}} f(\omega_n) = \frac{\Delta \omega}{2\pi}\sum_{\omega_n}  f(\omega_n)\stackrel{T \rightarrow 0}{\longrightarrow} \int_{-\infty}^{\infty} \frac{\mbox{d}\omega}{2\pi} f(\omega).
\end{align}

\section{Lattice magnets and continuum limit}
\label{AppIsing}

\noindent In this appendix, we apply the concepts of section \ref{SecFun} to the Ising model. The formulation of the system on a discrete and finite lattice allows for a transparent discussion of spontaneous symmetry breaking and the construction of the effective action. We show, why spontaneous symmetry breaking can only occur in an infinite system and in which sense the partition function and the effective action store the same physical information in a different manner. We then explain how our findings on the lattice can be translated to a continuum theory such as ultracold bosons and fermions. In particular, we give a brief overview onto functional differentiation and Gaussian functional integrals.

\vspace{5mm}
\begin{center}
 \emph{Spontaneous symmetry breaking: Magnet case study}
\end{center}

\noindent We consider spins $\vec{s}$ on a $d$-dimensional lattice. To each spin, we associate a magnetic moment $\vec{m}$ according to $\vec{m}/\mu_{\rm B} = g_{\rm L} \vec{s}/\hbar$, where $\mu_{\rm B}$ and  $g_{\rm L}$ are the Bohr magneton and the Land\'{e} factor, respectively. Choosing units such that $\vec{m}^2=1$ we obtain the classical Heisenberg model of order $n$, where $n$ is the number of components of $\vec{m}$. The corresponding Hamiltonian is given by
\begin{equation}
 \label{ising1} H = - \sum_{\langle i,j\rangle} J_{ij} \vec{m}_i \cdot \vec{m}_j + \sum_i \vec{h}_i \cdot \vec{m}_i
\end{equation}
with $i$ labeling the individual lattice sites and the summation $\langle i ,j \rangle$ being restricted to nearest neighbors. For $J_{ij} >0$, the first term favors magnets on neighboring sites $\vec{x}_i$ and $\vec{x}_j$ to be aligned by lowering the energy of such a configuration. The magnetic field $\vec{h}_i$ in the second term can be regarded as a \emph{source} of magnetization at the site $\vec{x}_i$.

For the purpose of our analysis, it is sufficient to restrict to the case of $n=1$ such that the orientation of the magnets can be either up or down. The system is then known as the Ising model. The Hamiltonian function of the latter is given by
\begin{equation}
 \label{eff1} H = - J \sum_{\langle i,j\rangle} m_i m_j - \sum_i h_i m_i,
\end{equation}
with $m_i$ being either $+1$ or $-1$. Moreover, we assumed $J_{ij}=J >0$ to be an overall constant. The first term can then be interpreted as a kinetic term. Indeed, assuming periodic boundary conditions we have
\begin{align}
 \nonumber &-J \sum_{\langle i,j \rangle} m_i m_j = -\frac{J}{2} \sum_i (m_i m_{i+1} + m_i m_{i-1}) \\
 \nonumber \label{ising2} &= -\frac{J}{2} a^2 \sum_i m_i \Bigl( \frac{m_{i+1}-2m_i+m_{i-1}}{a^2}\Bigr)-J \sum_i m_i^2\\
 &= \frac{J a^2}{2} \sum_i m_i (- \nabla^2) m_i - JN
\end{align}
with lattice spacing $a$, number of magnets $N$ and discrete Laplacian $\nabla^2$. The constant shift is irrelevant here. For cold atoms it can be absorbed into the definition of the chemical potential.

Assuming the system to be in equilibrium with a heat bath of temperature $T$, the orientation $m_i$ of the magnets on the individual lattice sites can be treated as a stochastic variable with a canonical probability distribution. We decompose $m_i$ into its mean and fluctuating part according to
\begin{equation}
 \label{eff1b} m_i = \bar{m}_i + \delta m_i
\end{equation}
with $\langle \delta m_i\rangle =0$. The mean magnetization is given by
\begin{equation}
 \label{eff2} \bar{m}_i = \langle m_i \rangle_h = \frac{1}{Z} \sum_{\{m_i\}} m_i e^{-\beta H} = \frac{1}{\beta} \frac{\partial }{\partial h_i}\log Z(\{h_i\}).
\end{equation}
Thus, taking a derivative of the partition function with respect to the $i$th component of the set $\{h_i\}$, we obtain the magnetization at site $i$. 

If we set $h_i \equiv 0$ after taking the derivative in Eq. (\ref{eff2}), we can calculate the spontaneous magnetization at vanishing field, $\langle m_i \rangle_{h=0}$. Since the Hamiltonian for $h_i \equiv 0$ is symmetric under $m_i \rightarrow -m_i$ ($\mathbb{Z}_2$-symmetry), we expect $\langle m_i \rangle_{h=0}$ to be zero, because each configuration with a certain number of spins up has a corresponding configuration with all spins reversed and they should cancel when summing over all configurations. This reasoning, however, can only be applied to finite systems (finite number of lattice sites). Indeed, the Ising model shows phases of nonvanishing spontaneous magnetization in the  infinite volume limit, most prominently in the exact solution of the two-dimensional Ising model by Onsager.

The magnetic field singles out a preferred direction at each site, because parallel alignment along the field minimizes the energy. If we first enlarge the system to infinite volume and then remove the field, we have
\begin{equation}
 \label{eff3} \lim_{h \rightarrow 0} \lim_{N \rightarrow \infty} \langle m \rangle_h = \left\{ \begin{array}{cc} m_{0} \neq 0 &(\text{``broken phase''})\\
                                             0 &(\text{``symmetric phase''}).
                                            \end{array}\right.
\end{equation}
Interchanging these limits we always get zero, because we can apply our reasoning from above. Note that Eq. (\ref{eff3}) already hints towards the answer to the deep question \emph{why} there actually are phase transitions in nature. The calculation of the partition function in a finite volume consists of a summation of a finite number of terms, all of them being nice analytic functions. Even if the Hamiltonian contains a symmetry breaking term like $\sum_i h_i m_i$ in the above example, the contribution of this terms to $Z(\{h_i\})$ can be removed at the end of the calculation, because $Z$ is analytic in the $h_i$. However, if we calculate the corresponding intensive potential $\frac{1}{N} \log Z$ (e.g. free energy density or pressure) in the thermodynamic limit $V,N \rightarrow \infty$ we may introduce singularities and non-analytical behavior in $Z(\{h_i\})$. In fact, the sum of infinitely many analytic functions will in general not be an analytic function again. Therefore, the thermodynamic potentials remember the field $h_i$, even if we set it to zero at the end of the calculation.

Leaving this aspect aside for the moment, we want to extract further information from Eq. (\ref{eff2}). Higher correlation functions can be obtained by taking higher derivatives with respect to the field. For instance, the connected two-point function
\begin{align}
 \nonumber \langle (m_i - \bar{m}_i)(m_j - \bar{m}_j)\rangle &= \langle m_i m_j \rangle - \bar{m}_i \bar{m}_j \\
 \label{eff4} &= \frac{\partial^2}{\partial h_i \partial h_j} \log Z(\{h_i\})
\end{align}
tells us how strongly deviations from the mean value at site $i$ are correlated to deviations at site $j$.

We now invert the magnetization problem. Given an arbitrary mean magnetization $\bar{m}_i$, how do we have to choose $\{h_i\}$ in order to obtain exactly this magnetization? The corresponding generating function is obtained from the Legendre transform of $\log Z(\{h_i\})$ in the variables $\{h_i\}$. We build
\begin{align}
 \nonumber \Gamma(\{\bar{m}_i\}) &= \sup_{\{h_j\}} \Bigl(\sum_i\bar{m}_i h_i  - \log Z(\{h_j\})\Bigr)\\
 \label{eff5} &= \sum_i \bar{m}_i h_i - \log Z(\{h_i\}).
\end{align}
In the second line, which is valid for continuously differentiable $\log Z$, the magnetic field is defined implicitly through the equation $\bar{m}_i \stackrel{!}{=} \frac{\partial \log Z(\{h_i\})}{\partial h_i}$. Note that the right hand side of Eq. (\ref{eff5}) does not depend on $\{h_i\}$. We arrive at the answer to our question, how the field has to be chosen for given $\bar{m}_j$, via
\begin{align}
 \nonumber \frac{\partial \Gamma}{\partial \bar{m}_j} &= \sum_i \underbrace{\frac{\partial \bar{m}_i}{\partial \bar{m}_j}}_{\delta_{ij}}h_i + \sum_i \bar{m}_i \frac{\partial h_i}{\partial \bar{m}_j} - \frac{\partial \log Z}{\partial \bar{m}_j}\\
 \nonumber &= h_j + \sum_i \bar{m}_i \frac{\partial h_i}{\partial \bar{m}_j} - \sum_i \underbrace{\frac{\partial \log Z}{\partial h_i}}_{\bar{m}_i} \frac{\partial h_i}{\partial \bar{m}_j}\\
 \label{eff6} &= h_j + \sum_i \bar{m}_i \frac{\partial h_i}{\partial \bar{m}_j} - \sum_i \bar{m}_i \frac{\partial h_i}{\partial \bar{m}_j} = h_j.
\end{align}

\vspace{5mm}
\begin{center}
 \emph{Functional differentiation}
\end{center}

\noindent Instead of studying fields which live on discrete lattice sites $\vec{x}_i$, we now turn our attention to a continuum theory with space-time variable $X=(\tau,\vec{x})$. This is of relevance for ultracold atoms considered in the main text. For a system of classical magnets as treated in this appendix, we do not have a time variable $\tau$. However, the functional methods developed here are not invalidated and we may replace $X \rightarrow \vec{x}$ in this case. The sets $\{h_i\}$ and $\{\bar{m}_i\}$ become functions $h(X)$ and $\bar{m}(X)$. Sums $\sum_i$ are replaced by integrals $\int_X$ and instead of partial derivatives $\partial/\partial \bar{m}_i$ we take functional derivatives $\delta/\delta \bar{m}(X)$. We formally write
\begin{equation}
 \label{eff6b} i \rightarrow X, \hspace{5mm} \partial \rightarrow \delta.
\end{equation}
More explicitly, we have
\begin{equation}
 \label{eff7} \frac{\delta \bar{m}(X)}{\delta \bar{m}(X')} = \delta(X-X') = \delta(\tau-\tau')\delta^{(d)}(\vec{x}-\vec{x}'),
\end{equation}
which mimics the well-known relation $\partial \bar{m}_i/\partial \bar{m}_j = \delta_{ij}$.

For a proper functional differentiation, we have to specify which
variables are treated as independent of each other. Certainly, the
field values on different space-time points (lattice sites) are
distinct variables. In addition, for a complex field $\bar{m}(X) =
\phi(X)$, as it is the case for ultracold bosons, we have to specify
whether $(\phi,\phi^*)$ or $(\phi_1,\phi_2)$ are treated as
independent variables. Here, the real fields $\phi_1$ and $\phi_2$ are
derived from the representation $\phi = \frac{1}{\sqrt{2}}(\phi_1 +
\rmi \phi_2)$.

Let us choose $\phi$ and $\phi^*$ as independent variables. As a generalization of
\begin{equation}
 \label{eff8} \frac{\partial}{\partial \phi_k^*}  \left( \phi_i^* \sum_{i,j} A_{ij}  \phi_j\right)   =   \sum_j A_{kj} \phi_j
\end{equation}
we have
\begin{equation}
 \label{eff9} \frac{\delta}{\delta \phi_Z^*}  \int_{X,Y} \phi^*_X A_{XY} \phi_Y = \int_Y A_{ZY} \phi_Y.
\end{equation}
The expression $A_{XY}$ is an operator kernel and thus similar to a matrix element. If the kernel contains derivative terms, we sometimes have to perform a partial integration before taking the functional derivative, e.g.
\begin{align}
  \nonumber \frac{\delta}{\delta \phi_X} \int_Z \phi^*_Z (\partial_\tau -\nabla^2)\phi_Z &= \frac{\delta}{\delta \phi_X} \int_Z \phi_Z (-\partial_\tau -\nabla^2)\phi^*_Z\\
 \label{eff11} &= (-\partial_\tau -\nabla^2)\phi^*_X.
\end{align}
The second functional derivative of this expression is found to be
\begin{align}
 \label{eff12} \frac{\delta^2}{\delta \phi^*_Y \delta \phi_X} \int_Z \phi^*_Z (\partial_\tau -\nabla^2)\phi_Z &= (-\partial_\tau -\nabla^2)^X \delta(X-Y).
\end{align}
We emphasize that this is a kernel and not an operator acting to the right. This may be more transparent from writing
\begin{align}
 \nonumber &\int_Y \phi^*_Y (\partial_\tau -\nabla^2)\phi_Y=\int_{X,Y} \phi^*_Y \delta(X-Y) (\partial_\tau-\nabla^2)^X \phi_X \\
 \label{eff13} &= \int_{X,Y} \phi^*_Y \left[ (-\partial_\tau-\nabla^2)^X \delta(X-Y)\right]  \phi_X,
\end{align}
which allows for reading off the second functional derivative immediately.

\vspace{5mm}
\begin{center}
 \emph{Gaussian functional integrals}
\end{center}

\noindent The partition function which corresponds to a Gaussian probability distribution 
\begin{equation}
 \label{ising3} \langle \mathcal{O} \rangle_{0} = \frac{1}{Z_0} \int \Bigl( \prod_{i=1}^N \mbox{d} m_i \Bigr) \mathcal{O}(\{m_i\}) e^{-\frac{1}{2}m_i A_{ij} m_j}
\end{equation}
can be computed for an arbitrary complex symmetric matrix $A$ with nonvanishing eigenvalues $\lambda_i\neq 0$ and $\mbox{Re}(A)\geq0$, i.e. $m^t \mbox{Re}(A) m' = m_i\mbox{Re}(A)_{ij} m_j' \geq0$ for all choices of $m_i$ and $m'_j$. We have
\begin{equation}
 \label{ising4} Z_0 = \int \Bigl( \prod_i \mbox{d} m_i\Bigr) e^{-\frac{1}{2}m_i A_{ij} m_j}.
\end{equation}
For real $A$, we can find an orthogonal transformation $O$ such that $O A O^t = \mbox{diag}(\lambda_1,\dots,\lambda_N)=B$. Writing $m^t A m = (O m)^t B (Om)$, we observe a transformation of the integration variable according to $m \mapsto O m$ to factorize the integral into $N$ Gaussian integrals. Indeed, the Jacobian of the transformation is unity. We arrive at
\begin{equation}
 \label{ising5} Z_0 = \frac{(2 \pi)^{N/2}}{\sqrt{\lambda_1 \cdots \lambda_N}} = \frac{(2\pi)^{N/2}}{\sqrt{\mbox{det}(A)}}.
\end{equation}
Given the above restrictions on $A$, we find both the left and the right hand side of this equation to be an analytic function in the complex coefficients $A_{ij}$ of $A$. Thus, by virtue of analytic continuation, Eq. (\ref{ising5}) is also valid in the case of complex $A$.

For complex integration variables $\phi=(\phi_1.\dots,\phi_N)$, the denominator of Eq. (\ref{ising5}) appears without the square root. Introducing in addition complex source fields $j=(j_1,\dots,j_N)$ on each lattice site, we find from completing the square that
\begin{align}
 \label{ising6}\int \Bigl( \prod_i \frac{\mbox{d}\phi^*_i\mbox{d}\phi_i}{2\pi\rmi}\Bigr) e^{-\phi_i^*A_{ij}\phi_i +j^*_i\phi_i + \phi_i^*j_i} =\frac{1}{\det(A)} e^{j^\dagger A^{-1} j}.
\end{align}
Here, $j^\dagger = (j^*)^t$ has the usual meaning. 

In practical calculations, the matrix $A$ (e.g. the inverse propagator) is often given in terms of real basis fields. One might then either work with real variables as in Eq. (\ref{ising4}) or with complex ones as in Eq. (\ref{ising6}). To see this, we write
\begin{equation}
 \label{ising6c} \int  \prod_i \mbox{d}\phi^*_i\mbox{d}\phi_i e^{-\phi^*_i A_{ij} \phi_j} = \int  \prod_i \mbox{d}\phi^*_i\mbox{d}\phi_i e^{-\frac{1}{2}(\phi_i,\phi^*_i) B_{ij} \binom{\phi_j}{\phi^*_j}}
\end{equation}
with
\begin{equation}
 \label{ising6d} B = \left(\begin{array}{cc} 0 & A^t \\ A & 0 \end{array}\right).
\end{equation}
The integral on the right hand side of Eq. (\ref{ising6c}) is of type (\ref{ising5}) with $m_i=(\phi_i,\phi^*_i)$ for $i=1,\dots,2N$. We then find 
\begin{align}
 \nonumber &\int \Bigl( \prod_i \frac{\mbox{d}\phi^*_i\mbox{d}\phi_i}{2\pi\rmi}\Bigr) e^{-\phi_i^*A_{ij}\phi_i}\\
 \nonumber &= \Bigl(\frac{1}{2\pi \rmi}\Bigr)^N \int \prod_i \mbox{d}\phi^*_i \mbox{d}\phi_i e^{-\frac{1}{2}(\phi_i,\phi^*_i) B_{ij} \binom{\phi_j}{\phi^*_j}}\\
 \label{ising6e} &= \Bigl(\frac{1}{2\pi \rmi}\Bigr)^N \frac{(2\pi)^N}{\sqrt{\mbox{det} B}} = \frac{1}{\rmi^N} \frac{1}{[(-1)^N(\mbox{det}A)^2]^{1/2}} =\frac{1}{\mbox{det A}}.
\end{align}
Thus, in both cases, the solution is found to be $(\mbox{det}A)^{-1}$ or $(\mbox{det}B)^{-1/2}$, respectively. In particular, applying the formula $\log \det A = \mbox{Tr} \log A$, which is easily seen by writing both sides of the relation in terms of the eigenvalues of $A$, we find up to an overall constant
\begin{equation}
 \label{ising6f} \log \int \Bigl( \prod_i \frac{\mbox{d}\phi^*_i\mbox{d}\phi_i}{2\pi\rmi}\Bigr) e^{-\phi_i^*A_{ij}\phi_i} = \mbox{Tr} \log A = \frac{1}{2} \mbox{Tr} \log B.
\end{equation}
For this reason, the factor of $1/2$ in the one-loop formula (\ref{Func19}) or the flow equation (\ref{FRG1}) for the effective action might be present or not, depending on the definition of $S^{(2)}$.

The generalization of these lattice formulae to the continuum are straightforward. For bosonic atoms, represented by the complex field $\varphi(X)$, which are coupled to a complex source field $j(X)$, we have
\begin{align}
 \nonumber &\int \mbox{D} \varphi^* \mbox{D} \varphi \exp\Bigl(-\int (\varphi^*_X A_{XY} \varphi_Y + j^*_X \varphi_X + \varphi^*_X j_X)\Bigr) \\
 \label{fun27} &= \frac{1}{\det(A)} \exp\Bigl(\int j^*_X (A^{-1})_{XY}j_Y\Bigr).
\end{align}
The functional measure can formally be written as
\begin{equation}
 \label{ising6b} \mbox{D}\vphi^*\mbox{D}\vphi = \int \Bigl( \prod_X \frac{\mbox{d}\vphi^*(X) \mbox{d}\vphi(X)}{2 \pi \rmi} \Bigr),
\end{equation}
the precise definition being given by the discretized version. Note that infinite constant prefactors such as $\mathcal{N} = \prod_X 2\pi \rmi$ are not of relevance for our purposes, because they always drop out in calculations of correlation functions. We can thus normalize them to be unity right from the beginning.

For Grassmannians, although they do not have concrete values, a complex conjugation can be defined. However, for all purposes of our interest, it is sufficient to work with two independent $N$-vectors $\psi=(\psi_1,\dots,\psi_N)$ and $\psi^*=(\psi_1^*,\dots,\psi_N^*)$. We still write $\psi^\dagger = (\psi^*)^t$, but $\psi$ is in no way related to $\psi^\dagger$. After introducing two source terms $\eta$ and $\eta^*$ we find
\begin{equation}
 \label{fun28} \int \mbox{D} \psi^* \mbox{D} \psi e^{-\psi^\dagger A \psi + \eta^\dagger \psi + \psi^\dagger\eta} = \det(A) e^{\eta^\dagger (A^{-1}) \eta}.
\end{equation}
Note that the determinant appears in the numerator. This peculiar feature of Gaussian integrals for Grassmannians is related to the definition of the exponential function. We used the functional measure
\begin{equation}
 \label{fun29} \mbox{D} \psi^* \mbox{D} \psi = \prod_n \mbox{d} \psi_n^* \mbox{d} \psi_n.
\end{equation}
Of course, formula (\ref{fun28}) immediately applies to the continuous case of Grassmann fields $\psi(X)$ and $\psi^*(X)$ with the usual replacements, for instance
\begin{equation}
 \label{ising7} \eta^\dagger A^{-1} \eta = \int_{X,Y} \eta^*_X (A^{-1})_{XY} \eta_Y.
\end{equation}

\section{One-loop effective potential for bosons}
\label{AppOneLoopBos}
\noindent This appendix provides computational details on the derivation of the one-loop effective potential for weakly interacting bosons. In particular, we derive the condensate depletion at zero temperature, which is a pure interaction effect.

\vspace{5mm}
\begin{center}
 \emph{Evaluation of the one-loop correction}
\end{center}

\noindent Here we perform the calculation of the effective action in the one-loop approximation. This computation provides some useful formulae and serves as a basis for extracting the condensate depletion present in weakly interacting Bose systems at zero temperature. It will also shed light on ultraviolet divergences, and how to cope with them. 

We start from
\begin{equation}
 \label{bog38b} \Gamma^{(\text{1-loop})}[\phi] = S[\phi] + \Delta \Gamma[\phi] = S[\phi] + \frac{1}{2} \mbox{Tr} \log S^{(2)}[\phi].
\end{equation}
The field $\phi(\tau,\vec{x})$ is arbitrary. By taking functional derivatives with respect to $\phi$, we obtain higher correlation functions $\Gamma^{(n)}$ to one-loop order. If we evaluate $\Gamma^{(\text{1-loop})}[\phi]$ for a constant field, we arrive at the effective potential $U^{(\text{1-loop})}(\phi)$. Analogous to Eqs. (\ref{bog7b}) and (\ref{bog7c}), we then find the gap equation and equation of state according to $(\partial U/\partial \phi)(\phi_0)=0$ and $(\partial U/\partial\mu)(\phi_0)=-n$, respectively.

We express the microscopic action in the basis of real fields $\vphi=\frac{1}{\sqrt{2}}(\vphi_1+\rmi \vphi_2)$ as
\begin{align}
 \nonumber &S[\vphi_1,\vphi_2] = \int_X U_{\rm \Lambda}(\rho_X)\\
 \label{bog9d} &+\frac{1}{2}\int_X (\vphi_{1,X},\vphi_{2,X}) \left(\begin{array}{cc}\frac{-\nabla^2}{2M} & \rmi \partial_{\tau}\\ -\rmi \partial_{\tau} & \frac{-\nabla^2}{2M} \end{array}\right)\left( \begin{array}{c} \vphi_{1,X} \\ \vphi_{2,X}\end{array}\right).
\end{align}
We introduced the microscopic or classical potential
\begin{equation}
 \label{bog10} U_{\Lambda}(\rho) = -\mu \rho + \frac{g}{2} \rho^2,
\end{equation}
which depends on $\rho = \vphi^*\vphi = \frac{1}{2}(\vphi_1^2+\vphi_2^2)$. It coincides with the classical effective potential from Eq. (\ref{bog6}). The label $\Lambda$ indicates that this expression for the effective potential is only valid at the energy scale of the UV cutoff, whereas it is changed due to the inclusion of fluctuations in the infrared regime.

We calculate $S^{(2)}[\phi^*,\phi]$ by expanding $S[\phi^* + \delta \vphi^*,\phi+\delta \vphi]$  to second order in $\delta \vphi$ around a \emph{real}, constant field $\phi=\sqrt{\rho}$. This allows for the implementation of spontaneous symmetry breaking and to distinguish between amplitude and phase fluctuations. We find
\begin{align}
 \nonumber &S[\sqrt{\rho}+\delta \vphi_1,\delta \vphi_2] \simeq S[\sqrt{\rho},0]\\
\label{bog11}  &+\frac{1}{2}\int_X (\delta\vphi_{1,X},\delta\vphi_{2,X}) \left(\begin{array}{cc}P_{11,X} & \rmi \partial_{\tau}\\ -\rmi \partial_{\tau} & P_{22,X} \end{array}\right)\left( \begin{array}{c}\delta\vphi_{1,X} \\ \delta\vphi_{2,X}\end{array}\right)
\end{align}
with
\begin{align}
 \label{bog10c} P_{11,X} &= - \frac{\nabla^2}{2M} + U_\Lambda'(\rho)+2\rho U_\Lambda''(\rho),\\
 \label{bog10d} P_{22,X} &= - \frac{\nabla^2}{2M} + U_\Lambda'(\rho).
\end{align}
Here, a prime denotes differentiation with respect to $\rho$. Note that the minimum $\rho_0$ of the \emph{full} effective potential $U(\rho)$ will usually not satisfy $U_\Lambda'(\rho_0)=0$, i.e. $\rho_0(\mu,T) \neq \mu/g$, because the inclusion of quantum and thermal fluctuations shifts the position of the minimum in field space. Otherwise, no phase transitions would occur in nature. However, in the case of a weakly interacting Bose gas at low temperatures, the shift is small and can be treated perturbatively.

In order to deal with the derivative terms in Eq. (\ref{bog11}), we transform our fields to momentum space $Q=(\omega_n,\vec{q})$. Note that the Fourier transformation of Eq. (\ref{bog11}) would be difficult, if the expansion point $\phi$ was not constant in space-time. We have $\delta\vphi_i^*(Q) = \delta\vphi_i(-Q)$, because $\delta\vphi_1$ and $\delta\vphi_2$ are real. This leads us to the expression
\begin{align}
 \nonumber & S[\sqrt{\rho}+\delta\vphi_1,\delta \vphi_2] \simeq S[\sqrt{\rho},0] \\
 \label{bog15}&+\frac{1}{2} \int_{Q} (\delta \vphi_{1,-Q},\delta \vphi_{2,-Q}) G^{-1}_\Lambda(Q) \left( \begin{array}{c} \delta \vphi_{1,Q} \\ \delta \vphi_{2,Q}\end{array}\right)
\end{align}
with classical inverse propagator
\begin{align}
 \nonumber G^{-1}_\Lambda(Q) &= \left(\begin{array}{cc}\vare_q + U_\Lambda'(\rho) + 2 \rho U_\Lambda''(\rho) & - \omega_n\\ \omega_n & \vare_q  +U_\Lambda'(\rho)\end{array}\right)\\
 \label{bog17} &=\left(\begin{array}{cc}\vare_q -\mu + 3 g \rho & - \omega_n\\ \omega_n & \vare_q  -\mu + g \rho\end{array}\right).
\end{align}
We also introduced the notation $\int_Q = T \sum_n \int \frac{{\rm d}^3q}{(2\pi)^3}$. For the corresponding zero temperature limit we refer to Eq. (\ref{Co32}).

The trace of the inverse propagator $G^{-1}_\Lambda(Q)$ consists of an integration over $Q$ and the usual trace of the $2\times2$-matrix. We have
\begin{align}
 \nonumber \mbox{Tr} \log S^{(2)}[\sqrt{\rho},0] &= \delta(P=0) \int_Q (\log G^{-1}_\Lambda(Q))_{ii} \\
 \label{bog22} &= \beta V \int_Q \log \det G^{-1}_\Lambda(Q).
\end{align}
We used here the matrix identity 
\begin{equation}
\mbox{tr} \log A = \log \det A,
\end{equation} 
which is easily seen by writing the left and right hand sides in terms of the eigenvalues of $A_{ij}$. Moreover, we used that
\begin{equation}
 \label{bog23} \delta(P=0) = \int_X \left.e^{iP\cdot X}\right|_{P=0} = \int_0^\beta \mbox{d}\tau \int \mbox{d}^3x = \beta V
\end{equation}
coincides with the volume of space-time.

With these formulae, we find for the one-loop effective potential 
\begin{align}
 \nonumber &U^{(\text{1-loop})}(\rho) = -\mu\rho + \frac{g}{2}\rho^2+\frac{1}{2}\int_Q \log \det G^{-1}_\Lambda(Q)\\
 \nonumber &= -\mu\rho + \frac{g}{2}\rho^2 + \frac{1}{2}\int \frac{\mbox{d}^3q}{(2 \pi)^3} T \sum_n \log\biggl(1+\frac{E_q(\rho)^2}{\omega_n^2}\biggr)\\
 \label{bog24} &=-\mu\rho + \frac{g}{2}\rho^2+T \int \frac{\mbox{d}^3q}{(2 \pi)^3} \log\sinh \Bigl(E_q(\rho)/2T\Bigr).
\end{align}
The bosonic Matsubara summation was evaluated according to $\sum_n \log(1+\frac{x^2}{(\pi n)^2})=2 \log \sinh(x)$. We dropped an overall constant in line two, which is irrelevant for the thermodynamics, and introduced the abbreviation
\begin{align}
 \label{bog25} E_q(\rho) = \sqrt{(\vare_q-\mu+3 g \rho)(\vare_q -\mu +g \rho)}.
\end{align}
For $\rho = \mu/g$ we find $E_q(\rho)=\sqrt{\vare_q(\vare_q+2g\rho_0)}=E_q$. The integral on the right hand side of Eq. (\ref{bog24}) is divergent for large momenta. Regularization via introduction of a UV cutoff $\Lambda$ yields terms proportional to $\Lambda^3$ and $\Lambda$. These unphysical divergences from perturbation theory are cured in a physically motivated ultraviolet (UV) renormalization scheme based on an investigation of the gap equation and the equation of state.

\vspace{5mm}
\begin{center}
 \emph{Thermodynamics, condensate depletion, and UV renormalization}
\end{center}

\noindent From the one-loop effective potential (\ref{bog24}) we obtain the phase structure and the equation of state. The order parameter $\rho_0(\mu,T)$ is found from the gap equation
\begin{align}
 \nonumber 0 &= \frac{\partial U^{(\text{1-loop})}}{\partial \rho}(\rho_0(\mu,T))\\
 \nonumber &= - \mu + g \rho_0 + \frac{1}{2} \int \frac{\mbox{d}^3q}{(2\pi)^3} \frac{\partial E_q}{\partial \rho}\biggr|_{\rho=\rho_0}\coth\biggl(\frac{E_q}{2T}\biggr)  \\
 \label{bog26} &= - \mu + g \rho_0 + g \int \frac{\mbox{d}^3q}{(2\pi)^3} \frac{\vare_q-\mu + \frac{3}{2} g \rho_0}{E_q}\coth\biggl(\frac{E_q}{2T}\biggr).
\end{align}
The integrand tends to $1$ for large momenta. The physical origin of this divergence can be understood from the fact that we have assumed interactions which are local in coordinate space, and hence constant in momentum space up to arbitrarily large momenta. It can be cured by a renormalization of the coupling constant $g_{\rm R} = g + \delta g(\Lambda)$, where $\delta g(\Lambda)$ is an explicitly cutoff dependent term, which replaces all the microscopic details we left out during our calculation. (The prescribed UV renormalization procedure is performed for the case of fermions within BCS theory in Eq. (\ref{bcs20d}).) We then observe from Eq. (\ref{bog26}) that $\rho_0 = \mu/g + O(g^0)$ at zero temperature.

Taking a $\mu$-derivative of Eq. (\ref{bog24}), we obtain
\begin{equation}
\label{bog29} \frac{\partial U}{\partial \mu}(\rho) = - \rho - \frac{1}{2} \int \frac{\mbox{d}^3q}{(2\pi)^3} \frac{\vare_q-\mu + 2 g \rho}{E_q}\coth\biggl(\frac{E_q(\rho)}{2T}\biggr).
\end{equation}
At zero temperature, the $\coth$ can  be replaced by unity. Inserting $\rho_0 \simeq \mu/g$ we then find
\begin{equation}
 \label{bog27} n(\mu) = -\frac{\partial U}{\partial \mu}(\rho_0) = \rho_0 + \frac{1}{2} \int \frac{\mbox{d}^3q}{(2\pi)^3} \frac{\vare_q+g \rho_0}{E_q},
\end{equation}
which again diverges due to high momenta in the integral. Writing the particle number as 
\begin{equation}
 \label{bog28} n = \phi^*_0\phi_0 + \int \frac{\mbox{d}^3q}{(2\pi)^3} \langle \delta \vphi^*_{\vec{q}} \delta \vphi_{\vec{q}} \rangle,
\end{equation}
we obtain a recipe to cure this divergence: it relates to the fact that the functional integral works with fields rather than operators and thus does not contain information on the operator ordering which is present in a second quantized formulation. Indeed, we have the relation $\langle \delta \vphi^* \delta \vphi \rangle = \langle \delta\vphi \delta \vphi^*\rangle$. We may contrast this with the particle number density defined in second quantization as $n = \phi^*_0\phi_0+\sum_{\vec{q}\neq0}\langle \hat{a}^\dagger_{\vec{q}}\,\hat{a}_{\vec{q}}\rangle$, where $\hat a^\dag_{\vec{q}}, \hat a_{\vec{q}}$ are bosonic operators. We can reconcile these two approaches by identifying permutation invariant combinations, which must be equal to each other: $\langle \delta \vphi^* \delta \vphi \rangle = \frac{1}{2} \langle \delta \vphi^* \delta \vphi + \delta\vphi \delta \vphi^*\rangle = \frac{1}{2} \langle \hat{a}^\dagger \hat{a} + \hat{a} \hat{a}^\dagger \rangle = \langle \hat{a}^\dagger \hat{a} \rangle + \frac{1}{2}$ for each momentum mode. We thus conclude that we overestimated the contribution of each mode by $1/2$ in the functional formulation, leading to a linear UV divergence in Eq. (\ref{bog27}) and a cubic one in the corresponding effective potential. Physically, we have thus identified this divergence as resulting from the quantum mechanical zero point shift for each oscillator mode $\vec q$ \cite{Diehl:2005ae}. Subtracting this term we finally arrive at
\begin{align}
 \nonumber n(\mu) &= \phi^*_0 \phi_0 + \int \frac{\mbox{d}^3q}{(2\pi)^3}\langle \hat{a}^\dagger_{\vec{q}}\hat{a}_{\vec{q}}\rangle\\
 \nonumber&= \phi^*_0 \phi_0 + \int \frac{\mbox{d}^3q}{(2\pi)^3}\Bigl(\langle \delta \vphi^*_{\vec{q}}\delta \vphi_{\vec{q}}\rangle - \frac{1}{2} \Bigr)\\
 \label{bog28b}&= \rho_0 + \frac{1}{2} \int \frac{\mbox{d}^3q}{(2\pi)^3} \biggl( \frac{\vare_q+g \rho_0}{E_q} - 1 \biggr).
\end{align}
The resulting \emph{condensate depletion} is not found in a noninteracting Bose gas. We can interpret this behavior as part of the bosons being kicked out of the condensate due to the repulsive interactions.  This is an observable effect of quantum fluctuations, which occurs in the absence of thermal fluctuations at $T=0$. In contrast, superfluid density and particle density are equal at zero temperature. This is ensured by the Ward identity which is related to Galilei symmetry of the effective action. Although the condensate spontaneously breaks Galilei symmetry, this does not invalidate the statement, because the Ward identity is a property the effective action.

\section{Symmetries of the effective action}
\label{SymmetriesEffAct}

\noindent A global continuous symmetry of the classical action yields a (classical) conserved charge -- this is Noether's theorem. In the
absence of anomalies, there is also a conserved charge for the full quantum theory. Here, we briefly
review the formalism for the construction of the conserved Noether charge for both the classical and the quantum case. 

The classical equations of motion are found from the action $S[\vphi]$ and the corresponding Lagrangian $\mathcal{L}$, the latter being defined by $S = \int_X \mathcal{L}$, from the variational principle
\begin{equation}
 \label{Symm1} 0 = \frac{\delta S}{\delta \vphi(X)}[\vphi_0] = \int_Y \frac{\delta \mathcal{L}}{\delta \vphi(X)}.
\end{equation}
In particular, for a Lagrangian $\mathcal{L} = \mathcal{L}(\vphi,\partial_\mu \vphi)$ which only depends on the field and its first derivatives, we have
\begin{equation}
 \label{Symm2} 0 = \left.\frac{\delta S}{\delta \vphi}\right|_{\rm EoM} =  \left(\frac{\partial \mathcal{L}}{\partial \vphi} - \partial_\mu \frac{\partial \mathcal{L}}{\partial(\partial_\mu \vphi)}\right)_{\vphi=\vphi_0}.
\end{equation}
These are the standard Euler--Lagrange equations. In general, the Lagrangian may also contain higher derivative terms of the field, which we indicate by writing $\mathcal{L}=\mathcal{L}(\vphi,\partial_\mu \vphi,\dots)$.

Eq. (\ref{Symm1}) tells us that the solution $\vphi_0$ is a stationary point of the action with respect to variations of the field. In particular, these variations $\vphi \to \vphi + \delta \vphi$ may be generated by a continuous transformation $\vphi \to \vphi^{(\alpha)}$, where $\alpha(X)$ is a continuous function quantifying the mapping such that $\alpha=0$ corresponds to the identity map.  We restrict the discussion to transformations, whose linear part in an expansion in powers of $\alpha$ is at most linear in the field $\vphi$. A typical example is provided by a ${\rm U}(1)$ transformation according to $\vphi \to \vphi^{(\alpha)}= e^{\rmi \alpha} \vphi\simeq (1 + \rmi \alpha) \vphi$ with real $\alpha$. Defining
\begin{equation}
 \label{Symm3} S^{(\alpha)}[\vphi] = S[\vphi^{(\alpha)}] = \int_X \mathcal{L}^{(\alpha)},
\end{equation}
which is a functional of $\alpha(X)$, we apply the chain rule for functional differentiation and specialize Eq. (\ref{Symm1}) to
\begin{equation}
 \label{Symm4} \left. \frac{\delta S^{(\alpha)}}{\delta \alpha(X)} \right|_{\alpha=0,\rm EoM} =0.
\end{equation}

An important class of microscopic actions $S[\vphi]$ are the ones possessing a \emph{global} symmetry of the type described above, i.e. $S^{(\alpha)}=S$ for \emph{constant} $\alpha(X) \equiv \alpha$. For a generic Lagrangian, we then find
\begin{align}
 \nonumber S^{(\alpha)}[\vphi] &= \int_X \mathcal{L}(\vphi^{(\alpha)},\partial_\mu \vphi^{(\alpha)},\dots)= \int_X \mathcal{L}^{(\alpha)}(\alpha,\partial_\mu \alpha, \dots)\\
 \label{Symm5}& \stackrel{\alpha = {\rm const.}}{=} \int_X \mathcal{L}^{(\alpha)}(\alpha) \stackrel{!}{=}  \int_X \mathcal{L}= S[\vphi].
\end{align}
Since this equation is valid to all orders in $\alpha$, we find for an action with global symmetry the relation
\begin{equation}
 \label{Symm6} \left.\frac{\partial \mathcal{L}^{(\alpha)}}{\partial \alpha}\right|_{\alpha =0} =0.
\end{equation}

Whereas Eq. (\ref{Symm4}) is valid for any action evaluated at the solution to the equations of motion, Eq. (\ref{Symm6}) is valid for all $\vphi$ but only actions with global continuous symmetry. We can now combine both findings to obtain
\begin{align}
 \nonumber 0 &= \left. \frac{\delta S^{(\alpha)}}{\delta \alpha}\right|_{\alpha=0,\rm EoM} = \left( \frac{\partial \mathcal{L}^{(\alpha)}}{\partial \alpha} - \partial_\mu \frac{\partial \mathcal{L}^{(\alpha)}}{\partial(\partial_\mu\alpha)} + \dots\right)_{\alpha=0,\rm EoM} \\
 \label{Symm7} &= - \partial_\mu \left.\frac{\partial \mathcal{L}^{(\alpha)}}{\partial(\partial_\mu \alpha)}\right|_{\alpha=0,\rm EoM} + \dots
\end{align}
The leading term, which is relevant for most cases of interest, has a divergence structure. In fact, it constitutes a local continuity equation $\partial_\mu J^\mu_{\rm cl}=0$  for the \emph{classical Noether current}
\begin{equation}
 \label{Symm8} J^{\mu}_{\rm cl} = -\left.\frac{\partial \mathcal{L}^{(\alpha)}}{\partial(\partial_\mu \alpha)}\right|_{\alpha=0}.
\end{equation}
It is related to the global conservation of the Noether charge
\begin{equation}
 \label{Symm9} Q_{\rm cl} = \int \mbox{d}^dx J^0_{\rm cl},
\end{equation}
where $J^0_{\rm cl}$ is the temporal component of the current.

The global continuous symmetry of the microscopic action yields a conserved quantity $Q$. For ultracold atoms with global ${\rm U}(1)$ symmetry, for instance, $Q=N$ is given by the particle number. Since the Noether charge is conserved globally, its local variations cannot be arbitrary, but rather must be such that the net change is locally zero. This fact is expressed by the continuity equation $\partial_\mu J^{\mu}_{\rm cl}=0$.

We now turn our attention to a full quantum theory which is described by a microscopic action $S[\vphi]$ and the partition function
\begin{equation}
 \label{Symm10} Z[j] = \int \mbox{D} \vphi e^{-S[\vphi]+j \cdot \vphi} = \int \mbox{D} \vphi e^{-S[\vphi^{(\alpha)}]+j \cdot \vphi^{(\alpha)}}.
\end{equation}
In the second line we performed a change of the integration variable and assumed the functional measure to be invariant with respect to this mapping. This is the case for unitary transformations, which we want to consider here. However, there are cases (so-called anomalies), where the assumption $\mbox{D} \vphi^{(\alpha)} =\mbox{D} \vphi$ is not valid. Since the left hand side of Eq. (\ref{Symm10}) does not depend on $\alpha$, we find from the right hand side to first order in $\alpha$ the relation
\begin{equation}
 \label{Symm11} \Bigl\langle \left. \frac{\delta S^{(\alpha)}}{\delta \alpha(X)} \right|_{\alpha=0}\Bigr\rangle_j = j\cdot \left.\frac{\delta \phi^{(\alpha)}}{\delta \alpha(X)}\right|_{\alpha=0}.
\end{equation}
Switching to the effective action via a Legendre transformation in the variables $(j,\phi)$, we have $j[\phi]=(\delta \Gamma/\delta \phi)[\phi]$ and deduce
\begin{equation}
 \label{Symm12} \Bigl\langle \left. \frac{\delta S^{(\alpha)}}{\delta \alpha(X)} \right|_{\alpha=0}\Bigr\rangle_j = \frac{\delta \Gamma}{\delta \phi} \cdot \left.\frac{\delta \phi^{(\alpha)}}{\delta \alpha(X)}\right|_{\alpha=0} = \left. \frac{\delta \Gamma^{(\alpha)}}{\delta \alpha(X)}\right|_{\alpha=0}.
\end{equation}
Analogous to the classical discussion, we introduced $\Gamma^{(\alpha)}[\phi] = \Gamma[\phi^{(\alpha)}]$.

From the quantum action principle, i.e. $(\delta \Gamma/\delta \phi)[\phi_0]=0$, we conclude that
\begin{equation}
 \label{Symm13} \left. \frac{\delta \Gamma^{(\alpha)}}{\delta \alpha(X)}\right|_{\alpha=0,\rm EoM}=0.
\end{equation}
This equation generalizes the classical relation (\ref{Symm4}). Together with the result in Eq. (\ref{Symm12}), we find for microscopic actions with global continuous symmetry
\begin{align}
 \label{Symm14} 0 = \Bigl\langle\left. \frac{\delta S^{(\alpha)}}{\delta \alpha(X)} \right|_{\alpha=0, \rm EoM}\Bigr\rangle =    - \partial_\mu \Bigl\langle\frac{\partial \mathcal{L}^{(\alpha)}}{\partial (\partial_\mu \alpha)}\Bigr\rangle_{\alpha=0,\rm EoM} + \dots 
\end{align}
We again applied Eq. (\ref{Symm6}). The ellipsis vanishes for a  Lagrangian which only depends on the first derivatives of the field. Eq. (\ref{Symm14}) is a continuity equation for the full \emph{quantum Noether current}
\begin{equation}
 \label{Symm15} J^\mu = - \partial_\mu \Bigl\langle\frac{\partial \mathcal{L}^{(\alpha)}}{\partial (\partial_\mu \alpha)}\Bigr\rangle_{\alpha=0}
\end{equation}
with conserved charge $Q=\int \mbox{d}^d x J^0$.

\section{Few-body physics in vacuum}
\label{AppVac}

\noindent In this appendix, we show how the physics of a few particles can be treated with the FRG. Special features of these vacuum calculations are diagrammatic simplifications and a closed hierarchy of flow equations of the $N$-body sector. For an introduction to this field see, for instance, Ref. \cite{Floerchinger:2011yv}. Therein, emphasis is on the Efimov effect, but general features of vacuum physics with the FRG and its relation to the many-body problem are covered as well.

The investigation of few-body problems with the FRG is motivated by
several points. Firstly, the calculation of vacuum observables from
first principles allows for comparison with experiments or exact
results from quantum mechanics and thus benchmarking the
technique. Moreover, the FRG approach reveals a different point of
view on well-known results like the Efimov effect. It also allows for
the computation of nonuniversal features away from resonance. Finally,
we already encountered the importance of the vacuum problem for the
many-body BCS-BEC crossover, where it provides the UV renormalization
of the microscopic couplings and influences the physics on the BEC
side.

We obtain the effective action in vacuum by choosing the parameters
such that $n=T=0$ with the temperature always being above
criticality. This implies condensation to be absent. In an FRG
setting, we then always remain in the symmetric phase with vanishing
density. For the BCS-BEC crossover, we had found $\mu_\psi = -
m^2_\psi \leq 0$ and $m^2_\phi \geq 0$ for the fermions and bosons,
respectively. Since the ground state has to be stable, the propagators
generically acquire a non-negative gap, $\Gamma^{(2)}(Q=0) \geq
0$. For the regularized propagator, we even have the stronger
statement that their gaps are strictly positive.

\vspace{5mm}
\begin{center}
 \emph{Simplification of the flow equations in vacuum}
\end{center}

\noindent \emph{All diagrams whose inner lines point into the same
  direction (thereby forming a closed tour), do not contribute to the
  flow in vacuum.\footnote{This holds under the mild assumption that
    the propagators are properly described in terms of a single
    pole. In other words, no additional non-analyticities are
    generated during the RG flow.}} This has important consequences
for the FRG, because from Eq. (\ref{FRG14}) we find the right hand
side of the flow equation to be constructed from one-loop
diagrams. For this reason, several quantities do not get renormalized
in vacuum (i.e. they do not receive contributions from quantum
fluctuations), because there are simply no diagrams. To intuitively
understand the above claim, we consider such a cyclic diagram. It
necessarily contains a line associated with a hole (or
anti-particle). However, such excitations are not contained in the
nonrelativistic quantum vacuum which conserves particle
number. Therefore, these processes cannot occur.

An example for a process which is described by such cyclic diagrams is
provided by the particle-hole fluctuations around the Fermi surface
which made up the Gorkov effect on the BCS side of the BCS-BEC
crossover. The corresponding screening effect is not found in vacuum,
because there is no Fermi surface.

Cyclic diagrams vanish in vacuum, because all poles of the propagators
lie in a definite half-plane of the complex $\omega$-plane. Therefore,
the loop integral yields zero by virtue of the residue
theorem. Indeed, consider a kinetic term $\int_Q \vphi^*(Q) P_{\rm
  B,F}(Q) \vphi(Q)$ in the effective action for either bosons or
fermions. In vacuum, there is no condensation and the corresponding
mean fields vanish. In the $(\vphi,\vphi^*)$-basis, we then find for
the inverse propagator $\Gamma^{(2)}(Q',Q)=\delta(Q+Q')G^{-1}(Q)$ the
simple expression
\begin{equation}
 \label{VAC1} G^{-1}_{\rm B,F}(Q) = \left(\begin{array}{cc} 0 & P_{\rm B,F}(-Q) \\ P_{\rm B,F}(Q) & 0 \end{array}\right).
\end{equation}
Connected lines pointing into the same direction are represented by
products of $P^{-1}_{\rm B}(Q)$ and $P^{-1}_{\rm F}(Q)$ with the same
sign of the frequency and spatial momentum variable $Q$. Without loss
of generality, we assume $\omega$ to be positive. As mentioned above,
for a nonvanishing cutoff the masslike term in the propagator $P(Q)$
will be strictly positive and the poles indeed lie in one distinct
half-plane of the complex $\omega$-plane. We close the integration
contour in the opposite half-plane and the loop integral
vanishes. Since the cutoff derivative $\tilde{\partial}_t$ only
increases the multiplicity of the poles but not their location, the
argument remains valid for the full flow equation.

We give three examples of this diagrammatic simplification in vacuum,
which are relevant for the BCS-BEC crossover. First, the fermion
propagator does not get renormalized, because the mixed diagram
containing a fermion line and a boson line is cyclic. Indeed, this is
a result of the form of the Yukawa coupling $\sim h (\vphi^*\psi_1
\psi_2 + {\rm h.c.})$. Second, the single box diagram which
contributes to the flow of the four-fermion coupling $\lambda_\psi$
vanishes in vacuum. Thus we have $\lambda_{\psi,k}=0$ on all scales if
$\lambda_{\psi,\Lambda}=0$. The latter may be obtained by a suitable
Hubbard--Stratonovich transformation of the microscopic action. This,
in turn, implies the Yukawa coupling to be protected from
renormalization, too, because the only diagram contributing to its
flow is proportional to $\lambda_\psi$ and thus zero. In particular,
$h_k(Q_1,Q_2)=h_\Lambda$ remains momentum independent.

As a further consequence, we have a strict hierarchy of flow equations
in vacuum. For instance, the two-body problem can be solved without
knowledge of the three-(or more)-body problem. In general, the
$N$-body problem only requires input only from processes involving $M
\leq N$ particles. This is easily understood in quantum
mechanical terms, where the solution of the $N$-body system is given
by the $N$-particle wave function.

We are thus led to a very different truncation scheme for vacuum
problems than we applied in the many-body sector. Indeed, it will turn
out to be useful to expand the most general $\Gamma_k[\phi]$ into
monomials of the field, i.e.
\begin{equation}
  \label{vac1} \Gamma_k[\phi] =  \Gamma_{k,2}[\phi] + 
  \Gamma_{k,3}[\phi] + \Gamma_{k,4}[\phi] + \dots,
\end{equation}
where $\Gamma_{k,N}$ is of $N$-th order in the field $\phi$. This a
particular case of a \emph{vertex expansion}.

Due to the translational invariance, we have for the momentum
representation of a particular vertex
\begin{equation}
 \label{vac2} \Gamma^{(n)}(Q_1,\dots,Q_n) = \gamma_n(Q_1,\dots,Q_{n-1}) \delta(Q_1+\dots+Q_n).
\end{equation}
The $n$-th momentum is determined by the remaining $n-1$ ones due to
momentum conservation. For example, we have $\Gamma^{(2)}(Q',Q) =
\delta(Q+Q')G^{-1}(Q)$, which we already encountered several times in
the earlier sections. To get quantitative precision, one has to keep
the full momentum dependence of the vertices $\gamma_n$. Because of
the simplifications which arise in vacuum, it might even in this case
be possible to solve the system of equations. However, for a
qualitative understanding, also the assumption of momentum
independence can be employed.

We dropped the zeroth order term in Eq. (\ref{vac1}), because it
represents the ground state energy of the vacuum. The linear term
vanishes for ${\rm U}(1)$-symmetric system, which are the only
relevant ones for our discussion. The higher order terms are
determined such that they respect ${\rm U}(1)$-symmetry and their
\emph{total} number of fields equals $N$.

 \vspace{5mm}
\begin{center}
 \emph{Two-body sectors and dimer binding energy for two-component fermions}
\end{center}

\noindent From our above considerations on the non-renormalization of
the fermion propagator, the four-fermion coupling and the Yukawa
coupling, we see that the most general truncation to third order for a
system of two-component fermions is given by
\begin{align}
  \nonumber &\Gamma_{k,2}[\phi,\psi] = \int_Q \biggl( \psi^\dagger(Q)(\rmi 
\omega-\vec{q}^2-\mu_\psi)\psi(Q) \\
  \nonumber &\hspace{15mm}+ \phi^*(Q) P_{\phi,k}(Q) \phi(Q)\biggr),\\
  \nonumber &\Gamma_{k,3}[\phi,\psi] = - \frac{h}{2} \int_{Q_1,Q_2,Q_3} 
\biggl(\phi^*(Q_3)\psi^t(Q_1)\vare \psi(Q_2) \\
\label{vac5} &\hspace{15mm} - \phi(Q_3)\psi^\dagger(Q_1)\vare
\psi^*(Q_2)\biggr)\delta(Q_1-Q_2-Q_3).
\end{align}
The UV condition on the boson propagator reads $P_{\phi,\Lambda}(Q) =
m^2_\Lambda$. Eq. (\ref{vac5}) contains everything that contributes to
the two-body problem. The flow equation
\begin{align}
 \nonumber &\partial_t P_{\phi,k}(P)\\
 \label{vac6} &= \tilde{\partial}_t \int_Q \frac{- h^2}{(P_\psi(Q)
   +R_{\psi,k}(Q))(P_\psi(P-Q)+R_{\psi,k}(P-Q))}
\end{align}
has been solved exactly for the cutoff $R_{\psi,k}= k^2$, and for the
Litim cutoff discussed in the context of the BCS-BEC crossover. After
UV renormalization, the physical dimer propagator is found to be
\begin{equation}
  \label{vac7} P_{\phi,k=0}(Q) = \frac{h^2}{8\pi} \left( - \frac{1}{a} + 
    \sqrt{\frac{\rmi \omega}{2}+\frac{\vec{q}^2}{4}-\mu_\psi}\right).
\end{equation}
The binding energy of the dimer is determined by the poles of the
analytically continued propagator. We use our expression for the
inverse propagator and compute
\begin{equation}
 \label{vac8} P_{\phi,k=0}(Q=0, \mu_\psi=\vare_{\rm B}/2) \stackrel{!}{=} 0.
\end{equation}
This yields
\begin{equation}
 \label{vac9} \vare_{\rm B} = - \frac{2}{a^2}
\end{equation}
as expected.
\vfill

\end{appendix}

\eject

\bibliographystyle{bibstyle}
\bibliography{ucoldbib}

\end{document}